%% file: main.tex
\documentclass[envcountsame,envcountchap]{svmono}
\usepackage{amsmath}
\usepackage{amssymb}
\usepackage{euscript}
\usepackage[latin1]{inputenc}
\usepackage{pstricks,pst-plot}
\usepackage{gastex}
\usepackage{url}
\usepackage{xspace}
\usepackage{makeidx} 
\usepackage{multind}

\makeindex{Gen}
\makeindex{Not}

\newcommand{\Nat}{\mathbb{N}}

\newcommand{\Q}{\mathbb{Q}}
\newcommand{\Z}{\mathbb{Z}}

\newcommand{\M}{\EuScript{M}}

\renewcommand{\E}{\EuScript{E}}

\newcommand{\Y}{\EuScript{Y}}

\renewcommand{\I}{\EuScript{I}}
\newcommand{\X}{\EuScript{X}}

\newcommand{\C}{\EuScript{C}}

\newcommand{\Lan}{\EuScript{L}}

\newcommand{\Hrond}{\EuScript{H}}
\renewcommand{\H}{\Hrond}

\newcommand{\classe}{\EuScript{C}}
\newcommand{\A}{\EuScript{A}}

\newcommand{\class}{\classe}

\renewcommand{\S}{\mathcal{S}}

\newcommand{\automaton}{\EuScript{A}}

\newcommand{\T}{\EuScript{T}}

\newcommand{\unit}{\textbf{e}}

\newcommand{\size}{\mathrm{size}}

\newcommand{\aff}{\mathrm{aff}}
\newcommand{\vecspace}{\mathrm{vec}}

\newcommand{\bound}[2]{\textrm{bound}_{#1}(#2)}

\renewcommand{\vec}[1]{\overrightarrow{#1}}
\newcommand{\saff}{\mathrm{saff}}
\newcommand{\vecsaff}{\vec{\mathrm{saff}}}
\newcommand{\comp}{\mathrm{comp}}

\newcommand{\finiteset}[2]{\{#1,\ldots,#2\}}
\newcommand{\cardinal}[1]{|#1|}
\newcommand{\relation}{\mathcal{R}}

\newcommand{\partie}{\mathcal{P}}
\newcommand{\moins}{\backslash}
\newcommand{\difference}[2]{#1\moins#2}
\newcommand{\scalar}[2]{\left<#1,#2\right>}

\newcommand{\fast}{\textsc{Fast}\xspace}
\newcommand{\lash}{\textsc{Lash}\xspace}

\newcommand{\brain}{\textsc{Brain}\xspace}

\newcommand{\mona}{\textsc{Mona}\xspace}

\newcommand{\omegatool}{\textsc{Omega}\xspace}

\newcommand{\norm}[2]{\left|\left|#2\right|\right|_{#1}}

\newcommand{\fo}[1]{\textrm{FO}\left(#1\right)}

\renewcommand{\gcd}[2]{\textrm{gcd}(#1,#2)}

\newcommand{\ind}{\rule{0.6cm}{0cm}}

\renewcommand{\P}{\EuScript{P}}

\newcommand{\group}{\textrm{group}}
\newcommand{\inv}{\textrm{inv}}

\begin{document}

\author{Jérôme Leroux}
\title{Least Significant Digit First\\ Presburger Automata}
\maketitle

\frontmatter

\include{dedic}

\include{pref}

\tableofcontents
\mainmatter




\input{chapter.introduction}
\input{chapter.notations}


\part{Logic and Automata}
\input{chapter.DVA}

\input{chapter.modifyingDVA}

\input{chapter.expressiveness}

\input{chapter.examplesFDVA}

\input{chapter.reduction}

\part{Geometry}
\input{chapter.linearsets}

\input{chapter.semilinearsets}

\input{chapter.degenerate}

\input{chapter.polyhedrons}

\input{chapter.presburgerdecomposition}

\part{From Automata to Presburger Formulas}
\input{chapter.stronglyconnected}

\input{chapter.geometricalextraction}
\input{chapter.finalalgorithm}




\backmatter

\printindex{Gen}{Index}
\printindex{Not}{Notations}

\end{document}

%% file: dedic.tex
%
%
%

\thispagestyle{empty}
\vspace*{3.5cm}
\begin{flushright}

{\large Aux Québécois}

\end{flushright}

%% file: chapter.introduction.tex
\chapter{Introduction}
Presburger arithmetic \cite{P-PCM29} is a decidable logic used in a large range of applications. As described in \cite{L-LICS04}, this logic is central in many areas including integer programming problems \cite{S-87}, compiler optimization techniques \cite{OMEGA}, program analysis tools \cite{BGP-TOPLAS99,FO-CONCUR97,F-WFLP00} and model-checking \cite{BFL-TACAS04,FAST,LASH}. Different techniques \cite{GBD-FMCAD02} and tools have been developed for manipulating \emph{the Presburger-definable sets} (the sets of integer vectors satisfying a Presburger formula): by working directly on the Presburger-formulas \cite{K-LICS04} (implemented in \omegatool\ \cite{OMEGA}), by using semi-linear sets \cite{GS-PACIF66} (implemented in \brain\ \cite{RV-AMAST02}), or automata (integer vectors being encoded as strings of digits) \cite{WB-SAS95,BC-CAAP96} (implemented in \fast\ \cite{BFLP-CAV03}, \lash\ \cite{LASH} and \mona\ \cite{KMS-IJFCS02}). Presburger-formulas and semi-linear sets lack canonicity. As a direct consequence, a set that possesses a simple representation could unfortunately be represented in an unduly complicated way. Moreover, deciding if a given vector of integers is in a given set, is at least \emph{NP-hard} \cite{B-FOCS77,GS-PACIF66}. On the other hand, a minimization procedure for automata provides a canonical representation. That means, the automaton that represents a given set only depends on this set and not on the way we compute it. For these reasons, autmata are well adapted for applications that require a lot of boolean manipulations such as model-checking.

Whereas there exist efficient algorithms for computing an automaton that represents the set defined by a given Presburger formula \cite{K-LICS04,WB-ICALP00,BC-CAAP96}, the inverse problem of computing a Presburger-formula from a Presburger-definable set represented by an automaton, called the \emph{Presburger synthesis problem}, was first studied in \cite{L-THESE03} and only \emph{partially solved in exponential time} (resp. \emph{doubly exponential time}) for \emph{convex integer polyhedrons} \cite{L-LICS04} (resp. for \emph{semi-linear sets with the same set of periods} \cite{L-CIAA04}). Presburger-synthesis has many applications. For example, in software verification, we are interested in computing the set of reachable states of an infinite state system by using automata and in analyzing the structure of these sets with a tool such as \cite{OMEGA} which manipulates Presburger-formulas. The Presburger-synthesis problem is also central to a new generation of constraint solvers for Presburger arithmetic that manipulate both automata and Presburger-formulas \cite{L-LICS04,K-LICS04}.

The Presburger-synthesis problem is naturally related to the problem of deciding whether an autamaton represents a Presburger-definable set, a well-known hard problem first solved by Muchnik in 1991 \cite{M-PREPRINT91} with a quadruple exponential time algorithm. To the best of our knowledge no better algorithm for the full class of Presburger-definable sets has been proposed since 1991.\\

In this paper, given an automaton that represents a set $X$ of integer vectors encoded by the least significant digit first decomposition, we prove that we can decide in \emph{polynomial time} whether $X$ is Presburger-definable. Moreover, in this case, we provide an algorithm that computes in \emph{polynomial time} a Presburger-formula that defines $X$. 

%% file: chapter.notations.tex
\chapter{Notations}
We provide in this chapter notations used in the sequel. 

\section{Sets, Functions, and Relations}
We denote by $\Q$\index{Gen}{rational numbers}\index{Not}{$\Q$}, $\Q_+$\index{Not}{$\Q_+$}, $\Z$\index{Gen}{integers}\index{Not}{$\Z$} and $\Nat$\index{Gen}{non-negative integers}\index{Not}{$\Nat$} respectively the set of \emph{rational numbers}, \emph{non-negative rational numbers}, \emph{integers}, and \emph{non-negative integers}.

The \emph{intersection}\index{Gen}{intersection}, \emph{union}\index{Gen}{union}, \emph{difference}\index{Gen}{difference}, and \emph{symmetric difference}\index{Gen}{symmetric difference} of two sets $A$ and $B$ are written $A\cap B$\index{Not}{$A\cap B$}, $A\cup B$\index{Not}{$A\cup B$}, $\difference{A}{B}$\index{Not}{$A\moins B$}, and $A\Delta B=(\difference{A}{B})\cup(\difference{B}{A})$\index{Not}{$A\Delta B$}. 

The \emph{class of subsets}\index{Gen}{class of subsets}  (resp. \emph{the class of finite subsets}\index{Gen}{class of finite subsets}) of a set $E$ is denoted by $\partie(E)$\index{Not}{$\partie(E)$} (resp. $\partie_f(E)$\index{Not}{$\partie_f(E)$}). The \emph{cardinal}\index{Gen}{cardinal} of a finite set $X$ is written $\cardinal{X}\in\Nat$\index{Not}{$\cardinal{X}$}. A \emph{partition}\index{Gen}{partition} $\class$ of a set $E$ is a class of non-empty subsets of $E$ such that $X_1\cap X_2=\emptyset$ for any $X_1,X_2\in\class$ and $E=\bigcup_{X\in \class}X$.

The \emph{Cartesian product}\index{Gen}{Cartesian product} of two sets $A$ and $B$ is written  $A\times B$\index{Not}{$A\times B$}. The set $X^m$\index{Not}{$X^m$} is called the \emph{set of vectors}\index{Gen}{vector} with $m\in\Nat$ \emph{components}\index{Gen}{component} in a set $X$. Given an integer $i\in\finiteset{1}{m}$ and a vector $x\in X^m$, the $i$-th component of $x$ is written $x[i]\in X$\index{Not}{$x[i]$}.

The set of \emph{functions}\index{Gen}{function} $f:X\rightarrow Y$, also called \emph{sequences}\index{Gen}{sequence} of elements in $Y$ indexed by $X$ is written $Y^X$\index{Not}{$Y^X$}. A function $f\in X^Y$ is said \emph{injective}\index{Gen}{injective} if $f(x_1)\not=f(x_2)$ for any $x_1\not=x_2\in X$, \emph{surjective}\index{Gen}{surjective} if for any $y\in Y$ there exists $x\in X$ such that $y=f(x)$, and \emph{bijective}\index{Gen}{bijective} or \emph{one-to-one}\index{Gen}{one-to-one} if it is both injective and surjective. A function $f\in Y^X$ is either denoted by $f:X\rightarrow Y$\index{Not}{$f:X\rightarrow Y$}, or it is denoted by $f=(f_x)_{x\in X}$\index{Not}{$(f_x)_{x\in X}$} and in this last case $Y$ is implicitly known. Given a function $f:X\rightarrow Y$ and two sets $A$ and $B$, we define $f(A)$\index{Not}{$f(A)$} and $f^{-1}(B)$\index{Not}{$f^{-1}(B)$} respectively the \emph{image}\index{Gen}{image} and the \emph{inverse image}\index{Gen}{inverse image} of $A$ and $B$ by $f$, given by $f(A)=\{f(x);\;x\in X\cap A\}$ and $f^{-1}(B)=\{x\in X;\;f(x)\in B\}$ (remark that $A$ is not necessary a subset of $X$ and $B$ is not necessary a subset of $Y$). 

An \emph{enumeration}\index{Gen}{enumeration} of a set $E$ is an injective function $f:\Nat\rightarrow E$. A \emph{countable set} $E$ is a set $E$ that has an enumeration. Recall that a finite set is countable and the class of finite subsets of a countable set is countable. 

Let $V$ be a countable set of \emph{boolean variables}. A \emph{boolean formula}\index{Gen}{boolean formula} $\phi$ over the boolean variables $V$ is a formula in the grammar $\phi:=v|\phi\cap\phi|\phi\cup\phi|\phi\moins\phi|\phi\Delta\phi$ where $v\in V$. A \emph{boolean valuation}\index{Gen}{boolean valuation} $\rho$ is a function that maps each boolean variable $v$ to a set $\rho(v)$. Observe that a boolean valuation $\rho$ can be naturally extended to any boolean formula $\phi$. Given a boolean formula $\phi(v_1,\ldots,v_n)$ where $v_1$, ..., $v_n$ are the boolean variables \emph{occurring} in $\phi$ and some sets $E_1$, ..., $E_n$, we denote by $\phi(E_1,\ldots,E_n)$ the unique set $\rho(\phi)$ where $\rho$ is any valuation such that $\rho(v_i)=E_i$. A set $E$ is called a \emph{boolean combination}\index{Gen}{boolean combination} of sets in a class $\class$ of sets if there exists a boolean formula $\phi(v_1,\ldots,v_n)$ and some sets $E_1$, ..., $E_n$ in $\class$ such that $E=\phi(E_1,\ldots,E_n)$.
\begin{lemma}\label{lem:bool}
  We can decide in polynomial time if a finite set $E$ is a boolean combination of sets in a finite class $\class$ of finite sets. Moreover, in this case we can compute in polynomial time a boolean formula $\phi(v_1,\ldots,v_n)$ and a sequence $E_1$, ..., $E_n$ of sets in $\class$ such that $E=\phi(E_1,\ldots,E_n)$.
\end{lemma}
\begin{proof}
  Let us consider an enumeration $E_1$, ..., $E_n$ of the sets in $\class$ and let $X=\bigcup_{i=1}^n E_i$. Let us consider the function $f:X\times\finiteset{1}{n}\rightarrow \partie(E)$ such that $f(x,i)$ is the unique set in $\{E_i,X\Delta E_i\}$ that contains $x$. Let us also consider the set $K_x=\bigcap_{i=1}^nf(x,i)$ and observe that $E$ is a boolean combination of sets in $\class$ if and only $E=\bigcup_{e\in E}K_e$.
  \qed 
\end{proof}

A \emph{relation}\index{Gen}{relation} $\relation$ is a subset of $S_1\times S_2$ where $S_1$ and $S_2$ are two sets. We denote by $s_1\relation s_2$ if $(s_1,s_2)\in\relation$. Such a relation is said \emph{one-to-one}\index{Gen}{relation!one-to-one} if there exists a unique $s_2\in S_2$ such that $s_1\relation s_2$ for any $s_1\in S_1$, and if there exists a unique $s_1\in S_1$ such that $s_1\relation s_2$ for any $s_2\in S_2$. The \emph{concatenation}\index{Gen}{relation!concatenation}\index{Gen}{concatenation!of two relations} $\relation_1.\relation_2$\index{Not}{$\relation_1.\relation_2$} of two relations $\relation_1\subseteq S_1\times S_2$ and $\relation_2\subseteq S_2\times S_3$ is the relation $\relation_1.\relation_2\subseteq S_1\times S_3$ defined by $s_1\relation_1.\relation_2 s_3$ if and only if there exists $s_2\in S_2$ such that $s_1\relation_1 s_2$ and $s_2\relation_2 s_3$. A \emph{binary relation}\index{Gen}{relation!binary} $\relation$ over a set $S$ is a relation $\relation\subseteq S_1\times S_2$ such that $S_1=S=S_2$. Recall that a binary relation $\relation$ over $S$ is an \emph{equivalence}\index{Gen}{relation!equivalence} if $\relation$ is \emph{reflexive}\index{Gen}{relation!reflexive} ($s\relation s$ for any $s$), \emph{symmetric}\index{Gen}{relation!symmetric} ($s_1\relation s_2$ if and only if $s_2\relation s_1$ for any $s_1,s_2\in S$), and \emph{transitive}\index{Gen}{relation!transitive} ($s_1\relation s_2$ and $s_2\relation s_3$ implies $s_1\relation s_3$ for any $s_1,s_2,s_3\in S$). Given an equivalence binary relation $\relation$ over $S$, the \emph{equivalence class}\index{Gen}{relation!equivalence class} of an element $s\in S$ is the set of $s'\in S$ such that $s\relation s'$. Recall that equivalence classes provide a partition of $S$.

\section{Linear Algebra}
The \emph{unit vector}\index{Gen}{unit vector} $\unit_{j,m}\in\Q^m$\index{Not}{$\unit_{j,m}$} where $j\in\finiteset{1}{m}$  is defined by $\unit_{j,m}[j]=1$ and $\unit_{j,m}[i]=0$ for any $i\in\finiteset{1}{m}\moins\{j\}$. The \emph{zero vector}\index{Gen}{zero vector} $\unit_{0,m}\in\Q^m$ is defined by $\unit_{0,m}=(0,\ldots,0)$. 

Vectors $x+y$ and $t.x$ are defined by $(x+y)[i]=(x[i])+(y[i])$ and $(t.x)[i]=t.(x[i])$ for any $i\in\finiteset{1}{m}$, $x,y\in\Q^m$, $t\in\Q$. We naturally define $A+B=\{a+b;\;(a,b)\in A\times B\}$ and $T.A=\{t.a;\;(t,a)\in T\times A\}$ for any $A,B\subseteq\Q^m$ and $T\subseteq\Q$. For any $a,b\in\Q^m$ and $t\in\Q$, let us define $a+B=\{a\}+B$, $A+b=A+\{b\}$, $t.A=\{t\}.A$ and $T.a=T.\{a\}$.

The \emph{infinite norm}\index{Gen}{infinite norm} of a vector $x\in\Q^m$ is defined by $\norm{\infty}{x}=\max_{i}|x[i]|$\index{Not}{$\norm{\infty}{x}$} where $|x[i]|$\index{Not}{$|x|$} is the \emph{absolute value}\index{Gen}{absolute value} of $x[i]$. 

The \emph{dot product}\index{Gen}{dot product} of two vectors $x,y\in\Q^m$ is denoted by $\scalar{x}{y}=\sum_{i=1}^mx[i].y[i]$\index{Not}{$\scalar{x}{y}$}.

The \emph{greatest common divisor (gcd)}\index{Gen}{greatest common divisor} of $m\in\Nat\moins\{0\}$ integers $x_1$, ..., $x_m$ is denoted by $\textrm{gcd}(x_1,\ldots,x_m)$. Recall that the gcd of of some integers can be efficiently computed in polynomial time thanks to an Euclidean algorithm.

A rational number $q\in\Q$ can be \emph{canonically represented} as a tuple $(n,d)\in\Z\times(\Nat\moins\{0\})$ such that $q=\frac{n}{d}$ and $\gcd{n}{d}=1$.  The integer $\size(q)\in\Nat$\index{Gen}{sizes!of a rational number} is defined as the least (for $\leq$) integer such that $n,d\leq 2^{\size(q)}$. The integer $\size(x)\in\Nat$\index{Gen}{sizes!of a vector of rational numbers} where $x\in\Q^m$  is defined by $\size(x)=\sum_{i=1}^m\size(x[i])$. The integer $\size(X)\in\Nat$\index{Gen}{sizes! of a finite set of vectors of rational numbers} where $X\in\partie_f(\Q^m)$ is defined by $\size(X)=\sum_{x\in X}\size(x)$. 

A function $f:\Q^m\rightarrow\Q^{m'}$ is said \emph{affine}\index{Gen}{affine function} if for any $i\in\finiteset{1}{m}$, there exists $v_i\in\Q^{m}$ and $c_i\in\Q$ such that $f(x)[i]=c_i+\scalar{v_i}{x}$ for any $x\in\Q^m$.

The set of \emph{matrices}\index{Gen}{matrix} with $n\in\Nat$ \emph{rows}\index{Gen}{rows} and $m\in\Nat$ \emph{columns}\index{Gen}{columns} with \emph{coefficients} in a set $X\subseteq \Q$ is denoted by $\M_{m,n}(X)$. Its elements are denoted by $M[i,j]\in X$ where $1\leq i\leq n$ and $1\leq j\leq m$.

\section{Alphabets, Graphs, and Automata}
An \emph{alphabet}\index{Gen}{alphabet} $\Sigma$\index{Not}{$\Sigma$} is a non-empty finite set. Given an alphabet $\Sigma$, we denote by $\Sigma^+$\index{Not}{$\Sigma^+$} the set of non-empty \emph{words}\index{Gen}{word} over $\Sigma$. Given a non-empty word $\sigma=b_1\ldots b_k$ of $k\in\Nat\moins\{0\}$ elements $b_i\in \Sigma$, and an integer $i\in\finiteset{1}{k}$, we denote by $\sigma[i]$\index{Not}{$\sigma[i]$} the element $\sigma[i]=b_i$. We denote by $\epsilon$\index{Not}{$\epsilon$} the empty word\index{Gen}{empty word}. As usual $\Sigma^*$\index{Not}{$\Sigma^*$} denotes the set of words $\Sigma^+\cup\{\epsilon\}$ and a \emph{language}\index{Gen}{language} $\Lan$\index{Not}{$\Lan$} is a subset of $\Sigma^*$. The \emph{concatenation}\index{Gen}{concatenation!of two words}\index{Gen}{concatenation!of two languages} of $\sigma_1\in\Sigma^*$ and $\sigma_2\in\Sigma^*$ (resp. $\Lan_1\subseteq\Sigma^*$ and $\Lan_2\subseteq \Sigma^*$) is denoted by $\sigma_1.\sigma_2$\index{Not}{$\sigma_1.\sigma_2$} (resp. $\Lan_1.\Lan_2=\{\sigma_1.\sigma_2;\;(\sigma_1,\sigma_2)\in\Lan_1\times\Lan_2\}$\index{Not}{$\Lan_1.\Lan_2$}). Given a word $\sigma\in\Sigma^*$, we define as usual $\sigma^i$\index{Not}{$\sigma^i$} where $i\in\Nat$ and $\sigma^*=\{\sigma^i;\;i\in\Nat\}$\index{Not}{$\sigma^*$}. The \emph{length}\index{Gen}{length of a word} of a word $\sigma\in\Sigma^*$ is denoted by $|\sigma|\in\Nat$\index{Not}{$|\sigma|$}. The \emph{residue}\index{Gen}{residue of languages} $\sigma^{-1}.\Lan$\index{Not}{$\sigma^{-1}.\Lan$} of a language $\Lan\subseteq\Sigma^*$ by a word $\sigma\in\Sigma^*$ is the language $\sigma^{-1}.\Lan=\{w\in\Sigma^*;\;\sigma.w\in\Lan\}$.

A \emph{graph $G$ labelled by $\Sigma$}\index{Gen}{graph} is a tuple $G=(Q,\Sigma,\delta)$ such that $Q$ is the non empty \emph{set of states}\index{Gen}{states}\index{Gen}{states!of a graph}, $\Sigma$ is an alphabet\index{Gen}{alphabet} and $\delta:Q\times\Sigma\rightarrow Q$\index{Not}{$\delta$} is the \emph{transition function}\index{Gen}{transition function}\index{Gen}{transition function}. Two graphs $G_1=(Q_1,\Sigma,\delta_1)$ and $G_2=(Q_2,\Sigma,\delta_2)$ labelled by $\Sigma$ are said \emph{isomorph}\index{Gen}{isomorph}\index{Gen}{isomorph!graph} by a one-to-one relation $\relation\subseteq Q_1\times Q_2$, if we have $\delta_1(q_1,b)\relation\delta_2(q_2,b)$ for any $q_1\relation q_2$ and for any $b\in\Sigma$. As usual, the transition function $\delta$ is uniquely extended into a function $\delta:Q\times\Sigma^*\rightarrow Q$ such that $\delta(q,\epsilon)=q$ for any $q\in Q$ and such that $\delta(q,\sigma_1.\sigma_2)=\delta(\delta(q,\sigma_1),\sigma_2)$. Given a word $\sigma\in\Sigma^*$, we denote by $\xrightarrow{\sigma}$\index{Not}{$\xrightarrow{\sigma}$} the binary relation over $Q$ defined by $q\xrightarrow{\sigma}q'$ if and only if $q'=\delta(q,\sigma)$. In this case, we say that there exists a \emph{path}\index{Gen}{path} from a state $q$ to a state $q'$ labelled by $\sigma$. Such a path is called a \emph{cycle on $q$}\index{Gen}{cycle} if $q=q'$ and $\sigma\not=\epsilon$. Given a language $\Lan\subseteq \Sigma^*$, the binary relation $\xrightarrow{\Lan}$\index{Not}{$\xrightarrow{\Lan}$} is defined by $\xrightarrow{\Lan}=\bigcup_{\sigma\in \Lan}\xrightarrow{\sigma}$. The binary relation $\rightarrow$\index{Not}{$\rightarrow$} is defined by $\rightarrow=\xrightarrow{\Sigma^*}$.  A state $q'$ is said \emph{reachable}\index{Gen}{reachable} from a state $q_0$ if $q_0\rightarrow q'$. The notion of reachability is naturally extended to the subsets of $Q$: a subset $Q'\subseteq Q$ is said \emph{reachable} from a subset $Q_0\subseteq Q$ if the exists a state $q'\in Q'$ reachable from a state $q_0\in Q_0$. In this case the set $Q'$ is said \emph{co-reachable}\index{Gen}{co-reachable} from $Q$. A \emph{strongly connected component}\index{Gen}{component!strongly connected} $Q'$ is an equivalence class for the equivalence binary relation $\rightleftarrows$\index{Not}{$\rightleftarrows$} defined over $Q$ by $q\rightleftarrows q'$ if and only if $q\rightarrow q'$ and $q'\rightarrow q$. A graph $G$ is said \emph{finite}\index{Gen}{graph!finite} if $Q$ is finite. In this case $|G|=|Q|$\index{Not}{$|G|$} denotes the number of states of $G$, and the integer $\size(G)\in\Nat$\index{Gen}{sizes!of a graph} is defined by $\size(G)=|\Sigma|.|Q|$.

An \emph{automaton $\automaton$ labelled by $\Sigma$}\index{Gen}{automaton} is a tuple $\automaton=(k_0,K,\Sigma,\delta,K_F)$ such that $(K,\Sigma,\delta)$ is a graph labelled by $\Sigma$, $k_0\in K$ is the \emph{initial state}\index{Gen}{automaton!initial state} and $K_F\subseteq K$ is the set of \emph{final states}\index{Gen}{automaton!final states}. Two automata $\automaton_1=(k_{0,1},K_1,\Sigma,\delta_1,K_{F,1})$ and $\automaton_2=(k_{0,2},K_2,\Sigma,\delta_2,K_2)$ labelled by $\Sigma$ are said \emph{isomorph}\index{Gen}{isomorph!automaton} by a one-to-one relation $\relation\subseteq K_1\times K_2$ if $(K_1,\Sigma,\delta_1)$ and $(K_2,\Sigma,\delta_2)$ are isomorph by $\relation$, $(k_{0,1},k_{0,2})\in\relation$, and we have $k_1\in K_{F,1}$ if and only if $k_2\in K_{F,2}$ for any $(k_1,k_2)\in\relation$. An automaton with a finite set of states $K$ is said \emph{finite}\index{Gen}{automaton!finite}. In this case, we denote by $|\automaton|$\index{Not}{$|\automaton|$} the number of states $|K|$ and the integer $\size(\automaton)$\index{Gen}{sizes!of an automaton} is defined by $\size(\automaton)=|\Sigma|.|K|$. The language $\Lan(\automaton)\subseteq\Sigma^*$\index{Not}{$\Lan(\automaton)$} \emph{recognized}\index{Gen}{recognized} by an automaton $\automaton$ labelled by $\Sigma$ is defined by $\Lan(\automaton)=\{\sigma\in\Sigma^*;\;\delta(q_0,\sigma)\in K_F\}$. A language $\Lan\subseteq \Sigma^*$ is said \emph{regular}\index{Gen}{regular language} if it can be recognized by a finite automaton. Recall that a language $\Lan\subseteq\Sigma^*$ is regular if and only if the set of residues $\{\sigma^{-1}.\Lan;\;\sigma\in\Sigma^*\}$ is finite. In this case the automaton $(\Lan,K,\Sigma,\delta,K_F)$ defined by the set of states $K=\{\sigma^{-1}.\Lan;\;\sigma\in\Sigma^*\}$, the transition function $\delta(k,b)=b^{-1}.k$ which is in $K$ since $b^{-1}.\sigma^{-1}.\Lan=(\sigma.b)^{-1}.\Lan$ and the final set of states $K_F=\{k\in K;\;\epsilon\in k\}$ is the unique (up to isomorphism) \emph{minimal}\index{Gen}{automaton!minimal} (for the number of states) automaton labelled by $\Sigma$ that recognizes $\Lan$.

%% file: chapter.DVA.tex
\chapter{Finite Digit Vector Automata}
In this chapter, the \emph{Finite Digit Vector Automata (FDVA)} representation, a state-based representation of set of integer vectors is presented.

\section{Digit Vector Decomposition}
In this paper, $r$\index{Not}{$r$} denotes an integer in $\Nat\moins\{0,1\}$ called \emph{basis of decomposition}\index{Gen}{basis of decomposition}. The set $\Sigma_r=\finiteset{0}{r-1}$\index{Not}{$\Sigma_r$} is called the set of \emph{$r$-digits}\index{Gen}{digits} and the set $S_r=\{0,r-1\}\subseteq \Sigma_r$\index{Not}{$S_r$} is called the set of \emph{$r$-signs}\index{Gen}{signs}. Given an integer $m\in\Nat\moins\{0\}$\index{Not}{$m$} called \emph{dimension}\index{Gen}{dimension}, we intensively used the alphabets $\Sigma_{r,m}=\Sigma_r^m$\index{Not}{$\Sigma_{r,m}$} and $S_{r,m}=S_r^m$\index{Not}{$S_{r,m}$} whose the elements are respectively called the \emph{$(r,m)$-digit vectors}\index{Gen}{digit vectors} and the \emph{$(r,m)$-sign vectors}\index{Gen}{sign vectors}. Naturally, a word over the alphabet $\Sigma_{r,m}$ can also be seen as a word over the alphabet $\Sigma_r$ with a length multiple of $m$. In order to simplify notations, these words are identified. Moreover, given a word $\sigma\in\Sigma_{r,m}^*$, we denote by $|\sigma|_m$\index{Not}{$|\sigma|_m$} the length of $\sigma$ seens as a word over the alphabet $\Sigma_{r,m}$ and defined by $|\sigma|_m=\frac{|\sigma|}{m}$, and given a word $\sigma=b_1\ldots b_k$ of $k\in\Nat\moins\{0\}$ $(r,m)$-digit vectors $b_i\in\Sigma_{r,m}$ and an integer $i\in\finiteset{1}{k}$, we denote by $\sigma[i]_m$\index{Not}{$\sigma[i]_m$} the $(r,m)$-digit vector $\sigma[i]_m=b_i$.

A \emph{$(r,m)$-decomposition}\index{Gen}{decomposition} $(\sigma,s)$ of an integer vector $x\in\Z^m$ is a couple $(\sigma,s)\in \Sigma_{r,m}^*\times S_{r,m}$ corresponding to \emph{a least significant digit first decomposition}\index{Gen}{least significant digit first decomposition} of $x$ in basis $r$. More formally, we have $x=\rho_{r,m}(\sigma,s)$\index{Not}{$\rho_{r,m}(\sigma,s)$} where $\rho_{r,m}:\Sigma_{r,m}^*\times S_{r,m}\rightarrow \Z^m$ is defined by the following equality:
$$\rho_{r,m}(\sigma,s)=r^{|\sigma|_m}.\frac{s}{1-r}+\sum_{i=1}^{|\sigma|_m}r^{i-1}.\sigma[i]_m$$

\begin{example}
  $(011,0)$ is a $(2,1)$-decomposition of $6=2^1+2^2$.
\end{example}

\begin{example}\label{ex:0andminus1}
  $(\epsilon,1)$, $(1,1)$, $(11,1)$, ..., $(1\ldots 1,1)$ are the $(2,1)$-decompositions of $-1$ and $(\epsilon,0)$, $(0,0)$, ..., $(0\ldots 0,0)$ are the $(2,1)$-decompositions of $0$.
\end{example}

Following notations introduced in \cite{L-INFINITY03}, function $\rho_{r,m}$ can be defined thanks to the unique sequence $(\gamma_{r,m,\sigma})_{\sigma\in\Sigma_r^*}$\index{Not}{$\gamma_{r,m,\sigma}$} of functions $\gamma_{r,m,\sigma}:\Z^m\rightarrow\Z^m$ such that $\gamma_{r,m,\sigma_1.\sigma_2}=\gamma_{r,m,\sigma_1}\circ\gamma_{r,m,\sigma_2}$ for any $\sigma_1,\sigma_2\in\Sigma_{r}^*$, $\gamma_{r,m,\epsilon}$ is the identity function, and such that $\gamma_{r,m,b}(x)$ is defined for any $(b,x)\in\Sigma_r\times\Z^m$ by the following equality:
$$\gamma_{r,m,b}(x[1],\ldots,x[m])=(r.x[m]+b,x[1],\ldots, x[m-1])$$
In fact, we deduce that for any $(r,m)$-decomposition $(\sigma,s)$, we have the following equality since $\gamma_{r,m,w}(x)=r.x+w$ for any $(w,x)\in\Sigma_{r,m}\times\Z^m$:
$$\rho_{r,m}(\sigma,s)=\gamma_{r,m,\sigma}(\frac{s}{1-r})$$

Function $\rho_{r,m}$ can be used to associate to any language $\Lan\subseteq\Sigma_{r,m}^*\times S_{r,m}$, the set of integer vectors $X=\rho_{r,m}(\Lan)$. Remark that $\rho_{r,m}$ is a surjective function (we have $\rho_{r,m}(\Sigma_{r,m}^*\times S_{r,m})=\Z^m$) because any vector $x\in\Z^m$ owns at least one $(r,m)$-decomposition. Hence, for any subset $X\subseteq\Z^m$, there exists at least one language $\Lan$ such that $X=\rho_{r,m}(\Lan)$. However, intersection of languages does not correspond to intersection of sets of integer vectors: for instance, consider $\Lan_1=\{(0,0)\}$ and $\Lan_2=\{(0.0,0)\}$ and remark that $\{0\}=\rho_{r,1}(\Lan_1)\cap\rho_{r,1}(\Lan_2)\not=\rho_{r,1}(\Lan_1\cap \Lan_2)=\emptyset$. In order to avoid this problem, we introduce the notion of \emph{saturated languages}.

A language $\Lan\subseteq \Sigma_{r,m}^*\times S_{r,m}$ is said \emph{$(r,m)$-saturated}\index{Gen}{saturated!language} if for any $(r,m)$-decompositions $(\sigma_1,s_1)$ and $(\sigma_2,s_2)$ of the same vector, we have $(\sigma_1,s_1)\in\Lan$ if and only if $(\sigma_2,s_2)\in\Lan$. Remark that $\Sigma_{r,m}^*\times S_{r,m}$ is a $(r,m)$-saturated language such that $\rho_{r,m}(\Sigma_{r,m}^*\times S_{r,m})=\Z^m$, and $\Lan_1\#\Lan_2$ is a $(r,m)$-saturated language such that $\rho_{r,m}(\Lan_1\#\Lan_2)=\rho_{r,m}(\Lan_1)\#\rho_{r,m}(\Lan_2)$ for any pair $(\Lan_1,\Lan_2)$ of $(r,m)$-saturated languages, and for any $\#\in\{\cup,\cap,\moins,\Delta\}$.

The $(r,m)$-decompositions of the same integer vector are characterized by the following lemma \ref{lem:carasame}.
\begin{lemma}\label{lem:carasame}
  Two $(r,m)$-decompositions $(\sigma_1,s_1)$ and $(\sigma_2,s_2)$ represent the same integer vector if and only if $s_1=s_2$ and $\sigma_1.s_1^*\cap \sigma_2.s_2^*\not=\emptyset$.
\end{lemma}
\begin{proof}
  Consider two $(r,m)$-decompositions $(\sigma_1,s_1)$ and $(\sigma_2,s_2)$ such that there exists $s\in S_{r,m}$ and $k_1,k_2\in\Nat$ satisfying $s_1=s=s_2$ and $\sigma_1.s_1^{k_1}=\sigma_2.s_2^{k_2}$, and let us prove that $(\sigma_1,s_1)$ and $(\sigma_2,s_2)$ represent the same vector. Just remark that $\gamma_{r,m,s}(\frac{s}{1-r})=\frac{s}{1-r}$ for any $s\in S_{r,m}$. Hence, an immediate induction (over $k_1$ and $k_2$) shows that $(\sigma_1,s_1)$ and $(\sigma_2,s_2)$ represent the same vector.
  
  For the converse, consider two $(r,m)$-decompositions $(\sigma_1,s_1)$ and $(\sigma_2,s_2)$ that represent the same vector. Remark that for any $(r,m)$-decomposition $(\sigma,s)$ of an integer vector $x\in\Z^m$, we have $s[i]=0$ if $x[i]\in\Nat$ and $s[i]=r-1$ if $x[i]\in\Z\moins\Nat$ for any $i\in\finiteset{1}{m}$. Therefore, as $(\sigma_1,s_1)$ and $(\sigma_2,s_2)$ represents the same vector, we deduce that there exists $s\in S_{r,m}$ such that $s_1=s=s_2$. Consider $k_1$ and $k_2$ such that $|\sigma_1|+k_1=|\sigma_2|+k_2$. From the first paragraph, we deduce that $(w_1,s)$ and $(w_2,s)$ represent the same vector where $w_1=\sigma_1.s_1^{k_1}$ and $w_2=\sigma_2.s_2^{k_2}$. By uniqueness of the $(r,m)$-decompositions with a fixed length, we deduce that $w_1=w_2$.
  \qed
\end{proof}

\section{State-based Decomposition}
A language of $(r,m)$-decompositions can be naturally represented by a state-based representation. Our representation is obtain by considering the natural one-to-one function from the set of $(r,m)$-decompositions to the set of words in $\Sigma_{r,m}^*.\Diamond.S_{r,m}$ that associate to a $(r,m)$-decomposition $(\sigma,s)$ the word $\sigma.\Diamond.s$ where $\Diamond$ is an additional letter not in $\Sigma_r$. 

Observe that an automaton $\automaton$ recognizing a language included in $\Sigma_{r,m}^*.\Diamond.S_{r,m}$ can be decomposed into (1) a graph called \emph{Digit Vector Graph} corresponding to the part of $\automaton$ before a $\Diamond$ letter, and the part of $\automaton$ after a $\Diamond$ letter called a \emph{final function}.
\begin{definition}
  A \emph{Digit Vector Graph (DVG)}\index{Gen}{DVG}\index{Gen}{digit vector graph} is a tuple $G=(Q,m,K,\Sigma_r,\delta)$ where $Q$ is the non empty set of \emph{principal states}\index{Gen}{principal states}\index{Gen}{principal state}, $r\in\Nat\moins\{0,1\}$ is the \emph{basis of decomposition}, $m\in\Nat\moins\{0\}$ is the \emph{dimension}, and $(K,\Sigma_r,\delta)$ is a graph such that $Q\subseteq K$ and $\delta(Q,\Sigma_{r,m})\subseteq Q$.
\end{definition}
A \emph{Finite Digit Vector Graph (FDVG) $G$}\index{Gen}{FDVG}\index{Gen}{digit vector graph!finite} is a DVG with a finite set of states $K$. Given a FDVG $G$, the integer $\size(G)\in\Nat$\index{Gen}{sizes!finite digit vector graph} is defined by $\size(G)=r.|K|$. The \emph{parallelization}\index{Gen}{parallelization!of a digit vector graph} $[G]$\index{Not}{$[G]$} of a DVG $G=(Q,m,K,\Sigma_r,\delta)$ is the graph $[G]=(Q,\Sigma_{r,m},\delta)$. We introduce DVG rather than graph labelled by $\Sigma_{r,m}$ in order to establish fine polynomial time complexity results that should be useless with an exponential size in $m$ of the alphabet $\Sigma_{r,m}$. Naturally any graph labelled by $\Sigma_{r,m}$ is equal to the parallelization of at least one DVG in basis $r$ and in dimension $m$. 

\begin{definition}
  A \emph{final function}\index{Gen}{final function} is a tuple $F=(Q,f,m,K,S_r,\delta,K_F)$ where $Q$ is the non empty set of \emph{principal states}\index{Gen}{state!principal}, $r\in\Nat\moins\{0,1\}$ is the \emph{basis of decomposition}, $m\in\Nat\moins\{0\}$ is the \emph{dimension}, $(K,S_r,\delta)$ is a \emph{finite} graph, $f:Q\rightarrow K$ is a function mapping principal states to states in $K$, and $K_F\subseteq K$ is the set of \emph{final states} such that the language recognized by the automaton $(f(q),K,S_r,\delta,K_F)$ is a subset of $S_{r,m}$ for any principal state $q\in Q$.
\end{definition}
A final function $F$ is said \emph{finite}\index{Gen}{final function!finite} if the set of principal states $Q$ is \emph{finite} (observe that $K$ is finite by definition). Given a finite final function $F$, the integer $\size(F)\in\Nat$\index{Gen}{sizes!of a finite final function} is defined by $\size(F)=|Q|+|K|$. The \emph{parallelization}\index{Gen}{parallelization!of a final function} $[F]$\index{Not}{$[F]$} of a final function $F=(Q,f,m,K,S_r,\delta,K_F)$ is the function $[F]:Q\rightarrow\partie(S_{r,m})$ such that $[F](q)$ is the language recognized by the automaton $(f(q),K,S_r,\delta,K_F)$.

A DVG $G$ and a final function $F$ are said \emph{compatible}\index{Gen}{final function!compatible}\index{Gen}{digit vector graph!compatible} if they are defined over the same set of principal states with the same basis $r$ and the same dimension $m$. Given a tuple $(q,G,F)$ where $q$ is a principal state, $G$ is a DVG and $F$ is a final function compatible, we denote by $\Lan((q,G,F))$\index{Not}{$\Lan(\automaton)$} the following language of $(r,m)$-decompositions:
$$\Lan((q,G,F))=\{(w,s)\in\Sigma_{r,m}^*\times S_{r,m};\;s\in [F](\delta(q,w))\}$$
Recall that we are interested in recognizing $(r,m)$-saturated languages. A final function $F$ is said \emph{saturated}\index{Gen}{saturated!final function} for a DVG $G$ if it is compatible with $G$ and if $\Lan((q,G,F))$ is $(r,m)$-saturated for any principal states $q\in Q$. 
\begin{proposition}\label{prop:satcara}
  A final function $F$ is saturated for a DVG $G$ if and only if $F$ and $G$ are compatible and $[F](q_1)\cap\{s\}=[F](q_2)\cap \{s\}$ for any $q_1\xrightarrow{s}q_2$ with $(q_1,s,q_2)\in Q\times S_{r,m}\times Q$.
\end{proposition}
\begin{proof}
  Assume first that $\Lan((q,G,F))$ is $(r,m)$-saturated for any state $q\in Q$, and let us prove that $s\in [F](q_1)$ if and only if $s\in [F](q_2)$ for any $q_1\xrightarrow{s}q_2$ with $(q_1,s,q_2)\in Q\times S_{r,m}\times Q$. Assume first that $s\in [F](q_1)$. Lemma \ref{lem:carasame} proves that $\rho_{r,m}(\epsilon,s)=\rho_{r,m}(s,s)$. As $\Lan((q_1,G,F))$ is $(r,m)$-saturated, we deduce that $(s,s)\in \Lan((q_1,G,F))$. From $q_2=\delta(q_1,s)$ we get $s\in [F](q_2)$. Next assume that $s\in [F](q_2)$. We get $(s,s)\in\Lan((q_1,G,F))$. As this language is $(r,m)$-saturated and $\rho_{r,m}(s,s)=\rho_{r,m}(\epsilon,s)$, we deduce that $(\epsilon,s)\in \Lan((q_1,G,F))$. Therefore $s\in [F](q_1)$. 
  
  Next, assume that $[F](q_1)\cap\{s\}=[F](q_2)\cap\{s\}$ for any $q_1\xrightarrow{s}q_2$ with $(q_1,s,q_2)\in Q\times S_{r,m}\times Q$, and let us prove that $\Lan((q,G,F))$ is $(r,m)$-saturated for any state $q\in Q$. Let us consider two $(r,m)$-decomposition $(\sigma,s)$ and $(\sigma',s')$ of the same integer vector such that $(\sigma',s')\in\Lan((q,G,F))$ and let us prove that $(\sigma,s)\in\Lan((q,G,F))$. From lemma \ref{lem:carasame}, we deduce that $s=s'$ and there exists $k,k'\in\Nat$ such that $\sigma.s^k=\sigma'.s^{k'}$. As $s\in \Lan((q_1,G,F))$ if and only if $s\in\Lan(q_2,G,F)$ for any $q_1\xrightarrow{s}q_2$ with $q_1,q_2\in Q$, an immediate induction shows that $(\sigma',s')\in\Lan((q,G,F))$ implies $(\sigma,s)\in\Lan((q,G,F))$. Therefore $\Lan((q,G,F))$ is $(r,m)$-saturated for any $q\in Q$.
  \qed
\end{proof}

We can now introduce our definition of digit vector automata.
\begin{definition}
  A \emph{Digit Vector Automaton (DVA)}\index{Gen}{DVA}\index{Gen}{digit vector automaton} is a tuple $\automaton=(q_0,G,F_0)$ where $q_0\in Q$ is the \emph{initial state}, $G$ is a DVG and $F_0$ is a final function saturated for $G$.
\end{definition}
A \emph{Finite Digit Vector Automaton (FDVA)}\index{Gen}{FDVA}\index{Gen}{digit vector automaton!finite} $\automaton$ is a DVA with a finite DVG $G$ and a finite final function $F$. Given a FDVA $\automaton$, the integer $\size(\automaton)$\index{Gen}{sizes!of a FDVA} is defined by $\size(\automaton)=\size(G)+\size(F)$. Given a DVA $\automaton=(q_0,G,F_0)$, the $(r,m)$-saturated language $\Lan(\automaton)=\Lan((q_0,G,F_0))$ is called the \emph{recognized language}\index{Gen}{recognized} of $\automaton$. The set $X=\rho_{r,m}(\Lan(\automaton))$ is called the set of integer vectors \emph{represented}\index{Gen}{represented} by $\automaton$. 

Let us show that any set $X\subseteq\Z^m$ can be represented by a DVA by introducing the DVG $G_{r,m}(X)=(Q_{r,m}(X),m,K_{r,m}(X),\Sigma_r,\delta_{r,m})$\index{Not}{$G_{r,m}(X)$} where $K_{r,m}(X)=\{\gamma_{r,m,w}^{-1}(X);\;w\in\Sigma_r^*\}$, $Q_{r,m}(X)=\{\gamma_{r,m,w}^{-1}(X);\;w\in\Sigma_{r,m}^*\}$, and $\delta_{r,m}$ is defined by $\delta_{r,m}(Y,b)=\gamma_{r,m,b}^{-1}(Y)$ for any $Y\in K_{r,m}(X)$ and $b\in\Sigma_r$. Finally, let us consider the tuple $\automaton_{r,m}(X)=(X,G_{r,m}(X),F_{r,m})$\index{Not}{$\automaton_{r,m,}(X)$} where $F_{r,m}$ is any final function such that $[F_{r,m}](Y)=S_{r,m}\cap (1-r).Y$ for any $Y\in Q_{r,m}(X)$.
\begin{proposition}\label{prop:autox}
  The tuple $\automaton_{r,m}(X)$ is a DVA in basis $r$ and in dimension $m$ that represents $X$.
\end{proposition}
\begin{proof}
  Let us first prove that $\automaton_{r,m}(X)$ is a DVA in basis $r$ and in dimension $m$. It is sufficient to show that $[F_{r,m}](q_1)\cap\{s\}=[F_{r,m}](q_2)\cap\{s\}$ for any $q_1\xrightarrow{s}q_2$ where $(q_1,s,q_2)\in Q\times S_{r,m}\times Q$. As $q_1\xrightarrow{s}q_2$, we get $q_2=\gamma_{r,m,s}^{-1}(q_1)$. Remark that $[F_{r,m}](q_1)=S_{r,m}\cap (1-r).q_1$ and $[F_{r,m}](q_2)=S_{r,m}\cap (1-r).q_2$. As $\gamma_{r,m,s}(\frac{s}{1-r})=\frac{s}{1-r}$, we deduce that $[F_{r,m}](q_1)\cap\{s\}=[F_{r,m}(q_2)]\cap\{s\}$. We are done.

  Now, let $X'$ be the set represented by the DVA $\automaton_{r,m}(X)$, and let us prove that $X'=X$. Let $x\in X'$. There exists a $(r,m)$-decomposition $(\sigma,s)$ of $x$ such that $(\sigma,s)\in\Lan(\automaton_{r,m}(X))$. Let $q=\delta_{r,m}(q_0,\sigma)$. We get $q=\gamma_{r,m,\sigma}^{-1}(X)$. From $s\in [F_{r,m}](q)$, we deduce $s\in S_{r,m}\cap (1-r).q$. Hence $\frac{s}{1-r}\in q=\gamma_{r,m,\sigma}^{-1}(X)$ and we obtain $\gamma_{r,m,\sigma}(\frac{s}{1-r})\in X$. As $\rho_{r,m}(\sigma,s)=x$, we get $x\in X$ and we have proved the inclusion $X'\subseteq X$. For the converse inclusion, let $x\in X$. Let us consider a $(r,m)$-decomposition $(\sigma,s)$ of $x$. As $x=\rho_{r,m}(\sigma,s)$ and $\rho_{r,m}(\sigma,s)=\gamma_{r,m,\sigma}^{-1}(X)$, we get $s\in S_{r,m}\cap (1-r).q$ where $q=\gamma_{r,m,\sigma}^{-1}(X)$. Therefore $q_0\xrightarrow{\sigma}q$ and $s\in[F_{r,m}](q)$. That means $\rho_{r,m}(\sigma,s)\in X'$ and we have proved the other inclusion $X\subseteq X'$.
  \qed
\end{proof}

%% file: chapter.modifyingDVA.tex
\chapter{Modifying a DVA}
The sets obtained by \emph{moving the initial state} of a DVA are geometrically characterized in section \ref{sub:moveinit} and the set obtained by \emph{modifying the final function} of a DVA are studied in section \ref{sub:replacingfinal}.

\section{Moving the initial state}\label{sub:moveinit}
The DVA obtained from $\automaton$ by replacing the initial state $q_0$ by another principal state $q\in Q$ is denoted by $\automaton_q$\index{Not}{$\automaton_q$}. Given a set $X$ implicitly represented by a DVA $\automaton$ with a set of principal states $Q$, we denote by $X_q$\index{Not}{$X_q$} the set represented by the DVA $\automaton_q$. In this section the set $X_{q_2}$ is geometrically characterized in function of $X_{q_1}$ for any path $q_1\xrightarrow{w}q_2$ where $(q_1,w,q_2)\in Q\times\Sigma_{r,m}^*\times Q$. 

\begin{proposition}\label{prop:moveinit}
  Let $X$ be a set represented by a DVA in basis $r$ and in dimension $m$ with a set $Q$ of principal states. We have $X_{q_2}=\gamma_{r,m,w}^{-1}(X_{q_1})$ for any path $q_1\xrightarrow{w}q_2$ where $(q_1,w,q_2)\in Q\times\Sigma_{r,m}^*\times Q$.
\end{proposition}
\begin{proof}
  Consider $x\in X_{q_2}$. There exists $(\sigma,s)\in\Lan(\automaton_{q_2})$ such that $x=\rho_{r,m}(\sigma,s)$. From $(w.\sigma,s)\in\Lan(\automaton_{q_1})$, we deduce that $\rho_{r,m}(w.\sigma,s)\in X_{q_2}$. Just remark that $\rho_{r,m}(w.\sigma,s)=\gamma_{r,m,w}(\rho_{r,m}(\sigma,s))=\gamma_{r,m,w}(x)$. We have proved that $X_{q_2}\subseteq \gamma_{r,m,w}^{-1}(X_{q_1})$. For the converse, consider $x\in \gamma_{r,m,w}^{-1}(X_{q_1})$. As any vector owns at least one $(r,m)$-decomposition, there exists a $(r,m)$-decomposition $(\sigma,s)$ such that $x=\rho_{r,m}(\sigma,s)$. From $x\in\gamma_{r,m,w}^{-1}(X_{q_1})$, we get $\gamma_{r,m,w}(x)\in X_{q_1}$. Just remark that $\gamma_{r,m,w}(x)=\rho_{r,m}(w.\sigma,s)$. As $\Lan(\automaton_{q_1})$ is $(r,m)$-saturated, we get $(w.\sigma,s)\in\Lan(\automaton_{q_1})$. In particular $(\sigma,s)\in\Lan(\automaton_{q_2})$. Hence $x=\rho_{r,m}(\sigma,s)\in X_{q_2}$. We have proved the other inclusion $\gamma_{r,m,w}^{-1}(X_{q_1})\subseteq X_{q_2}$.
  \qed
\end{proof}

\begin{theorem}
  Let $X$ be a Presburger-definable set represented by a DVA $\automaton=(q_0,G,F_0)$. The set $X_{q}$ is Presburger-definable for any reachable (for $[G]$) principal state $q\in Q$.
\end{theorem}
\begin{proof}
  The proof is immediate because if $X$ is Presburger-definable, there exists a formula $\phi$ in $\fo{\Z,\Nat,+}$ that defines $X$. Consider a reachable (for $[G]$) principal state $q\in Q$. There exists a path $q_0\xrightarrow{\sigma}q$ with $\sigma\in\Sigma_{r,m}^*$. From proposition \ref{prop:moveinit}, we deduce that $X_{q}$ is defined by the Presburger formula $\phi_\sigma(x):=\exists y;\;(y=\gamma_{r,m,\sigma}(x)\wedge \phi(y))$. Therefore $X_q$ is Presburger-definable.
  \qed
\end{proof}

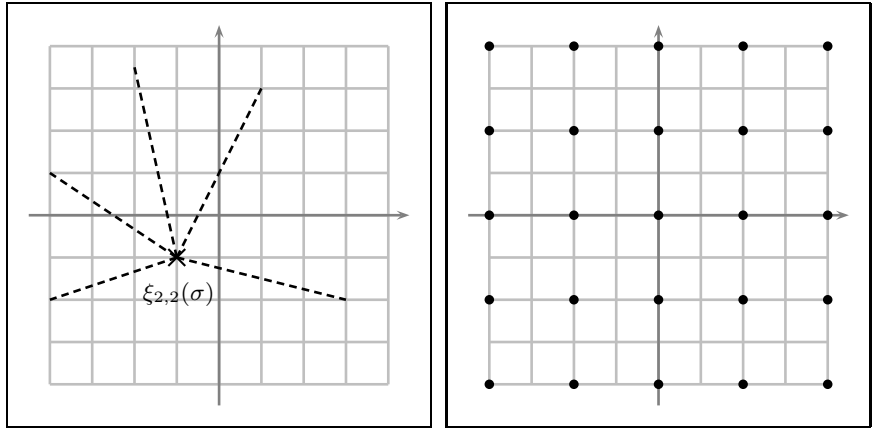
\begin{figure}[htbp]
  \begin{center}
    \setlength{\unitlength}{16pt}
    \pssetlength{\psunit}{16pt}
    \pssetlength{\psxunit}{16pt}
    \pssetlength{\psyunit}{16pt}    
    \begin{picture}(10,10)(-5,-5)
      \put(-5,-5){\framebox(10,10){}}
        \multido{\iii=-4+1}{9}{\psline[linecolor=lightgray,linewidth=1pt](-4,\iii)(4,\iii)}\multido{\iii=-4+1}{9}{\psline[linecolor=lightgray,linewidth=1pt](\iii,-4)(\iii,4)}\psline[linecolor=gray,arrowscale=1,linewidth=1pt]{->}(0,-4.5)(0,4.5)\psline[linecolor=gray,arrowscale=1,linewidth=1pt]{->}(-4.5,0)(4.5,0)
      
      \psline(-1.2,-1.2)(-0.8,-0.8)
      \psline(-1.2,-0.8)(-0.8,-1.2)
      \put(-1.8,-2){$\xi_{2,2}(\sigma)$}
      \psline[linewidth=1pt,linestyle=dashed,dash=3pt 2pt](-1,-1)(1,3)
      \psline[linewidth=1pt,linestyle=dashed,dash=3pt 2pt](-1,-1)(3,-2)
      \psline[linewidth=1pt,linestyle=dashed,dash=3pt 2pt](-1,-1)(-4,1)
      \psline[linewidth=1pt,linestyle=dashed,dash=3pt 2pt](-1,-1)(-2,3.5)
      \psline[linewidth=1pt,linestyle=dashed,dash=3pt 2pt](-1,-1)(-4,-2)
    \end{picture}
    \setlength{\unitlength}{16pt}
    \pssetlength{\psunit}{16pt}
    \pssetlength{\psxunit}{16pt}
    \pssetlength{\psyunit}{16pt}    
    \begin{picture}(10,10)(-5,-5)
      \put(-5,-5){\framebox(10,10){}}
       \multido{\iii=-4+1}{9}{\psline[linecolor=lightgray,linewidth=1pt](-4,\iii)(4,\iii)}\multido{\iii=-4+1}{9}{\psline[linecolor=lightgray,linewidth=1pt](\iii,-4)(\iii,4)}\psline[linecolor=gray,arrowscale=1,linewidth=1pt]{->}(0,-4.5)(0,4.5)\psline[linecolor=gray,arrowscale=1,linewidth=1pt]{->}(-4.5,0)(4.5,0)      
      \multido{\ix=-4+2}{5}{\multido{\iy=-4+2}{5}{\psdot(\ix,\iy)}}
    \end{picture}
  \end{center}
  \caption{On the left, $\Gamma_{2,2,\sigma}^{-1}$ with its fix-point $\xi_{2,2}(\sigma)$. On the right $\Gamma_{2,2,(0,0)}(\Z^2)$ \label{fig:mask}}
\end{figure}

Previous proposition \ref{prop:moveinit} provides a characterization of the sets obtained by moving the initial state of a DVA to another principal state. This characterization can be translated into a geometrical one by considering the unique sequence $(\Gamma_{r,m,w})_{w\in\Sigma_r^*}$\index{Not}{$\Gamma_{r,m,\sigma}$} of affine functions $\Gamma_{r,m,w}:\Q^m\rightarrow\Q^m$ such that $\Gamma_{r,m,w_1.w_2}=\Gamma_{r,m,w_1}\circ\Gamma_{r,m,w_2}$ for any $(w_1,w_2)\in\Sigma_r^*$, such that $\Gamma_{r,m,\epsilon}$ is the identity function and such that $\Gamma_{r,m,b}(x)$ is defined for any $(b,x)\in\Sigma_r\times\Q^m$ by the following equality:
$$\Gamma_{r,m,b}(x[1],\ldots,x[m])=(r.x[m]+b,x[1],\ldots, x[m-1])$$
As $\gamma_{r,m,\sigma}(x)=\Gamma_{r,m,\sigma}(x)$ for any $x\in\Z^m$, we deduce that $\gamma_{r,m,\sigma}^{-1}(X)=\Gamma_{r,m,\sigma}^{-1}(X\cap \Gamma_{r,m,\sigma}(\Z^m))$. Now, just remark that given $\sigma\in\Sigma_{r,m}^*$, $\Gamma_{r,m,\sigma}(x)=r^{|\sigma|_m}.x+\gamma_{r,m,\sigma}(\unit_{0,m})$ is simply a \emph{scaling function}\index{Gen}{scaling function} (an affine function of the form $x\rightarrow \mu.x+v$ where $\mu\in\Q\moins\{0\}$ and $v\in\Q^m$) and $\Gamma_{r,m,\sigma}(\Z^m)=r^{|\sigma|}.\Z^m+\gamma_{r,m,\sigma}(\unit_{0,m})$ is a \emph{pattern} (see figure \ref{fig:mask} and section \ref{sec:pattern}).

\begin{remark}
  Function $\Gamma_{r,m,w}$ is the unique affine function that \emph{extends} $\gamma_{r,m,w}$: there exists a unique affine function $f:\Q^m\rightarrow\Q^m$ such that $f(x)=\gamma_{r,r,w}(x)$ for any $x\in\Z^m$.
\end{remark}

The following lemma introduces the geometrically characterized vectors $\xi_{r,m}(\sigma)$\index{Not}{$\xi_{r,m}(\sigma)$} that will be useful in the sequel.
\begin{lemma}
  The function $\xi_{r,m}:\Sigma_{r,m}^+\rightarrow\Q^m$ defined by $\xi_{r,m}(\sigma)=\frac{\gamma_{r,m,\sigma}(\unit_{0,m})}{1-r^{|\sigma|_m}}$ is the unique function such that $\xi_{r,m}(\sigma)$ is a fix-point of $\Gamma_{r,m,\sigma}$ for any $\sigma\in\Sigma_{r,m}^+$.
\end{lemma}
\begin{proof}
  Remark that $\xi_{r,m}(\sigma)$ is a fix-point of $\Gamma_{r,m,\sigma}$, and if $x$ is a fix-point of $\Gamma_{r,m,\sigma}$, then $r^{|\sigma|_m}.x+\gamma_{r,m,\sigma}(\unit_{0,m})=x$ and we deduce that $x=\xi_{r,m}(\sigma)$.
  \qed
\end{proof}

In the sequel the sets $X\subseteq\Z^m$ such that there exists $\sigma\in\Sigma_{r,m}^+$ satisfying $\gamma_{r,m,\sigma}^{-1}(X)=X$ are useful since intuitively $\xi_{r,m}(\sigma)$ is a fix point of these sets. Such a set is said \emph{$(r,m,\sigma)$-cyclic}\index{Gen}{cyclic set}.

\section{Replacing the final function}\label{sub:replacingfinal}
Given a set $X$ implicitly represented by a DVA $\automaton=(q_0,G,F_0)$ and given a final function $F$ saturated for $G$, we denote by $X^{F}$\index{Not}{$X^F$} the set represented by the DVA $\automaton^F$\index{Not}{$\automaton^F$} obtained from $\automaton$ by replacing $F_0$ by $F$.

\subsection{Detectable sets}
A set $X'\subseteq\Z^m$ is said \emph{$(r,m)$-detectable}\index{Gen}{detectable} in a set $X\subseteq\Z^m$ if $\gamma_{r,m,\sigma_1}^{-1}(X')=\gamma_{r,m,\sigma_2}^{-1}(X')$ for any words $\sigma_1,\sigma_2\in\Sigma_{r,m}^*$ such that $\gamma_{r,m,\sigma_1}^{-1}(X)=\gamma_{r,m,\sigma_2}^{-1}(X)$. The following theorem \ref{thm:detectable} shows that these sets characterize the sets $X'\subseteq \Z^m$ such that for any DVA $\automaton=(q_0,G,F_0)$ that represents $X$, there exists a final function $F$ saturated for $G$ such that $X'=X^F$.
\begin{theorem}\label{thm:detectable}
  A set $X'\subseteq\Z^m$ is $(r,m)$-detectable in a set $X\subseteq\Z^m$ if and only if for any DVA $\automaton$ that represents $X$, there exists a final function $F$ saturated for $G$ such that $X'=X^F$.
\end{theorem}
\begin{proof}
  Assume first that for any DVA $\automaton=(q_0,G,F_0)$ that represents $X$, there exists a final function $F$ saturated for $G$ such that $X'=X^F$. Let us consider the DVA $\automaton_{r,m}(X)=(X,G_{r,m}(X),F_{r,m})$ where $G_{r,m}(X)=(Q_{r,m}(X),m,K_{r,m}(X),\Sigma_r,\delta_{r,m})$. There exists $F:Q_{r,m}(X)\rightarrow \partie(S_{r,m})$ such that $X'$ is represented by the DVA $(X,G_{r,m}(X),F)$. Consider $\sigma_1,\sigma_2\in\Sigma_{r,m}^*$ such that $\gamma_{r,m,\sigma_1}^{-1}(X)=\gamma_{r,m,\sigma_2}^{-1}(X)$. By definition of $\automaton_{r,m}(X)$, there exists $Y\in Q_{r,m}(X)$ such that $\delta_{r,m}(X,\sigma_1)=Y=\delta_{r,m}(X,\sigma_2)$. Proposition \ref{prop:moveinit} proves that $\gamma_{r,m,\sigma_1}^{-1}(X')=X^F_Y=\gamma_{r,m,\sigma_2}^{-1}(X')$. Therefore $X'$ is $(r,m)$-detectable in $X$.
  
  Next, assume that $X'$ is $(r,m)$-detectable in $X$ and let us consider a DVA $\automaton=(q_0,G,F_0)$ that represents $X$ where $G=(Q,m,K,\Sigma_r,\delta)$. Let $F$ be a final function over $Q$ such that $[F](q)=\{s\in S_{r,m};\exists \sigma\in\Sigma_{r,m}^*;\;\delta(q_0,\sigma)\in\delta(q,s^*)\wedge \rho_{r,m}(\sigma,s)\in X'\}$.
  
  Let us first prove that $F$ is saturated for $G$. Consider a transition $q\xrightarrow{s}q'$ with $s\in S_{r,m}$, and let us prove that $s\in [F](q)$ if and only if $s\in [F](q')$. Assume first that $s\in [F](q)$. We deduce that there exists $\sigma\in\Sigma_{r,m}^*$, and integer $k\in\Nat$ such that $\delta(q_0,\sigma)=\delta(q,s^k)$ and $\rho_{r,m}(\sigma,s)\in X'$. From $\delta(q_0,\sigma.s)=\delta(q',s^k)$ and $\rho_{r,m}(\sigma.s,s)=\rho_{r,m}(\sigma,s)\in X'$, we deduce that $s\in [F](q')$. Let us prove the converse and assume now that $s\in [F](q')$. There exists a word $\sigma\in\Sigma_r^*$, an integer $k\in\Nat$ such that $\delta(q_0,\sigma)=\delta(q',s^k)$ and such that $\rho_{r,m}(\sigma,s)\in X'$. Just remark that $\delta(q_0,\sigma.s)=\delta(q,s^{k+1})$ and $\rho_{r,m}(\sigma.s,s)=\rho_{r,m}(\sigma,s)\in X'$. Hence $s\in [F](q)$. We have proved that $F$ is saturated for $G$.
  
  By construction of $F$, we have $X'\subseteq X^F$. Let us prove the converse inclusion. Consider a vector $x\in X^F$. There exists a $(r,m)$-decomposition $(w,s)\in\Lan_{q_0}^{F}$ such that $\rho_{r,m}(w,s)=x$. Let $q=\delta(q_0,w)$. We get $s\in [F](q)$. That means there exists $\sigma\in\Sigma_{r,m}^*$ such that $\delta(q_0,\sigma)\in\delta(q,s^*)$ and such that $\rho_{r,m}(\sigma,s)\in X'$. By replacing $w$ by a word in $w.s^*$, we can assume that $\delta(q_0,\sigma)=q$. From $\delta(q_0,\sigma)=\delta(q_0,w)$, proposition \ref{prop:moveinit} shows that $\gamma_{r,m,\sigma}^{-1}(X)=\gamma_{r,m,w}^{-1}(X)$. As $X'$ is detectable in $X$, we get $\gamma_{r,m,\sigma}^{-1}(X')=\gamma_{r,m,w}^{-1}(X')$. Moreover, as $\rho_{r,m}(\sigma,s)\in X'$, we deduce from the previous equality that $x=\rho_{r,m}(w,s)\in X'$. We have proved the other inclusion $X^F\subseteq X'$.
  \qed
\end{proof}

The following proposition will be useful for deciding if a set $X'$ is $(r,m)$-detectable in a set $X$ represented by a DVA $\automaton$ in basis $r$. 
\begin{proposition}\label{prop:critdetectable}
  Let us consider a FDVA $\automaton$ in dimension $m$ in basis $r$ with $n$ states. We can compute in polynomial time a set $U$ of at most $r.m.n$ pairs $(\sigma_1,\sigma_2)$ of words in $\Sigma_r^{\leq n}$ satisfying $|\sigma_1|+m.\Z=|\sigma_2|+m.\Z$ for any $(\sigma_1,\sigma_2)\in U$, and such that for any set $X'\subseteq\Z^m$, there exists a final function $F$ such that $X'$ is represented by $\automaton^F$ if and only if $\gamma_{r,m,\sigma_1}^{-1}(X')=\gamma_{r,m,\sigma_2}^{-1}(X')$ for any $(\sigma_1,\sigma_2)\in U$.
\end{proposition}
\begin{proof}
  We first show that for any $z\in\Nat$ and for any $X\subseteq\Z^m$ we have $\bigcup_{\sigma\in\Sigma_r^z}\gamma_{r,m,\sigma}(\gamma_{r,m,\sigma}^{-1}(X))=X$. Naturally $\gamma_{r,m,\sigma}(\gamma_{r,m,\sigma}^{-1}(X))\subseteq X$ for any word $\sigma\in\Sigma_r^*$ and in particular we get the inclusion $\bigcup_{\sigma\in\Sigma_r^z}\gamma_{r,m,\sigma}(\gamma_{r,m,\sigma}^{-1}(X))\subseteq X$. For the converse inclusion, let $x\in X$. There exists a $(r,m)$-decomposition $(w,s)$ of $x$ and by replacing $w$ by a word in $w.s^*$, we can assume that $|w|\geq z$. In particular there exists a decomposition of $w$ into $w=\sigma.w'$ where $\sigma\in\Sigma_r^z$. Since $\rho_{r,m}(\sigma.w',s)=\gamma_{r,m,\sigma}(\rho_{r,m}(w',s))$ and $\rho_{r,m}(\sigma.w',s)=x\in X$, we deduce that $\rho_{r,m}(w',s)\in \gamma_{r,m,\sigma}^{-1}(X)$ and hence $x\in\gamma_{r,m,\sigma}(\gamma_{r,m,\sigma}^{-1}(X))$. We have proved the converse inclusion.  
  
  Let $S$ be the set of couples $s=(k,Z)\in K\times \Z/m.\Z$ such that there exists a word $\sigma_s\in\Sigma_r^*$ satisfying $s=(\delta(q_0,\sigma_s),|\sigma_s|+m.\Z)$, and let $(\sigma_s)_{s\in S}$ be a sequence of words satisfying the previous condition, $\sigma_{(q_0,m.\Z)}=\epsilon$ and $|\sigma_s|<n$ for any $s\in S$. Observe that such a sequence $(\sigma_s)_{s\in S}$ is computable in polynomial time. Let us consider the set $U$ of pairs $(\sigma_{s_1}.b,\sigma_{s_2})$ where $s_1=(k_1,Z_1)$, $s_2=(k_2,Z_2)$ are in $S$ and $b\in\Sigma_r$ satisfies $s_2=(\delta(k_1,b),Z_1+1)$. Note that $U$ is computable in polynomial time and it contains at most $r.m.n$ pairs $(\sigma_1,\sigma_2)$ of words in $\Sigma_r^{\leq n}$ satisfying $|\sigma_1|+m.\Z=|\sigma_2|+m.\Z$ for any $(\sigma_1,\sigma_2)\in U$.
  
  Assume first that there exists a final function $F$ such that $X'$ is represented by $\automaton^F$ and let us prove that $\gamma_{r,m,\sigma_1}^{-1}(X')=\gamma_{r,m,\sigma_2}^{-1}(X')$ for any $(\sigma_1,\sigma_2)\in U$. Remark that it sufficient to prove that $\gamma_{r,m,\sigma_1}^{-1}(X')=\gamma_{r,m,\sigma_2}^{-1}(X')$ for any pair $(\sigma_1,\sigma_2)$ of words in $\Sigma_r^*$ such that there exists $s=(k,Z)\in U$ satisfying $(\delta(q_0,\sigma_1),|\sigma_1|+m.\Z)=s=(\delta(q_0,\sigma_2),|\sigma_2|+m.\Z)$. There exists $z\in\finiteset{0}{m-1}$ such that $Z+z=m.\Z$. Since $\delta(q_0,\sigma_1)=\delta(q_0,\sigma_2)$ we deduce that $\delta(q_0,\sigma_1.\sigma)=\delta(q_0,\sigma_2.\sigma)$ for any word $\sigma\in\Sigma_r^z$. As $\sigma_1.\sigma$ and $\sigma_2.\sigma$ are both in $\Sigma_{r,m}^*$, proposition \ref{prop:moveinit} shows that $\gamma_{r,m,\sigma_1.\sigma}^{-1}(X')=\gamma_{r,m,\sigma_2.\sigma}^{-1}(X')$. Thus $\gamma_{r,m,\sigma}^{-1}(X_1')=\gamma_{r,m,\sigma}^{-1}(X_2')$ for any $\sigma\in\Sigma_r^z$ where $X_1'=\gamma_{r,m,\sigma_1}^{-1}(X')$ and $X_2'=\gamma_{r,m,\sigma_2}^{-1}(X')$. We have proved that $\bigcup_{\sigma\in\Sigma_{r,m}^z}\gamma_{r,m,\sigma}(\gamma_{r,m,\sigma}^{-1}(X_1'))=\bigcup_{\sigma\in\Sigma_r^z}\gamma_{r,m,\sigma}(\gamma_{r,m,\sigma}^{-1}(X_2'))$. From the first paragraph we get $X_1'=X_2'$.
  
  Next assume that $\gamma_{r,m,\sigma_1}^{-1}(X')=\gamma_{r,m,\sigma_2}^{-1}(X')$ for any $(\sigma_1,\sigma_2)\in U$ and let us prove that there exists a final function $F$ such that $X'$ is represented by $\automaton^F$. As previously, it is sufficient to prove that $\gamma_{r,m,\sigma_1}^{-1}(X')=\gamma_{r,m,\sigma_2}^{-1}(X')$ for any pair $(\sigma_1,\sigma_2)$ of words in $\Sigma_r^*$ such that there exists $s=(k,Z)\in S$ satisfying $(\delta(q_0,\sigma_1),|\sigma_1|+m.\Z)=s=(\delta(q_0,\sigma_2),|\sigma_2|+m.\Z)$. Let us remark that it is sufficient to prove that $\gamma_{r,m,\sigma}^{-1}(X')=\gamma_{r,m,\sigma_s}^{-1}(X')$ for any $\sigma\in\Sigma_r^*$ where $s=(\delta(q_0,\sigma),|\sigma|+m.\Z)$. Let us consider a sequence $b_1$, ..., $b_i$ of $r$-digits $b_j\in \Sigma_r$ such that $\sigma=b_1\ldots b_i$ and let $s_j=(\delta(q_0,b_1\ldots b_j),j+m.\Z)\in S$ for any $j\in\finiteset{0}{i}$. By hypothesis, we have $\gamma_{r,m,\sigma_{s_{j-1}}.b_j}^{-1}(X')=\gamma_{r,m,\sigma_{s_j}}^{-1}(X')$. In particular $\gamma_{r,m,\sigma_{s_{j-1}}.b_j\ldots b_i}^{-1}(X')=\gamma_{r,m,\sigma_{s_j}.b_{j+1}\ldots b_i}^{-1}(X')$ for any $j\in\finiteset{1}{i}$. We deduce that $\gamma_{r,m,\sigma_{s_0}.b_1\ldots b_i}^{-1}(X')=\gamma_{r,m,\sigma_{s_i}}^{-1}(X')$. Since $\sigma_{s_0}=\epsilon$, $\sigma=b_1\ldots b_j$ and $s_i=s$, we have proved that $\gamma_{r,m,\sigma}^{-1}(X')=\gamma_{r,m,\sigma_s}^{-1}(X')$.
  \qed
\end{proof}


Let $Z_{r,m,s}$\index{Not}{$Z_{r,m,s}$} be the set of vectors $x\in\Z^m$ having a $(r,m)$-decomposition of the form $(\sigma,s)$ where $\sigma\in\Sigma_{r,m}^*$. This set is defined by the follwoing Presburger-formula:
$$(\bigwedge_{i;\;s[i]=0}x[i]\geq 0)\wedge(\bigwedge_{i;\;s[i]=r-1}x[i]<0)$$
The sets $Z_{r,m,s}$ naturally appear as $(r,m)$-detectable sets as shown by the following proposition \ref{prop:detectableinall} that characterize these sets.
\begin{proposition}\label{prop:detectableinall}
  A set is $(r,m)$-detectable in any set $X\subseteq\Z^m$ if and only if it is equal to a union of $Z_{r,m,s}$.
\end{proposition}
\begin{proof}
  Let us consider a finite set $\Lan\subseteq S_{r,m}$ and a DVA $\automaton$ that represents a set $X$ and just remark that $\bigcup_{s\in \Lan}Z_{r,m,s}$ is represented by the DVA $\automaton^F$ where $F$ is a final function such that $[F](q)=\Lan$ for any $q\in Q$. Therefore $\bigcup_{s\in \Lan}Z_{r,m,s}$ is $(r,m)$-detectable in any set $X\subseteq \Z^m$. Conversely, let us consider a set $X'$ that is $(r,m)$-detectable in any set $X$. As $\emptyset$ is represented by a DVA with one unique principal state $q_0$, and $X'$ is $(r,m)$-detectable in $\emptyset$, we deduce that there exists a final function $F$ such that $X'$ is represented by $\automaton^F$. Therefore $X'=\bigcup_{s\in [F](q_0)}Z_{r,m,s}$.
  \qed
\end{proof}

\begin{example}\label{rem:detectable}
  The set $X_1\# X_2$ is $(r,m)$-detectable in $X$ for any $(r,m)$-detectable sets $X_1$, $X_2$ in $X$, and for any  $\#\in\{\cup,\cap,\moins,\Delta\}$. Thus, any boolean combination of sets $(r,m)$-detectable in $X$ is $(r,m)$-detectable in $X$.
\end{example}

\subsection{Eyes and kernel}
Consider a FDVG $G=(Q,m,K,\Sigma_r,\delta)$. Given a $(r,m)$-sign vector $s\in S_{r,m}$, let us consider the equivalence relation $\sim_s$\index{Not}{$\sim_s$} over the principal states $Q$ defined by $q_1\sim_s q_2$ if and only if $\delta(q_1,s^*)\cap\delta(q_2,s^*)\not=\emptyset$. An equivalence class $Y\subseteq Q$ for $\sim_s$ is called an \emph{$s$-eye}\index{Gen}{eye} (or just an \emph{eye}). Given an $s$-eye $Y$, we denote $F_{s,Y}:Q\rightarrow\partie(S_{r,m})$\index{Not}{$F_{s,Y}$} a final function defined by $[F_{s,Y}](q)=\{s\}$ if $q\in Y$ and defined by $[F_{s,Y}](q)=\emptyset$ otherwise. Notice that a final function $F:Q\rightarrow\partie(S_{r,m})$ is saturated for $G$ if and only if $[F]$ is a finite union of final functions $[F_{s,Y}]$.

\begin{figure}[htbp]
  \begin{center}
    \setlength{\unitlength}{8pt}
    \pssetlength{\psunit}{8pt}
    \pssetlength{\psxunit}{8pt}
    \pssetlength{\psyunit}{8pt}
    \begin{picture}(20,20)(-3,-3)%
      \put(-3,-3){\framebox(20,20){}}
      \node[Nadjust=wh](n0)(0,10){}\node[Nadjust=wh](n1)(2,10){}\node[Nadjust=wh](n2)(2,12){}\node[Nadjust=wh](n3)(4,8){}\node[Nadjust=wh](nz)(6,10){}\node[Nadjust=wh](n4)(2,4){}\node[Nadjust=wh](n5)(4,6){}\node[Nadjust=wh](n6)(6,2){}\node[Nadjust=wh](n7)(6,4){}\node[Nadjust=wh](n8)(10,2){}\node[Nadjust=wh](n9)(8,4){}\node[Nadjust=wh](na)(12,-1){}\node[Nadjust=wh](nb)(12.5,2){}\node[Nadjust=wh](nc)(12,4){}\node[Nadjust=wh](ne)(14,6){}\node[Nadjust=wh](nf)(12,10){}\node[Nadjust=wh](ng)(10,8){}\node[Nadjust=wh](nh)(14,10){}\node[Nadjust=wh](ni)(12,12){}\node[Nadjust=wh](nj)(9,14){}\node[Nadjust=wh](nk)(13,14){}\node[Nadjust=wh](nl)(8,12){}\node[Nadjust=wh](nm)(8,10){}\node[Nadjust=wh](no)(6.2,14){}\node[Nadjust=wh](np)(5,12){}\node[Nadjust=wh](nq)(10,6){}%
      \drawedge(n0,n1){$s$}%
      \drawedge(n2,n1){$s$}%
      \drawedge(n1,n3){$s$}%
      \drawedge(n4,n5){$s$}%
      \drawedge(n6,n7){$s$}%
      \drawedge[ELside=r](n8,n9){$s$}%
      \drawedge[ELside=r](na,n8){$s$}%
      \drawedge[ELside=r](nb,n8){$s$}%
      \drawedge[ELside=r](nc,n8){$s$}%
      \drawedge[ELside=r](ne,nc){$s$}%
      \drawedge(nf,ng){$s$}%
      \drawedge(nh,nf){$s$}%
      \drawedge(ni,nf){$s$}%
      \drawedge(nj,nf){$s$}%
      \drawedge[ELside=r](nk,ni){$s$}%
      \drawedge(nl,nm){$s$}%
      \drawedge(no,nl){$s$}%
      \drawedge(np,nl){$s$}%
      \drawedge[ELside=r](n3,n5){$s$}%
      \drawedge[ELside=r](n5,n7){$s$}%
      \drawedge[ELside=r](n7,n9){$s$}%
      \drawedge[ELside=r](n9,nq){$s$}%
      \drawedge[ELside=r](nq,ng){$s$}%
      \drawedge[ELside=r](ng,nm){$s$}%
      \drawedge[ELside=r](nm,nz){$s$}%
      \drawedge[ELside=r](nz,n3){$s$}%
    \end{picture}
    \begin{picture}(20,20)(-3,-3)%
      \put(-3,-3){\framebox(20,20){}}
      \node[Nadjust=wh](n3)(4,8){}%
      \node[Nadjust=wh](nz)(6,10){}%
      \node[Nadjust=wh](n5)(4,6){}%
      \node[Nadjust=wh](n7)(6,4){}%
      \node[Nadjust=wh](n9)(8,4){}%
      \node[Nadjust=wh](nq)(10,6){}%
      \node[Nadjust=wh](ng)(10,8){}%
      \node[Nadjust=wh](nm)(8,10){}%
      \drawedge[ELside=r](n3,n5){$s$}%
      \drawedge[ELside=r](n5,n7){$s$}%
      \drawedge[ELside=r](n7,n9){$s$}%
      \drawedge[ELside=r](n9,nq){$s$}%
      \drawedge[ELside=r](nq,ng){$s$}%
      \drawedge[ELside=r](ng,nm){$s$}%
      \drawedge[ELside=r](nm,nz){$s$}%
      \drawedge[ELside=r](nz,n3){$s$}%
    \end{picture}
  \end{center}
\caption{On the left an $s$-eye. On the right its $s$-kernel.\label{fig:eye}}
\end{figure}
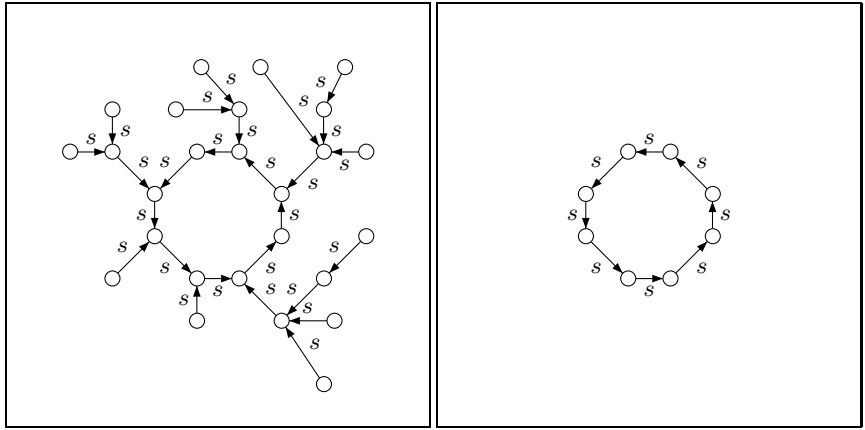
The \emph{$s$-kernel}\index{Gen}{kernel} $\ker_s(Y)$\index{Not}{$\ker_s(Y)$} of an $s$-eye $Y\subseteq Q$ is defined by $\ker_s(Y)=\bigcap_{i\in\Nat}\delta(Y,s^i)$. Remark that the $s$-kernel of an $s$-eye $Y$ is a non empty set of the form $\ker_s(Y)=\{q_1,\ldots, q_{k}\}$ such that $q_1\xrightarrow{s}q_2\ldots q_{k}\xrightarrow{s}q_1$ (see figure \ref{fig:eye}).

%% file: chapter.expressiveness.tex
\chapter{Expressiveness}\label{sub:expressiveness}
The expressiveness of the FDVA representation is studied in this section. We first prove in section \ref{subsub:rdef} that a subset of $\Z^m$ can be represented by a FDVA if and only if it is $r$-definable \cite{BHMV-BMS94}. Next in section  section \ref{subsub:ndd}, we show that the \emph{Number Decision Diagram (NDD)} \cite{WB-ICALP00} representation, an other state-based symbolic representation for subsets of $\Z^m$ is slightly equivalent (up to polynomial time translation) to the FDVA.

\section{Sets $r$-definable}\label{subsub:rdef}
Recall \cite{BHMV-BMS94} that a set $X\subseteq\Z^m$ is said \emph{$r$-definable}\index{Gen}{definable} if it can be defined in the first order theory $\fo{\Z,\Nat,+,V_r}$ where $V_r:\Z\rightarrow\Z$\index{Not}{$V_r$}\index{Gen}{valuation} is the $r$-valuation function defined by $V_r(0)=0$ and $V_r(x)$ is the greatest power of $r$ that divides $x\in\Z\moins\{0\}$. Note \cite{BHMV-BMS94} that a subset $X\subseteq \Nat^m$ is definable in $\fo{\Nat,+,V_r}$ if and only if the language $\{\sigma\in\Sigma_{r,m}^*;\;\rho_{r,m}(\sigma,\unit_{0,m})\}$ is regular. We are going to prove that a set $X\subseteq\Z^m$ can be represented by a FDVA in basis $r$ if and only if it is $r$-definable by decomposing such a set into sets of the form $f_{r,m,s}(X_s)$ where $X_s\subseteq\Nat^m$, $s\in S_{r,m}$ is a $(r,m)$-sign vector, and $f_{r,m,s}$\index{Not}{$f_{r,m,s}$} is the function given in the following definition.
\begin{definition}
  Given a $(r,m)$-sign vector $s\in S_{r,m}$, we denote by $f_{r,m,s}:\Z^m\rightarrow\Z^m$ the function defined for any $x\in\Z^m$ and for any $i\in\finiteset{1}{m}$ by:
  $$f_{r,m,s}(x)[i]=
  \begin{cases}
    x[i] & \text{if }s[i]=0\\
    -1-x[i] & \text{otherwise}
  \end{cases}$$
\end{definition}
Remark that $X=\bigcup_{s\in S_{r,m}}f_{r,m,s}(X_s)$ where $X_s=\Nat^m\cap f_{r,m,s}(X)$\index{Not}{$X_s$}. The following two propositions \ref{prop:ZtoN} and \ref{prop:ZtoN2} shows that a FDVA that represents $X_s$ is computable in linear time from a FDVA that represents $X$.
\begin{proposition}\label{prop:ZtoN}
  For any $(r,m)$-sign vectors $s\in S_{r,m}$, a FDVA that represents $f_{r,m,s}(X)$ in basis $r$ is computable in time $O(m.\textrm{size}(\automaton))$ from a FDVA $\automaton$ that represents a set $X\subseteq \Z^m$ in basis $r$. 
\end{proposition}
\begin{proof}
  Let us consider a FDVA $\automaton=(q_0,G,F_0)$ that represents $X$ in basis $r$. Without loss of generality, we can assume that $G$ and $F_0$ share the same set of states $K$ and the same transition function $\delta$. That means $G=(Q,m,K,S_r,\delta)$ and $F=(Q,f,m,K,S_r,\delta,K_F)$. 

  Let us first assume that there exists a function $l:K\rightarrow \Z/m.\Z$ such that $l(k')=l(k)+1$ for any transition $k\xrightarrow{b}k'$ where $(k,b,k')\in K\times \Sigma_r\times K$, such that $l(q_0)=1$ and $l(f(q))=1$ for any $q\in Q$. Let us consider the two bijective functions $t_{r,0},t_{r,r-1}:\Sigma_r\rightarrow\Sigma_r$ where $t_{r,0}$ is the identity function and $t_{r,r-1}(b)=r-1-b$ for any $b\in\Sigma_r$. By replacing the function $\delta$ in $G$ and $F_0$ by the function $\delta'$ given by $\delta'(k,b)=\delta(k,t_{r,s[l(k)]}(b))$ we deduce a DVG $G'$ and a final function $F'$ such that the DVA $\automaton'=(q_0,G',F')$ represents $f_{r,m,s}(X)$ in basis $r$. This result is well know and the proof is left to the reader.
  
  In the general case, if the labeling function $l$ does not exist, by multiplying the size of $\automaton$ by $m$, a DVA $\automaton''$ that represents $X$ in basis $r$ and owns a labelling function $l$ can be easily obtained. Hence, we are done.
\qed
\end{proof}

\begin{proposition}\label{prop:ZtoN2}
  A FDVA that represents $\Nat^m\cap X$ in basis $r$ is computable in linear time from a FDVA that represents a set $X\subseteq \Z^m$ in basis $r$. 
\end{proposition}
\begin{proof}
  Let us consider a FDVA $\automaton=(q_0,G,F_0)$ that represents $X$. Remark that in linear time we can compute a final function $F$ with the set $Q$ of principal states such that $[F](q)=\{\unit_{0,m}\}$ if $\unit_{0,m}\in [F_0](q)$ and $[F](q)=\emptyset$ otherwise. Now, just remark that $\Nat^m\cap X$ is represented by the FDVA $(q_0,G,F)$.
  \qed
\end{proof}

We can easily deduce the following theorem \ref{thm:hdef}.
\begin{theorem}\label{thm:hdef}
  A set $X\subseteq\Z^m$ can be represented by a FDVA in basis $r$ if and only if it is $r$-definable.
\end{theorem}
\begin{proof}
  Assume first that $X$ is $r$-definable and let us prove that $X$ can be represented by a FDVA in basis $r$. As $X$ is $r$-definable, the set $X_s=\Nat^m\cap f_{r,m,s}(X)$ is $r$-definable for any $s\in S_{r,m}$. As $X_s\subseteq \Nat^m$, from \cite{BHMV-BMS94} we deduce that $\{\sigma\in \Sigma_{r,m}^*;\;\rho_{r,m}(\sigma,\unit_{0,m})\in X_s\}$ is regular. Therefore $X_s$ can be represented by a FDVA in basis $r$. From proposition \ref{prop:ZtoN} we deduce that $f_{r,m,s}(X_s)$ can be represented by a FDVA in basis $r$. Therefore $X=\bigcup_{s\in S_{r,m}}f_{r,m,s}(X_s)$ can be represented by a FDVA in basis $r$. For the converse, assume that $X$ is represented by a FDVA in basis $r$ and let us prove that $X$ is $r$-definable. From propositions \ref{prop:ZtoN} and \ref{prop:ZtoN2} we deduce that $X_s=\Nat^m\cap f_{r,m,s}(X)$ can be represented by a FDVA in basis $r$. As $X_s\subseteq \Nat^m$, from \cite{BHMV-BMS94} we deduce that $X_s$ is $r$-definable. As $X=\bigcup_{s\in S_{r,m}}f_{r,m,s}(X_s)$, we deduce that $X$ is $r$-definable.
  \qed
\end{proof}

\begin{remark}
  We can easily prove that for any set $X\subseteq \Z^m$, the set $X$ is $r$-definable if and only if the DVA $\automaton_{r,m}(X)$ is finite and moreover in this case it is the unique (up to isomorphism) minimal (for the total number of states) FDVA that represents $X$ in basis $r$. 
\end{remark}

\section{Number Decision Diagrams (NDD)}\label{subsub:ndd}
Recall \cite{WB-ICALP00} that a \emph{Number Decision Diagram (NDD)}\index{Gen}{NDD}\index{Gen}{number decision diagram} $\automaton$ in basis $r$ and in dimension $m$ that represents a $r$-definable set $X\subseteq \Z^m$ is a finite automaton over the alphabet $\Sigma_r$ that recognizes the regular language $\{\sigma.s;\;(\sigma,s)\in\rho_{r,m}^{-1}(X)\}$. We do not consider NDD in this paper because (1) the class of regular languages included in $\Sigma_{r,m}^*.S_{r,m}$ is not stable by residue which means the automaton obtained by moving the initial state of a NDD is not a NDD anymore, and (2) rather than replacing the final function $F_0$ of a FDVA $\automaton$ by another final function $F$ is structurally obvious, the corresponding operation over NDD is not immediate since the FDVG $G$ and the finite final function $F_0$ are encoded into a single automaton. Nevertheless, polynomial time algorithms provided in this paper can be applied to NDD thanks to the following translation proposition \ref{prop:nddfdva}.
\begin{proposition}\label{prop:nddfdva}
  A NDD that represents $X$ in a basis $r$ is computable in quadratic time from a FDVA that represents a set $X$ in basis $r$. Conversely, a FDVA that represents $X$ in basis $r$ is computable in linear time from a NDD that represents a set $X$ in basis $r$.
\end{proposition}
\begin{proof}
  Let us consider a letter $\Diamond$ not in $\Sigma_r$ and let us consider the one-to-one function $f:\Sigma_{r,m}^*.\Diamond.S_{r,m}\rightarrow \Sigma_{r,m}^*.S_{r,m}$. It is sufficient to show that (1) a finite automaton that recognizes $\Lan'=f(\Lan)$ is computable in quadratic time from a finite automaton that recognizes a language $\Lan\subseteq \Sigma_{r,m}^*.\Diamond.S_{r,m}$, and (2) a finite automaton that recognizes $\Lan=f^{-1}(\Lan')$ is computable in linear time from a finite automaton that recognizes a language $\Lan'\subseteq \Sigma_{r,m}^*.S_{r,m}$. These two computations are immediate.
  \qed
\end{proof}


%% file: chapter.examplesFDVA.tex
\chapter{Some Examples of FDVA}\label{sub:examples}
The FDVA $\automaton_{r,1}(\Z)$, $\automaton_{r,1}(\Nat)$, $\automaton_{r,3}(+)$ and $\automaton_{r,2}(V_r)$, are given in figures \ref{fig:DVAobvious}, \ref{fig:dvaplus} and \ref{fig:DVAval}. Remark that a principal state $q\in Q$ is labelled by the set $X_q$ (in fact a formula in $\fo{\Z,\Nat,+,V_r}$ defining $X_q$), and a dot-edge from $q$ to $[F_0](q)$ is drawn for each state $q\in Q$ such that $[F_0](q)\not=\emptyset$.

\begin{figure}[htbp]
  \begin{center}
    \begin{picture}(20,40)(-2,-10)
      \put(-2,-10){\framebox(20,40){}}
      \node[Nmarks=i,iangle=180](A)(8,10){$\Z$}
      \node[Nadjust=wh,Nframe=n](B)(8,-5){$\scriptstyle S_r$}
      \drawloop(A){$\scriptstyle\Sigma_r$}
      \drawedge[dash={1.0 1.0 1.0 1.0}{0.0}](A,B){}
    \end{picture}
    \begin{picture}(20,40)(-2,-10)
      \put(-2,-10){\framebox(20,40){}}
      \node[Nmarks=i,iangle=180](A)(8,10){$\Nat$}
      \node[Nadjust=wh,Nframe=n](B)(8,-5){$\scriptstyle \{0\}$}
      \drawloop(A){$\scriptstyle \Sigma_r$}
      \drawedge[dash={1.0 1.0 1.0 1.0}{0.0}](A,B){}
    \end{picture}
  \end{center}
  \caption{On the left, FDVA $\automaton_{r,1}(\Z)$. On the right, FDVA $\automaton_{r,1}(\Nat)$\label{fig:DVAobvious}}
\end{figure}

\begin{figure}[htbp]
  \begin{center}
    \begin{picture}(120,60)(-15,-105)
      \put(-15,-105){\framebox(120,60){}}
      \node[NLangle=0.0,Nadjust=w,Nmarks=i,iangle=180](n0)(8.0,-74.74){$x[1]+x[2]=x[3]$}
      \node[NLangle=0.0,Nadjust=w](n1)(82.0,-74.74){$x[1]+x[2]+1=x[3]$}
      \node[NLangle=0.0](n2)(45.0,-87.74){$\emptyset$}
      \node[Nframe=n,Nadjust=wh](n28)(8.0,-96.0){$\begin{array}{c}\scriptstyle s\in S_{r,3};\\[-0pt]\scriptstyle s[1]+s[2]=s[3]\end{array}$}
      \node[Nframe=n,Nadjust=wh](n29)(82.0,-96.0){$\begin{array}{c}\scriptstyle  s\in S_{r,3};\\[-0pt]\scriptstyle s[1]+s[2]+1=s[3]+r\end{array}$}
      \drawedge[curvedepth=12.0](n0,n1){$\begin{array}{c}\scriptstyle b\in\Sigma_{r,3};\\[-0pt]\scriptstyle b[1]+b[2]=b[3]+r\end{array}$}
      \drawedge[ELside=r](n1,n0){$\begin{array}{c}\scriptstyle b\in\Sigma_{r,3};\\[-0pt]\scriptstyle b[1]+b[2]+1=b[3]\end{array}$}
      \drawloop(n1){$\begin{array}{c}\scriptstyle b\in\Sigma_{r,3};\\[-0pt]\scriptstyle b[1]+b[2]+1=b[3]+r\end{array}$}
      \drawloop(n0){$\begin{array}{c}\scriptstyle b\in\Sigma_{r,3};\\[-0pt]\scriptstyle b[1]+b[2]=b[3]\end{array}$}
      \drawedge[ELside=r,ELpos=60,curvedepth=5.0,ELdist=-1](n0,n2){$\begin{array}{c}\scriptstyle b\in\Sigma_{r,3};\\[-0pt]\scriptstyle b[1]+b[2]\not\in b[3]+r.\Z\end{array}$}
      \drawedge[ELpos=60,curvedepth=-5.0,ELdist=-1](n1,n2){$\begin{array}{c}\scriptstyle b\in\Sigma_{r,3};\\[-0pt]\scriptstyle b[1]+b[2]+1\not\in b[3]+r.\Z\end{array}$}
      
      \drawedge[dash={1.0 1.0 1.0 1.0}{0.0},curvedepth=0](n0,n28){}
      \drawedge[dash={1.0 1.0 1.0 1.0}{0.0},curvedepth=0](n1,n29){}
      \put(-5,-68){$q_0$}
      \put(95,-68){$q_1$}
      \put(44,-81){$q_\perp$}
      \drawloop[loopangle=-90](n2){$ \scriptstyle  \Sigma_{r,3}$}
    \end{picture}
  \end{center}
  \caption{The FDVA $\automaton_{r,3}(\{x\in\Z^3;\;x[1]+x[2]=x[3]\})$\label{fig:dvaplus}}
\end{figure}

\begin{figure}[htbp]
  \begin{center}
    \begin{picture}(95,68)(25,-12)
      \put(25,-12){\framebox(95,68){}}
      \node[Nadjust=w,Nmarks=i,iangle=180](A)(46,38){$V_r(x[1])=x[2]$}
      \node[Nadjust=wh,Nframe=n](Ap)(46,18){$\scriptstyle \{(0,0)\}$}
      \drawedge[dash={1.0 1.0 1.0 1.0}{0.0}](A,Ap){}
      \node[Nadjust=w](B)(108,38){$x[2]=0$}
      \node[Nadjust=wh,Nframe=n](Bp)(108,18){$\scriptstyle S_r\times\{0\}$}
      \drawedge[dash={1.0 1.0 1.0 1.0}{0.0}](B,Bp){}
      \node(C)(77,6){$\emptyset$}
      \drawloop[loopangle=-90](C){$\scriptstyle\Sigma_{r,2}$}
      \drawedge(A,B){$\scriptstyle(\Sigma_{r}\moins\{0\})\times\{1\}$}
      \drawloop(A){$\scriptstyle \{(0,0)\}$}
      \drawloop(B){$\scriptstyle \Sigma_{r}\times\{0\}$}
      \drawedge[ELpos=65](B,C){$\scriptstyle \Sigma_{r}\times (\Sigma_r\moins \{0\})$}
      \drawedge[ELpos=60,ELside=r](A,C){$\begin{array}{@{}c@{}}\scriptstyle b\in\Sigma_{r,2} \\[-0pt] \scriptstyle (\neg(b[1]=0\wedge b[2]=0))\\[-0pt]\scriptstyle \wedge(\neg(b[1]\not=0\wedge b[2]=1))\end{array}$}
      \put(33,44){$q_0$}
      \put(115,44){$q_1$}
      \put(75,12){$q_\perp$}
    \end{picture}
  \end{center}
  \caption{FDVA $\automaton_{r,2}(\{x\in\Z^2;\;V_r(x[1])=x[2]\})$\label{fig:DVAval}}
\end{figure}

%% file: chapter.reduction.tex
\chapter{Reductions}
In this section, we prove that the problem of deciding if the set $X$ represented by a FDVA $\automaton$ is Presburger-definable and in this case the problem of computing a Presburger formula that defines $X$ can be reduced in polynomial time to:
\begin{itemize}
\item the cyclic case\index{Gen}{reduction!cyclic}: there exists a loop on the initial state $q_0$. In particular the set $X$ represented by $\automaton$ is cyclic from proposition \ref{prop:moveinit}.
\item the positive case\index{Gen}{reduction!positive}: the final function $F_0$ is such that $[F_0](q)\in\{\emptyset,\{\unit_{0,m}\}\}$. In particular $X\subseteq\Nat^m$.
\end{itemize}

\section{Cyclic reduction}
Given a word $\sigma\in\Sigma_{r,m}^+$, a set $X\subseteq\Z^m$ is said \emph{$(r,m,\sigma)$-cyclic}\index{Gen}{cyclic} (or just \emph{cyclic}) if $\gamma_{r,m,\sigma}^{-1}(X)=X$ and a DVA $\automaton$ is said \emph{$(r,m,\sigma)$-cyclic} (or just \emph{cyclic}) if $\delta(q_0,\sigma)=q_0$. From proposition \ref{prop:moveinit}, we deduce that the set represented by a $(r,m,\sigma)$-cyclic DVA $\automaton$ is $(r,m,\sigma)$-cyclic. Conversely, remark that if a set $X$ is $(r,m,\sigma)$-cyclic then the DVA $\automaton_{r,m}(X)$ is $(r,m,\sigma)$-cyclic. The notion of cyclic sets is useful in the sequel for reducing some problems to the special cyclic case since a cyclic Presburger-definable set can be defined by a Presburger formula of a very particular form (see lemma \ref{lem:relprime}).

\begin{remark}
  The first application of the cyclic reduction is the positive reduction given in section \ref{sub:positivereduction}. 
\end{remark}

\begin{lemma}\label{lem:relprime}
  For any $(r,m,\sigma)$-cyclic Presburger-definable set $X$, there exists an integer $n\in\Nat\moins\{0\}$ relatively prime with $r$ such that $X$ can be defined by a formula equal to a boolean combination of formulas of the form $\scalar{\alpha}{x-\xi_{r,m}(\sigma)}<0$ where $\alpha\in\Z^m$ and formulas of the form $x\in b+n.\Z^m$ where $b\in\Z^m$.
\end{lemma}
\begin{proof}
  A quantification elimination shows that there exists an integer $n_0\in\Nat\moins\{0\}$ and a finite set $D_0\subseteq\Z^m\times \Z$ such that $X$ can be defined by a formula equal to a boolean combination of formulas of the from $\scalar{\alpha}{x}<c$ where $(\alpha,c)\in D_0$ and $x\in b+n_0.\Z^m$ where $b\in\Z^m$. Remark that there exists an integer $k\in\Nat$ enough larger such that $n=\frac{n_0}{\gcd{n_0}{r^{|\sigma^k|_m}}}$ is relatively prime with $r$ and such that the rational number $\beta_{\alpha,c}=r^{-|\sigma^k|_m}.((r-1).c-\scalar{\alpha}{\rho_{r,m}(\sigma,\unit_{0,m})})$ satisfies $|\beta_{\alpha,c}|<1$ for any $(\alpha,c)\in D_0$. As $\gamma_{r,m,\sigma^k}^{-1}(X)=X$, we deduce that $X$ can be defined by a formula equal to a boolean combination of formulas of the form $\scalar{\alpha}{\gamma_{r,m,\sigma^k}(x)}<c$ where $(\alpha,c)\in D_0$ and $\gamma_{r,m,\sigma^k}(x)\in b+n_0.\Z^m$ where $b\in\Z^m$. Now remark that $\scalar{\alpha}{\gamma_{r,m,\sigma^k}(x)}<c$ is equivalent to $\scalar{\alpha}{(r-1).x+\rho_{r,m}(\sigma,\unit_{0,m})}<\beta_{\alpha,c}$. Since $\scalar{\alpha}{(r-1).x+\rho_{r,m}(\sigma,\unit_{0,m})}\in\Z$ and $|\beta_{\alpha,c}|<1$, we have proved that $\scalar{\alpha}{\gamma_{r,m,\sigma^k}(x)}<c$ is equivalent to $\scalar{\alpha}{x-\xi_{r,m}(\sigma)}<0$ if $\beta_{\alpha,c}<0$ and it is equivalent to $\neg \scalar{-\alpha}{x-\xi_{r,m}(\sigma)}< 0$ if $\beta_{\alpha,c}\geq 0$. Finally, remark that $\gamma_{r,m,\sigma}(x)\in b+n_s.\Z^m$ is either false if $b\not\in n.\Z^m$ or equivalent to a formula of the form $x\in b'+n.\Z^m$ where $b'\in\Z^m$ otherwise.
  \qed
\end{proof}

\begin{lemma}\label{lem:fini}
  From an automaton $\automaton$ over $\Sigma_r$ that represents a finite language $\Lan\subseteq\Sigma_{r,m}^*$, we can compute in polynomial time a Presburger formula $\phi$ that defines $\rho_{r,m}(\Lan,\unit_{0,m})$.
\end{lemma}
\begin{proof}
  Let us consider a finite automaton $\automaton=(q_0,Q,\Sigma_r,\delta,Q_F)$ that recognizes $\Lan$. We denote by $\automaton_q$ the automaton obtained from $\automaton$ by replacing the initial state $q_0$ by an other state $q\in Q$. Let us remark that $\Lan\subseteq\Sigma_r^{\leq |Q|}$ since otherwise $\Lan$ is infinite thanks to the \emph{pumping lemma}. For any $k\in\finiteset{0}{|Q|}$, we can compute in polynomial time a finite automaton that recognizes $\Lan\cap \Sigma_r^k$. Hence, without loss of generality, we can assume that there exists $k\in\Nat$ such that $\Lan\subseteq\Sigma_r^k$. The cases $k=0$ or $\Lan=\emptyset$ are left to the reader. Since $\Lan\subseteq\Sigma_{r,m}^*$ and $\Lan$ is not empty and included in $\Sigma_{r,m}^k$, we deduce that $m$ divides $k$. Let $n=\frac{k}{m}$ and remark that $x\in\rho_{r,m}(\Lan,\unit_{0,m})$ if and only if there exists a sequence $b_1$, ..., $b_k$ of integers in $\Sigma_r$ such that $\bigwedge_{j=1}^m x[j]=\sum_{i=0}^{n-1} b_{j+m.i}.r^{i}$ and such that $\delta(q_0,b_1\ldots b_k)\in Q_F$. Now remark that this last property can be translated into a Presburger formula in polynomial time.
  \qed
\end{proof}

\begin{proposition}\label{prop:NDDtocyclic}
  Let $X\subseteq \Z^m$ be a set represented by a FDVA $\automaton$ in basis $r$ and let $Q_c$ be the set of principal states reachable for $[G]$ that have a loop. The set $X$ is Presburger-definable if and only if $X_{q_c}$ is Presburger-definable for any $q_c\in Q_c$. Moreover, from a sequence of Presburger formulas $(\phi_{q_c})_{q_c\in Q_c}$ such that $\phi_{q_c}$ defines $X_{q_c}$, we can compute in polynomial time a Presburger formula $\phi$ that defines $X$.
\end{proposition}
\begin{proof}
  Assume first that $X$ is Presburger-definable. Recall that we have proved that $X_q$ is Presburger-definable for any principal state $q$ reachable for $[G]$. In particular $X_{q_c}$ is Presburger-definable for any $q_c\in Q_c$. Next, assume that $X_{q_c}$ is defined by a Presburger formula $\phi_{q_c}$ for any $q_c\in Q_c$ and let us prove that we can compute in polynomial time a Presburger formula $\phi$ that defines $X$. For any $k\in\finiteset{0}{|Q|-1}$ and for any $q\in Q$, we can compute in polynomial time an automaton $\automaton_{k,q}$ over $\Sigma_r$ that recognizes $\Lan_{k,q}=\{\sigma\in\Sigma_{r,m}^k;\;\delta(q_0,\sigma)=q\}$. From lemma \ref{lem:fini} we can compute in polynomial time a Presburger formula $\phi_{k,q_c}$ that defines the set $X_{k,q_c}=\rho_{r,m}(\Lan_{k,q_c},\unit_{0,m})$. Let us prove that $X$ is defined by the Presburger formula $\phi(x):=\bigvee_{q_c\in Q_c}\bigvee_{k=0}^{|Q|-1}(\exists y\;\exists z\;((x=r^k.y+z)\wedge\phi_{q_c}(y)\wedge\phi_{k,q_c}(z)))$. Let $x\in X$. There exists a $(r,m)$-decomposition $(w,s)$ of $x$ such that $|w|_m\geq m.|Q|$. In this case, $w$ can be decomposed in $w=\sigma.w'$ where $\sigma\in\Sigma_{r,m}^{\leq |Q|}$ is such that there exists a loop on $q_c=\delta(q_0.\sigma)$ and $w'\in\Sigma_r^*$. From $x=\gamma_{r,m,\sigma}(x')$ where $x'=\rho_{r,m}(w',s)$ and $x\in X$, we deduce that $x'\in\gamma_{r,m,\sigma}^{-1}(X)=X_{q_c}$. Let $k=|\sigma|_m$. From $x=r^k.x'+\rho_{r,m}(\sigma,\unit_{0,m})\in r^k.X_{q_C}+X_{k,q_c}$ we deduce that $\phi(x)$ is true. For the converse, consider $x\in\Z^m$ such that $\phi(x)$ is true. There exists  $(q_c,k)\in Q_C\times \finiteset{0}{|Q|-1}$, $x'\in X_{q_c}$ and a word $\sigma\in\Lan(\automaton_{k,q_c})$ such that $x=r^{k}.x'+\rho_{r,m}(\sigma,\unit_{0,m})$. Let us consider a $(r,m)$-decomposition $(w',s)$ of $x'$. As $|\sigma|_m=k$, we deduce that $x'=\gamma_{r,m,\sigma}(x)$. As $q_0\xrightarrow{\sigma}q_c$, we have $X_{q_c}=\gamma_{r,m,\sigma}^{-1}(X)$. Hence $x\in\gamma_{r,m,\sigma}(\gamma_{r,m,\sigma}^{-1}(X))\subseteq X$. We have proved that $x\in X$. 
  \qed
\end{proof}

\section{Positive reduction}\label{sub:positivereduction}
The following proposition \ref{prop:Sdetection} and proposition \ref{prop:ZtoN} provide the positive reduction since a set $S$ satisfying the following proposition \ref{prop:Sdetection} is computable in quadratic time. 
\begin{proposition}\label{prop:Sdetection}
  Let $\automaton$ be a FDVA that represents a set $X\subseteq \Z^m$. Let us consider a set $S$ of $(r,m)$-sign vectors such that $S\cap(F_0(q)\Delta F_0(q'))\not=\emptyset$ for any state $q,q'\in Q$ such that $F_0(q)\Delta F_0(q')\not=\emptyset$. The set $X$ is Presburger-definable if and only if the set $\Nat^m\cap f_{r,m,s}(X)$ is Presburger-definable for any $s\in S$. Moreover from a sequence of Presburger formulas $(\phi_s)_{s\in S}$ such that $\phi_s$ defines $X_s=\Nat^m\cap f_{r,m,s}(X)$, we can compute in polynomial time a Presburger formula $\phi$ that defines $X$.
\end{proposition}
\begin{proof}
  Naturally, if $X$ is Presburger-definable, then $X_s=\Nat^m\cap f_{r,m,s}(X)$ is Presburger-definable for any $s\in S$. Let us prove the converse. Form proposition \ref{prop:NDDtocyclic}, we can assume that there exists a loop on the initial state. Consider a sequence $(\phi_s)_{s\in S}$ of Presburger formulas $\phi_s$ that defines $X_s$. Let us consider the function $\textrm{sign}:\Z^m\rightarrow S_{r,m}$ that associate to any vector $x\in\Z^m$ the unique $(r,m)$-sign vector $s\in S_{r,m}$ such that there exists $x\in Z_{r,m,s}$.

  Let us consider the following Presburger formula $\theta_s(x,k)$ and remark that $\theta_s(x,k)$ is true if and only if $x+k.\frac{s-\textrm{sign}(x)}{1-r}\in Z_{r,m,s}\cap X$. We denote by $K_{s,x}$ the Presburger-definable set $K_{s,x}=\{k\in\Z;\;\theta_s(x,k)\}$. Since $K_{s,x}$ is a Presburger definable set included in $\Z$, there exists a unique minimal integer $n_{s,x}\in\Nat\moins\{0\}$ such that there exists a finite set $B_{s,x}\subseteq \finiteset{0}{n_{s,x}-1}$ and an integer $k_{s,x}\in\Z$ such that $K_{s,x}\cap (k_{s,x}+\Nat)=k_{s,x}+B_{s,x}+n_{s,x}.\Nat$. Let us prove that $n_{s,x}$ is relatively prime with $r$. From lemma \ref{lem:relprime}, we deduce that there exists an integer $n_s$ relatively prime with $r$ such that $Z_{r,m,s}\cap X$ can be defined by a formula equal to a boolean combination of formulas of the form $\scalar{\alpha}{x}<c$ and $x\in b+n_s.\Z^m$. Now, just remark that $n_{s,x}$ divides $n_s$. We deduce that $n_{s,x}$ is relatively prime with $r$.
  $$\theta_s(x,k):=\exists y\;\phi_s\circ f_{r,m,s}(y)\wedge \bigwedge_{i=1}^m\left(\begin{array}{ll}&(x[i]\geq 0\Longrightarrow y[i]=x[i]+k.\frac{s[i]}{1-r})\\ \vee &(x[i]<0\Longrightarrow y[i]=x[i]+k.\frac{s[i]-(r-1)}{1-r})\end{array}\right)$$

  Let us consider the Presburger formula $W_s(x,n):=n\geq 1\wedge \exists k_0\;\forall k\geq k_0;\ \theta_s(x,k)\Longleftrightarrow\theta_s(x,k+n)$. Remark that $W_s(x,n)$ is true if and only if $n\in n_{s,x}.(\Nat\moins\{0\})$.
  
  Next, let us denote by $Q_s$ the set of principal states $q\in Q$ such that $s\in [F_0](q)$. Observe that we can compute in polynomial time the partition $\class$ of $Q$ corresponding to the equivalence relation $\sim$ defined by $q_1\sim q_2$ if and only if $F_0[q_1]=F_0[q_2]$. Given $C\in\class$, remark that $[F_0](q)$ does not depend on $q\in C$ and we can denote by $[F_0](C)$ the unique subset of $S_{r,m}$ such that $[F_0](C)=[F_0](q)$ for any $q\in C$. From lemma \ref{lem:bool}, we deduce that for any $C\in\class$ there exists a boolean formula $\relation_C$ computable in polynomial time such that $C$ is defined by $\relation_C( [F_0](q)_{s\in S})$. 
  
  We are going to prove that $X$ is defined by the following Presburger formula $\phi(x)$:
  $$\phi(x):=\bigvee_{C\in\class}(\textrm{sign}(x)\in [F_0](C)\wedge \forall N\;\exists n\geq N\;\relation_C(\theta_s(x,1+n)\wedge W_s(x,n))_{s\in S})$$

  Let us consider $x\in \Z^m$ such that $\phi(x)$ is satisfied and let us prove that $x\in X$. There exists $C\in\class$ such that $\textrm{sign}(x)\in [F_0](C)$ and for any $N$ there exists $n\geq N$ such that $\relation_C(\theta_s(x,1+n))_{s\in S}$ and $W_s(x,n)$ are true. Let us consider $N=k_{s,x}-1$ and let $n\geq N$ be such that $\relation_C(\theta_s(x,1+n))_{s\in S}$ and $W_s(x,n)$ are true. Since $W_s(x,n)$ is true, we deduce that $n\in n_{s,x}.(\Nat\moins\{0\})$. Let us consider a $(r,m)$-decomposition $(\sigma_0,s_0)$ of $x_0$ such that $r^{|\sigma_0|_m}\geq k_{s,x}$ for any $s\in S$. Since $n_{s,x}$ is relatively prime with $r$, by replacing $\sigma_0$ by a word in $\sigma_0.s_0^*$, we can assume that $r^{|\sigma_0|_m}\in 1+n_{s,x}.\Z$. Since $1+n$ and $r^{|\sigma_0|_m}$ are both greater than $k_{s,x}$ and the difference of these two integers $(1+n)-(r^{|\sigma_0|_m})$ is in $n_{s,x}.\Z$, we deduce that $\theta_{s}(x,1+n)$ is equivalent to $\theta_s(x,r^{|\sigma_0|_m})$. Therefore $\relation_C(\theta_s(x,1+n))_{s\in S}$ is true. Remark that $\theta_s(x,r^{|\sigma_0|_m})$ is true if and only if $x+r^{|\sigma_0|_m}.\frac{s-s_0}{1-r}\in Z_{r,m,s}\cap X$. Remark that $x+r^{|\sigma_0|_m}.\frac{s-s_0}{1-r}=\rho_{r,m}(\sigma_0,s)$. Therefore $\theta_s(x,r^{|\sigma_0|_m})$ is equivalent to $s\in [F_0](q)$ where $q=\delta(q_0,\sigma_0)$. We deduce that $\relation_C(s\in [F_0](q))_{s\in S}$ is true. Hence $q\in C$ and from $s_0\in [F_0](C)$ we get $s_0\in [F_0](q)$. We have proved that $x\in X$.
  
  Now, let us consider $x\in X$ and let us prove that $\phi(x)$ is true. Since $Q$ is finite and  $\prod_{s\in S}n_{s,x}$ is relatively prime with $r$, there exists a $(r,m)$-decomposition $(\sigma_0,s_0)$ of $x$ and an integer $d_0\in\Nat\moins\{0\}$ such that $q=\delta(q_0,\sigma_0)$ satisfies $\delta(q,s_0^{d_0})=q$ and such that $r^{|\sigma_0|_m}$ and $r^{d_0}$ are in $1+n_{s,x}.\Z$. Since $\class$ is a partition of $Q$, there exists $C\in\class$ such that $q\in\class$. Let us consider $N\in\Z$. There exists $k\in\Nat$ such that the integer $n=r^{|\sigma_0|_m+k.d_0}-1$ is greater than or equal to $N$ and $1$. Remark that $n\in n_{s,x}.(\Nat\moins\{0\})$. Therefore $W_s(x,n)$ is true. Moreover, as $x\in X$ we deduce that $s_0\in [F_0](q)$ and hence $\textrm{sign}(x)\in [F_0](C)$. Moreover, as $q\in C$ we get $\relation_C(s\in [F_0](q))_{s\in S}$ is true. Remark that $s\in [F_0](q)$ if and only if $\rho_{r,m}(\sigma_0.s_0^{k.d_0},s)\in Z_{r,m,s}\cap X$ if and only if $x+r^{|\sigma_0|_m+k.d_0}.\frac{s-s_0}{1-r}\in Z_{r,m,s}\cap X$ if and only if $\theta_s(x,1+n)$ is true. Therefore $\phi(x)$ is true.
  \qed
\end{proof}

%% file: chapter.linearsets.tex
\chapter{Linear Sets}

\section{Vector spaces}
\begin{figure}[htbp]
  \begin{center}
    \setlength{\unitlength}{16pt}
    \pssetlength{\psunit}{16pt}
    \pssetlength{\psxunit}{16pt}
    \pssetlength{\psyunit}{16pt}    
    \begin{picture}(10,10)(-5,-5)
      \put(-5,-5){\framebox(10,10){}}
      \multido{\iii=-4+1}{9}{\psline[linecolor=lightgray,linewidth=1pt](-4,\iii)(4,\iii)}\multido{\iii=-4+1}{9}{\psline[linecolor=lightgray,linewidth=1pt](\iii,-4)(\iii,4)}\psline[linecolor=gray,arrowscale=1,linewidth=1pt]{->}(0,-4.5)(0,4.5)\psline[linecolor=gray,arrowscale=1,linewidth=1pt]{->}(-4.5,0)(4.5,0)
      \put(-3,-0.8){$V$}
      
      \psline[linecolor=black,linewidth=1pt](-4,-2)(4,2)
      \psline[linecolor=black,linewidth=2pt,arrowscale=1]{|->}(0,0)(2,1)
    \end{picture}
  \end{center}
  \caption{The vector space $V=\Q.(2,1)$\label{fig:vectorspace}} 
\end{figure}

A \emph{vector space}\index{Gen}{vector space} $V$ of $\Q^m$ is a non empty subset of $\Q^m$ such that $\lambda.V\subseteq V$ for any $\lambda\in\Q$ and such that $V+V\subseteq V$. As any finite or infinite intersection of vector spaces of $\Q^m$ remains a vector space and we deduce that any set $X\subseteq\Q^m$ is included into a unique minimal (for $\subseteq$) vector space denoted by $\vecspace(X)$\index{Not}{$\vecspace(X)$} and called the \emph{vector hull of $X$} or the \emph{vector space generated} by $X$\index{Gen}{vector hull}\index{Gen}{vector space generated}. A \emph{basis}\index{Gen}{basis!of a vector space} of a vector space $V$ is a sequence $v_1$, .., $v_d$ of vectors in $V$ such that for any $x\in V$ there exists a unique sequence $\lambda_1$, ..., $\lambda_d$ of rational numbers such that $x=\sum_{i=1}^d \lambda_i.v_i$. Recall that any vector space has a basis and the number of elements of a basis only depends on $V$ and it is called the \emph{dimension}\index{Gen}{dimension} of $V$, and it is denoted by $\dim(V)\in\finiteset{0}{m}$\index{Not}{$\dim(V)$}. 

There exists unduly complicated basis of vector spaces. For instance consider the vector space $V=\Q^2$ and for each $n\in\Nat$ let $v^n_1$, $v^n_2$ be the basis of $V$ given by $v^n_1=(2.n+1,n)$ and $v^n_2=(2,1)$. That means complex basis of simple vector spaces (for instance $\Q^2$) can be computed if vector spaces are symbolically manipulated by basis. In order to overcome this problem, we are going to associate to any vector space a canonical basis.

A \emph{set of indices} $I\subseteq \finiteset{1}{m}$ is said \emph{full rank}\index{Gen}{full rank set of indices} for a vector space $V$ if for any $x\in \Q^I$ there exists a \emph{unique} $v\in V$ such that $v[i]=x[i]$ for any $i\in I$.
\begin{proposition}\label{prop:fullrow}
  Any vector space has a full rank set of indices.
\end{proposition}
\begin{proof}
  Let us consider subset $I\subseteq \finiteset{1}{m}$ maximal for the inclusion amongst the subset $J\subseteq \finiteset{1}{m}$ satisfying for any $x\in \Q^J$, there exists a unique $v\in V$ such that $v[j]=x[j]$ for any $j\in J$. Remark that such a set $I$ exists since $J=\emptyset$ satisfies the condition. Let us consider two vectors $v_1,v_2\in V$ such that $v_1[i]=v_2[i]$ for any $i\in I$ and let $w=v_1-v_2$. Assume by contradiction that $w\not=\unit_{0,m}$. There exists $j_0\in \finiteset{1}{m}\moins I$ such that $w[j_0]\not=0$. Let $J=I\cup\{j_0\}$ and let us prove that for any $x\in \Q^J$ there exists $v\in V$ such that $v[j]=x[j]$ for any $j\in J$. By definition of $I$, there exists $v_0\in V$ such that $v_0[i]=x[i]$ for any $i\in I$. Let $v=v_0+\frac{x[j_0]-v_0[j_0]}{w[j_0]}.w$ and remark that $v[i]=x[i]$ for any $i\in I$ since $w[i]=0$ and $v[j_0]=0$. Therefore $I$ is not maximal and we get a contradiction. Thus $w=\unit_{0,m}$ and we have proved that for any $x\in \Q^I$, there exists a unique $v\in V$ such that $v[i]=x[i]$ for any $i\in I$.
  \qed
\end{proof}


A \emph{vector $I$-representation}\index{Gen}{representation!of a vector space} of a vector space $V$ where $I$ is a full rank set of indices for $V$ is a sequence $(v_{i})_{i\in I}$ of vectors in $V$ satisfying $v_{i}[i]=1$ and $v_{i}[j]=0$ for any $j\in I\moins \{i\}$. Observe that such a sequence $(v_i)_{i\in I}$ is a basis of $V$ and given a full rank set $I$, there exists a unique vector $I$-representation of $V$. The integer $\textrm{size}(V)\in\Nat$\index{Gen}{sizes!of a vector space} of a vector space $V$ is defined by $\size(V)=\max_{I}(\sum_{i\in I}\size(v_i))$ where $(v_i)_{i\in I}$ is the unique vector $I$-representation of $V$. 

The following proposition provides a simple way for computing incrementally a vector $I$-representation of a vector space $V$.
\begin{proposition}\label{prop:manipulate}
  Let $I$ be a full rank set of indices for a vector space $V$, let $(v_i)_{i\in I}$ be the vector $I$-representation of $V$ and let $V'$ be the vector space $V'=V+\Q.x$ where $x$ is any vector in $\Q^m$. The vector spaces $V$ and $V'$ are equal if and only if the vectors $y=x-\sum_{i\in I} x[i].v_i$ and $\unit_{0,m}$ are equal. Moreover, if $V'$ is not equal to $V$ then given $j_0$ such that $y[j_0]\not=0$, the set of indices $J=I\cup\{j_0\}$ is full rank for $V'$ and the vector $J$-representation of $V'$ is the following sequence $(v'_j)_{j\in J}$:
  $$v_j'=
  \begin{cases}
    v_j-v_j[j_0].\frac{y}{y[j_0]} & \text{if } j\in I\\
    \frac{y}{y[j_0]} & \text{if } j=j_0\\
  \end{cases}$$
\end{proposition}
\begin{proof}
  Assume first that $y=\unit_{0,m}$ and let us prove that $V=V'$. Since $y=\unit_{0,m}$, we get $x=\sum_{i\in I} x[i].v_i\in V$ and we deduce $V=V'$. Otherwise, if $V=V'$ we deduce that $y\in V$. Since $(v_i)_{i\in I}$ is a basis of $V$, there exists a sequence $(\lambda_i)_{i\in I}$ of rational numbers such that $y=\sum_{i\in I}\lambda_i.v_i$. From this last equality, we get $y[i]=\lambda_i$ and from $y=x-\sum_{i\in I} x[i].v_i$, we get $y[i]=x[i]-x[i]=0$. Thus $\lambda_i$ for any $i$ and we have proved that $y=\unit_{0,m}$. We have proved that the vector spaces $V$ and $V'$ are equal if and only if the vectors $y=x-\sum_{i\in I} x[i].v_i$ and $\unit_{0,m}$ are equal.
  
  Now, assume that $V'$ is not equal to $V$ and observe that $J$ is a set of indices full rank for $V'$ and the sequence $(v'_j)_{j\in J}$ is a vector $I$-representation of $V'$.
  \qed
\end{proof}

Our representation is motivated by the following corollary.
\begin{corollary}\label{cor:canonicalaffine}
  The size of a vector space $V$ is at most polynomially larger than the size of any finite subset $V_0\subseteq \Q^m$ that generates $V$.
\end{corollary}
\begin{proof}
  Assume fixed a full row set of indices $I$ of $V$. Let us consider a finite set $V_0$ of vectors that generates $V$. It is sufficient to show that we can compute in polynomial time a sequence $(v_i)_{i\in I}$ from $V_0$. By applying the polynomial time algorithm given in proposition \ref{prop:manipulate} and adding one by one the vector $v_0$ in $V$ and by selecting $j_0$ in $I$, we deduce that the sequence $(v_i)_{i\in I}$ is computable in polynomial time.
  \qed
\end{proof}

\section{Affine spaces}\label{sub:aff}
\begin{figure}[htbp]
  \begin{center}
    \setlength{\unitlength}{16pt}
    \pssetlength{\psunit}{16pt}
    \pssetlength{\psxunit}{16pt}
    \pssetlength{\psyunit}{16pt}    
    \begin{picture}(10,10)(-5,-5)
      \put(-5,-5){\framebox(10,10){}}
        \multido{\iii=-4+1}{9}{\psline[linecolor=lightgray,linewidth=1pt](-4,\iii)(4,\iii)}\multido{\iii=-4+1}{9}{\psline[linecolor=lightgray,linewidth=1pt](\iii,-4)(\iii,4)}\psline[linecolor=gray,arrowscale=1,linewidth=1pt]{->}(0,-4.5)(0,4.5)\psline[linecolor=gray,arrowscale=1,linewidth=1pt]{->}(-4.5,0)(4.5,0)
      \psline[linecolor=black,linewidth=1pt](-4,-1)(4,3)
      \psline[linecolor=black,linewidth=2pt,arrowscale=1]{|->}(0,1)(2,2)
      \put(-3,0.2){$A$}
    \end{picture}
    \begin{picture}(10,10)(-5,-5)
      \put(-5,-5){\framebox(10,10){}}
        \multido{\iii=-4+1}{9}{\psline[linecolor=lightgray,linewidth=1pt](-4,\iii)(4,\iii)}\multido{\iii=-4+1}{9}{\psline[linecolor=lightgray,linewidth=1pt](\iii,-4)(\iii,4)}\psline[linecolor=gray,arrowscale=1,linewidth=1pt]{->}(0,-4.5)(0,4.5)\psline[linecolor=gray,arrowscale=1,linewidth=1pt]{->}(-4.5,0)(4.5,0)      
      \psline[linecolor=black,linewidth=1pt](-4,-2)(4,2)
      \psline[linecolor=black,linewidth=2pt,arrowscale=1]{|->}(0,0)(2,1)
      \put(-3,-0.8){$\vec{A}$}
    \end{picture}
  \end{center}
  \caption{On the left an affine space $A=(0,1)+\Q.(2,1)$. On the right its direction.\label{fig:affinespace}} 
\end{figure}
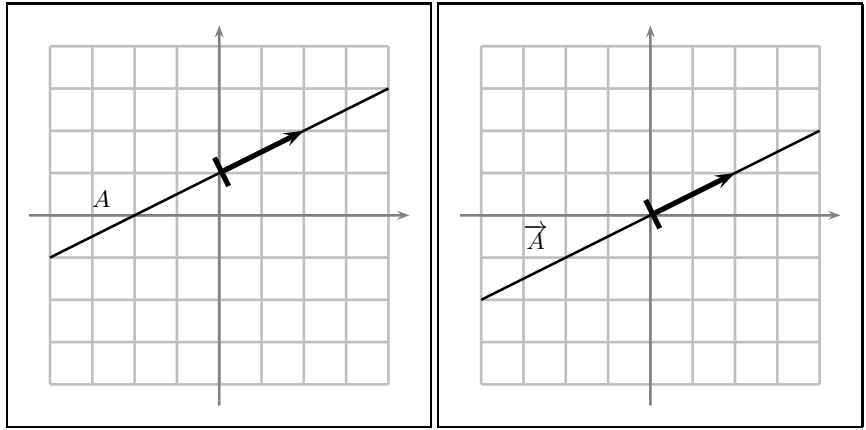

An \emph{affine space}\index{Gen}{affine space} $A$ of $\Q^m$ is either the empty-set, or a set of the form $A=a_0+V$ where $a_0\in\Q^m$ and $V$ is a vector space of $\Q^m$. This vector space $V$ is unique, denoted by $\vec{A}$ and called the \emph{direction}\index{Gen}{direction!of an affine space} of $A$ (see figure \ref{fig:affinespace}). If $A=\emptyset$, we denote by $\vec{A}=\emptyset$\index{Not}{$\vec{A}$} the \emph{direction} of $A$. A non-empty affine space $A$ is called a $V$-affine space if $\vec{A}$ is equal to a vector space $V$.

An \emph{affine $I$-representation}\index{Gen}{representation!of an affine space} of a $V$-affine space $A$ where $I$ is a full rank set of indices of $V$ is a couple $(a,(v_i)_{i\in I})$ where $a$ is a vector in $A$ such that $a[i]=0$ for any $i\in I$ and $(v_i)_{i\in I}$ is the $I$-vector representation of $V$. Observe that such a couple is unique. The integer $\textrm{size}(A)\in\Nat$\index{Gen}{sizes!of an affine space} of a non-empty affine space $A$ is defined by $\size(A)=\max_{I}(\size(a))+\size(V)$ where $(a,(v_i)_{i\in I})$ is the unique $I$-affine representation of $A$. The integer $\size(\emptyset)$ is defined by $\size(\emptyset)=0$. Notice that $\size(A)=\size(V)$ if the affine space $A$ is a vector space $A=V$ since in this case $a=\unit_{0,m}$.

The direction of affine spaces, has an interesting application intensively used in the sequel and given by the following lemma. 
\begin{lemma}[Comparable affine lemma]\label{lem:comparableaffine}
  Two comparable (for $\subseteq$) affine spaces that have the same direction are equal.
\end{lemma}
\begin{proof}
  Consider two affine spaces $A_1$ and $A_2$ such that $A_1\subseteq A_2$ and such that $\vec{A_1}=\vec{A_2}$. Naturally, if $A_1=\emptyset$, as $\vec{A_1}=\vec{A_2}$ we deduce that $A_2=\emptyset$ and we are done. Assume that $A_1\not=\emptyset$. Consider $a_1\in A_1$. As $a_1\in A_1\subseteq A_2$, we deduce that $A_2=a_1+\vec{A_2}$. From $\vec{A_1}=\vec{A_2}$, we get $A_2=a_1+\vec{A_1}=A_1$.
  \qed
\end{proof}

Recall that any finite or infinite intersection of affine spaces of $\Q^m$ remains an affine space, and we deduce that any set $X\subseteq\Q^m$ is included into a unique minimal (for $\subseteq$) affine space denoted by $\aff(X)$\index{Not}{$\aff(X)$} and called the \emph{affine hull of~$X$} or the affine space generated by $X$\index{Gen}{affine hull}\index{Gen}{affine space generated}. The direction of $\aff(X)$ is denoted by $\vec{\aff}(X)=\vec{\aff(X)}$\index{Not}{$\vec{\aff}(X)$}.

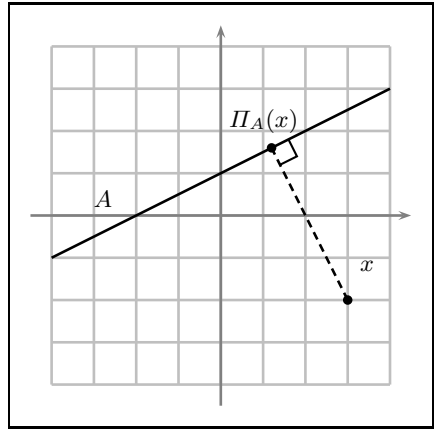
\begin{figure}[htbp]
  \begin{center}
   \setlength{\unitlength}{16pt}
    \pssetlength{\psunit}{16pt}
    \pssetlength{\psxunit}{16pt}
    \pssetlength{\psyunit}{16pt}    
    \begin{picture}(10,10)(-5,-5)
      \put(-5,-5){\framebox(10,10){}}
      \multido{\iii=-4+1}{9}{\psline[linecolor=lightgray,linewidth=1pt](-4,\iii)(4,\iii)}\multido{\iii=-4+1}{9}{\psline[linecolor=lightgray,linewidth=1pt](\iii,-4)(\iii,4)}\psline[linecolor=gray,arrowscale=1,linewidth=1pt]{->}(0,-4.5)(0,4.5)\psline[linecolor=gray,arrowscale=1,linewidth=1pt]{->}(-4.5,0)(4.5,0)
      \psline[linecolor=black,linewidth=1pt](-4,-1)(4,3)
      \psline[linewidth=1pt,linestyle=dashed,dash=3pt 2pt](3,-2)(1.2,1.6)
      \psdot*(3,-2)\psdot*(1.2,1.6)
      \put(3.3,-1.3){$x$}
      \put(0.2,2.1){$\Pi_A(x)$}
      \put(-3,0.2){$A$}
      \psline(1.6,1.8)(1.8,1.4)(1.4,1.2)
    \end{picture}
  \end{center}
  \caption{Orthogonal projection $\Pi_A(x)=\frac{(6,3)}{5}$ of $x=(3,-2)$ over $A=(0,1)+\Q.(2,1)$.\label{fig:projection}} 
\end{figure}


Finally, recall that the \emph{orthogonal}\index{Gen}{orthogonal} $X^\perp$\index{Not}{$X^\perp$} of a subset $X\subseteq\Q^m$ is the vector space $X^\perp=\{y\in\Q^m;\;\forall x\in X \scalar{y}{x}=0\}$. Recall that $(X^\perp)^\perp=\vecspace(X)$. In particular, $X=V$ is a vector space if and only if $(V^\perp)^\perp=V$. The \emph{orthogonal projection over a non-empty affine space $A$}\index{Gen}{orthogonal projection} is the unique function $\Pi_A:\Q^m\rightarrow A$\index{Not}{$\Pi_A$} such that $\Pi_A(x)-x\in (\vec{A})^\perp$ for any $x\in \Q^m$ (see figure \ref{fig:projection}). Recall that $\Pi_A$ is an affine function that satisfies $\Pi_A(x)=(1-\sum_{i=1}^mx[i]).\Pi_A(\unit_{0,m})+\sum_{i=1}^mx[i].\Pi_{A}(\unit_{i,m})$.


\section{Vector lattices}
An \emph{additive group}\index{Gen}{group} $M$ of $\Q^m$ is a non-empty finite subset of $\Q^m$ such that $-M\subseteq M$ and $M+M\subseteq M$. As any finite or infinite intersection of additive groups remains an additive group and $\Q^m$ is a group, any set $X\subseteq \Q^m$ is included into a minimal (for $\subseteq$) additive group, denoted by $\group(X)$\index{Not}{$\group(X)$} and called the \emph{group generated}\index{Gen}{group generated} by $X$. An additive group $M$ such that there exists a finite set $X$ satisfying $M=\group(X)$ is called a \emph{vector lattice}\index{Gen}{vector lattice}. Lattices are characterized by introducing \emph{discrete sets}\index{Gen}{discrete set}. A set $Z\subseteq \Q^m$ is said \emph{discrete} if for any $x\in M$, there exists a rational number $\epsilon>0$ such that $\norm{\infty}{x-y}\geq\epsilon$ for any $y\in M\moins\{x\}$.
\begin{proposition}[\cite{T-92-maths}]
  A group is discrete if and only if it is a vector lattice.
\end{proposition}
\begin{proof}
  Assume first that $M$ is a discrete group and let us prove that $M$ is a vector lattice. Since $\unit_{0,m}\in M$, there exists $\epsilon>0$ such that $\norm{\infty}{x}>\epsilon$ for any $x\in M$. Let $V$ be the vector space generated by $M$ and let $v_1$, ..., $v_d$ be a basis of $V$ formed by vectors in $M$. Let us denote by $B=\{\sum_{i=1}^d\lambda[i].v_i;\;0\leq \lambda[i]\leq 1\}$. The rational $k=\sum_{i=1}^d\norm{\infty}{v_i}$ satisfies $\norm{\infty}{b}\leq k$ for any $b\in B$. Assume by contradiction that $M\cap B$ contains more than $(\frac{2.k+1}{\epsilon})^n$ elements. Hence, there exists $x_1,x_2\in M\cap B$ such that $x_1\not=x_2$ and such that $\norm{\infty}{x_1-x_2}\leq \epsilon$. By definition of $\epsilon$ we deduce that $x_1-x_2=\unit_{0,m}$ and we get a contradiction. Thus $M\cap B$ is finite. For any $x\in M$, there exists $\lambda\in\Q^d$ such that $x=\sum_{i=1}^d\lambda[i].v_i$. Let us consider a vector $z\in \Z^d$ such that $0\leq \lambda[i]-z[i]\leq 1$ and remark that $x-\sum_{i=1}^dz[i].v_i\in M\cap B$. Thus $M=\group(\{v_1,\ldots,v_d\}\cup (M\cap B))$ and we have proved that there exists a finite set $X$ of vectors such that $M=\group(X)$. For the converse, assume that $M$ is a vector lattice and let us prove that $M$ is discrete. There exists a finite set $X$ of vectors such that $M=\group(X)$. Let us consider an integer $d\in\Nat\moins\{0\}$ such that $d.X\subseteq \Z^m$ and let us remark that for any $x,y\in M$ such that $x\not=y$, we have $\norm{\infty}{x-y}\geq \frac{1}{d}$. Thus $M$ is discrete.
  \qed
\end{proof}
Thanks to this characterization, we deduce that any group included in a vector lattice is a vector lattice since any set included in a discrete set remains discrete. Given a vector space $V$, a vector lattice $M$ such that $V=\vecspace(M)$ is called a $V$-vector lattice. The previous proposition also proves that $\Z^m\cap V$ is a $V$-vector lattice since it is a discrete group such that $\vecspace(\Z^m\cap V)=V$.

\subsection{Hermite representation}
We are going to provide a canonical (up to a full rank set of indices $I$ for $V$) representation of any $V$-vector lattice. 

An \emph{Hermite matrix}\index{Gen}{Hermite matrix} $B$ of order $d$ is a lower triangular (we have $B[i,j]=0$ for any $j>i$), non-negative square matrix $B\in\M_{d,d}(\Q_+)$, in which each row has a unique maximal entry which is located on the main diagonal of $B$. Given a full row set of indices $I=\{i_1<\cdots<i_d\}$ of a vector space $V$, an \emph{Hermite $I$-representation}\index{Gen}{representation!of a vector lattice} $B$ of a $V$-vector lattice $M$ is an Hermite matrix $B$ of order $d$ such that we have the following equality where $(v_i)_{i\in I}$ is the vector $I$-representation of $V$:
$$M=\group\{\sum_{k=1}^dB[k,j].v_{i_k};\;j\in\finiteset{1}{d}\}$$
The integer $\size(M)\in\Nat$ of a $V$-vector lattice $M$ is defined by $\size(M)=\max_I(\size(B))+\size(V)$.

The following theorem shows that the Hermite $I$-representation provides a canonical representation that is polynomially bounded by the size of any finite set $X$ such that $M=\group(X)$. 
\begin{theorem}[Theorem 4.1, 4.2 and 5.3 of \cite{S-87}]
  Given a full rank set of indices $I$ of a vector space $V$, any $V$-vector lattice $M$ owns a unique Hermite $I$-representation. Moreover, this representation is computable in polynomial time from any finite set of vectors that generates $M$.
\end{theorem}
This theorem also proves that for any $V$-vector lattice, there exists a basis $v_1$, ..., $v_d$ of $V$ such that $M=\sum_{j=1}^d\Z.v_j$ (for instance take $v_j=\sum_{k=1}^dB[k,j].v_{i_k}$). Such a sequence $v_1$, ..., $v_d$ is called a \emph{$\Z$-basis}\index{Gen}{basis!of a vector lattice} of $M$.

The following proposition will be useful in the sequel.
\begin{proposition}[Corollary 5.3b and 5.3c of \cite{S-87}]\label{prop:hermitevector}
  From an $I$-representation of a vector space $V$, we can compute in polynomial time the Hermite $I$-representation of the $V$-vector lattice $\Z^m\cap V$.
\end{proposition}

\subsection{Stability by intersection} 
Naturally, any intersection of vector lattices\index{Gen}{intersection!of two vector lattices} remains a vector lattice. The following lemma \ref{lem:Vgroupinter} shows that the class of $V$-vector lattice is stable by finite intersection (remark \ref{rem:Vgroupinter} shows that this class is not stable by infinite intersection).
\begin{lemma}\label{lem:Vgroupinter}
  The class of $V$-vector lattices is stable by finite intersection. Moreover, given a finite sequence $M_1$, ..., $M_n$ of $V$-vector lattices, we can compute in polynomial time the $V$-vector lattice $\bigcap_{j=1}^n M_j$.
\end{lemma}
\begin{proof}
  Let $I$ be a full rank set of indices. Recall that from an Hermite $I$-representation of $M_j$, we get a $\Z$-basis $v_{1,j}$, ..., $v_{d,j}$ of $M_j$. Now, remark that $x\in M$ where $M=\bigcap_{j=1}^nM_j$ if and only if there exists $z_1$, ..., $z_n$ in $\Z^{d}$ such that $x=\sum_{i=1}^dz_j.v_{i,j}$ for any $j\in\finiteset{1}{n}$. Let us consider the vector space $W=\{(x,z_1,\ldots,z_n)\in \Q^n\times\Q^d\times\cdots\Q^d;\;\bigcap_{j=1}^nx=\sum_{i=1}^dz_j.v_{i,j}\}$. From proposition \ref{prop:hermitevector} we deduce in polynomial time a $\Z$-basis of $\Z^m\cap W$ of the form $(x_1,z_{1,1},\ldots,z_{1,n})$, ..., $(x_d,z_{d,1},\ldots,z_{d,n})$. Let us remark that $\bigcap_{j=1}^dM_j$ is the $V$-vector lattice generated by $x_1$, ..., $x_d$. We deduce the $I$-representation of $M$ in polynomial time.
  \qed
\end{proof}

\begin{remark}\label{rem:Vgroupinter}
  The class of $V$-vector lattices is not stable by infinite intersection. In fact, let $M_n$ be the $V$-vector lattice $M_n=(n+1).\Z^m$ where $V=\Q^m$, and just remark that $\bigcap_{n\in\Nat}M_n=\{\unit_{0,m}\}$ is naturally a group as any intersection of groups, but it is not a $V$-vector lattice if $m\geq 1$.
\end{remark}

\subsection{Sub-lattice}
The \emph{quotient}\index{Gen}{quotient} $M'/M$\index{Not}{$M/M'$} of two $V$-vector lattices $M\subseteq M'$ is defined by $M/M'=\{m'+M;\;m'\in M\}$. The following theorem \ref{thm:inv1} proves that this set is finite.
\begin{theorem}[\cite{T-92-maths}]\label{thm:inv1}
  Given two vector lattices $M\subseteq M'$, there exists a unique sequence $n_1$, ..., $n_{d}$ of integers in $\Nat\moins\{0\}$ such that $n_{i}$ divides $n_{i+1}$ for any $i$ and such that there exists a $\Z$-basis $v_1$, ..., $v_{d}$ of $M'$ satisfying $n_1.v_1$, ..., $n_d.v_d$ is a $\Z$-basis of $M$. Moreover such a sequence $(n_1,v_1)$, ..., $(n_d,v_d)$ is computable in polynomial time. 
\end{theorem}
The unique sequence $n_1$, ..., $n_d$ is called the \emph{characteristic sequence}\index{Gen}{characteristic sequence} of $M$ in $M'$. 

The following lemma will be useful in the sequel.
\begin{lemma}[\cite{T-92-maths}]\label{lem:MMpMpp}
  Given three $V$-vector lattices $M\subseteq M'\subseteq M''$, we have the following equality:
  $$|M''/M'|.|M'/M|=|M''/M|$$
\end{lemma}

\subsection{Vector lattices included in $\Z^m$}
In the sequel we denote by $h_r:\Nat\moins\{0\}\rightarrow\Nat\moins\{0\}$\index{Not}{$h_r$} the function defined by $h_r(n)=\frac{n}{\gcd{n}{r}}$, and we denote by $\theta_m$\index{Not}{$\theta_m$} is the function $\theta_m\in\finiteset{1}{m}\rightarrow\finiteset{1}{m}$ defined by $\theta_m(i)\in (i-1+m.\Z)\cap\finiteset{1}{m}$.

\subsubsection{Inverse image by $\gamma_{r,m,0}$}
Theorem \ref{thm:inv1} proves that any $V$-vector lattice $M$ included in $\Z^m$ is a set of the form $M=\sum_{i=1}^dn_i.\Z.v_i$ where $v_1$, ..., $v_d$ is a $\Z$-basis of $\Z^m\cap V$ and $n_1$, ..., $n_d$ are integers in $\Nat\moins\{0\}$. Thus, the following lemma \ref{lem:invmask0} shows that the class of $V$-vector lattices included in $\Z^m$ is stable by inverse image by $\gamma_{r,m,\unit_{0,m}}$.
\begin{lemma}\label{lem:invmask0}
  Given a $\Z$-basis $v_1$, ..., $v_d$ of $\Z^m\cap V$ where $V$ is a vector space and a sequence $n_1$, ..., $n_d$ of integers in $\Nat\moins\{0\}$, we have:
  $$\gamma_{r,m,\unit_{0,m}}^{-1}(\sum_{i=1}^d n_i.\Z.v_i)=\sum_{i=1}^dh_r(n_i).\Z.v_i$$
\end{lemma}
\begin{proof}
  Let $x\in \gamma_{r,m,\unit_{0,m}}^{-1}(\sum_{i=1}^dn_i.\Z.v_i)$. There exists $z_1$, ..., $z_d$ in $\Z$ such that $r.x=\sum_{i=1}^dn_i.z_i.v_i$. In particular $x\in\Z^m\cap V$ and there exists $t_1$, ..., $t_d$ in $\Z$ such that $x=\sum_{i=1}^dt_i.v_i$. As $v_1$, ..., $v_d$ is a $\Z$-basis, we get $r.t_i=n_i.z_i$ for any $i$. Therefore $r_i.t_i=h_r(n_i).z_i$ where $r_i=\frac{r}{\gcd{n_i}{r}}$. As $r_i$ and $h_r(n_i)$ are relatively prime, there exists $u_i$, $u_i'$ in $\Z$ such that $u_i.r_i+u_i'.h_r(n_i)=1$. From $u_i.r_i.t_i=h_r(n_i).u_i.z_i$, we get $t_i=h_r(n_i).(u_i.z_i+u_i'.t_i)$. Therefore, $x\in \sum_{i=1}^dh_r(n_i).\Z.v_i$ and we have proved the inclusion $\gamma_{r,m,\unit_{0,m}}^{-1}(\sum_{i=1}^dn_i.\Z.v_i)\subseteq \sum_{i=1}^dh_r(n_i).\Z.v_i$. Let us prove the other inclusion. Consider $x\in  \sum_{i=1}^d h_r(n_i).\Z.v_i$. There exists a sequence $z_1$, ..., $z_d$ in $\Z$ such that $x=\sum_{i=1}^dh_r(n_i).z_i.v_i$. Hence $\gamma_{r,m,\unit_{0,m}}(x)=\sum_{i=1}^nr.h_r(n_i).z_i.v_i$. As $n_i$ divides $r.h_r(n_i)$, we deduce that $\gamma_{r,m,\unit_{0,m}}(x)\in \sum_{i=1}^dn_i.\Z.v_i$. Therefore $x\in\gamma_{r,m,\unit_{0,m}}^{-1}(\sum_{i=1}^dn_i.\Z.v_i)$ and we have proved the other inclusion.
  \qed
\end{proof}

The stability of vector lattices by inverse image by $\gamma_{r,m,0}$ is provided by the following proposition \ref{prop:invmask1}. 
\begin{proposition}\label{prop:invmask1}
  The set $M_z=\gamma_{r,m,0}^{-z}(M)$ is a $V_z$-vector lattice included in $\Z^m$ for any $V$-vector lattice $M$ included in $\Z^m$ where $V_z$ is the vector space $V_z=\Gamma_{r,m,0}^{-z}(V)$ and for any $z\in\Nat$. Moreover, from an Hermite $I$-representation of $M$, we can compute in polynomial time the Hermite $I_z$-representation of $M_z$ where $I_z=\theta_m^z(I)$.
\end{proposition}
\begin{proof}
  Recall that form the $I$-representation of $M$, we immediately deduce a $\Z$-basis $v_1$, ..., $v_d$ of $M$. Let us remark that $\gamma_{r,m,0}^{-z}(M)$ is the set of vectors $x\in\Z^m$ such that there exists a vector $k\in\Z^d$ satisfying $\Gamma_{r,m,0}^z(x)=\sum_{i=1}^dk[i].v_i$. Let us consider the vector space $W=\{(k,x)\in\Q^d\times \Q^m;\;\Gamma_{r,m,0}^z(x)=\sum_{i=1}^dk[i].v_i\}$. Remark that $W$ is a vector space and $J=\finiteset{1}{d}$ is a full rank set of indices of $W$. From proposition \ref{prop:hermitevector} we deduce that we can compute in polynomial time the $J$-representation of $W$. That means we can compute in polynomial time a $\Z$-basis of $\Z^m\cap W$ denoted by $(k_1,x_1)$, ..., $w_d=(k_d,x_d)$ where $k_i\in\Z^d$ and $x_i\in\Z^m$. Now, just remark that $\gamma_{r,m,0}^{-z}(M)$ is a the $V_z$-vector lattice generated by the vectors $x_1$, ..., $x_d$. Therefore, the $I_z$-representation of $\gamma_{r,m,0}^{-z}(M)$ is computable in polynomial time for any $z\in\finiteset{0}{m-1}$. Observe that in general, an integer $z\in\Nat$ can be decomposed into $z=z'+m.k$ where $k\in\Nat$ and $z'\in\finiteset{0}{m-1}$. Observe that $\gamma_{r,m,\unit_{0,m}}^{-k}(M)$ can be computed in polynomial time thanks to lemma \ref{lem:invmask0}.
  \qed
\end{proof}

\subsubsection{Relatively prime properties}
A $V$-vector lattice $M$ included in $\Z^m$ is said \emph{relatively prime with a basis of decomposition $r$}\index{Gen}{relatively prime} if the integer $|\Z^m\cap V/M|$ is relatively prime with $r$. 

Thanks to lemma \ref{lem:MMpMpp}, we deduce that the class of $V$-vector lattices included in $\Z^m$ and relatively prime with $r$ is stable by finite intersection. In fact given two relatively prime $V$-vector lattices $M_1$ and $M_2$ included in $\Z^m$, from $M_1\cap M_2\subseteq n.\Z^m\cap V\subseteq \Z^m\cap V$ where $n=|\Z^m\cap V/M_1|.|\Z^m\cap V/M_2|$, we deduce that $|\Z^m\cap V/M_1\cap M_2|.|M_1\cap M_2/n.\Z^m\cap V|=|\Z^m\cap V/n.\Z^m\cap V|=n^{\dim(V)}$. In particular $|\Z^m\cap V/M_1\cap M_2|$ divides an integer relatively prime with $r$. That means it is relatively prime with $r$.

We are going to show that the $V$-vector lattices included in $\Z^m$ and relativelly prime with $r$ naturally appear when computing inverse images of a $V$-vector lattice by $\gamma_{r,m,0}$.

As $h_r(n)\leq n$ for any integer $n\in\Nat\moins\{0\}$ we deduce that $(h_r^k(n))_{k\in\Nat}$ is a non increasing sequence ultimely stationary: there exists $k_n\in\Nat$ such that $h_r^{k}(n)=h_r^{k_n}(n)$ for any $k\geq k_n$. We denote by $h_r^\infty(n)$ this limit. Remark that $h_r^\infty(n)$ is relatively prime with $r$ and $h_r^\infty(n)=n$ if and only if $n$ is relatively prime with $r$. The previous lemma \ref{lem:invmask0} shows that $(\gamma_{r,m,\unit_{0,m}}^{-k}(M))_{k\in\Nat}$ is a non decreasing sequence of $V$-vector lattices ultimately stationary. The limit is denoted by $\gamma_{r,m,\unit_{0,m}}^{-\infty}(M)$\index{Not}{$\gamma_{r,m,\unit_{0,0}}^{-\infty}(M)$} and naturally satisfies the following equality:
$$\gamma_{r,m,\unit_{0,m}}^{-\infty}(M)=\bigcup_{k\in\Nat}\gamma_{r,m,\unit_{0,m}}^{-k}(M)$$
From the previous lemma \ref{lem:invmask0} we deduce that $\gamma_{r,m,\unit_{0,m}}^{-\infty}(M)$ is relatively prime with $r$ and if $M$ is relatively prime with $r$ then $\gamma_{r,m,\unit_{0,m}}^{-\infty}(M)=M$. In particular the class of $V$-vector lattices relatively prime with $r$ is stable by inverse image by $\gamma_{r,m,\unit_{0,m}}$.

Let us remark that the elements in $\gamma_{r,m,\unit_{0,m}}^{-\infty}(M)$ are geometrically characterized by the following lemma \ref{lem:caraGinfty}
\begin{lemma}\label{lem:caraGinfty}
  Given a vector lattice $M$ included in $\Z^m$ and a vector $x\in\Z^m$, we have $x\in\gamma_{r,m,\unit_{0,m}}^{-\infty}(M)$ if and only if there exists $k\in\Nat$ such that $r^k.x\in M$.
\end{lemma}
\begin{proof}
  Let $V=\vecspace(M)$. There exists a $\Z$-basis of $M$ of the form $n_1.v_1$, ..., $n_d.v_d$ where $n_1$, ..., $n_d$ are integers in $\Nat\moins\{0\}$ and $v_1$, ..., $v_d$ is a $\Z$-basis of $\Z^m\cap V$. From lemma \ref{lem:invmask0} we deduce that $h_r^\infty(n_1).v_1$, ..., $h_r^\infty(n_d).v_d$ is a $\Z$-basis of $\gamma_{r,m,\unit_{0,m}}^{-\infty}(M)$. Remark that there exists an integer $k_0\in\Nat$ such that $r^{k_0}h_r^\infty(n_i)$ divides $n_i$ for any $i\in\finiteset{1}{d}$.

  First, let us first prove that there exists $k\in\Nat$ satisfying $r^k.x\in M$ for any $x\in\gamma_{r,m,\unit_{0,m}}^{-\infty}(M)$. There exists $z\in\Z^d$ such that $x=\sum_{i=1}^dh_r^\infty(n_i).z[i].v_i$. In particular $r^{k_0}.x=\sum_{i=1}^d r^{k_0}.h_r^\infty(n_i).z[i].v_i\in M$ and we have proved that there exists an integer $k\in\Nat$ such that $r^k.x\in M$.

  Next, let us show that $x\in\gamma_{r,m,\unit_{0,m}}^{-\infty}(M)$ for any $x\in\Z^m$ such that there exists $k\in\Nat$ satisfying $r^k.x\in M$. As $r^k.x\in M$, we deduce that $x\in\Z^m\cap V$. Hence, there exists $z\in\Z^d$ such that $x=\sum_{i=1}^dz[i].v_i$. Hence $r^k.x=\sum_{i=1}^dz[i]r^k.z[i].v_i$. Moreover, as $r^k.x\in M$, there exists $t\in\Z^d$ such that $r^{k}.x=\sum_{i=1}^dn_i.t[i].v_i$. As $v_1$, ..., $v_d$ is a $\Z$-base, we get $r^k.z[i]=n_i.t[i]$. As $h_r^\infty(n_i)$ divides $n_i$, we deduce that $n_i'=\frac{n_i}{h_r^\infty(n_i)}$ is $\Nat$. Hence $r^k.z[i]=h_r^\infty(n_i).n_i'.t[i]$. As $h_r^\infty(n_i)$ is relatively prime with $r$, then $h_r^\infty(n_i)$ is relatively prime with $r^k$, and we deduce that $r^k$ divides $n_i'.t[i]$. Hence $z[i]\in h_r^\infty(n_i).\Z$. We deduce that $x\in \gamma_{r,m,\unit_{0,m}}^{-1}(M)$.
  \qed
\end{proof}

\section{Affine lattices}
An \emph{affine lattice $P$}\index{Gen}{affine lattice} is a subset of $\Q^m$ of the form $P=a+M$ where $a\in\Q^m$ and $M$ is a lattice. A \emph{$V$-affine lattice $P$} is an \emph{affine lattice $P$} of the form $P=a+M$ where $M$ is a $V$-vector lattice.

Given a $V$-affine space $A$, observe that $\Z^m\cap A$ is either empty or a $V$-affine lattice of the form $a+(\Z^m\cap V)$ where $a$ is any vector in $\Z^m\cap A$. The following proposition will be useful for computing a vector in $\Z^m\cap A$ when such a vector exists.

\begin{proposition}[Corollary 5.3b and 5.3c of \cite{S-87}]\label{prop:heher}
  Given an affine space $A$, we can decide in polynomial time if $\Z^m\cap A$ is non empty and in this case, we can compute in polynomial time a vector $a$ in this set.
\end{proposition}

\begin{corollary}\label{cor:intersectionaffinelattice}
  Given two affine lattices $P_1=b_1+M_1$ and $P_2=b_2+M_2$ where $b_1$, $b_2$ are two vectors in $\Q^d$ and $M_1$, $M_2$ are two vectors lattices, we can decide in polynomial time if $(b_1+M_1)\cap (b_2+M_2)\not=\emptyset$. Moreover, in this case we can compute in polynomial time a vector $a$ in this set. Observe that we have $P_1\cap P_2=a+(M_1\cap M_2)$. 
\end{corollary}
\begin{proof}
  From the vector $I_1$-representation of $M_1$, we deduce in linear time a $\Z$-basis $v_{1,1}$, ..., $v_{1,d_1}$ of $M_1$, and from the vector $I_2$-representation of $M_2$, we get in linear time a $\Z$-basis $v_{2,1}$, ..., $v_{2,d_2}$ of $M_2$. Observe that $(b_1+M_1)\cap (b_2+M_2)\not=\emptyset$ if and only if $\Z^m\cap A$ is non empty where $A$ is the affine space $A=\{(x_1,x_2)\in \Q^{d_1}\times \Q^{d_2};\; b_1+\sum_{i=1}^{d_1}x_1[i].v_{1,i}=b_2+\sum_{i=1}^{d_2}x_2[i].v_{2,i}\}$. Note that proposition \ref{prop:heher} provides a polynomial time algorithm for deciding if $\Z^m\cap A$ is non-empty and in this case it provides in polynomial time a vector $(x_1,x_2)\in\Z^m\cap A$. Note that $a=b_1+\sum_{i=1}^{d_1}x_1[i].v_{1,i}$ is a vector in $P_1\cap P_2$.
  \qed
\end{proof}

%% file: chapter.semilinearsets.tex
\chapter{Semi-linear Sets}

\section{Semi-linear Spaces}
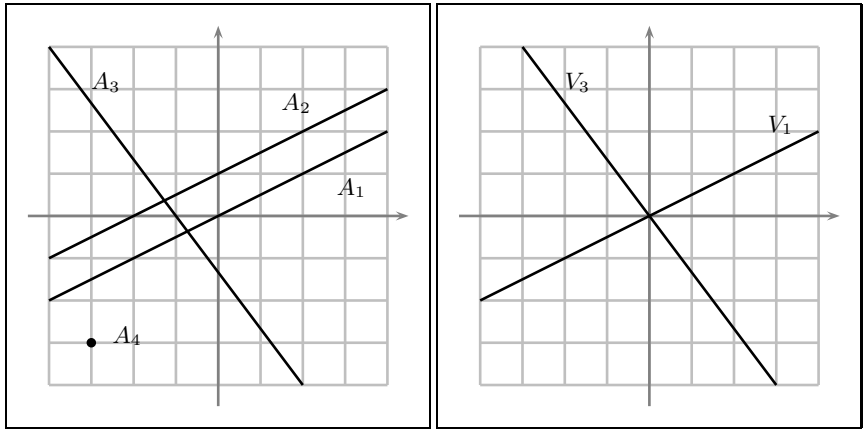
\begin{figure}[htbp]
  \begin{center}
    \setlength{\unitlength}{16pt}
    \pssetlength{\psunit}{16pt}
    \pssetlength{\psxunit}{16pt}
    \pssetlength{\psyunit}{16pt}
    \begin{picture}(10,10)(-5,-5)
      \put(-5,-5){\framebox(10,10){}}
      \multido{\iii=-4+1}{9}{\psline[linecolor=lightgray,linewidth=1pt](-4,\iii)(4,\iii)}\multido{\iii=-4+1}{9}{\psline[linecolor=lightgray,linewidth=1pt](\iii,-4)(\iii,4)}\psline[linecolor=gray,arrowscale=1,linewidth=1pt]{->}(0,-4.5)(0,4.5)\psline[linecolor=gray,arrowscale=1,linewidth=1pt]{->}(-4.5,0)(4.5,0)
      \psdot*(-3,-3)%
      \psline[linecolor=black,linewidth=1pt](-4,-1)(4,3)
      \psline[linecolor=black,linewidth=1pt](-4,-2)(4,2)
      \psline[linecolor=black,linewidth=1pt](-4,4)(2,-4)
      \put(2.8,0.5){$A_1$}
      \put(1.5,2.5){$A_2$}
      \put(-3,3){$A_3$}
      \put(-2.5,-3){$A_4$}
    \end{picture}
    \begin{picture}(10,10)(-5,-5)
      \put(-5,-5){\framebox(10,10){}}
      \multido{\iii=-4+1}{9}{\psline[linecolor=lightgray,linewidth=1pt](-4,\iii)(4,\iii)}\multido{\iii=-4+1}{9}{\psline[linecolor=lightgray,linewidth=1pt](\iii,-4)(\iii,4)}\psline[linecolor=gray,arrowscale=1,linewidth=1pt]{->}(0,-4.5)(0,4.5)\psline[linecolor=gray,arrowscale=1,linewidth=1pt]{->}(-4.5,0)(4.5,0)
      \psline[linecolor=black,linewidth=1pt](-4,-2)(4,2)
      \psline[linecolor=black,linewidth=1pt](-3,4)(3,-4)
      \put(2.8,2){$V_1$}
      \put(-2,3){$V_3$}
    \end{picture}
  \end{center}
  \caption{On the left a semi-affine space $S$. On the right its direction.\label{fig:semiaffinespace}} 
\end{figure}
A \emph{semi-affine space}\index{Gen}{semi-affine space} (resp. a \emph{semi-vector space}\index{Gen}{semi-vector space}) $S$ of $\Q^m$ is a finite union of affine spaces (resp. vector spaces) of $\Q^m$ (see figure \ref{fig:semiaffinespace}). Given a vector space $V$, a finite union of $V$-affine spaces is called a \emph{semi-$V$-affine space}. In this section we show that a semi-affine space can be canonically decomposed into maximal affine spaces, called \emph{affine components}. Moreover, by proving that any finite or infinite intersection of semi-affine spaces remains a semi-affine space, we define the notion of \emph{semi-affine hull}.

\subsection{Affine components}
\begin{definition}
  An \emph{affine component}\index{Gen}{affine component} $A$ of a semi-affine space $S$ is a maximal (for $\subseteq$) affine space included in $S$. The set of affine components is denoted by $\comp(S)$\index{Not}{$\comp(S)$}.
\end{definition}
We are going to prove that $\comp(S)$ provides a canonical representation of $S$. We first prove the following lemma, intensively used in the sequel.
\begin{lemma}[Insecable lemma]\label{lem:insecable}
  Let $\class$ be a non-empty finite class of affine spaces and $A_0$ be an affine space such that $A_0\subseteq \bigcup_{A\in\class}A$. There exists $A\in\class$ such that $A_0\subseteq A$.
\end{lemma}
\begin{proof}
  Let us consider an affine space $A_0$ and let us prove by induction over $n\in\Nat\moins\{0\}$ that for any finite class $\class$ of affine spaces such that $|\class|=n$ and $A_0\subseteq \bigcup_{A\in\class}A$, there exists $A\in\class$ such that $A_0\subseteq A$. Naturally the case $n=1$ is immediate. Assume that the induction hypothesis is true for an integer $n\in\Nat\moins\{0\}$ and let us consider a finite class $\class$ of affine spaces such that $|\class|=n+1$ and $A_0\subseteq \bigcup_{A\in\class}A$. Let us consider $A'\in\class$. The case $A_0\subseteq A'$ is also immediate so we can assume that $A_0\not\subseteq A'$. Let us consider $\class'=\class\moins\{A'\}$. As $A_0\not\subseteq A'$, there exists $a_0\in \difference{A_0}{A'}$. Let $a_1\in A_0$ and remark that $a_t=a_0+t.(a_1-a_0)\in A_0$ for any $t\in\Q$ because $A_0$ is an affine space. From $A_0\subseteq \bigcup_{A\in\class}A$, we deduce that for any $t\in\Q$, there exists $A\in\class$ such that $a_t\in A$. As $\Q$ is infinite whereas $\class$ is finite, there exists $A\in\class$ and at least two different $t\in\Q$ satisfying $a_t\in A$. As $A$ is an affine space, we deduce that $a_t\in A$ for every $t\in\Q$. From $a_0\in A$ and $a_0\not\in A'$, we deduce that $A\in \class'$. We get $a_1\in\bigcup_{A\in\class'}A$. We have proved that $A_0\subseteq \bigcup_{A\in\class'}A$. From $|\class'|=n$, we deduce that there exists $A''\in\class'$ such that $A_0\subseteq A''$. We have proved the induction hypothesis for $\class$.
  \qed
\end{proof}

\begin{proposition}\label{prop:affinecomponent}
  The set $\comp(S)$ of a semi-affine space $S$ is finite and $S$ is equal to the finite union of its affine components $S=\bigcup_{A\in\comp(S)}A$. Moreover, from any finite class $\class$ of affine spaces such that $S=\bigcup_{A\in\class}A$, we can compute in polynomial time $\comp(S)$.
\end{proposition}
\begin{proof}
  Let us consider a semi-affine space $S=\bigcup_{A\in\class}A$ where $\class$ is a finite class of affine spaces. 

  Consider the class $\class'$ of non-empty affine spaces in $\class$ maximal for $\subseteq$. Let us first prove that $S=\bigcup_{A'\in\class'}A'$. Naturally, from $\class'\subseteq \class$, we deduce that $\bigcup_{A'\in\class'}A'\subseteq S$. For any $A\in\class$, either $A=\emptyset$ and in this case $A\subseteq \bigcup_{A'\in\class'}A'$, or $A\not=\emptyset$, and in this case there exists $A'\in\class'$ such that $A\subseteq A'$. Hence $A\subseteq \bigcup_{A'\in\class'}A'$. Therefore, $S=\bigcup_{A'\in\class'}A'$.

  By replacing $\class$ by $\class'$, we can assume without loss of generality that $\class$ is a finite class of non-empty affine spaces such that $A_1\subseteq A_2$ implies $A_1=A_2$ for any $A_1$, $A_2$ in $\class$.
  
  Let us now prove that $\comp(S)=\class$. Let $A_0\in\class$ and consider an affine space $A'$ such that $A_0\subseteq A'\subseteq S$. Insecable lemma \ref{lem:insecable} proves that $A'\subseteq S=\bigcup_{A\in\class}A$ implies that there exists $A\in\class$ such that $A'\subseteq A$. From $A_0\subseteq A$ and $A$, $A_0$ in $\class$, we get $A_0=A$. We deduce that $A_0=A'$. Hence $A_0$ is a maximal (for $\subseteq$) non-empty affine space such that $A_0\subseteq S$. That means $A_0\in\comp(S)$ and we have proved that $\class\subseteq \comp(S)$. Let us prove the converse inclusion. Let $A_0\in\comp(S)$. As $A_0\subseteq S=\bigcup_{A\in\class}A$, insecable lemma \ref{lem:insecable} shows that there exists $A\in\class$ such that $A_0\subseteq A$. From $A_0\subseteq A\subseteq S$, we deduce by maximality of $A_0$ that $A_0=A$. Hence $A_0\in\class$ and we have proved that $\comp(S)\subseteq\class$.
  \qed
\end{proof}

\subsection{Size}
The set of affine components provides a natural way for \emph{canonically} representing semi-affine spaces as finite set of affine spaces. The integer $\size(S)\in\Nat$\index{Gen}{sizes!of a semi-affine space} where $S$ is a semi-affine space is naturally defined by $\size(S)=\sum_{A\in\comp(S)}\size(A)$.

\subsection{Direction}
\begin{definition}
  The \emph{direction}\index{Gen}{direction of a semi-affine space} $\vec{S}$ of a semi-affine space $S$ is defined by $\vec{S}=\bigcup_{A\in\comp(S)}\vec{A}$\index{Not}{$\vec{S}$}.
\end{definition}
Remark that the semi-affine space direction definition extends the affine space direction definition because if $S=A$ is a non-empty affine space then $\comp(S)=\{A\}$, and if $S=\emptyset$ then $\comp(S)=\emptyset$. Remark also that insecable lemma \ref{lem:insecable} shows that for any class $\class$ of affine spaces such that $S=\bigcup_{A\in\class}A$, we have $\vec{S}=\bigcup_{A\in\class}\vec{A}$ even if $\class$ is not equal to $\comp(S)$. That shows in particular that a semi-affine space $S$ is a semi-vector space if and only if $\vec{S}=S$.

\begin{example}\label{ex:direx}
  Let us consider the semi-affine space $S=A_1\cup A_2\cup A_3\cup A_4$ where $A_1=\Q.(2,1)$, $A_2=(0,1)+\Q.(2,1)$, $A_3=(-1,0)+\Q.(3,-4)$ and $A_4=\{(-3,-3)\}$ given in figure \ref{fig:semiaffinespace}. We have $\vec{S}=V_1\cup V_3$ where $V_1=\Q.(2,1)$ and $V_3=\Q.(3,-4)$. Remark that $S$ owns $4$ affine components $\comp(S)=\{A_1,A_2,A_3,A_4\}$ and $\vec{S}$ owns only $2$ affine components $\comp(\vec{S})=\{V_1,V_3\}$. 
\end{example}

\subsection{Semi-affine hull}
Following proposition \ref{prop:intersectionsemiaffine} proves that any finite or infinite intersection of semi-affine spaces remains a semi-affine space. In particular for any subset $X\subseteq \Q^m$, there exists a minimal (for $\subseteq$) semi-affine space written $\saff(X)$\index{Not}{$\saff(X)$} that contains $X$. This semi-affine space is called the \emph{semi-affine hull}\index{Gen}{semi-affine hull} of $X$. The semi-vector space $\vec{\saff(X)}$\index{Not}{$\vecsaff(X)$} is written $\vecsaff(X)$.

\begin{proposition}\label{prop:intersectionsemiaffine}
  Any finite or infinite intersection of semi-affine spaces remains a semi-affine space.
\end{proposition}
\begin{proof}
  Observe that a semi-affine space is a finite union of affine spaces that can be represented by a finite set of vectors in $\Q^m$. Hence the class of semi-affine spaces is \emph{countable}. In order to prove the lemma, it is therefore sufficient to prove that $\bigcap_{n\in\Nat}S_n$ is a semi-affine space for any sequence $(S_n)_{n\in\Nat}$ of semi-affine spaces. As the class of semi-affine spaces is stable by finite intersection, we can also assume that $(S_n)_{n\in\Nat}$ is non-increasing. Let us prove by induction over the dimension $k\in\Nat\cup\{-1\}$ that any non-increasing sequence of semi-affine spaces $(S_n)_{n\in\Nat}$ such that $\dim(\vec{\aff}(S_0))\leq k$, is ultimately stationary. Case $k=-1$ is immediate because in this case $S_n=\emptyset$ for any $n\in\Nat$. Now, assume the induction true for $k\geq -1$ and let us consider a non-increasing sequence of semi-affine spaces $(S_n)_{n\in\Nat}$ such that the dimension of $\vec{\aff}(S_0)$ is equal to $k+1$. Remark that if $S_n$ is an affine space for any $n\geq 0$, then $(S_n)_{n\geq 0}$ is a non-increasing sequence of affine spaces. In particular, this sequence is ultimately constant. So, we can assume that there exists an integer $n_0\geq 0$ such that $S_{n_0}$ is not an affine space. There exists a finite class $\class$ of affine spaces such that $S_{n_0}=\bigcup_{A\in\class}A$. Let $A\in\class$. From $A\subseteq S_{n_0}\subseteq S_0\subseteq\aff(S_0)$, we deduce that the dimension of $\vec{A}$ is less than or equal to $k+1$. Moreover, if it is equal to $k+1$, from $A\subseteq \aff(S_0)$, we deduce $A=\aff(S_0)$ and we get $S_{n_0}=A$ is an affine space which is a contradiction. As the sequence $(S_n\cap A)_{n\geq 0}$ is a non-increasing sequence of semi-affine spaces such that the dimension of $\vec{\aff}(S_n\cap A)\subseteq \vec{A}$ is less than or equal to $k$, the induction hypothesis proves that there exists $n_A\geq 0$ such that $S_n\cap A=S_{n_A}\cap A$ for any $n\geq n_A$. Let us consider $N=\max_{A\in\class}(n_0,n_A)$. For any $n\geq N$, we have $S_n\subseteq S_{n_0}=\bigcup_{A\in\class}A$ and $S_n\cap A=S_{N}\cap A$. Hence $S_n=S_n\cap (\bigcup_{A\in\class}A)=\bigcup_{A\in\class}(S_n\cap A)=\bigcup_{A\in\class}(S_N\cap A)=S_N\cap(\bigcup_{A\in\class}A)=S_N$ for any $n\geq N$ and we have proved the induction.
  \qed
\end{proof}

\begin{example}
  The semi-affine hull of a \emph{finite} subset $X\subseteq\Q^m$ is equal to $X$ because $X$ is the finite union over $x\in X$ of the affine spaces $\{x\}$. The semi-affine hull of an \emph{infinite} subset  $X\subseteq\Q$ (remark that $m=1$) is equal to $\Q$. In fact, the class of affine spaces of $\Q$ is equal to $\{\Q,\emptyset\}\cup\{\{x\};\;x\in\Q\}$.
\end{example}

\begin{remark}
  As $\aff(X)$ is an affine space and in particular a semi-affine space that contains $X$, we deduce that $\saff(X)\subseteq\aff(X)$. This last inclusion can be strict as shown by the example $X=\{\unit_{0,m},\ldots,\unit_{m,m}\}$. In fact, in this case, we have $\saff(X)=X$ and $\aff(X)=\Q^m$.
\end{remark}

The following lemma will be useful to compute the semi-affine hull of some subsets of $\Q^m$ (see example \ref{ex:nm}).
\begin{lemma}[Covering lemma\index{Gen}{covering lemma}]\label{lem:cov}\hfill
  \begin{itemize}
  \item For any affine function $f:\Q^{m}\rightarrow\Q^{m'}$ and for any subset $X\subseteq\Q^{m}$, we have $\saff(f(X))=f(\saff(X))$.
  \item For any subsets $X,X'\subseteq\Q^m$, we have 
    \begin{itemize}
    \item $\saff(X\times X')=\saff(X)\times\saff(X')$,
    \item $\saff(X\cup X')=\saff(X)\cup\saff(X')$, and
    \item $\saff(X+X')=\saff(X)+\saff(X')$.
    \end{itemize}
  \end{itemize}
\end{lemma}
\begin{proof}
  Let us consider an affine function $f$. From $X\subseteq \saff(X)$, we deduce $f(X)\subseteq f(\saff(X))$. As $f(\saff(X))$ is a semi-affine space that contains $f(X)$ (observe that $f(A)$ is an affine space for any affine space $A$ and for any affine function $f$), by minimality of the semi-affine hull, we deduce $\saff(f(X))\subseteq f(\saff(X))$. Let us prove the converse inclusion. As $f(X)\subseteq\saff(f(X))$, we have $X\subseteq f^{-1}(\saff(f(X)))$. As $f^{-1}(\saff(f(X)))$ is a semi-affine space (observe that $f^{-1}(A)$ is an affine space for any affine space $A$ and for any affine function $f$), by minimality of the semi-affine hull, we get $\saff(X)\subseteq f^{-1}(\saff(f(X)))$. Hence $f(\saff(X))\subseteq f(f^{-1}(\saff(f(X))))$. Recall that for any function $g:A\rightarrow B$, and for any subset $Y\subseteq B$, we have $g(g^{-1}(Y))=g(A)\cap Y$. Hence $f(f^{-1}(\saff(f(X))))=f(\Q^m)\cap\saff(f(X))$. From $f(X)\subseteq f(\Q^m)$, we also deduce $\saff(f(X))\subseteq f(\Q^m)$ and we get $f(\Q^m)\cap\saff(f(X))=\saff(f(X))$. Therefore $f(\saff(X))\subseteq\saff(f(X))$.
  
  Let us consider $X,X'\subseteq\Q^m$ and let us prove that $\saff(X\cup X')=\saff(X)\cup\saff(X')$. From $X\cup X'\subseteq\saff(X)\cup\saff(X')$, we deduce by minimality of the semi-affine hull $\saff(X\cup X')\subseteq\saff(X)\cup\saff(X')$. Moreover, from $X\subseteq X\cup X'\subseteq \saff(X\cup X')$, we get $\saff(X)\subseteq\saff(X\cup X')$ and symmetrically $\saff(X')\subseteq\saff(X\cup X')$. We have shown $\saff(X)\cup\saff(X')\subseteq\saff(X\cup X')$.
  
  Let us consider $X,X'\subseteq\Q^m$ and let us prove that $\saff(X\times X')=\saff(X)\times\saff(X')$. From $X\times X'\subseteq \saff(X)\times\saff(X')$, we deduce that $\saff(X\times X')\subseteq \saff(X)\times\saff(X')$. By considering the affine function $f_{1,x}:\Q^m\rightarrow\Q^{2m}$ defined by $f_{1,x}(x')=(x,x')$, we get $\saff(\{x\}\times X')=\{x\}\times\saff(X')$ for any $x\in X$. From $\{x\}\times X'\subseteq X\times X'$, we deduce $\saff(\{x\}\times X')\subseteq\saff(X\times X')$. So $X\times \saff(X')\subseteq\saff(X\times X')$. In particular, for any $x'\in\saff(X')$, we have $X\times\{x'\}\subseteq\saff(X\times X')$. Affine function $f_{2,x'}:\Q^m\rightarrow\Q^{2m}$ defined by $f_{2,x'}(x)=(x,x')$ proves that $\saff(X)\times\{x'\}\subseteq\saff(X\times X')$ for any $x'\in\saff(X')$. So, we have proved $\saff(X)\times\saff(X')\subseteq\saff(X\times X')$.
  
  Let us consider $X,X'\subseteq\Q^m$ and let us prove that $\saff(X+ X')=\saff(X)+\saff(X')$. By considering the affine function $f:\Q^{2m}\rightarrow \Q^m$ defined by $f(x,x')=x+x'$, we deduce that $\saff(X)\times\saff(X')=f(\saff(X)\times\saff(X'))=f(\saff(X\times X'))=\saff(f(X\times X'))=\saff(X+X')$.
  \qed
\end{proof}
  
\begin{example}\label{ex:nm}
  The semi-affine hull of $\Nat^m$ is equal to $\Q^m$. In fact, from covering lemma \ref{lem:cov}, we deduce $\saff(\Nat^m)=\sum_{i=1}^{m}\saff(\Nat.\unit_{i,m})= \sum_{i=1}^{m}\saff(\Nat).\unit_{i,m}=\sum_{i=1}^{m}\Q.\unit_{i,m}=\Q^m$.
\end{example}

\subsection{Cyclic sets}
Recall that a $(r,m,\sigma)$-cyclic set $X$ where $\sigma\in\Sigma_{r,m}^*$ is a subset of $\Z^m$ such that $\gamma_{r,m,\sigma}^{-1}(X)=X$. The following proposition \ref{prop:saffcyclic} shows that the semi-affine hull of a $(r,m,\sigma)$-cyclic set $X\subseteq\Z^m$ is a finite union of affine spaces of the form $\xi_{r,m}(\sigma)+V$ where $V$ is a vector space.
\begin{proposition}\label{prop:saffcyclic}
  We have $\saff(X)=\xi_{r,m}(\sigma)+\vecsaff(X)$ for any $(r,m,\sigma)$-cyclic set $X\subseteq \Z^m$.
\end{proposition}
\begin{proof}
  It is sufficient to prove that for any affine component $A$ of $\saff(X)$, we have $\xi_{r,m}(s)\in A$. Consider $x\in X$. As $\gamma_{r,m,\sigma}^{-1}(X)=X$ then $\gamma_{r,m,\sigma}^k(x)=r^{k.|\sigma|}.(x-\xi(\sigma))+\xi(\sigma)\in X$ for any $k\in\Nat$. Covering lemma \ref{lem:cov} proves that $\Q.(x-\xi_{r,m}(\sigma))+\xi_{r,m}(\sigma)\subseteq\saff(X)$. In particular, for any $\lambda\in\Q$, we have $\lambda.(X-\xi_{r,m}(\sigma))+\xi_{r,m}(\sigma)\subseteq\saff(X)$. From covering lemma \ref{lem:cov}, we also prove that $\lambda.(\saff(X)-\xi_{r,m}(\sigma))+\xi_{r,m}(\sigma)\subseteq\saff(X)$. Let $A$ be an affine component of $\saff(X)$. We have proved that $\Q.(A-\xi_{r,m}(\sigma))+\xi_{r,m}(\sigma)\subseteq\saff(X)$. From $A\subseteq \Q.(A-\xi_{r,m}(\sigma))+\xi_{r,m}(\sigma)\subseteq\saff(X)$, we deduce by maximality of the affine component $A$, the equality $A=\Q.(A-\xi(\sigma))+\xi_{r,m}(\sigma)$. In particular $\xi_{r,m}(\sigma)\in A$.
  \qed
\end{proof}

\section{Semi-affine lattices}
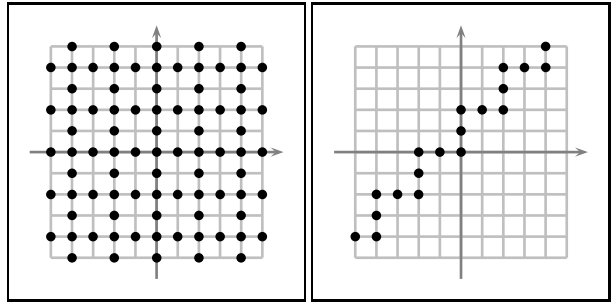
\begin{figure}[htbp]
 \setlength{\unitlength}{8pt}
 \pssetlength{\psunit}{8pt}
 \pssetlength{\psxunit}{8pt}
 \pssetlength{\psyunit}{8pt}
 \begin{center}
   \begin{picture}(14,14)(-7,-7)%
     \put(-7,-7){\framebox(14,14){}}
     \multido{\iii=-5+1}{11}{\psline[linecolor=lightgray,linewidth=1pt](-5,\iii)(5,\iii)}\multido{\iii=-5+1}{11}{\psline[linecolor=lightgray,linewidth=1pt](\iii,-5)(\iii,5)}\psline[linecolor=gray,arrowscale=1,linewidth=1pt]{->}(0,-6)(0,6)\psline[linecolor=gray,arrowscale=1,linewidth=1pt]{->}(-6,0)(6,0)
     \multido{\ix=-5+1}{11}{\multido{\iy=-4+2}{5}{\psdots*(\ix,\iy)}}
     \multido{\ix=-4+2}{5}{\multido{\iy=-5+2}{6}{\psdots*(\ix,\iy)}}
   \end{picture}
   \begin{picture}(14,14)(-7,-7)%
     \put(-7,-7){\framebox(14,14){}}
     \multido{\iii=-5+1}{11}{\psline[linecolor=lightgray,linewidth=1pt](-5,\iii)(5,\iii)}\multido{\iii=-5+1}{11}{\psline[linecolor=lightgray,linewidth=1pt](\iii,-5)(\iii,5)}\psline[linecolor=gray,arrowscale=1,linewidth=1pt]{->}(0,-6)(0,6)\psline[linecolor=gray,arrowscale=1,linewidth=1pt]{->}(-6,0)(6,0)
     \multido{\ix=-4+2}{5}{\psdots*(\ix,\ix)}
     \psdot(-5,-4)\psdot(-4,-3)\psdot(-3,-2)\psdot(-2,-1)\psdot(-1,0)\psdot(0,1)\psdot(1,2)\psdot(2,3)\psdot(3,4)\psdot(4,5)
     \psdot(-4,-2)\psdot(-2,0)\psdot(0,2)\psdot(2,4)
   \end{picture}
   \caption{On the left a semi-$\Q^2$-affine lattice $P_1$. On the right a semi-$\Q.(1,1)$-affine lattice $P_2$.\label{fig:modularspace}}
 \end{center}
\end{figure}

A \emph{semi-$V$-affine lattice $P$}\index{Gen}{semi-affine lattice} is a finite union of $V$-affine lattices. Observe that the class of semi-$V$-affine lattice is stable by boolean combinations.

\begin{lemma}\label{lem:uniformmodul}
  For any non-empty semi-$V$-affine lattice, there exists a non-empty finite set $B\subseteq \Q^m$ and a $V$-vector lattice $M$ such that $P=B+M$.
\end{lemma}
\begin{proof}
  There exists a non-empty finite sequence $(a_j,M_j)_{j\in J}$ where $a_j\in\Q^m$ and $M_j$ is a $V$-vector lattice such that $P=\bigcup_{j\in J}(a_j+M_j)$. From lemma \ref{lem:Vgroupinter}, we deduce that $M=\bigcap_{j\in J}M_j$ is a $V$-vector lattice. Since $M\subseteq M_j$, theorem \ref{thm:inv1} shows that there exists a finite set $B_j\subseteq M_j$ such that $M_j=B_j+M$. We have proved that $P=B+M$ where $B$ is the finite set $B=\bigcup_{j\in J}(a_j+B_j)$.
  \qed
\end{proof}

The \emph{group of invariants}\index{Gen}{invariants} $\inv(X)$\index{Not}{$\inv(X)$} of a subset $X\subseteq \Q^m$ is the group of vectors $v\in\Q^m$ that let $X$ invariant: we have $X-v=X$. 
\begin{lemma}\label{lem:invVgroup}
  The group of invariants of a non empty semi-$V$-affine lattice is a $V$-vector lattice.
\end{lemma}
\begin{proof}
  Let $P$ be a non-empty semi-$V$-affine lattice. Lemma \ref{lem:uniformmodul} proves that there exists a non-empty finite set $B\subseteq \Q^m$ and a $V$-vector lattice $M$ such that $P=B+M$. Let us show that $\inv(P)\subseteq (V\cap (B-B))+M$. Consider a vector $v\in\inv(P)$. Let $b\in B$. Since $P-k.v=P$ for any $k\in\Nat$, there exists $b_k\in B$ and $m_k\in M$ such that $b-k.v=b_k+m_k$. Since $B$ is finite, there exists $k_1<k_2$ such that $b_{k_1}=b_{k_2}$. We deduce that $(k_1-k_2).v=m_{k_2}-m_{k_1}$. In particular $v\in V$ since $M\subseteq V$. Moreover, from $v=b-b_1+m_1$ and $m_1\in M\subseteq V$, we get $b-b_1\in V$. We have proved that $v\in (V\cap (B-B))+M$. Thus $\inv(P)$ is included in the discrete set $(V\cap (B-B))+M$ and we have proved that $\inv(P)$ is a vector lattice. Let us prove that $\vecspace(\inv(P))=V$. From $\inv(P)\subseteq (V\cap (B-B))+M$ we get $\vecspace(\inv(P))\subseteq V$. Moreover, from $M\subseteq \inv(P)$ we get $V=\vecspace(M)\subseteq \vecspace(\inv(P))$. Therefore $\inv(P)$ is a $V$-vector lattice.
  \qed
\end{proof}

The $V$-vector lattice of invariants of a non-empty semi-$V$-affine lattice is geometrically characterized by the following proposition \ref{prop:invgeo}.
\begin{proposition}\label{prop:invgeo}
  Let $P$ be a non-empty semi-$V$-affine lattice and let $M$ be a $V$-vector lattice. There exists a finite subset $B\subseteq \Z^m$ such that $P=B+M$ if and only if $M\subseteq \inv(P)$.
\end{proposition}
\begin{proof}
  Observe that if there exists a finite set $B\subseteq \Z^m$ such that $P=B+M$, we deduce that $M\subseteq \inv(P)$. Let us now prove the converse. Assume that $M$ is a $V$-vector lattice such that $M\subseteq \inv_V(P)$ and let us prove that there exists a finite set $B\subseteq \Z^m$ such that $P=B+M$. Lemma \ref{lem:uniformmodul} proves that there exists a non-empty finite set $B_0\subseteq \Q^m$ and a $V$-vector lattice $M_0$ such that $P=B_0+M_0$. As $M\subseteq \inv(P)$, we deduce that $P=B_0+M_0+M$. Since $M\subseteq M_0+M$, theorem \ref{thm:inv1} proves that there exists a finite set $B_1\subseteq M_0+M$ such that $M_0+M=B_1+M$. Therefore $P=B+M$ where $B=B_0+B_1$.
  \qed
\end{proof}

\begin{proposition}\label{prop:invcomputation}
  Let $M$ be a $V$-vector lattice and let $B$ be a non-empty finite subset of $\Q^m$. We can compute in polynomial time the $V$-vector lattice of invariants of $P=B+M$.
\end{proposition}
\begin{proof}
  Let us fix a vector $b_0\in B$ and let us prove that the $V$-vector lattice of invariant $\inv(P)$ is equal to the $V$-vector lattice $M'$ generated by $M$ and the vectors $v\in B-b_0$ such that $v+B+M=B+M$. Observe that $M'\subseteq \inv(P)$. Conversely, let $x\in\inv(P)$. We have $x+B+M=B+M$. In particular $x+b_0\in B+M$ and we deduce that there exists $v\in B-b_0$ and $m\in M$ such that $x=v+m$. Observe that $x+B+M=B+M$ implies $v+B+M=B+M$. Thus $x\in M'$ and we have proved that $\inv(P)=M'$. Note that a vector $v\in \Q^m$ satisfies $v+B+M=B+M$ if and only if for any $b\in B$ there exists $b'\in B$ such that $v+b-b'\in M$. Since we can decide in polynomial time if a vector is in $M$, we are done.
  \qed
\end{proof}

\begin{corollary}\label{cor:semiaffinelatticeequal}
  Given two semi-affine lattice $P_1=B_1+M_1$ and $P_2=B_2+M_2$ where $B_1$, $B_2$ are two finite subsets of $\Q^m$, and $M_1$, $M_2$ are two vector lattices, we can decide in polynomial time if $B_1+M_1=B_2+M_2$.  
\end{corollary}
\begin{proof}
  Naturally if $B_1$ and $B_2$ are both empty then $P_1=P_2$ and if only one of then is empty then $P_1\not=P_2$. Thus, without loss of generality, we cam assume that $B_1$ and $B_2$ are non empty. From proposition \ref{prop:invcomputation}, we deduce that $\inv(P_1)$ and $\inv(P_2)$ are computable in polynomial time. Observe that if $\inv(P_1)\not=\inv(P_2)$ then $P_1\not=P_2$. Hence we can assume that there exists a vector lattice $M$ such that $\inv(P_1)=M=\inv(P_2)$. We have reduced our problem to decide if $B_1+M_1=B_2+M_2$ where $M_1$ and $M_2$ are equal to a $V$-vector lattice $M$. Let $S_1$ and $S_2$ be the semi-$V$-affine spaces $S_i=\bigcup_{b\in B_i}(b+V)$. If $S_1\not=S_2$ then $P_1\not=P_2$. So, we can assume that there exists a semi-$V$-vector space $S$ such that $S_1=S=S_2$. Remark that $P_1=P_2$ if and only if $(B_1\cap A)+M=(B_2\cap A)+M$ for any affine component $A$ of $S$. Thus we can assume that $B_1$ and $B_2$ are included into a $V$-affine space $A$. Let $a_0\in A$ (for instance take $a_0\in B_1$) and notice that $P_1=P_2$ if and only if $(B_1-a_0)+M=(B_2-a_0)+M$. Hence, we can assume that $B_1$ and $B_2$ are included in $V$. From an Hermite $I$-representation of $M$, we get in linear time a $\Z$-basis $v_1$, ..., $v_d$ of $M$. Let us consider the function $\lambda\in V\rightarrow \Q^d$ defined by $\lambda(v)$ is the unique $x\in \Q^d$ such that $0\leq x[i]<1$ and such that there exists $k\in \Z^d$ satisfying $v=\sum_{i=1}^d(x+k)[i].v_i$. Note that $\lambda(v)$ is computable in polynomial time and $B_1+M=B_2+M$ if and only if $\lambda(B_1)=\lambda(B_2)$. Thus, we can decide in polynomial time if $B_1+M=B_2+M$.
  \qed
\end{proof}

\begin{example}
  Let $P_1$ be the semi-$\Q^2$-affine lattice $P_1=\{(0,0),(1,0),(0,1)\}+2.\Z^2$ and let $P_2$ be the semi-$\Q.(1,1)$-affine lattice $P_2=\{(0,0),(0,1),(0,2),(1,2)\}+\Z.(2,2)$ given in figure \ref{fig:modularspace}. We have $\inv(P_1)=\Z.(2,0)+\Z.(0,2)$ and $\inv(P_2)=\Z.(2,2)$.
\end{example}

\section{Semi-patterns}\label{sec:pattern}
A \emph{$V$-pattern}\index{Gen}{pattern} is a $V$-affine lattice included in $\Z^m$ and a \emph{semi-$V$-pattern}\index{Gen}{semi-pattern} is a semi-$V$-affine lattice included in $\Z^m$. 

Observe that the the $V$-lattice $\inv(P)$ of a non-empty semi-$V$-pattern $P$ is included in $\Z^m\cap V$ and if $P$ is empty then $\inv(P)=V$. We denote by $\inv_V(P)$\index{Not}{$\inv_V(X)$} the $V$-vector lattice $\inv_V(P)=\Z^m\cap V\cap \inv(X)$ for any (empty or non-empty) semi-$V$-pattern $P$.

\subsection{Inverse image by $\gamma_{r,m,\sigma}$}
Proposition \ref{prop:invsemipattern} proves that the class of semi-patterns is stable by inverse image by $\gamma_{r,m,\sigma}$ where $\sigma\in\Sigma_r^*$.

\begin{proposition}\label{prop:invsemipattern}
  Let $B$ be a finite subset of $\Z^m$ and let $M$ be a $V$-vector lattice included in $\Z^m$. For any word $\sigma\in\Sigma_r^*$, we can compute in polynomial time a finite set $B_\sigma\subseteq \Z^m$ such that $|B_\sigma|\leq |B|$ and $\gamma_{r,m,\sigma}^{-1}(B+M)=B_\sigma+\gamma_{r,m,0}^{-|\sigma|}(M)$.
\end{proposition}
\begin{proof}
  Let us consider for each $b\in B$ such that $\gamma_{r,m,\sigma}(\Z^m)\cap (b+M)\not=\emptyset$, a vector $b'\in\Z^m$ such that $\gamma_{r,m,\sigma}(b')\in b+M$. We denote by $B'$ the set of $b'\in\Z^m$ obtained. Note that corollary \ref{cor:intersectionaffinelattice} provides a polynomial time algorithm for computing $B'$. Let us prove that $\gamma_{r,m,\sigma}^{-1}(B+M)=B'+\gamma_{r,m,0}^{-|\sigma|}(M)$. Let $x\in B'+\gamma_{r,m,0}^{-|\sigma|}(M)$. That means, there exists $b'\in B'$ such that $\gamma_{r,m,0}^{|\sigma|}(x-b')\in M$. Moreover, by definition of $b'$, there exists $b\in B$ such that $\gamma_{r,m,\sigma}(b')\in b+M$. From $\gamma_{r,m,0}^{|\sigma|}(x-b')=\gamma_{r,m,\sigma}(x)-\gamma_{r,m,0}^{-|\sigma|}(b')$, we get $\gamma_{r,m\sigma}(x)\in B+M$. Therefore $x\in\gamma_{r,m,\sigma}^{-1}(B+M)$, and we have proved the inclusion $B'+\gamma_{r,m,0}^{-|\sigma|}(M)\subseteq \gamma_{r,m,\sigma}^{-1}(B+M)$. For the converse inclusion, consider $x\in  \gamma_{r,m,\sigma}^{-1}(B+M)$. There exists $b\in B$ such that $\gamma_{r,m,\sigma}(x)\in b+M$. By construction, there exists $b'\in\Z^m$ such that $\gamma_{r,m,\sigma}(b')\in b+M$. Hence $\gamma_{r,m,\sigma}(x)-\gamma_{r,m,\sigma}(b')\in M$. From $\gamma_{r,m,0}^{|\sigma|}(x-b')=\gamma_{r,m,\sigma}(x)-\gamma_{r,m,\sigma}(b')$, we get $\gamma_{r,m,0}^{|\sigma|}(x-b')\in M$. Therefore $x\in b'+\gamma_{r,m,0}^{-|\sigma|}(M)$ and we have proved the other inclusion.
   \qed
\end{proof}

\subsection{Relatively prime properties}
A semi-$V$-pattern $P$ is said \emph{relatively prime}\index{Gen}{relatively prime} with $r$ if the $V$-lattice $\inv_V(P)$ is relatively prime with $r$. From lemma \ref{lem:MMpMpp} we deduce that the class of relatively prime semi-$V$-patterns is stable by boolean combinations. In fact, consider two semi-$V$-patterns $P_1$ and $P_2$ and $\#\in\{\cup,\cap,\moins,\Delta\}$. Observe that $\inv_V(P_1)\cap\inv_V(P_2)\subseteq \inv_V(P_1\#P_2)\subseteq \Z^m\cap V$. From these inclusions, lemma \ref{lem:MMpMpp} proves $|\Z^m\cap V/\inv_V(P_1\#P_2)|.|\inv_V(P_1\#P_2)/\inv_V(P_1)\cap\inv_V(P_2)|$ is equal to the integer $|\Z^m\cap V/\inv_V(P_1)\cap\inv_V(P_2)|$. As $\inv_V(P_1)$ and $\inv_V(P_2)$ are two $V$-lattices relatively prime with $r$, we deduce that $\inv_V(P_1)\cap\inv_V(P_2)$ is relatively prime with $r$. In particular $|\Z^m\cap V/\inv_V(P_1\#P_2)|$ divides an integer relatively prime with $r$ and we deduce that this integer is relatively prime with $r$. Hence $P_1\#P_2$ is relatively prime with $r$.

The following lemma provides a geometrical characterization of these semi-$V$-patterns. This characterization and proposition \ref{prop:invsemipattern} prove that the class of semi-$V$-patterns relatively prime with $r$ is stable by inverse image by $\gamma_{r,m\sigma}$ for any $\sigma\in\Sigma_{r,m}^*$. 
\begin{lemma}
  A semi-$V$-pattern is relatively prime with $r$ if and only if there exists a $V$-lattice $M$ relatively prime with $r$ and a finite set $B\subseteq\Z^m$ such that $P=B+M$.
\end{lemma}
\begin{proof}
  Remark that if $P$ is relatively prime with $r$ then there exists a finite subset $B\subseteq\Z^m$ such that $P=B+\inv_V(X)$. Conversely, assume that there exists a $V$-lattice $M$ relatively prime with $r$ and a finite set $B\subseteq\Z^m$ such that $P=B+M$ and let us prove that $P$ is relatively prime with $r$. Since $M\subseteq\inv_V(X)\subseteq\Z^m\cap V$, lemma \ref{lem:MMpMpp} shows that $|\Z^m\cap V/\inv_V(X)|.|\inv_V(X)/M|=|\Z^m\cap V/M|$. As $|\Z^m\cap V/M|$ is relatively prime with $r$, we deduce that $|\Z^m\cap V/\inv_V(X)|$ is relatively prime with $r$. Thus $P$ is relatively prime with $r$.
  \qed
\end{proof}

The class of semi-$V$-patterns relatively prime with $r$ that are also included into a $V$-affine space naturally appear when computing the inverse image of a semi-$V$-pattern by $\gamma_{r,m,\sigma}$ when $\sigma$ is a word enough longer in $\Sigma_{r,m}^*$ as proved by the following proposition \ref{prop:relaffpat}.
\begin{lemma}\label{lem:cyclicpattern}
  Any $(r,m,w)$-cyclic semi-$V$-pattern $P$ is relatively prime with $r$ and included in the $V$-affine space $A=\xi_{r,m}(w)+V$.
\end{lemma}
\begin{proof}
  As $P$ is $(r,m,w)$-cyclic, we deduce that $P=\gamma_{r,m,w^k}^{-1}(P)$ for any $k\in\Nat$. From proposition \ref{prop:invsemipattern}, we deduce that $P$ is relatively prime with $r$. Moreover, from proposition \ref{prop:saffcyclic}, we get $\saff(P)=\xi_{r,m}(w)+\vecsaff(P)$. As $P$ is a semi-$V$-pattern, we deduce that $\vecsaff(P)$ is either empty or equal to $V$. Hence $\saff(P)\subseteq \xi_{r,m}(w)+V$. From $P\subseteq \saff(P)$, we are done.
  \qed
\end{proof}

\begin{proposition}\label{prop:relaffpat}
  The class of semi-$V$-pattern relatively prime with $r$ and included into a $V$-affine space is stable by inverse image by $\gamma_{r,m,\sigma}$ for any $\sigma\in\Sigma_{r,m}^*$. Moreover, given a general semi-$V$-pattern $P$, there exists an integer $k\in\Nat$ such that $\gamma_{r,m,\sigma}^{-1}(P)$ is a semi-$V$-pattern relatively prime with $r$ and included into a $V$-affine space for any word $\sigma\in\Sigma_{r,m}^{\geq k}$.
\end{proposition}
\begin{proof}
  Let us first consider a semi-$V$-pattern $P$ relatively prime with $r$ and included into a $V$-affine space $A$, let $\sigma\in\Sigma_{r,m}^*$ and let us prove that $\gamma_{r,m,\sigma}^{-1}(P)$ is a semi-$V$-pattern relatively prime with $r$ and included into a $V$-affine space. Recall that we have previously proved that $\gamma_{r,m,\sigma}^{-1}(P)$ is a semi-$V$-pattern relatively prime with $r$. Since $P\subseteq A$, we deduce that $\gamma_{r,m,\sigma}^{-1}(P)\subseteq A'$ where $A'$ is the $V$-affine space $A'=\Gamma_{r,m,\sigma}^{-1}(A)$. We are done.
  
  Now, let us consider a general semi-$V$-pattern there exists an integer $k\in\Nat$ such that $\gamma_{r,m,\sigma}^{-1}(P)$ is a semi-$V$-pattern relatively prime with $r$ and included into a $V$-affine space for any word $\sigma\in\Sigma_{r,m}^{\geq k}$. Since $P$ is Presburger-definable, there exists a FDVA $\automaton$ that represents $P$ in basis $r$. Let us consider the integer $k=|\automaton|$ the number of principal states of $\automaton$. Now consider $\sigma\in\Sigma_{r,m}^*$. Since $|\sigma|\geq |\automaton|$, the word $\sigma$ can be decomposed in $\sigma=\sigma_1.\sigma_2$ such that there exists a loop $q\xrightarrow{w}q$ where $w\in\Sigma_{r,m}^+$ and $q=\delta(q_0,\sigma_1)$. As $P_q=\gamma_{r,m,\sigma_1}^{-1}(P)$ this set is a semi-$V$-pattern. Moreover, as $\gamma_{r,m,w}^{-1}(P_q)=P_q$, lemma \ref{lem:cyclicpattern} proves that $P_q$ is relatively prime $r$ and included in a $V$-affine space. Finally, as $\gamma_{r,m,\sigma}^{-1}(P)=\gamma_{r,m,\sigma_2}^{-1}(P_q)$, the previous paragraph shows that $\gamma_{r,m,\sigma}^{-1}(P)$ is relatively prime with $r$ and included into a $V$-affine space.
  \qed
\end{proof}

Given a non-empty semi-$V$-pattern $P$ included into a $V$-affine space $A$, we naturally deduce that $\gamma_{r,m,\sigma}^{-1}(A)=\emptyset$ implies $\gamma_{r,m,\sigma}^{-1}(P)=\emptyset$. The class of semi-$V$-pattern relatively prime $r$ that are included into a $V$-affine space plays an important role since the following corollary \ref{cor:densepattern} intensively used in the sequel proved that for this class, the converse is true: $\gamma_{r,m,\sigma}^{-1}(P)=\emptyset$ implies $\gamma_{r,m,\sigma}^{-1}(A)=\emptyset$.

\begin{lemma}\label{lem:invdensepattern}
  Let $P$ be a semi-$V$-pattern relatively prime with $r$ and included into a $V$-affine space $A$. We have $\gamma_{r,m,\sigma}^{-1}(P)=\xi_{r,m}(s)+P-\rho_{r,m}(\sigma,s)$ for any semi-$V$-pattern $P$ relatively prime with $r$ and included into a $V$-affine space $A$ and for any $(r,m)$-decomposition $(\sigma,s)$ such that $\rho_{r,m}(\sigma,s)\in A$ and such that $r^{|\sigma|_m}\in 1+|\Z^m\cap V/\inv_V(P)|.\Z$.
\end{lemma}
\begin{proof}
  Let us consider $x\in\gamma_{r,m,\sigma}^{-1}(P)$. We have $\gamma_{r,m,\sigma}(x)=r^{|\sigma|_m}.x+\rho_{r,m}(\sigma,s)$. Hence $r^{|\sigma|_m}.(x-\xi_{r,m}(s))\in P-\rho_{r,m}(\sigma,s)$. In particular, from $P\subseteq A$ and $\rho_{r,m}(\sigma,s)\in A$, we deduce that $r^{\sigma}.(x-\xi_{r,m}(s))\in V$. Hence $x-\xi_{r,m}(s)\in\Z^m\cap V$. From $r^{|\sigma|_m}\in 1+|\Z^m\cap V/\inv_V(P)|$, we deduce that $(r^{|\sigma|_m}-1).(x-\xi_{r,m}(s))\in \inv_V(P)$. As $x-\xi_{r,m}(s)\in P-(r^{|\sigma|_m}-1).(x-\xi_{r,m}(s))-\rho_{r,m}(\sigma,s)$, we get $x\in \xi_{r,m}(s)+P-\rho_{r,m}(\sigma,s)$ and we have proved the inclusion $\gamma_{r,m,\sigma}^{-1}(P)\subseteq \xi_{r,m}(s)+P-\rho_{r,m}(\sigma,s)$. For the converse inclusion, let $x\in \xi_{r,m}(s)+P-\rho_{r,m}(\sigma,s)$. From $\gamma_{r,m,\sigma}(x)=r^{|\sigma|_m}.(x-\xi_{r,m}(s))+\rho_{r,m}(\sigma,s)$, we deduce that there exists $p\in P$ such that $\gamma_{r,m,\sigma}(x)=r^{|\sigma|_m}.(p-\rho_{r,m}(\sigma,s))+\rho_{r,m}(\sigma,s)$. Hence $\gamma_{r,m,\sigma}(x)=p+(r^{|\sigma|_m}-1).(p-\rho_{r,m}(\sigma,s))$. As $\rho_{r,m}(\sigma,s)$ and $p$ are both in $A$, we deduce that $p-\rho_{r,m}(\sigma,s)\in\Z^m\cap V$. Moreover, as $r^{|\sigma|_m}-1\in |\Z^m\cap V/\inv_V(X)|.\Nat$, we deduce that $(r^{|\sigma|_m}-1).(p-\rho_{r,m}(\sigma,s))\in\inv_V(P)$. From $p\in P$, we get $\gamma_{r,m,\sigma}(x)\in P$ and we have proved the other inclusion $\xi_{r,m}(s)+P-\rho_{r,m}(\sigma,s)\subseteq\gamma_{r,m,\sigma}^{-1}(P)$.
  \qed
\end{proof}

\begin{corollary}[Dense pattern corollary]\label{cor:densepattern}
  Let $P$ be a non-empty semi-$V$-pattern relatively prime with $r$ and included into a $V$-affine space $A$. The set $\gamma_{r,m,\sigma}^{-1}(P)$ is a non-empty semi-$\Gamma_{r,m,0}^{-|\sigma|}(V)$-pattern relatively prime with $r$ and included into the $\Gamma_{r,m,0}^{-|\sigma|}(V)$-affine space $\Gamma_{r,m,\sigma}^{-1}(A)$ for any word $\sigma\in\Sigma_r^*$ such that $\gamma_{r,m,\sigma}^{-1}(A)\not=\emptyset$. 
\end{corollary}
\begin{proof}
  As $\gamma_{r,m,\sigma}^{-1}(A)$ is non empty, there exists a couple  $(w,s)$ such that $\rho_{r,m}(w,s)\in \gamma_{r,m,\sigma}^{-1}(A)$ and such that $|\sigma|+|w|\in m.\Z$. By replacing $w$ by a word in $w.s^*$, we can assume without loss of generality that $r^{|\sigma.w|_m}\in 1+|\Z^m\cap V/\inv_V(P)|.\Z$. From lemma \ref{lem:invdensepattern}, we deduce that $\gamma_{r,m,\sigma.w}^{-1}(P)=\xi_{r,m}(s)+P-\rho_{r,m}(\sigma.w,s)$. As $\gamma_{r,m,\sigma.w}^{-1}(P)=\gamma_{r,m,w}^{-1}(\gamma_{r,m,\sigma}^{-1}(P))$ and $\gamma_{r,m,\sigma.w}^{-1}(P)\not=\emptyset$, we deduce that $\gamma_{r,m,\sigma}^{-1}(P)\not=\emptyset$. From proposition \ref{prop:invsemipattern} we deduce that $\gamma_{r,m,\sigma}^{-1}(P)$ is a semi-$\Gamma_{r,m,0}^{-|\sigma|}(V)$-pattern. Let us now show that $\gamma_{r,m,\sigma}^{-1}(P)$ is relatively prime with $r$. Since $P\subseteq A$, we deduce that $\gamma_{r,m,\sigma}^{-1}(P)$ is included in the $\Gamma_{r,m,0}^{-|\sigma|}(V)$-affine space $\Gamma_{r,m,\sigma}^{-1}(A)$. Now, let us prove that $\gamma_{r,m,\sigma}^{-1}(P)$ is relatively prime with $r$. From proposition \ref{prop:invsemipattern} we deduce that $\gamma_{r,m,0}^{-|\sigma|}(\inv_V(P))\subseteq \inv_V(\gamma_{r,m,\sigma}^{-1}(P))$. As $\inv_V(P)$ is relatively prime with $r$ we get $\gamma_{r,m,0}^{-\infty}(\inv_V(P))=\inv_V(P)$. Hence $\gamma_{r,m,0}^{-\infty}(\gamma_{r,m,0}^{-|\sigma|}(\inv_V(P)))=\gamma_{r,m,0}^{-|\sigma|}(\gamma_{r,m,0}^{-\infty}(\inv_V(P)))=\gamma_{r,m,0}^{-|\sigma|}(\inv_V(P))$ and we have proved that $\gamma_{r,m,0}^{-|\sigma|}(\inv_V(P))$ is relatively prime with $r$. From the inclusion $\gamma_{r,m,0}^{-|\sigma|}(\inv_V(P))\subseteq \inv_V(\gamma_{r,m,\sigma}^{-1}(P))$ and lemma \ref{lem:MMpMpp}, we deduce that $\inv_V(\gamma_{r,m,\sigma}^{-1}(P))$ is relatively prime with $r$. We are done.
  \qed
\end{proof}

%% file: chapter.degenerate.tex
\chapter{Degenerate Sets}
Given a vector space $V$, a subset $X\subseteq \Q^m$ is said \emph{$V$-degenerate}\index{Gen}{degenerate} if $V$ is not included in $\vecsaff(\Z^m\cap X)$. The following lemma \ref{lem:simVequi} shows that the binary relation $\sim^V$\index{Not}{$\sim^V$} defined over the subsets of $\Q^m$ by $X_1\sim^V X_2$ if and only if $X_1\Delta X_2$ is $V$-degenerate, is an equivalence relation. The equivalence class for $\sim^V$ of a subset $X\subseteq \Q^m$ is denote by $[X]^V$\index{Not}{$[X]^V$}.
\begin{lemma}\label{lem:simVequi}
  The binary relation $\sim^V$ is an equivalence.
\end{lemma}
\begin{proof}
  The binary relation $\sim^V$ is an equivalence relation. Naturally $\sim^V$ is reflexive and symmetric. So, it is sufficient to prove that $\sim^V$ is transitive. Consider $X_1,X_2,X_3\subseteq \Q^m$ such that $X_1\sim^VX_2$ and $X_2\sim^V X_3$ and let us prove that $X_1\sim^V X_3$. We have $\Z^m\cap(X_1\Delta X_3)\subseteq (\Z^m\cap (X_1\Delta X_2))\cup (\Z^m\cap(X_2\Delta X_3))$ and from insecable lemma \ref{lem:insecable}, we deduce that $V$ is not included in $\vecsaff(\Z^m\cap (X_1\Delta X_3))$. Hence $X_1\sim^V X_3$.
  \qed
\end{proof}
  
Given two equivalence classes $\X_1$ and $\X_2$ and a boolean operation $\#\in\{\cup,\cap,\moins,\Delta\}$, the following lemma \ref{lem:simVX1X2} shows that $[X_1\# X_2]^V$ is independent of $X_1\in\X_1$ and $X_2\in\X_2$. This equivalence class is naturally denoted by $\X_1\#^V\X_2$\index{Not}{$\cup^V$}\index{Not}{$\cap^V$}\index{Not}{$\moins^V$}\index{Not}{$\Delta^V$}.
\begin{lemma}\label{lem:simVX1X2}
  We have $[X_1\#X_2]^V=[X_1'\#X_2']^V$ for any $X_1,X_1',X_2,X_2'\subseteq\Q^m$ such that $X_1\sim^V X_1'$ and $X_2\sim^V X_2'$ and for any $\#\in\{\cup,\cap,\moins,\Delta\}$. 
\end{lemma}
\begin{proof}
  Let us prove that $(X_1\#X_2)\Delta (X_1'\#X_2')\subseteq (X_1\Delta X'_1)\#(X_2\Delta X'_2)$ for any $X_1,X_1',X_2,X_2'\subseteq \Q^m$ and for any $\#\in\{\cup,\cap,\moins,\Delta\}$. 

  Case $\#$ equals to $\Delta$: in this case, we have the equality $(X_1\#X_2)\Delta (X_1'\#X_2')=(X_1\Delta X'_1)\#(X_2\Delta X'_2)$ and we are done.

  Case $\#$ equals to $\cap$: we have $(X_1\#X_2)\Delta (X_1'\#X_2')=((X_1\cap X_2)\moins (X_1'\cap X_2'))\cup ((X_1'\cap X_2')\moins (X_1\cap X_2))$. Remark that $(X_1\cap X_2)\moins (X_1'\cap X_2')=((X_1\cap X_2)\moins X_1')\cup  ((X_1\cap X_2)\moins X_2')$. From  $(X_1\cap X_2)\moins X_1'\subseteq X_1\moins X_1'$ and $(X_1\cap X_2)\moins X_2'\subseteq X_2\moins X_2'$, we deduce that $(X_1\cap X_2)\moins (X_1'\cap X_2')\subseteq (X_1\moins X_1')\cup (X_2\moins X_2')\subseteq (X_1\Delta X_1')\cup (X_2\Delta X_2')$. By symmetry, we also get $(X_1'\cap X_2')\moins (X_1\cap X_2)\subseteq (X_1\Delta X_1')\cup (X_2\Delta X_2')$. We are done.

  Case $\#$ equals to $\moins$: this case can be reduced to the previous case $\cap$. In fact, if $\#$ is equal to $\moins$ then $(X_1\#X_2)\Delta (X_1'\#X_2')=(X_1\cap (\Q^m\moins X_2))\Delta (X_1'\cap (\Q^m\moins X_2'))$. From the previous case $\cap$, we deduce that $ (X_1\cap (\Q^m\moins X_2))\Delta (X_1'\cap (\Q^m\moins X_2'))\subseteq (X_1\Delta X_1')\cup ((\Q^m\moins X_2)\Delta (\Q^m\cap X_2')$. As $(\Q^m\moins X_2)\Delta (\Q^m\cap X_2')=X_2\Delta X_2'$, we are done.
  
  Case $\#$ equals to $\cup$: we have $(X_1\#X_2)\Delta (X_1'\#X_2')=((X_1\cup X_2)\moins (X_1'\cup X_2'))\cup ((X_1'\cup X_2')\moins (X_1\cup X_2))$. Remark that $(X_1\cup X_2)\moins (X_1'\cup X_2')=(X_1\moins (X_1'\cup X_2'))\cup  (X_2\moins (X_1'\cup X_2'))$. From  $X_1\moins (X_1'\cup X_2')\subseteq X_1\moins X_1'$ and $X_2\moins (X_1'\cup X_2')\subseteq X_2\moins X_2'$, we deduce that $(X_1\cup X_2)\moins (X_1'\cup X_2')\subseteq (X_1\moins X_1')\cup (X_2\moins X_2')\subseteq (X_1\Delta X_1')\cup (X_2\Delta X_2')$. By symmetry, we also get $(X_1'\cup X_2')\moins (X_1\cup X_2)\subseteq (X_1\Delta X_1')\cup (X_2\Delta X_2')$. We are done.
  
  From insecable lemma \ref{lem:insecable}, we deduce that if $X_1\sim^V X_1'$ and $X_2\sim^V X_2'$ then $X_1\#X_2\sim^V X_1'\# X_2'$ for any $\#\in\{\cup,\cap,\moins,\Delta\}$.
  \qed
\end{proof}

For any equivalence class $\X$ and for any word $\sigma\in\Sigma_{r,m}^*$, following lemma \ref{lem:simVgamma} shows that the equivalence class $[\gamma_{r,m,\sigma}^{-1}(X)]^V$ does not depend on $X\in\X$. This equivalence class is denoted by $\gamma_{r,m,\sigma}^{-1}(\X)$\index{Not}{$\gamma_{r,m,\sigma}^{-1}(\X)$}.
\begin{lemma}\label{lem:simVgamma}
  We have $\gamma_{r,m,\sigma}^{-1}(X)\sim^V\gamma_{r,m,\sigma}^{-1}(X')$ for any $X,X'\subseteq\Q^m$ such that $X\sim^V X'$, and for any $\sigma\in\Sigma_{r,m}^*$.
\end{lemma}
\begin{proof}
  Consider $X,X'\subseteq\Q^m$ such that $X\sim^V X'$. We denote by $Z=\Z^m\cap (X\Delta X')$. As $X\sim^V X'$, the vector space $V$ is not included in $\vecsaff(Z)$. We have $\Z^m\cap (\gamma_{r,m,\sigma}^{-1}(X_1)\Delta\gamma_{r,m,\sigma_2}^{-1}(X_2))=\gamma_{r,m,\sigma}^{-1}(Z)$. From covering lemma we get $\saff(\gamma_{r,m,\sigma}^{-1}(Z))\subseteq \Gamma_{r,m,\sigma}^{-1}(\saff(Z))$. By considering the direction of the previous inclusion, we get $\vecsaff(\gamma_{r,m,\sigma}^{-1}(Z))\subseteq \vecsaff(Z)$ since $\Gamma_{r,m,\sigma}(x)=r^{|\sigma|}.x+\gamma_{r,m,\sigma}(\unit_{0,m})$. As $V$ is not included in $\vecsaff(Z)$, we deduce that $V$ is neither included in $\vecsaff(\Z^m\cap (\gamma_{r,m,\sigma}^{-1}(X_1)\Delta\gamma_{r,m,\sigma_2}^{-1}(X_2)))$. Therefore $\gamma_{r,m,\sigma}^{-1}(X)\sim^V\gamma_{r,m,\sigma}^{-1}(X')$.
\end{proof}

The following lemma \ref{lem:simVgammadi} provides a commutativity result.
\begin{lemma}\label{lem:simVgammadi}
  We have $\gamma_{r,m,\sigma}^{-1}(\X_1\#^V\X_2)=\gamma_{r,m,\sigma}^{-1}(\X_1)\#^V\gamma_{r,m,\sigma}^{-1}(\X_2)$ for any equivalence class $\X_1$ and $\X_2$, for any $\#\in\{\cup,\cap,\moins,\Delta\}$, and for any $\sigma\in\Sigma_{r,m}^*$.
\end{lemma}
\begin{proof}
  Consider $X_1\in \X_1$ and $X_2\in\X_2$. We have $\gamma_{r,m,\sigma}^{-1}(\X_1\#^V\X_2)=[\gamma_{r,m,\sigma}^{-1}(X_1\# X_2)]^V=[\gamma_{r,m,\sigma}^{-1}(X_1)\#\gamma_{r,m,\sigma}^{-1}(X_2)]^V=[\gamma_{r,m,\sigma}^{-1}(X_1)]^V\#^V[\gamma_{r,m,\sigma}^{-1}(X_2)]^V=\gamma_{r,m,\sigma}^{-1}(\X_1)\#^V\gamma_{r,m,\sigma}^{-1}(\X_2)$.
  \qed
\end{proof}

%% file: chapter.polyhedrons.tex
\chapter{Polyhedrons}\label{sub:polyhedron}
In this section, we recall the definition of a polyhedron and associate to a polyhedron $C$ included into a vector space $V$, a boundary that only depends on the equivalence class $[C]^V$.

\begin{figure}[htbp]
  \begin{center}
    \setlength{\unitlength}{16pt}
    \pssetlength{\psunit}{16pt}
    \pssetlength{\psxunit}{16pt}
    \pssetlength{\psyunit}{16pt}
    \begin{picture}(10,10)(-5,-5)%
      \put(-5,-5){\framebox(10,10){}}
      \multido{\iii=-4+1}{9}{\psline[linecolor=lightgray,linewidth=1pt](-4,\iii)(4,\iii)}\multido{\iii=-4+1}{9}{\psline[linecolor=lightgray,linewidth=1pt](\iii,-4)(\iii,4)}
      \begin{psclip}{\pspolygon*[linecolor=lightgray,fillcolor=lightgray](-4,-4)(-4,4)(4,-4)}%
        \multido{\iii=-4+1}{9}{\psline[linecolor=gray,linewidth=1pt](-4,\iii)(4,\iii)}\multido{\iii=-4+1}{9}{\psline[linecolor=gray,linewidth=1pt](\iii,-4)(\iii,4)}%
      \end{psclip}%
      \psline[linecolor=gray,arrowscale=1,linewidth=1pt]{->}(0,-4.5)(0,4.5)\psline[linecolor=gray,arrowscale=1,linewidth=1pt]{->}(-4.5,0)(4.5,0)
      \psline[linecolor=black,linewidth=2pt,arrowscale=1]{|->}(0,0)(1,1)
      \psline[linecolor=black,linewidth=1pt](-4,4)(4,-4)
      \put(0.1,0.8){$\alpha$}
      \put(-3,-1.8){$\scalar{\alpha}{x}<0$}
    \end{picture}
    \begin{picture}(10,10)(-5,-5)%
      \put(-5,-5){\framebox(10,10){}}
      \multido{\iii=-4+1}{9}{\psline[linecolor=lightgray,linewidth=1pt](-4,\iii)(4,\iii)}\multido{\iii=-4+1}{9}{\psline[linecolor=lightgray,linewidth=1pt](\iii,-4)(\iii,4)}
      \begin{psclip}{\pspolygon*[linecolor=lightgray,fillcolor=lightgray](4,4)(-4,4)(4,-4)}%
        \multido{\iii=-4+1}{9}{\psline[linecolor=gray,linewidth=1pt](-4,\iii)(4,\iii)}\multido{\iii=-4+1}{9}{\psline[linecolor=gray,linewidth=1pt](\iii,-4)(\iii,4)}%
      \end{psclip}%
      \psline[linecolor=gray,arrowscale=1,linewidth=1pt]{->}(0,-4.5)(0,4.5)\psline[linecolor=gray,arrowscale=1,linewidth=1pt]{->}(-4.5,0)(4.5,0)
      \psline[linecolor=black,linewidth=2pt,arrowscale=1]{|->}(0,0)(1,1)
      \psline[linecolor=black,linewidth=1pt](-4,4)(4,-4)
      \put(0.1,0.8){$\alpha$}
      \put(0.5,2.2){$\scalar{\alpha}{x}>0$}
    \end{picture}
  \end{center}
  \caption{Let $\alpha=(1,1)$. On the left $\{x\in\Q^2;\;\scalar{\alpha}{x}< 0\}$. On the right $\{x\in\Q^2;\;\scalar{\alpha}{x}> 0\}$.\label{fig:halfspace}}
\end{figure}

\section{Orientation}
A \emph{$V$-hyperplane}\index{Gen}{hyperplane} $H$, where $V$ is a vector space, is a set of the form $\{x\in V;\;\scalar{\alpha}{x}=c\}$ where $(\alpha,c)\in (V\moins\{\unit_{0,m}\})\times\Q$. A $V$-hyperplane $H$ provides a partition of $V\moins H$ into two \emph{open $V$-half spaces}\index{Gen}{half-space} $\{x\in V;\;\scalar{\alpha}{x}<c\}$ and $\{x\in V;\;\scalar{\alpha}{x}>c\}$ that only depends on the $V$-hyperplane $H$ (see figures \ref{fig:halfspace}).

An \emph{orientation}\index{Gen}{orientation} $o$ is a function that associate to any couple $(V,H)$ where $H$ is a $V$-hyperplane, one of these two open $V$-half spaces. Given an implicit orientation $o$, we denote by $(V,H)^>$\index{Not}{$(V,H)^\#$} and $(V,H)^<$ the open $V$-half spaces $(V,H)^>=o(V,H)$ and $(V,H)^<=(V\moins H)\moins o(V,H)$. We denote by $(V,H)^=$ the hyperplane $H$ and the \emph{closed $V$-half spaces} $(V,H)^{\geq}$ and $(V,H)^{\leq}$ are naturally defined by $(V,H)^{\geq}=H\cup (V,H)^>$ and $(V,H)^{\leq}=H\cup (V,H)^<$.

Remark that a $V$-hyperplane $H$ is an affine space and in particular $\vec{H}$ is well defined. Moreover, $\vec{H}$ is also a $V$-hyperplane. Remark that $H+(V,\vec{H})^>$ is an open half space of the form $(V,H)^\#$ where $\#\in\{<,>\}$ depends on $H$. A \emph{uniform orientation}\index{Gen}{uniform orientation} is an orientation that only depends on the direction of the $V$-hyperplane $\vec{H}$: we have $(V,H)^\#=H+(V,\vec{H})^\#$ for any $\#\in\{\leq,<,=,>,\geq\}$.

In the remaining of this paper, we assume fixed a uniform orientation (see remark \ref{rem:unif} for the existence of such an \emph{effective and efficient} orientation). Moreover when $V$ is implicit, the set $(V,H)^\#$ is simply written $H^\#$\index{Not}{$H^\#$}.
\begin{remark}\label{rem:unif}
  Consider the function $o$ that associate to any $(V,H)$ where $H$ is a $V$-hyperplane, the open $V$-half space $H+(\Q_+\moins\{0\}).\Pi_{V}(\unit_i)$ where $i\in\finiteset{1}{m}$ is the least (for $\leq$) integer such that $\Pi_{V}(\unit_i)\not\in \vec{H}$. Remark that such an integer $i$ exists because if $\Pi_{V}(\unit_i)\in \vec{H}$ for any $i\in\finiteset{1}{m}$, then $V\subseteq \vec{H}$ which is impossible. Remark that $o$ is an uniform orientation \emph{computable in polynomial time}. 
\end{remark}

\section{$V$-polyhedral equivalence class}
Recall that a \emph{polyhedron}\index{Gen}{polyhedron} $C$ of $\Q^m$ is a boolean combination in $\Q^m$ of sets $H^\#$ where $H$ is a $\Q^m$-hyperplane and $\#\in\{\leq,<,=,>,\geq\}$. A \emph{$V$-polyhedron} $C$ is a polyhedron included into a vector space $V$. A polyhedron $C$ is said \emph{$(V,\Hrond)$-definable}\index{Gen}{definable polyhedron}, where $\Hrond$ is a finite set of $V$-hyperplanes if $C$ is a boolean combination in $V$ of sets in $\{H^\#;\;(H,\#)\in\Hrond\times\{\leq,<,=,>,\geq\}\}$.
\begin{lemma}\label{lem:vpoly}
  A polyhedron is a $V$-polyhedron if and only if it is $(V,\Hrond)$-definable for a finite set $\Hrond$ of $V$-hyperplanes.
\end{lemma}
\begin{proof}
  Naturally, if $C$ is $(V,\Hrond)$-definable then $C$ is a $V$-polyhedron. For the converse, consider a $V$-polyhedron $C$. By definition $C\subseteq V$ and there exists $D_0\in\partie_f(\Q^m\moins\{\unit_{0,m}\})$ and $K\in\partie_f(\Q)$ such that $C$ is a boolean combination in $\Q^m$ of sets $\{x\in\Q^m;\;\scalar{\alpha_0}{x}\#c\}$ where $(\alpha_0,c)\in D_0\times K$ and $\#\in \{\leq,<,=,>,\geq\}$. From $C\subseteq V$ we deduce that $C=C\cap V$ and in particular $C$ is a boolean combination in $V$ of sets $\{x\in V;\;\scalar{\alpha_0}{x}\#c\}=\{x\in V;\;\scalar{\Pi_{V}(\alpha_0)}{x}\#c\}$. Let $D=\Pi_V(D_0)\moins\{\unit_{0,m}\}$ and 
  consider the set of $V$-hyperplanes $\Hrond=\{\{x\in V;\;\scalar{\alpha}{x}=c\};\;(\alpha,c)\in D\times K\}$ and let us prove that $C$ is $(V,\Hrond)$-definable. Let $(\alpha_0,c)\in D_0\times K$. Remark 
  that $\{x\in  V;\;\scalar{\Pi_{V}(\alpha_0)}{x}\#c\}$ is either  
  empty or equal to $V$ in the case
  $\Pi_V(\alpha_0)=\unit_{0,m}$, 
  or it is in the class $\{H^\#;\;(H,\#)\in\Hrond\times\{\leq,<,=,>,\geq\}\}$ 
  if $\Pi_V(\alpha_0)\not=\unit_{0,m}$.
  \qed
\end{proof}

\begin{definition}
  A \emph{$V$-polyhedral equivalence class}\index{Gen}{polyhedral equivalence class} $\class$ is the equivalence class for $\sim^V$ of a $V$-polyhedron.
\end{definition}

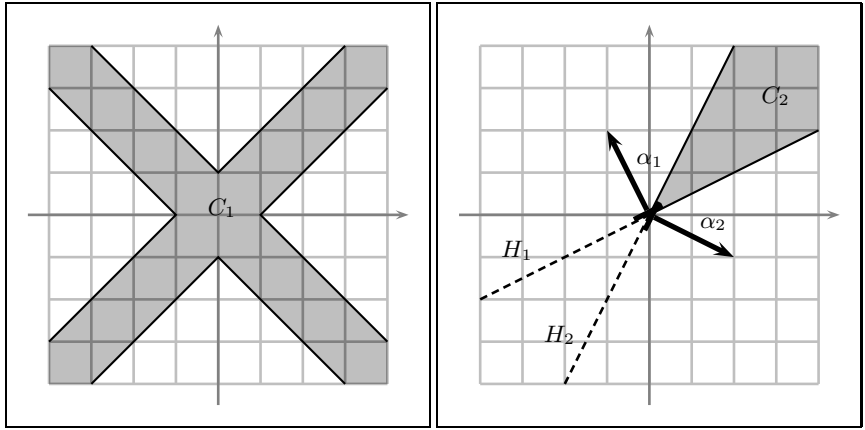
\begin{figure}[htbp]
  \begin{center}
    \setlength{\unitlength}{8pt}
    \pssetlength{\psunit}{8pt}
    \pssetlength{\psxunit}{8pt}
    \pssetlength{\psyunit}{8pt}
    \begin{picture}(20,20)(-10,-10)%
      \put(-10,-10){\framebox(20,20){}}
      \multido{\iii=-8+2}{9}{\psline[linecolor=lightgray,linewidth=1pt](-8,\iii)(8,\iii)}\multido{\iii=-8+2}{9}{\psline[linecolor=lightgray,linewidth=1pt](\iii,-8)(\iii,8)}\psline[linecolor=gray,arrowscale=1,linewidth=1pt]{->}(0,-9)(0,9)\psline[linecolor=gray,arrowscale=1,linewidth=1pt]{->}(-9,0)(9,0)
      \begin{psclip}{\pspolygon*[linecolor=lightgray,fillcolor=lightgray](-8,-8)(-8,-6)(6,8)(8,8)(8,6)(-6,-8)}%
        \multido{\iii=-8+2}{9}{\psline[linecolor=gray,linewidth=1pt](-8,\iii)(8,\iii)}\multido{\iii=-8+2}{9}{\psline[linecolor=gray,linewidth=1pt](\iii,-8)(\iii,8)}%
      \end{psclip}%
      \begin{psclip}{\pspolygon*[linecolor=lightgray,fillcolor=lightgray](-8,8)(-8,6)(6,-8)(8,-8)(8,-6)(-6,8)}%
        \multido{\iii=-8+2}{9}{\psline[linecolor=gray,linewidth=1pt](-8,\iii)(8,\iii)}\multido{\iii=-8+2}{9}{\psline[linecolor=gray,linewidth=1pt](\iii,-8)(\iii,8)}%
      \end{psclip}%
      \psline(-8,-6)(-2,0)(-8,6)
      \psline(8,-6)(2,0)(8,6)
      \psline(-6,-8)(0,-2)(6,-8)
      \psline(-6,8)(0,2)(6,8)
      \put(-0.5,0){$C_1$}
    \end{picture}
     \begin{picture}(20,20)(-10,-10)%
      \put(-10,-10){\framebox(20,20){}}
      \multido{\iii=-8+2}{9}{\psline[linecolor=lightgray,linewidth=1pt](-8,\iii)(8,\iii)}\multido{\iii=-8+2}{9}{\psline[linecolor=lightgray,linewidth=1pt](\iii,-8)(\iii,8)}%
      \begin{psclip}{\pspolygon*[linecolor=lightgray,fillcolor=lightgray](0,0)(8,4)(8,8)(4,8)}%
       \multido{\iii=-8+2}{9}{\psline[linecolor=gray,linewidth=1pt](-8,\iii)(8,\iii)}\multido{\iii=-8+2}{9}{\psline[linecolor=gray,linewidth=1pt](\iii,-8)(\iii,8)}%
      \end{psclip}%
      \psline[linecolor=gray,arrowscale=1,linewidth=1pt]{->}(0,-9)(0,9)\psline[linecolor=gray,arrowscale=1,linewidth=1pt]{->}(-9,0)(9,0)
      \psline(4,8)(0,0)(8,4)
      \psline[linewidth=1pt,linestyle=dashed,dash=3pt 2pt](-8,-4)(0,0)
      \psline[linewidth=1pt,linestyle=dashed,dash=3pt 2pt](-4,-8)(0,0)
      \psline[linecolor=black,arrowscale=1,linewidth=2pt]{|->}(0,0)(-2,4)
      \psline[linecolor=black,arrowscale=1,linewidth=2pt]{|->}(0,0)(4,-2)    
      \put(5.3,5.3){$C_2$}
      \put(-7,-2){$H_1$}
      \put(-5,-6){$H_2$}
      \put(-0.6,2.4){$\alpha_1$}
      \put(2.4,-0.6){$\alpha_2$}
     \end{picture}
  \end{center}
  \caption{Let $V=\Q^2$. On the left a $V$-degenerate $V$-polyhedron $C_1$. On the right a non $V$-degenerate $V$-polyhedron $C_2$.\label{fig:polyhedron}}
\end{figure}

\section{Open convex polyhedrons}
A $V$-polyhedron $C$ is said \emph{open convex}\index{Gen}{open convex} in $V$ (or just open convex when $V$ is implicitly known) if it is equal to a finite intersection of open $V$-half spaces (in particular $V$ is an open convex). 

\begin{definition}
  Given a finite set $\H$ of $V$-hyperplanes and a sequence $\#\in\{<,>\}^\H$, we denote by $C_{V,\#}$\index{Not}{$C_{V,\#}$} the open convex $V$-polyhedron $C_{V,\#}=\bigcap_{H\in\H}H^{\#_H}$ (if $\H=\emptyset$, then $C_{V,\#}=V$).
\end{definition}

Given a $(V,\Hrond)$-definable polyhedron $C$, remark that $C\moins(\bigcup_{H\in\Hrond}H)$ is a finite union of open convex polyhedrons $C_{V,\#}$ where $\#\in\{<,>\}^\Hrond$. As $[C]^V=[C\moins(\bigcup_{H\in\Hrond}H)]^V$, this property will be useful for decomposing $V$-polyhedrons.

\section{Degenerate polyhedrons}
We geometrically characterize the $V$-degenerate $V$-polyhedrons (see figure \ref{fig:polyhedron}) thanks to the following proposition \ref{prop:vdegpoly}.

We first prove the following two lemmas \ref{lem:degex} and \ref{lem:vdegpolyconv}.
\begin{lemma}\label{lem:degex}
  For any $V$-hyperplanes $H_1,H_2$ such that $\vec{H_1}=\vec{H_2}$, the open convex $V$-polyhedron $H_1^>\cap H_2^<$ is $V$-degenerate. 
\end{lemma}
\begin{proof}
  Let $\alpha\in \Z^m\cap V\moins\{\unit_{0,m}\}$ and $c_1,c_2\in\Q$ such that $H_1^>=\{x\in V;\;\scalar{\alpha}{x}>c_1\}$ and such that $H_2^<=\{x\in V;\;\scalar{\alpha}{x}<c_2\}$. Let us prove that $C=H_1^>\cap H_2^<$ is $V$-degenerate. Let $K=\{k\in\Z;\; c_1<k<c_2\}$ and remark that for any $x\in \Z^m\cap C$ we have $c_1<\scalar{\alpha}{x}<c_2$ and $\scalar{\alpha}{x}\in\Z$. Hence, there exists $k\in K$ such that $\scalar{\alpha}{x}\in K$. We deduce that $\Z^m\cap C\subseteq\bigcup_{k\in K}H_k$ where $H_k$ is the $V$-hyperplane $H_k=\{x\in V;\;\scalar{\alpha}{x}=k\}$. Hence $\vecsaff(\Z^m\cap C)\subseteq \{x\in V;\;\scalar{\alpha}{x}=0\}$. As $\alpha$ is in $V$ but not in this semi-vector space, we deduce that $V$ is not included in $\vecsaff(\Z^m\cap C)$. Hence $C$ is $V$-degenerate.
  \qed
\end{proof}

\begin{lemma}\label{lem:vdegpolyconv}
  We have $[C_{V,\#}]^V\not=[\emptyset]^V$ if and only if $\bigcap_{H\in\Hrond}\vec{H}^{\#_H}\not=\emptyset$, for any $\#\in\{<,>\}^{\H}$ where $\H$ is a finite set of $V$-hyperplanes. 
\end{lemma}
\begin{proof}
  Let us consider a sequence $(\alpha_H,c_H)_{H\in\Hrond}$ of elements in $(V\moins\{\unit_{0,m}\})\times\Q$ such that $H^{\#_H}=\{x\in V;\;\scalar{\alpha_H}{x}>c_H\}$, and let $C=\bigcap_{H\in\Hrond}H^{\#_H}$.
  
  Assume first that $\bigcap_{H\in\Hrond}\vec{H}^{\#_H}\not=\emptyset$ and let us prove that $C$ is non $V$-degenerate. Consider a vector $v$ in this open convex $V$-polyhedron and remark that $\scalar{\alpha_H}{v}> 0$ for every $H\in\Hrond$. By replacing $v$ by a vector in $(\Nat\moins\{0\}).v$, we can assume that $v\in\Z^m\cap V$. Let us first show that there exists $x_0\in\Z^m\cap C$. In fact, there exists $k\in\Nat$ enough larger such that $\scalar{\alpha_H}{k.v}>c_H$ for any $H\in \Hrond$. For such a $k$, just remark that $x_0=k.v\in \Z^m\cap C$. Next, let us prove that there exists a finite set $V_0$ of vectors in $\Z^m$ that generates $V$ and such that $\scalar{\alpha_H}{v_0}>0$ for any $(v_0,H)\in V_0\times\Hrond$. We know that there exists a finite set $V_0$ of vectors in $\Z^m$ that generates $V$. By replacing $V_0$ by $V_0+k.v$ where $k\in\Nat$ is enough larger, we can assume that $\scalar{\alpha_H}{v_0}>0$ for any $(v_0,H)\in V_0\times\Hrond$. We have proved that $x_0+\sum_{v_0\in V_0}\Nat.v_0\subseteq \Z^m\cap C$. From covering lemma \ref{lem:cov}, we get $\saff(x_0+\sum_{v_0\in V_0}\Nat.v_0)=x_0+V$. Hence $V\subseteq \vecsaff(\Z^m\cap C)$. Therefore $C$ is non $V$-degenerate.
  
  Now, assume that $\bigcap_{H\in\Hrond}\vec{H}^{\#_H}=\emptyset$. Hence, for any $v\in V$, there exists $H\in\Hrond$ such that $\scalar{\alpha_H}{v}\leq 0$. In particular for any $v\in C$, there exists $H\in \Hrond$ such that $c_H<\scalar{\alpha_H}{v}\leq 0$. Lemma \ref{lem:degex} shows that $C$ is $V$-degenerate.
  \qed
\end{proof}

\begin{proposition}\label{prop:vdegpoly}
  A $V$-polyhedron is $V$-degenerate if and only if it is included into a finite union of $H_1^>\cap H_2^<$ where $H_1$ and $H_2$ are two $V$-hyperplanes with the same direction. 
\end{proposition}
\begin{proof}
  As a finite union of $V$-degenerate subsets of $V$ remains $V$-degenerate, we deduce from lemma \ref{lem:degex} that if a $V$-polyhedron is included into a finite union of $H_1^>\cap H_2^<$ where $H_1$ and $H_2$ are two $V$-hyperplanes with the same direction, then it is $V$-degenerate.
  
  For the converse consider a $V$-polyhedron $C$ such that for any finite set $D\subseteq V\moins\{\unit_{0,m}\}$, the $V$-polyhedron $C$ is not included in $\bigcup_{\alpha\in D}\{x\in V;\;-1< \scalar{\alpha}{x}< 1\}$. Let $\Hrond$ be a finite set of $V$-hyperplanes such that $C$ is $(V,\Hrond)$-definable. Recall that $C'=C\moins(\bigcup_{H\in\Hrond}H)$ is a finite union of open convex definable polyhedron $C_{V,\#}$ where $\#\in\{<,>\}^\H$ and it satisfies $[C]^V=[C']^V$. So, we can assume without loss of generality that $C=C_{V,\#}$. Consider a sequence $(\alpha_H,c_H)_{H\in\Hrond}$ of elements in $(V\moins\{\unit_{0,m}\})\times\Q$ such that $H^{\#_H}=\{x\in V;\;\scalar{\alpha_H}{x}>c_H\}$. Naturally, $C\not=\emptyset$ (otherwise we obtain a contradiction). Hence, there exists $x_0\in C$. Let us consider $c\in\Q$ such that $c\geq 1$, $c\geq \scalar{\alpha_H}{x_0}$ and $c\geq -c_H$ for any $H\in \Hrond$. As $C$ is not included in $\bigcup_{H\in\Hrond}\{x\in V;\;-1< \scalar{\frac{\alpha_H}{c}}{x}< 1\}$, there exists $x_1\in C$ and such that for any $H\in\Hrond$ either $\scalar{\alpha_H}{x_1}>c$ or $\scalar{\alpha_H}{x_1}<-c$. As $x_1\in C$, recall that $\scalar{\alpha_H}{x_1}> c_\alpha$. Hence $\scalar{\alpha_H}{x_1}<-c$ implies $c<-c_\alpha$ which is impossible. Therefore $\scalar{\alpha_H}{x_1}>c$ for any $H\in\Hrond$. Consider $v=x_1-x_0$ and remark that $\scalar{\alpha_H}{v}>0$ for any $H\in\Hrond$. Hence $v$ is in $\bigcap_{H\in\Hrond}\vec{H}^{\#_H}$. From lemma \ref{lem:alldir}, we deduce that $C$ is non $V$-degenerate.
  \qed
\end{proof}

\begin{example}
  The $\Q^2$-polyhedrons $C_1=\{x\in\Q^2;(-1\leq x[1]+x[2]\leq 1)\vee (-1\leq x[1]-x[2]\leq 1)\}$ and $C_2=\{x\in\Q^2;\;-x[1]+2.x[2]\geq 0\wedge 2.x[1]-x[2]\geq 0\}$ are given in figure \ref{fig:polyhedron}. Remark that $C_1$ is $\Q^2$-degenerate because $\vecsaff(\Z^m\cap C_1)=V_1\cup V_2$ where $V_1=\{x\in\Q^2;\;x[1]=x[2]\}$ and $V_2=\{x\in\Q^2;\;x[1]+x[2]=0\}$, and $C_2$ is non $\Q^2$-degenerate because $\vecsaff(\Z^m\cap C_2)=\Q^2$.
\end{example}

\section{Boundary}
We are interested in associating to a $V$-polyhedral equivalence class $\class$, a set of $V$-hyperplanes that intuitively corresponds to the ``constraints of $\class$''.\\

A \emph{possible $V$-boundary $\Hrond$}\index{Gen}{possible boundary} of a $V$-polyhedral equivalence class $\class$ is a finite set of $V$-hyperplanes such that there exists a $(V,\Hrond)$-definable polyhedron in $\class$. Following lemma shows that a possible $V$-boundary can be translated, and in particular the \emph{direction of any possible $V$-boundary remains a possible $V$-boundary}.
\begin{lemma}\label{lem:movebound}
  For any possible $V$-boundary $\Hrond$ of a $V$-polyhedral equivalence class $\class$ and for any sequence $(V_{H})_{H\in\Hrond}$ of non-empty finite subset of $V$, the set $\{v+H;\;H\in\Hrond;v\in V_H\}$ is a possible $V$-boundary of $\class$.
\end{lemma}
\begin{proof}
  There exists a $(V,\Hrond)$-definable polyhedron $C\in \class$. That means $C$ is a boolean combination in $V$ of sets in $\{H^\leq,H^<,H^=,H^>,H^\geq;\;H\in\Hrond\}$. Lemma \ref{lem:degex} proves that $[(v+H)^\#]^V=[H^\#]^V$ for any $(H,\#)\in\Hrond\times\{\leq,<,=,>,\geq\}$ and for any $v\in V_H$.
  \qed
\end{proof}

\begin{lemma}\label{lem:alldir}
  Let $C$ be an open convex $V$-polyhedron and $H_1$ be a $V$-hyperplane such that $[C\cap H_1^<]^V\not=[\emptyset]^V$ and $[C\cap H_1^>]^V\not=[\emptyset]^V$. For any $V$-hyperplane $H_0$ such that $\vec{H_0}\not=\vec{H_1}$, 
  there exist $\#_0\in\{<,>\}$ such that 
  $[C\cap H_0^{\#_0}\cap H_1^<]^V\not=[\emptyset]^V$
  and $[C\cap H_0^{\#_0}\cap H_1^>]^V\not=[\emptyset]^V$.
\end{lemma}
\begin{proof}
  As $C$ is an open convex set, there exists a finite set $\Hrond$ of $V$-hyperplanes and $\#\in\{<,>\}^\Hrond$ such that $C=C_{V,\#}$. Let us consider a sequence $(\alpha_{H},c_{H})_{H\in\Hrond}$ of elements in $(V\moins\{\unit_{0,m}\})\times\Q$ such that $H^{\#_H}=\{x\in V;\;\scalar{\alpha_{H}}{x} \#_H c_{H}\}$. Let us also consider $(\alpha_0,c_0)$ and $(\alpha_1,c_1)$ in $(V\moins\{0\})\times\Q$ such that $H_0^{\#_0}=\{x\in V;\;\scalar{\alpha_0}{x} \#_0 c_0\}$ and $H_1^{\#_1}=\{x\in V;\;\scalar{\alpha_1}{x}\#_1 c_1\}$. As $[C\cap H_1^{\#_1}]^V\not=[\emptyset]^V$, lemma \ref{lem:vdegpolyconv} shows that there exists $v_{\#_1}\in \Q^m$ such that $\scalar{\alpha_1}{v_{\#_1}}\#_1 0$ and such that $\scalar{\alpha_{H}}{v_{\#_1}} \#_H 0$ for any $H\in\Hrond$.
  
  Let us first prove that there exists a finite set $V_1$ of vectors in $\bigcap_{H\in\Hrond}\vec{H}^{\#_H}$ that generates $\vec{H_1}$. There exist $\mu_<$ and $\mu_>$ in $\Q_+\moins\{0\}$ such that the vector $v_==\mu_<.v_<+\mu_>.v_>$ satisfies $\scalar{\alpha_1}{v_=}=0$. Remark that $v_=\in \vec{H_1}$ and satisfies $\scalar{\alpha_{H}}{v_=} \#_H 0$ for any $H\in\Hrond$. Let us consider a finite set of vectors $V_1$ that generate $\vec{H_1}$ and just remark that there exists $\mu\in\Q_+$ enough larger such that  $\scalar{\alpha_{H}}{v} \#_H 0$ for any $(H,v)\in \Hrond\times (V_1+\mu.v_=)$. Finally, as $V_1$ generates $\vec{H_1}$ and $v_=\in\vec{H_1}$, the set $V_1+\mu.v_=$ also generates $\vec{H_1}$. By replacing $V_1$ by $V_1+\mu.v_=$, we are done.
  
  Naturally, if $V_1\subseteq \vec{H_0}$ then $\vec{H_1}=\vec{H_0}$ which is impossible. Hence, there exists $v_1\in V_1$ such that $\scalar{\alpha_0}{v_1}\not=0$. Let $\#_0\in\{<,>\}$ such that $\scalar{\alpha_0}{v_1}\#_00$. Remark that there exists $\mu\in\Q_+$ enough larger such that $v_{\#_1}+\mu.v_1\in \bigcap_{H\in\Hrond}\vec{H}^{\#_H}\cap \vec{H_{0}}^{\#_0}\cap \vec{H_1}^{\#_1}$ for any $\#_1\in\{<,>\}$. Lemma \ref{lem:vdegpolyconv} shows that $[C\cap H_{0}^{\#_0}\cap H_1^<]^V\not=[\emptyset]^V$ and $[C\cap H_0^{\#_0}\cap H_1^>]^V\not=[\emptyset]^V$.
  \qed
\end{proof}

\begin{lemma}\label{lem:Pleqgeq}
  Let $C$ be an open convex $V$-polyhedron and $H$ be a $V$-hyperplane such that $[C\cap H^<]^V\not=[\emptyset]^V$ and $[C\cap H^>]^V\not=[\emptyset]^V$. The set $\vec{C\cap H}$ is non $\vec{H}$-degenerate open convex $\vec{H}$-polyhedron.
\end{lemma}
\begin{proof}
  Without loss of generality, we can assume that $\vec{C}=X$ and $\vec{H}=H$. Since $C\cap H^\#$ is an open convex non $V$-degenerate $V$-polyhedron, there exists a vector $v_\#$ in this set. Let us remark that there exists two rational numbers $x_<,x_>$ in $\Q_+\moins\{0\}$ such that $x=x_<.v_<+x_>.v_>\in H$. Since $x_<,x_>$ are both in $C$ and $x_<,x_>$ are strictly positive rational numbers, we deduce that $x\in C$. Hence $x\in H\cap C$ and from lemma \ref{lem:vdegpolyconv} we deduce that $H\cap C$ is non-$H$-degenerate.
  \qed
\end{proof}

\begin{proposition}\label{prop:bound}
  Let $\class$ be a $V$-polyhedral equivalence class and $\Hrond_V(\class)$\index{Not}{$\Hrond_V(\class)$} be the set of $V$-hyperplanes $H$ such that there exists an open convex $V$-polyhedron $C_H$ such that $[C_H\cap H^<]^V\not=[\emptyset]^V$ and $[C_H\cap H^>]^V\not=[\emptyset]^V$, and such that $[C_H]^V\cap^V \class$ is equal to one of these two equivalence classes. The set $\vec{\Hrond_V(\class)}$ is a possible $V$-boundary of $\class$ included into the direction of any possible $V$-boundary of $\class$.
\end{proposition}
\begin{proof}
  Let us first consider a possible $V$-boundary $\Hrond$ of $\class$ and let us prove that for any $H_0\in\Hrond\moins\Hrond_V(\class)$, the set $\Hrond\moins\{H_0\}$ is a possible $V$-boundary of $\class$. Let $\Hrond'=\Hrond\moins\{H_0\}$. As $\Hrond$ is a possible $V$-boundary of $\class$, there exists a $(V,\Hrond)$-definable polyhedron $C$ in $\class$. We have the following equality:
  \begin{align*}
    \class 
    &=\left[C\moins \left(\bigcup_{H\in\Hrond}H\right)\right]^V\\
    &=\bigcup_{\#\in\{<,>\}^{\Hrond'}}^V\left[C_{V,\#}\cap H_0^<\cap C\right]^V\cup^V\left[C_{V,\#}\cap H_0^>\cap C\right]^V
  \end{align*}
  As $C$ is $(V,\Hrond)$-definable, we deduce that $C_{V,\#}\cap H_0^{\#_0}\cap C$ is either empty or equal to $C_{V,\#}\cap H_0^{\#_0}$. Let us prove that $[C_{V,\#}\cap C]^V$ is either equal to $[\emptyset]^V$ or equal to $[C_{V,\#}]^V$. Naturally, if $C_{V,\#}\cap H_0^<$ or $C_{V,\#}\cap H_0^{>}$ is $V$-degenerate, we are done. Otherwise, $[C_{V,\#}\cap H_0^<]^V\not=[\emptyset]^V$ and $[C_{V,\#}\cap H_0^>]^V\not=[\emptyset]^V$. As $C_{V,\#}$ is an open convex $V$-polyhedron and $H_0\not\in \Hrond_V(\class)$, we deduce that $[C_{V,\#}]^V\cap^V\class$ is neither equal to $[C_{V,\#}\cap H_0^<]^V$ nor equal to $[C_{V,\#}\cap H_0^>]^V$. However, $(C_{V,\#}\cap C)\moins H_0$ is either equal to $\emptyset$, $C_{V,\#}\moins H_{0}$, $C_{V,\#}\cap H_0^<$ or $C_{V,\#}\cap H_0^>$. As the two last cases are impossible, we deduce that $[C_{V,\#}]^V\cap^V\class$ is either equal to $[\emptyset]^V$ in the first case, or equal to $[C_{V,\#}]^V$ in the second case. We have proved that the following $(V,\Hrond')$-definable polyhedron $C'$ is in $\class$. That means $\Hrond'$ is a possible $V$-boundary.
  $$C'=\bigcup_{\#\in\{<,>\}^{\Hrond'};\; [C_{V,\#}]^V\cap^V\class\not=[\emptyset]^V}C_{V,\#}$$
  Finally, let us now consider a possible $V$-boundary $\Hrond$ of $\class$ and $H_0\in\Hrond_V(\class)$, and let us prove that $\vec{H_0}\in\vec{\Hrond}$. Lemma \ref{lem:movebound} shows that we can assume that $\vec{\Hrond}=\Hrond$. As $H_0\in\Hrond_V(\class)$, there exists an open convex $V$-polyhedron $C_{H_0}$ such that $[C_{H_0}\cap H_0^<]^V\not=[\emptyset]^V$ and $[C_{H_0}\cap H_0^>]^V\not=[\emptyset]^V$ and such that $[C_{H_0}]^V\cap^V\class$ is equal to one of these two equivalence classes. Assume by contradiction that $\vec{H_0}\not\in\vec{\Hrond}$. From lemma \ref{lem:alldir}, an immediate induction proves there exists $\#\in\{<,>\}^\Hrond$ such that $[C_{H_0}\cap C_{V,\#}\cap H_0^<]^V\not=[\emptyset]^V$ and $[C_{H_0}\cap C_{V,\#}\cap H_0^>]^V\not=[\emptyset]^V$. As $\Hrond$ is a possible $V$-boundary of $\class$, we deduce that $[C_{V,\#}]^V\cap^V\class$ is either equal to $[\emptyset]^V$ or equal to $[C_{V,\#}]^V$. In particular $[C_{H_0}\cap C_{V,\#}]^V\cap^V\class$ is either equal to $[\emptyset]^V$ or equal to $[C_{H_0}\cap C_{V,\#}]^V$. Moreover, as $[C_{H_0}]^V\cap^V\class$ is equal to $[C_{H_0}\cap H_0^<]^V$ or $[C_{H_0}\cap H_0^>]^V$, we also deduce that $[C_{H_0}\cap C_{V,\#}]^V\cap^V\class$ is either equal to $[C_{H_0}\cap C_{V,\#}\cap H_0^<]^V$ or equal to $[C_{H_0}\cap C_{V,\#} \cap H_0^>]^V$. Hence there exists $\#_0\in\{<,>\}$ such that $[C_{H_0}\cap C_{V,\#} \cap H_0^{\#_0}]^V$ is either equal to $[\emptyset]^V$ or equal to $[C_{H_0}\cap C_{V,\#}]^V$. The first case is impossible and the second case implies $[C_{H_0}\cap C_{V,\#} \cap H_0^{\#_0'}]^V=[\emptyset]^V$ where $\#_0'\in\{<,>\}\moins\{\#_0\}$. We obtain a contradiction. Therefore $\vec{H_0}\in\vec{\Hrond}$.
  \qed
\end{proof}

The previous proposition \ref{prop:bound} shows in particular that the set of directions of possible $V$-boundaries of a $V$-polyhedron $C$, owns a minimal elements for $\subseteq$. 
\begin{definition}
  The finite class $\vec{\Hrond_V(\class)}$ is denoted by $\bound{V}{\class}$\index{Not}{$\bound{V}{\class}$} and called the \emph{$V$-boundary}\index{Gen}{boundary} of $\class$.
\end{definition}

\begin{example}\label{ex:vbound}
  Let $C_2=\{x\in\Q^2;\;\scalar{\alpha_1}{x}\geq 0\wedge\scalar{\alpha_2}{x}\geq 0\}$ be the $\Q^2$-polyhedron given in figure \ref{fig:polyhedron} where $\alpha_1=(-1,2)$ and $\alpha_2=(2,-1)$. Let $H_1$ and $H_2$ be the $\Q^2$-hyperplanes defined by $H_1=\{x\in\Q^2;\;\scalar{\alpha_1}{x}=0\}$ and $H_2=\{x\in\Q^2;\;\scalar{\alpha_2}{x}=0\}$. Naturally, as $C_2$ is $(\Q^2,\{H_1,H_2\})$-definable, we deduce that $\{H_1,H_2\}\subseteq \bound{\Q^2}{[C_2]^{\Q^2}}$. Let us show the converse inclusion. Consider the open convex $\Q^2$-polyhedron $C_{H_1}=\{x\in \Q^2;\; \scalar{\alpha_2}{x}>0\wedge x[2]>0\}$. Remark that $[C_{H_1}\cap H_1^{<}]^{\Q^2}$ and $[C_{H_1}\cap H_1^{>}]^{\Q^2}$ are not equal to $[\emptyset]^{\Q^2}$ and $[C_{H_1}\cap C_2]^{\Q^2}$ is equal to one of this two classes. We deduce that $H_1\in\bound{\Q^2}{C_2}$. Symmetrically, we get $H_2\in\bound{\Q^2}{[C_2]^{\Q^2}}$. Therefore $\bound{\Q^2}{[C_2]^{\Q^2}}=\{H_1,H_2\}$.
\end{example}

\section{Polyhedrons of the form $C+V^\perp$}
In the sequel, we often consider $\Q^m$-polyhedrons of the form $C+V^\perp$ where $C$ is a $V$-polyhedron. In this section, we provide some properties satisfied by these sets.\\

Given a $V$-polyhedral equivalence class $\class$, following lemma \ref{lem:simVVperp} shows that the equivalence class $[C+V^\perp]^V$ does not depend on the $V$-polyhedron $C\in\class$. This equivalence class $[C+V^\perp]^V$ is naturally denoted by $\class+V^\perp$\index{Not}{$\class+V^\perp$}.
\begin{lemma}\label{lem:simVVperp}
  We have $C+V^\perp\sim^V C'+V^\perp$ for any $V$-polyhedrons $C$ and $C'$ such that $C\sim^V C'$ 
\end{lemma}
\begin{proof}
  We have $\Z^m\cap ((C+V^\perp)\Delta (C'+V^\perp))=\Z^m\cap (C_0+V^\perp)$ where $C_0=C\Delta C'$. As $C\sim^V C'$, we deduce that $C_0$ is $V$-degenerate. In order to prove the lemma, we have to show that $V$ is not included in $\vecsaff(\Z^m\cap (C_0+V^\perp))$. Proposition \ref{prop:vdegpoly} proves that there exists a finite set $D\subseteq \Z^m\cap V\moins\{\unit_{0,m}\}$ and an integer $k\in\Nat$ such that $C_0\subseteq \bigcup_{\alpha\in D}\{x\in V;\;|\scalar{\alpha}{x}|\leq k\}$. Let $K=\finiteset{-k}{k}$ and remark that we get $\Z^m\cap (C_0+V^\perp)\subseteq\bigcup_{(\alpha,k)\in D\times K}\{x\in\Q^m;\;\scalar{\alpha}{x}=k\}$. Hence $\vecsaff(\Z^m\cap (C_0+V^\perp))\subseteq \bigcup_{\alpha\in D}\alpha^\perp$. As $\alpha\in V$ for any $\alpha\in D$, we deduce that $V$ is not included in $\alpha^\perp$ for any $\alpha\in D$. From insecable lemma \ref{lem:insecable} we deduce that $V$ is not included in $\bigcup_{\alpha\in D}\alpha^\perp$. In particular $V$ is not included in $\vecsaff(\Z^m\cap (C_0+V^\perp))$. Therefore $C+V^\perp\sim^V C'+V^\perp$.
  \qed
\end{proof}
Remark that even if $[C+V^\perp]^V$ does not depends on a $V$-polyhedron $C\in\class$, there exist subsets $X\subseteq V$ in $\class$ such that $[X+V^\perp]^V\not=[C+V^\perp]^V$ as shown by the following example \ref{ex:CplusVperp}. That explains why our definition of $\class+V^\perp$ is limited to $V$-polyhedral equivalence classes $\class$.
\begin{example}\label{ex:CplusVperp}
  Assume that $m=2$, let $V=\{x\in\Q^2;\;x[1]=x[2]\}$. Let us consider the $V$-polyhedron $C=\emptyset$ and the set $X=(\frac{1}{2},\frac{1}{2})+(\Z^m\cap V)$. Remark that $[C]^V=[X]^V$. However $[C+V^\perp]^V=[\emptyset]^V$ whereas $[X+V^\perp]^\perp\not=[\emptyset]^V$ since $\Z^m\cap (X+V^\perp)=(0,1)+2.\Z^2$.  
\end{example}

Let us finally proves that $\gamma_{r,m,\sigma}^{-1}(\class+V^\perp)=\class+V^\perp$ for any $V$-polyhedral equivalence class $\class$ and for any word $\sigma\in\Sigma_{r,m}^*$. In fact, given a $V$-polyhedron $C\in\class$, we have the following equalities:

\begin{align*}
  \gamma_{r,m,\sigma}^{-1}(\class+V^\perp)
  &=\gamma_{r,m,\sigma}^{-1}([C+V^\perp]^V)\\
  &=[\gamma_{r,m,\sigma}^{-1}(C+V^\perp)]^V\\
  &=[\Gamma_{r,m,\sigma}^{-1}(C+V^\perp)]^V
\end{align*}

We can easily prove that $\Gamma_{r,m,\sigma}^{-1}(C+V^\perp)$ is a $\Q^m$-polyhedron of the form $C'+V^\perp$ by introducing the sequence $(\Gamma_{V,r,m,\sigma})_{\sigma\in\Sigma_{r,m}^*}$ of affine functions $\Gamma_{V,r,m,\sigma}:V\rightarrow V$\index{Not}{$\Gamma_{V,r,m,\sigma}$} defined by the following equality for any $x\in V$:
$$\Gamma_{V,r,m,\sigma}(x)=r^{|\sigma|}.x+\Pi_V(\gamma_{r,m,\sigma}(\unit_{0,m}))$$
Remark that $\Gamma_{V,r,m,\sigma_1.\sigma_2}=\Gamma_{V,r,m,\sigma_1}\circ\Gamma_{V,r,m,\sigma_2}$ for any word $\sigma_1,\sigma_2\in\Sigma_{r,m}^*$, $\Gamma_{V,r,m,\epsilon}$ is the identity function, and $\Gamma_{r,m,\sigma}^{-1}(C+V^\perp)=\Gamma_{V,r,m,\sigma}^{-1}(C)+V^\perp$ for any subset $C\subseteq V$.

Thanks to the following proposition \ref{prop:gammacone}, we deduce the following corollary \ref{cor:gammacone}.
\begin{proposition}\label{prop:gammacone}
  We have $[\Gamma_{r,m,\sigma}^{-1}(C)]^V=[C]^V$ for any $V$-polyhedron $C$ and for any $\sigma\in\Sigma_{r,m}^*$. 
\end{proposition}
\begin{proof}
  Let us consider a finite class $\Hrond$ of $V$-polyhedrons such that $C$ is $(V,\Hrond)$-definable. As $C$ is a boolean combination in $V$ of sets $H^\#$ where $H\in\Hrond$ and $\#\in\{<,>\}$, we can assume that $C$ is equal to such a set. As $H$ and $\Gamma_{V,r,m,\sigma}^{-1}(H)$ have the same direction, from lemma \ref{lem:degex}, we are done.
  \qed
\end{proof}

\begin{corollary}\label{cor:gammacone}
  We have $\gamma_{r,m,\sigma}^{-1}(\class+V^\perp)=\class+V^\perp$ for any $V$-polyhedral equivalence class and for any $\sigma\in\Sigma_{r,m}^*$.
\end{corollary}

%% file: chapter.presburgerdecomposition.tex
\chapter{Presburger Decomposition}
A subset $X\subseteq \Q^m$ can be naturally decomposed into $X=\bigcup_{V\in\comp(\vecsaff(X))}X_V$ where $X_V$\index{Not}{$X_V$} is defined by the following equality:
$$X_V=X\cap\left(\bigcup_{\begin{array}{@{}c@{}}\scriptstyle A\in\comp(\saff(X))\vspace{-.0cm}\\\scriptstyle \vec{A}\subseteq V\end{array}}A\right)$$
Obseve that $X_V$ is non empty and as shown by the following dense component lemma \ref{lem:dense}, the semi-affine hull direction $\vecsaff(X_V)$ is equal to $V$.
\begin{lemma}[Dense component lemma]\label{lem:dense}
  We have $\saff(X\cap A)=A$ for any subset $X\subseteq\Q^m$ and for any affine component $A$ of $\saff(X)$. 
\end{lemma}
\begin{proof}
  We have $\saff(X)=A\cup S$ where $S$ is the semi-affine space equal to the finite union of affine spaces $A'\in\comp(\saff(X))\moins\{A\}$. From $X\subseteq \saff(X)$, we deduce that $X\subseteq (X\cap A)\cup S\subseteq\saff(X\cap A)\cup S$. By minimality of the semi-affine hull, we get $\saff(X)\subseteq \saff(X\cap A)\cup S$. As $A\subseteq \saff(X)$, insecable lemma \ref{lem:insecable} shows that either $A\subseteq \saff(X\cap A)$ or $A\subseteq S$. In this last case, by definition of $S$, insecable lemma \ref{lem:insecable} proves that there exists $A'\in\comp(\saff(X))\moins\{A\}$ such that $A\subseteq A'$. As $A$ is an affine component of $\saff(X)$ and $A\subseteq A'\subseteq\saff(X)$, we get the equality $A=A'$ which is impossible. Therefore $A\subseteq \saff(X\cap A)$. Moreover, as $X\cap A\subseteq A$, we get the other inclusion $\saff(X\cap A)\subseteq A$.
  \qed
\end{proof}

We are going to prove that this decomposition of $X$ can be refined when $X$ is Presburger-definable. In fact, in this case, we show that $X_V$ can be decomposed (up to $V$-degenerate sets) into sets of the form $P\cap (C+V^\perp)$ where $P$ is a semi-$V$-pattern and $C$ is a $V$-polyhedron.\\

Naturally, a set $P\cap (C+V^\perp)$ is Presburger-definable. The semi-affine hull direction of such a set is characterized by the following lemma \ref{lem:dimdim}
\begin{lemma}\label{lem:dimdim}
  Let $P$ be a semi-$V$-pattern and $\class$ a $V$-polyhedral equivalence class. We have $[P]^V\cap^V (\class+V^\perp)\not=[\emptyset]^V$ if and only if $P\not=\emptyset$ and $\class\not=[\emptyset]^V$. 
\end{lemma}
\begin{proof}
  Naturally if $P=\emptyset$ or $\class=[\emptyset]^V$ then $[P]^V=[\emptyset]^V$ or $\class+V^\perp=[\emptyset]^V$ and in this case $[P]^V\cap^V (\class+V^\perp)=[\emptyset]^V$. Assume that $P\not=\emptyset$ and $\class$ is non $V$-degenerate and let us prove that $[P]^V\cap^V (\class+V^\perp)\not=[\emptyset]^V$. As $\class$ is polyhedral, there exists a $V$-polyhedron $C\in\class$. Let us consider a finite class $\Hrond$ of $V$-hyperplanes such that $V$ is $(V,\H)$-definable. As $C\moins(\bigcup_{H\in\Hrond}H)$ is a finite union of $V$-polyhedrons of the form $C_{V,\#}$ where $\#\in\{<,>\}^\Hrond$ and $[H]^V=[\emptyset]^V$, we can assume without loss of generality that there exists $\#$ such that $\class=[C_{V,\#}]^V$. Moreover,as a semi-$V$-pattern is a finite union of $V$-pattern, we can also assume without loss of generality that there exists $a\in\Z^m$ and a $V$-group $M$ such that $P=a+M$. We have to prove that $[P]^V\cap^V (\class+V^\perp)\not=[\emptyset]^V$. That means $V$ is included in $\vecsaff((a+M)\cap (V_{V,\#}+V^\perp))$. Let $(\alpha_H,c_H)_{H\in\Hrond}$ be a sequence of elements in $(V\moins\{\unit_{0,m}\})\times\Q$ such that $H^{\#_H}=\{x\in V;\;\scalar{\alpha_H}{x}>c_H\}$ for any $H\in\Hrond$. Lemma \ref{lem:vdegpolyconv} proves that there exists $v\in V$ such that $\scalar{\alpha_H}{v}>0$ for any $H\in\Hrond$. By replacing $v$ by a vector in $(\Nat\moins\{0\}).v$, we can assume that $v\in M$. Let $a'=\Pi_V(a)$ be the orthogonal projection of $a$ over $V$. Vector $v'=a-a'\in V^\perp$. There exists an integer $k\in\Nat$ enough larger such that $\scalar{\alpha}{a'+k.v}>c_H$ for any $H\in \Hrond$. In particular $a'+k.v\in C$. As $k.v\in M$, we deduce that $a+k.v\in P$. From $a+k.v=(a'+k.v)+v'$ we get $a+k.v\in C+V^\perp$. Hence $x_0=a+k.v\in P\cap (C+V^\perp)$. Let us now consider a finite set $V_0$ of $\dim(V)$ vectors in $\Q^m$ that generates $V$. By replacing $V_0$ by $k.V_0$ where $k\in\Nat\moins\{0\}$ is enough larger, we can assume that $V_0\subseteq M$. Moreover, by replacing $V_0$ by $V_0+k.v$ where $k\in\Nat$ is enough larger, we can assume that $\scalar{\alpha_H}{v_0}>0$ for every $(H,v_0)\in\Hrond\times V_0$. We deduce that $x_0+\sum_{v_0\in V_0}\Nat.v_0\subseteq P\cap (C+V^\perp)$. Covering lemma \ref{lem:cov} proves that $\vecsaff(x_0+\sum_{v_0\in V_0}\Nat.v_0)=V$. In particular from $x_0+\sum_{v_0\in V_0}\Nat.v_0\subseteq P\cap (C+V^\perp)$ we get $V\subseteq \vecsaff(P\cap (C+V^\perp))$.
  \qed
\end{proof}

\begin{definition}
  A \emph{$V$-polyhedral partition}\index{Gen}{polyhedral partition} $(\class_i)_{i\in I}$ is a non empty finite sequence of $V$-polyhedral equivalence classes such that $\class_{i_1}\cap^V\class_{i_2}=[\emptyset]^V$ if and only if $i_1\not=i_2$ and such that $[V]^V=\bigcup^V_{i\in I}\class_i$.
\end{definition}

\begin{theorem}[Decomposition theorem\index{Gen}{decomposition theorem}]\label{theo:decomposition}\label{thm:decomposition}
  Let $X\subseteq\Z^m$ be a Presburger-definable set and $V$ be an affine component of $\vecsaff(X)$. There exists a unique $V$-polyhedral partition $(\class_{V,P}(X))_{P\in\P_V(X)}$\index{Not}{$\class_{V,P}(X)$} indexed by a non-empty finite class $\P_V(X)$\index{Not}{$\P_V(X)$} of semi-$V$-patterns such that:
  $$[X_V]^V=\bigcup_{P\in\P_V(X)}^V([P]^V\cap^V(\class_{V,P}(X)+V^\perp))$$
\end{theorem}
\begin{proof}
  Let us first prove that two $V$-polyhedral partitions $(\class_{V,P})_{P\in\P_V}$ and $(\class'_{V,P'})_{P'\in\P_V'}$ that satisfies $[X_V]^V=\bigcup_{P\in\P_V}^V([P]^V\cap^V(\class_{V,P}+V^\perp))$ and $[X_V]^V=\bigcup_{P'\in\P_V'}^V([P']^V\cap^V(\class'_{V,P'}+V^\perp))$ are equal. Consider $P\in\P_V$. As $[V]^V=\bigcup^V_{P'\in\P_V'}\class'_{V,P'}$, we deduce that $\class_{V,P}=\bigcup^V_{P'\in\P_V'}(\class_{V,P}\cap^V \class'_{V,P'})$. In particular, there exists $P'\in\P_V'$ such that $\class_{V,P}\cap^V \class'_{V,P'}\not=[\emptyset]_V$. Consider such a $P'\in\P_V'$. By intersecting the equality $\bigcup_{P\in\P_V}^V([P]^V\cap^V(\class_{V,P}(X)+V^\perp))=\bigcup_{P'\in\P_V'}^V([P']^V\cap^V(\class'_{V,P'}(X)+V^\perp))$ with $\class_{V,P}\cap^V \class'_{V,P'}$, we get $[P\Delta P']^V\cap^V((\class_{V,P}\cap^V \class'_{V,P'})+V^\perp)=[\emptyset]^V$. Lemma \ref{lem:dimdim} proves that $P\Delta P'=\emptyset$. Hence $P=P'$ and we have proved the inclusion $\P_V\subseteq\P_V'$ and by symmetry the equality $P_V=\P_V'$. Remark that we have also proved that for any $P'\in\P_V'\moins\{P\}$ we have $\class_{V,P}\cap^V \class'_{V,P'}=[\emptyset]^V$. Therefore $\class_{V,P}\cap^V(\bigcup_{P'\in\P_V'\moins\{P\}}^V\class'_{V,P'})=[\emptyset]^V$. As $(\class'_{V,P'})_{P'\in\P_V'}$ is a $V$-polyhedral partition, we deduce that $\class_{V,P}\subseteq^V\class'_{V,P}$ and by symmetry $\class_{V,P}=\class'_{V,P}$. We have proved that $(\class_{V,P})_{P\in\P_V}$ and $(\class'_{V,P'})_{P'\in\P_V'}$ are equal.
  
  Next, let us prove that there exists a $V$-polyhedral partition $(\class_{V,P})_{P\in\P_V}$ satisfying $[X_V]^V=\bigcup_{P\in\P_V}^V([P]^V\cap^V(\class_{V,P}+V^\perp))$. Let us denote by $\A_V$ the set of $A\in\comp(\saff(X_V))$ such that $\vec{A}=V$ and let $X_V'=X_V\cap(\bigcup_{A\in\A_V}A)$. As $X_V'$ is Presburger-definable, a quantification elimination shows that $X_V'$ is a boolean combination in $\Z^m$ of sets of the form $\{x\in\Z^m;\;\scalar{\alpha}{x}\in c+n.\Z\}$ and of the form $\{x\in\Z^m;\;\scalar{\alpha}{x}\#c\}$ where $(\alpha,\#,c,n)\in (\Z^m\moins\{0\})\times\{<,>\}\times \Z\times(\Nat\moins\{0\})$. Remark that any boolean combination of sets of the form $\{x\in\Z^m;\;\scalar{\alpha}{x}\in c+n.\Z\}$ is a semi-$\Q^m$-pattern and any boolean combination in $\Q^m$ of $\{x\in\Q^m;\;\scalar{\alpha}{x}\#c\}$ is a polyhedron. Hence, there exists a finite sequence $(P_i,C_i)_{i\in I}$ where $P_i$ is a semi-$\Q^m$-pattern and $C_i$ is a polyhedron such that $X_V'=\bigcup_{i\in I}(P_i\cap C_i)$. Let us consider a sequence $(v'_A)_{A\in \A_V}$ of vectors $v'_A\in A$. For any $i\in I$ and $A\in\A_V$, we have $A\cap C_i=A\cap (C_{i,A}+V^\perp)$ where $C_{i,A}$ is the $V$-polyhedron $C_{i,A}=(A\cap C_i)-v'_A$. As $I\times\A_V$ is finite, there exists a finite set $\Hrond$ of $V$-hyperplanes such that $C_{i,A}$ is $(V,\Hrond)$-definable for any $(i,A)\in I\times\A_V$. We have:
  \begin{align*}
    X_V'\moins(\bigcup_{H\in\Hrond}(H+V^\perp))
    &=\bigcup_{\#\in\{<,>\}^\Hrond}(X_V'\cap (C_{V,\#}+V^\perp))\\
    &=\bigcup_{\#\in\{<,>\}^\Hrond}\bigcup_{(i,A)\in I\times\A_V}(P_i\cap A\cap (C_{i,A}+V^\perp)\cap (C_{V,\#}+V^\perp))\\
    &=\bigcup_{\#\in\{<,>\}^\Hrond} \bigcup_{(i,A)\in I\times\A_V}(P_i\cap A\cap ((C_{i,A}\cap C_{V,\#})+V^\perp))\\
    &=\bigcup_{\#\in\{<,>\}^\Hrond}P_\#\cap (C_{V,\#}+V^\perp)
  \end{align*}
  Where $P_\#$ is the semi-$V$-pattern $P_\#=\bigcup_{(i,A)\in I\times\A_V;\;C_{i,A}\cap C_{V,\#}\not=\emptyset}(P_i\cap A)$ (recall that $C_{i,A}\cap C_{V,\#}$ is either empty or equal to $C_{V,\#}$). Let us denote by $\P_V=\{P_\#;\;[C_{V,\#}]_V\not=[\emptyset]_V\}$ and consider the sequence $(C_{V,P})_{P\in\P_V}$ of $V$-polyhedrons defined by:
  $$C_{V,P}=\bigcup_{\#\in\{<,>\}^\Hrond;\;P_\#=P}C_{V,\#}$$
  Remark that $(\class_{V,P})_{P\in\P_V}$ where $\class_{V,P}=[C_{V,P}]^V$ is a $V$-polyhedral partition. Moreover, the set $Z_V=X_V\Delta(\bigcup_{P\in\P_V}(P\cap (C_{V,P}+V^\perp)))$ is included in the union of $\bigcup_{A\in\comp(\saff(X_V))\moins\A_V}A$, $\bigcup_{\#\in\{<,>\}^\Hrond;\;[C_{V,\#}]_V=[\emptyset]_V}(P_\#\cap (C_{V,\#}+V^\perp))$, and $\bigcup_{H\in\Hrond}(X_V\cap (H+V^\perp))$. Remark that for any $A\in\comp(\saff(X_V))\moins\A_V$, we have $[A]^V=[\emptyset]^V$, for any $\#\in\{<,>\}^\Hrond$ such that $[C_{V,\#}]_V=[\emptyset]_V$, lemma \ref{lem:dimdim} shows that $[P_\#\cap (C_{V,\#}+V^\perp)]^V=[\emptyset]^V$, and for any $H\in\Hrond$, we have $[X_V\cap (H+V^\perp)]^V=[X_V]^V\cap^V[H+V^\perp]^V=[X_V]^V\cap^V[\emptyset]^V=[\emptyset]^V$. We deduce that $[Z_V]^V=[\emptyset]^V$. Therefore $[X_V]^V=\bigcup_{P\in\P_V}^V([P]^V\cap^V(\class_{V,P}+V^\perp))$.
  \qed
\end{proof}

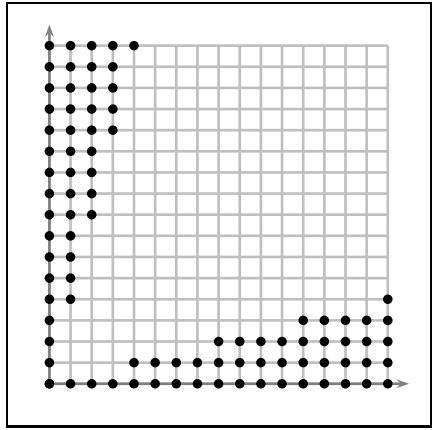
\begin{figure}[htbp]
  \setlength{\unitlength}{8pt}
  \pssetlength{\psunit}{8pt}
  \pssetlength{\psxunit}{8pt}
  \pssetlength{\psyunit}{8pt}
  \begin{center}
    \begin{picture}(20,20)(-2,-2)%
      \put(-2,-2){\framebox(20,20){}}
      \multido{\iii=0+1}{17}{\psline[linecolor=lightgray,linewidth=1pt](0,\iii)(16,\iii)}\multido{\iii=0+1}{17}{\psline[linecolor=lightgray,linewidth=1pt](\iii,0)(\iii,16)}\psline[linecolor=gray,arrowscale=1,linewidth=1pt]{->}(0,0)(0,17)\psline[linecolor=gray,arrowscale=1,linewidth=1pt]{->}(0,0)(17,0)
      \psdot*(0,1)\psdot*(0,2)\psdot(0,4)\psdot*(0,5)\psdot*(0,7)\psdot*(0,8)\psdot*(0,10)\psdot*(0,11)\psdot(0,13)\psdot*(0,14)\psdot*(0,16)\psdot(1,4)\psdot*(1,5)\psdot*(1,7)\psdot*(1,8)\psdot*(1,10)\psdot*(1,11)\psdot(1,13)\psdot*(1,14)\psdot*(2,8)\psdot*(2,10)\psdot*(2,11)\psdot(2,13)\psdot*(2,14)\psdot*(2,16)\psdot(3,13)\psdot*(3,14)\psdot*(3,16)\psdot*(4,16)\psdot*(0,0)\psdot*(0,3)\psdot*(0,6)\psdot*(0,9)\psdot*(0,12)\psdot*(0,15)\psdot*(1,6)\psdot*(1,9)\psdot*(1,12)\psdot*(1,15)\psdot*(2,9)\psdot*(2,12)\psdot*(2,15)\psdot*(3,12)\psdot*(3,15)\psdot*(1,16)
      
      \psdot*(1,0)\psdot*(2,0)\psdot(4,0)\psdot*(5,0)\psdot*(7,0)\psdot*(8,0)\psdot*(10,0)\psdot*(11,0)\psdot(13,0)\psdot*(14,0)\psdot*(16,0)\psdot(4,1)\psdot*(5,1)\psdot*(7,1)\psdot*(8,1)\psdot*(10,1)\psdot*(11,1)\psdot(13,1)\psdot*(14,1)\psdot*(8,2)\psdot*(10,2)\psdot*(11,2)\psdot(13,2)\psdot*(14,2)\psdot*(16,2)\psdot(13,3)\psdot*(14,3)\psdot*(16,3)\psdot*(16,4)\psdot*(0,0)\psdot*(3,0)\psdot*(6,0)\psdot*(9,0)\psdot*(12,0)\psdot*(15,0)\psdot*(6,1)\psdot*(9,1)\psdot*(12,1)\psdot*(15,1)\psdot*(9,2)\psdot*(12,2)\psdot*(15,2)\psdot*(12,3)\psdot*(15,3)\psdot*(16,1)

    \end{picture}
  \end{center}
  \caption{The Presburger-definable set $X=\{x\in\Nat^2;\;(x[2]\geq 4.x[1])\vee(x[1]\geq 4.x[2])\}$\label{fig:semilinear}}
\end{figure}

\begin{example}
  Let us consider the Presburger-definable set $X=\{x\in\Nat^2;\;(x[2]\geq 4.x[1])\vee(x[1]\geq 4.x[2])\}$ given in figure \ref{fig:semilinear}. We have $\saff(X)=\Q^2$. Hence $V=\Q^2$ is the only affine component of $\vecsaff(X)$. The $V$-polyhedral partition $([C_{V,P}]^V)_{P\in\P_V}$ defined by $\P_V=\{\Z^2,\emptyset\}$, $C_{V,\Z^2}=\{x\in\Q^2;\;0\leq x[1]\leq 4.x[2]\vee 0\leq x[2]\leq 4.x[1]\}$ and $C_{V,\emptyset}=V\moins C_{V,\Z^2}$ satisfies decomposition theorem.
\end{example}

The following proposition shows that the decomposition theorem can be also applied to $[X]^V$ since $[X_V]^V=[X]^V$.
\begin{proposition}\label{prop:XvtoX}
  We have $[X_V]^V=[X]^V$ for any set $X\subseteq \Z^m$ and for any affine component $V$ of $\vecsaff(X)$. 
\end{proposition}
\begin{proof}
  Let us consider the semi-affine space $S$ equal to the affine component $A$ of $\comp(\saff(X))$ such that $\vec{A}\subseteq V$. Recall that $X_V=X\cap S$. In order to prove that $[X_V]^V=[X]^V$, it is sufficient to show that $V$ is not included in $\vecsaff(\Z^m\cap (X\Delta X_V))$. Remark that $\Z^m\cap (X_V\Delta X)=X\moins S$. Moreover as $X\subseteq \bigcup_{A\in\comp(\saff(X))}A$, we deduce that $X\moins S\subseteq \bigcup_{A\in\comp(\saff(X))}(A\moins S)$. Naturally, if $\vec{A}\subseteq V$ then $A\subseteq S$ and in particular $A\moins S=\emptyset$. Hence $X\moins S$ is included into the finite union of affine component $A$ of $\saff(X)$ such that $\vec{A}\not\subseteq V$. Assume by contradiction that $V$ is included in $\vecsaff(X\moins S)$. From insecable lemma \ref{lem:insecable}, we deduce that there exists such an affine component $A$ such that $V\subseteq \vec{A}$. Hence $V\subseteq \vec{A}\subseteq\vecsaff(X)$ and as $V$ is an affine component of $\vecsaff(X)$, we deduce that $V=\vec{A}$ which is in contradiction with $\vec{A}\not\subseteq V$. Hence $V$ is not included in $\vecsaff(X\moins S)$ and we have proved that $[X_V]^V=[X]^V$.
  \qed
\end{proof}

%% file: chapter.stronglyconnected.tex
\chapter{Strongly Connected Components}
A \emph{component}\index{Gen}{component} $T$ of a FDVG $G$ is a strongly connected component of the parallelization $[G]$.

\section{Untransient strongly connected components}
A component $T$ is said \emph{untransient}\index{Gen}{untransient component} if there exists a loop $q\xrightarrow{\sigma}q$ where $q\in T$ and $\sigma\in\Sigma_{r,m}^+$. Otherwise, the component $T$ is said \emph{transient}\index{Gen}{transient component}.

In this section, we prove that for any untransient component $T$ of a FDVG $G$ there exists a unique vector space $V_G(T)$\index{Not}{$V_G(T)$} and a unique sequence $(a_G(q))_{q\in T}$\index{Not}{$a_G(q)$} of vectors in $V_G(T)^\perp$ such that we have the following equality:
$$\saff(\{\xi_{r,m}(w);\;q\xrightarrow{w\in \Sigma_{r,m}^+}q\})=a_G(q)+V_G(T)$$
Moreover, an algorithm for computing $V_G(T)$ and $(a_G(q))_{q\in T}$ in polynomial time is provided.

\begin{remark}
  The vector space $V_G(T)$ does not depend on $q\in T$.
\end{remark}

The polynomial time computation is based on a fix-point system provided by the following proposition \ref{prop:untransient}

\begin{proposition}\label{prop:untransient}
  Let $T$ be an untransient component of a FDVG $G$ and let $K_0$ be the set of states $k_0\in K$ reachable and co-reachable from $T$. There exists a unique minimal (for the point-wise inclusion) sequence of affine spaces $(A_{k_0})_{k_0\in K_0}$ not equal to $(\emptyset)_{k_0\in K_0}$ such that for any transition $k_0\xrightarrow{b}k_0'$ where $(k_0,b,k_0')\in K_0\times \Sigma_r\times K_0$, we have the following inclusion:
  $$\Gamma_{r,m,b}^{-1}(A_{k_0})\subseteq A_{k_0'}$$
  Moreover, this sequence satisfies $\saff(\{\xi_{r,m}(w);\;k_0\xrightarrow{w\in \Sigma_{r,m}^+}k_0\})=A_{k_0}$ for any $k_0\in K_0$.
\end{proposition}
\begin{proof}
  We denote by $Z_{k_0}$ the set of $Z_{k_0}=\{\xi_{r,m}(w);\;k_0\xrightarrow{w\in \Sigma_{r,m}^+}k_0\}$. By developing the expression $\xi_{r,m}(\sigma_1.w^k.\sigma_2)$ where $\sigma_1$, $\sigma_2$ are in $\Sigma_r^*$ such that $\sigma_1.\sigma_2\in\Sigma_{r,m}^+$, $w\in\Sigma_{r,m}^+$ and $n\in\Nat$, we obtain the following equality:
  $$\xi_{r,m}(\sigma_1.w^n.\sigma_2)=\frac{\Gamma_{r,m,\sigma_1}\circ\xi_{r,m}(w)}{1-r^{|\sigma_1.\sigma_2|_m+n.|w|_m}}+\Gamma_{r,m,\sigma_2}^{-1}\circ\xi_{r,m}(w)$$
  
  Let us first prove that $(\saff(Z_{k_0}))_{k_0\in K_0}$ satisfies the fix-point system. Consider a transition $k_0\xrightarrow{b}k_0'$ where $(k_0,b,k_0')\in K_0\times\Sigma_r\times K_0$. As $k_0$ and $k_0'$ and in the same strongly connected component, there exists a path $k_0'\xrightarrow{\sigma_1}k_0$. By replacing $\sigma_1$ by $\sigma_1.(b.\sigma_1)^{m-1}$, we can assume that $\sigma_1.b\in\Sigma_{r,m}^+$. Let us consider $x\in Z_{k_0}$. There exists a loop $k_0\xrightarrow{w}k_0$ where $w\in\Sigma_{r,m}^+$ such that $x=\xi_{r,m}(w)$. Remark that for any $n\in\Nat$, we have the loop $k_0'\xrightarrow{\sigma_1.w^n.b}k_0'$. Therefore $\xi_{r,m}(\sigma_1.w^n.b)\in Z_{k_0'}$. Thanks the the equality given in the first paragraph and covering lemma \ref{lem:cov}, we deduce that $\Q.\Gamma_{r,m,\sigma_1}(x)+\Gamma_{r,m,b}^{-1}(x)\subseteq \saff(Z_{k_0'})$. In particular $\Gamma_{r,m,b}^{-1}(x)\in \saff(Z_{k_0'})$. We have proved the inclusion $\Gamma_{r,m,b}^{-1}(Z_{k_0})\subseteq \saff(Z_{k_0'})$ and from covering lemma \ref{lem:cov}, we get $\Gamma_{r,m,b}^{-1}(\saff(Z_{k_0}))\subseteq \saff(Z_{k_0'})$. We have proved that $(\saff(Z_{k_0}))_{k_0\in K_0}$ satisfies the fix-point system.
  
  Now, let us prove that $\saff(Z_{k_0})$ is an affine space. Remark that this semi-affine space is not empty and in particular there exists at least one affine component $A$ of $\saff(Z_{k_0})$. Let $x\in Z_{k_0}$. Assume by contradiction that $Z_{k_0}\moins A$ is not empty. Let us consider a vector $x\in Z_{k_0}\moins A$. By definition of $Z_{k_0}$, there exists a loop ${k_0}\xrightarrow{w}k_0$ where $w\in\Sigma_{r,m}^+$ such that $x=\xi_{r,m}(w)$. From the previous paragraph, we deduce that $\Gamma_{r,m,w^n}^{-1}(\saff(Z_{k_0}))\subseteq \saff(Z_{k_0})$ for any $n\in\Nat$. In particular $\Gamma_{r,m,w^n}^{-1}(A)\subseteq \saff(Z_{k_0})$ for any $n\in \Nat$. Remark that $\Gamma_{r,m,w^n}^{-1}(A)=r^{-n.|w|}.(A-x)+x$ thanks to $x=\xi_{r,m}(w)$. Covering lemma \ref{lem:cov} shows that $\Q.(A-x)+x\subseteq \saff(Z_{k_0})$. As $A\subseteq \Q.(A-x)+x\subseteq \saff(Z_{k_0})$ and $A$ is an affine component of $\saff(Z_{k_0})$, we deduce the equality $A=\Q.(A-x)+x$. In particular $x\in A$ and we obtain a contradiction. We have proved that $Z_{k_0}\moins A=\emptyset$. Therefore $Z_{k_0}\subseteq A$. We get $\saff(Z_{k_0})=A$. Therefore $\saff(Z_{k_0})$ is an affine space (remark that even if these proof is similar to the one provided by proposition \ref{prop:saffcyclic}, we cannot apply this proposition since $Z_{k_0}$ is not necessary $(r,m,w)$-cyclic).
  
  Finally, let us consider a sequence of affine spaces $(A_{k_0})_{k_0\in K_0}$ not equal to $(\emptyset)_{k_0\in K_0}$ such that $\Gamma_{r,m,b}^{-1}(A_{k_0})\subseteq A_{k_0'}$ for any transition $(k_0\xrightarrow{b}k_0'$ with $(k_0,b,k_0')\in K_0\times\Sigma_r\times K_0$ and let us prove that $\saff(Z_{k_0})\subseteq A_{k_0}$ for any $k_0\in K_0$. An immediate induction shows that $\Gamma_{r,m,\sigma}^{-1}(A_{k_0})\subseteq A_{k_0'}$ for any path $k_0\xrightarrow{\sigma}k_0'$ where $(k_0,\sigma,k_0')\in K_0\times\Sigma_r^*\times K_0$. Since $\saff(Z_{k_0})$ is an affine space, it is sufficient to show that $Z_{k_0}\subseteq A_{k_0}$. Since $(A_{k_0})_{k_0\in K_0}$ is not equal to the empty sequence $(\emptyset)_{k_0\in K_0}$, there exists at least a state $k_1\in K_0$ such that $A_{k_1}\not\emptyset$. By definition of $K_0$, there exists a path $k_1\xrightarrow{\sigma}k_0$. From $\Gamma_{r,m,\sigma}^{-1}(A_{k_1})\subseteq A_{k_0}$ we deduce that $A_{k_0}\not=\emptyset$. Hence, there exists $a\in A_{k_0}$. Since $x\in Z_{k_0}$, there exists $w\in\Sigma_{r,m}^+$ such that $k_0\xrightarrow{w}k_0$. From the path $k_0\xrightarrow{w^n}k_0$, we get $\Gamma_{r,m,w^n}^{-1}(A_{k_0})\subseteq A_{k_0}$ for any $n\in\Nat$. Hence $\Gamma_{r,m,w^n}^{-1}(a)\in A_{k_0}$ for any $n\in \Nat$. Since $\Gamma_{r,m,w^n}^{-1}(a)=r^{-|w|_m}.(a-\xi_{r,m}(w))+\xi_{r,m}(w)$, from covering lemma \ref{lem:cov}, we get $\Q.(a-\xi_{r,m}(w))+\xi_{r,m}(w)\in A_{k_0}$. In particular $\xi_{r,m}(w)\in A_{k_0}$ and we have proved that $Z_{k_0}\subseteq A_{k_0}$. Thus $\saff(Z_{k_0})\subseteq A_{k_0}$ for any $k_0\in K_0$.
  
  Since $(\saff(Z_{k_0}))_{k_0\in K_0}$ is not equal to $(\emptyset)_{k_0\in K_0}$, we are done.
  \qed
\end{proof}

We deduce the following proposition \ref{prop:vgt} that shows that a characteristic vector space denoted by $V_G(T)$ is associated to any untransient component $T$ of a finite DVG $G$. This vector space is extremely useful in the sequel for extracting geometrical properties from a FDVA.

\begin{proposition}\label{prop:vgt}
  Let $T$ be an untransient component of a finite graph $G$ labelled by $\Sigma_{r,m}$. There exists a unique vector space $V_G(T)$ and a unique sequence $(a_G(q))_{q\in T}$ of vectors in $V_G(T)^\perp$ such that for any $q\in Q$:
  $$\saff(\{\xi_{r,m}(w);\;q\xrightarrow{w\in\Sigma_{r,m}^+}q\})=a_G(q)+V_G(T)$$
\end{proposition}
\begin{proof}
  Let $A_q=\saff(\{\xi_{r,m}(w);\;q\xrightarrow{w\in \Sigma_{r,m}^+}q\})$. The previous proposition \ref{prop:untransient} proves that $A_q$ is a non empty affine space. It is sufficient to show that the vector space $\vec{A_q}$ that does not depend on $q\in T$. By symmetry, it is sufficient to prove that $\vec{A_{q_1}}\subseteq \vec{A_{q_2}}$ for any $q_1,q_2\in T$. Since $T$ is strongly connected, there exists a path $q_1\xrightarrow{\sigma}q_2$ with $\sigma\in\Sigma_{r,m}^*$. Proposition \ref{prop:untransient} proves by an immediate induction that $\Gamma_{r,m,\sigma}^{-1}(A_{q_1})\subseteq A_{q_2}$. Since the affine space $\Gamma_{r,m,w}^{-1}(A_{q_1})$ is equal to $r^{-|w|_m}.(A_{q_1}-\rho_{r,m}(w,\unit_{0,m}))$, its direction is equal to $\vec{A_{q_1}}$. We deduce that $\vec{A_{q_1}}\subseteq \vec{A_{q_2}}$.  
  \qed
\end{proof}

\subsection{A polynomial time algorithm}
Thanks to the fix-point system provided by proposition \ref{prop:untransient}, we are going to show that $V_G(T)$ is computable in polynomial time from $G$. 

\begin{theorem}\label{thm:vgtpoly}
  Let $T$ be an untransient component of a FDVG $G$. The vector space $V_G(T)$ is computed in polynomial by the algorithm given in figure \ref{fig:algovgt}.
\end{theorem}
\begin{proof}
  Naturally, the algorithm terminates in polynomial time. Let us prove that the vector space $V$ returned by the algorithm is equal to $V_G(T)$. Let $(S_{k_0})_{k_0\in K_0}$ be the sequence of affine spaces $S_{k_0}=\saff(\{\xi_{r,m}(w);\;k_0\xrightarrow{w\in \Sigma_{r,m}^+}k_0\})$. For any state $k_0\in K_0$ let us consider the set $J_{k_0}=\lambda(k_0)-\lambda(k_0)$ the set of difference of two elements in $\lambda(k_0)$.

  Let us show that for any $k_0,k_0'\in K_0$, we have $J_{k_0}+m.\Z=J_{k_0'}+m.\Z$. It is sufficient to show the inclusion $J_{k_0}\subseteq J_{k_0'}+m.\Z$. Let $i_1,i_2\in J_{k_0}$. There exists two paths $q_1\xrightarrow{\sigma_1}k_0$ and $q_2\xrightarrow{\sigma_2}k_0$ where $|\sigma_1|\in i_1+m.\Z$, $|\sigma_2|\in i_2+m.\Z$ and $q_1,q_2\in T$. Since $T$ is strongly connected (for $[G]$), there exists a path $k_0\xrightarrow{w}k_0'$. From the path $q_1\xrightarrow{\sigma_1.w}k_0'$ and  $q_2\xrightarrow{\sigma_2.w}k_0'$ we deduce that $(|\sigma_1|+|w|)-(|\sigma_2|+|w|)\in J_{k_0'}+m.\Z$. Hence $i_1-i_2\in J_{k_0'}+m.\Z$. We have proved that for any $k_0,k_0'\in K_0$, we have $J_{k_0}+m.\Z=J_{k_0'}+m.\Z$. 
  
  Thanks to the previous paragraph, we deduce that $\Gamma_{r,m,0}^{-i_1}(V)=\Gamma_{r,m,0}^{-i_2}(V)$ for any $i_1,i_2\in \lambda(k_0)$ and for any $k_0\in K_0$ is an invariant of the algorithm. Thus, for any $k_0\in K_0$, there exists a vector space $V_{k_0}$ such that $V_{k_0}=\Gamma_{r,m,0}^{-i}(V)$ for any $i\in \lambda(k_0)$. For any transition $k_0\xrightarrow{b}k_0'$ such that $(k_0,b,k_0')\in K_0\times \Sigma_r\times K_0$, let $x_{k_0,b,k_0'}=\Gamma_{r,m,b}^{-1}(\xi_{r,m}(\sigma_{k_0}))-\xi_{r,m}(\sigma_{k_0'})$, and let $A_{k_0}=\xi_{r,m}(\sigma_{k_0})+V_{k_0}$. 
  
  Let us show that $V_G(T)\subseteq V$. Since for any transition $k_0\xrightarrow{b}k_0'$ where $(k_0,b,k_0')\in K_0\times \Sigma_r\times K_0$ and for any $i\in \lambda(k_0)$, we have $\Gamma_{r,m,0}^i(x_{k_0,b,k_0'})\in V$, we deduce that $\Gamma_{r,m,b}^{-1}(A_{k_0})=A_{k_0'}$ and in particular $(A_{k_0})_{k_0\in K_0}$ is a sequence of affine spaces satisfying the fix-point system provided by proposition \ref{prop:untransient} and not equal to $(\emptyset)_{k_0\in K_0}$. By minimality of the sequence $(S_{k_0})_{k_0\in K_0}$, we deduce that $S_{k_0}\subseteq A_{k_0}$. Taking the direction of the previous inclusion, we get $V_G(T)\subseteq V$. 
  
  Let us prove the converse inclusion $V\subseteq V_G(T)$. Remark that $V$ is generated by vectors $\Gamma_{r,m,0}^i(x_{k_0,b,k_0'})$ where $k_0\xrightarrow{b}k_0'$ is a transition such that $(k_0,b,k_0')\in K_0\times\Sigma_r\times K_0$, and $i\in\lambda(k_0)$. Since $V_G(T)$ is a vector space, it is sufficient to prove that $\Gamma_{r,m,0}^{i+1}(x_{k_0,b,k_0'})\in V_G(T)$. Remark that $\xi_{r,m}(\sigma_{k_0})\in S_{k_0}$ and since $\Gamma_{r,m,b}^{-1}(S_{k_0})\subseteq S_{k_0'}$, we get $\Gamma_{r,m,b}^{-1}(\xi_{r,m}(\sigma_{k_0}))\in S_{k_0'}$. Moreover, as $\xi_{r,m}(\sigma_{k_0'})\in S_{k_0'}$ and $S_{k_0'}$ is an affine space, we get $x_{k_0,b,k_0'}\in \vec{S}_{k_0'}$. By definition of $\lambda$, there exists a path $q\xrightarrow{\sigma}k_0$ such that $|\sigma|\in i+m.\Z$. As $T$ is strongly connected for $[G]$, there exists a path $k_0'\xrightarrow{w}q$ where $q\in T$ and $\sigma.b.w\in\Sigma_{r,m}^*$. As $\Gamma_{r,m,w}^{-1}(S_{k_0'})\subseteq S_q$, taking the direction of the previous inclusion provides $\Gamma_{r,m,0}^{-|w|}(\vec{S}_{k_0'})\subseteq V_G(T)$. From $x_{k_0,b,k_0'}\in \vec{S}_{k_0'}$ we get $x_{k_0,b,k_0'}\in \Gamma_{r,m,0}^{|w|}(V_G(T))$. As $\Gamma_{r,m,0}^m(V_G(T))=V_G(T)$ (in fact for any vector space $W$ we have $\Gamma_{r,m,0}^m(W)=W$), we deduce that $Gamma_{r,m,0}^{|w|}(V_G(T))=\Gamma_{r,m,0}^{-(i+1)}(V_G(T))$. Thus $\Gamma_{r,m,0}^{i+1}(x_{k_0,b,k_0'})\in V_G(T)$ and we have proved the other inclusion $V\subseteq V_G(T)$.
  \qed
\end{proof}

\begin{figure}[htbp]
  \begin{center}
    \psframebox{\footnotesize\begin{minipage}{13.5cm}
      \textbf{function} $V_G(T)$.\\
      \textbf{input}\\
      A FDVG $G=(Q,m,K,\Sigma_r,\delta)$ and an untransient component of $T$ of $G$.\\
      \textbf{output}\\
      $V_G(T)$.\\
      \textbf{begin}\\
      \ind \textbf{let} $K_0$ be the set of states $k_0\in K$ reachable and co-reachable from $T$.\\
      \ind \textbf{for} each state $k_0\in K_0$.\\
      \ind \ind \textbf{let} $\sigma_{k_0}\in \Sigma_{r,m}^+$ such that $k_0\xrightarrow{\sigma_{k_0}}k_0$.\\
      \ind \ind \textbf{let} $\lambda(k_0)\leftarrow\{i\in\finiteset{0}{m-1};\;T\xrightarrow{\Sigma_{r,m}^*.\Sigma_r^i}k_0\}$.\\
      \ind \textbf{end for}.\\
      \ind \textbf{let} $V\leftarrow\{\unit_{0,m}\}$.\\
      \ind \textbf{for} each transition $k_0\xrightarrow{b}k_0'$.\\
      \ind \ind \textbf{let} $x\leftarrow \Gamma_{r,m,b}^{-1}(\xi_{r,m}(\sigma_{k_0}))-\xi_{r,m}(\sigma_{k_0'})$.\\
      \ind \ind \textbf{let} $V\leftarrow V+\sum_{i\in\lambda(k_0)}\Q.\Gamma_{r,m,0}^{i+1}(x)$.\\
      \ind \textbf{end for}.\\
      \ind \textbf{return} $V$.\\
      \textbf{end}
    \end{minipage}}
  \end{center}
  \caption{An algorithm computing in polynomial time $V_G(T)$.\label{fig:algovgt}}
\end{figure}

\begin{figure}[htbp]
  \begin{center}
    \begin{picture}(78,42)(0,-10)
      \put(0,-10){\framebox(78,42){}}
      \node[Nmarks=i,iangle=180](A)(20,26){$\{1\}$}
      \node(B)(60,26){$\{0\}$}
      \node(C)(40,6){$\emptyset$}
      \node[Nframe=n](Bp)(60,6){$\scriptstyle \{0\}$}
      \drawedge[dash={1.0 1.0 1.0 1.0}{0.0}](B,Bp){}
      \drawloop[loopangle=-90](C){$\scriptstyle \Sigma_{r}$}
      \drawedge[curvedepth=0](A,B){$\scriptstyle 1$}
      \drawloop[loopangle=0](B){$\scriptstyle 0$}
      \drawedge[curvedepth=0,ELside=r](B,C){$\scriptstyle\Sigma_r\moins\{0\}$}
      \drawedge[curvedepth=0](A,C){$\scriptstyle\Sigma_r\moins\{1\}$}
    \end{picture}
  \end{center}
  \caption{The FDVA $\automaton_{r,1}(\{1\})$\label{fig:DVAun}}
\end{figure}

\begin{example}\label{exmp:dvaun}
  Let $\automaton_{r,1}(\{1\})$ be the FDVA given in figure \ref{fig:DVAun}. The two components $T_1=\{\{0\}\}$ and $T_\perp=\{\emptyset\}$ are untransient,and the component $T_0=\{\{1\}\}$ is transient.
\end{example}


\begin{example}
  Let $\automaton_{r,3}(+)$ be the FDVA representing $\{x\in\Z^3;\;x[1]+x[2]=x[3]\}$ and given in figure \ref{fig:dvaplus}. We denote by $q_0$, $q_1$ and $q_\perp$, the principal states $q_0=\{x\in\Z^3;\;x[1]+x[2]=x[3]\}$, $q_1=\{x\in\Z^m;\;x[1]+x[2]+1=x[3]\}$ and $q_\perp=\emptyset$. The two strongly connected components $T_0=\{q_0,q_1\}$ and $T_\perp=\{q_\perp\}$ are untransient. We have $V_{G}(T_\perp)=\Q^3$ and $V_{G}(T_0)=\{x\in\Q^3;\;x[1]+x[2]=x[3]\}$.
\end{example}

\begin{example}
  Let $\automaton_{r,2}(V_r)$ be the FDVA representing $\{x\in\Z^2;\;V_r(x[1])=x[2]\}$ and given in figure \ref{fig:DVAval}. We denote by $q_0$, $q_1$ and $q_\perp$ the principal states $q_0=\{x\in\Z^2;\;V_r(x[1])=x[2]\}$, $q_1=\Z\times\{0\}$ and $q_\perp=\emptyset$. The three strongly connected components $T_0=\{q_0\}$, $T_1=\{q_1\}$ and $T_\perp=\{q_\perp\}$ are untransient. Moreover, the vector spaces associated to $T_0$, $T_1$, $T_\perp$ are respectively equal to $\{\unit_{0,m}\}$, $\Q\times\{0\}$ and $\Q^2$.
\end{example}

\section{Detectable semi-$V$-patterns}
In this section, we prove that any semi-$V$-pattern $P\in\P_V(X)$ introduced by decomposition theorem \ref{theo:decomposition} is $(r,m)$-detectable in $X$ for any affine component $V$ of $\vecsaff(X)$ and for any Presburger-definable set $X$. That means, given a DVA $\automaton$ that represents $X$, there exists a final function $F$ such that $P$ is represented by $\automaton^F$. Independently, being given a semi-$V$-pattern $P$ and a FDVA $\automaton$ that represents a set $X$ not necessary Presburger-definable, a polynomial time algorithm for deciding if there exists a final function $F$ such that $P$ is represented by $\automaton^F$ is provided.

\begin{lemma}\label{lem:Xsigma}
  Given a Presburger definable set $X$, an affine component $V$ of $\vecsaff(X)$ and a word $\sigma\in\Sigma_{r,m}^*$, we have:
  $$[\gamma_{r,m,\sigma}^{-1}(X)]^V=\bigcup_{P\in\P_V(X)}^V([\gamma_{r,m,\sigma}^{-1}(P)]^V\cap^V(\class_{V,P}(X)+V^\perp))$$
\end{lemma}
\begin{proof}
  Recall that $[X]^V=\bigcup_{P\in\P_V(X)}^V([P]^V\cap^V(\class_{V,P}(X)+V^\perp))$ from decomposition theorem \ref{thm:decomposition} and proposition \ref{prop:XvtoX}. We deduce that $[\gamma_{r,m,\sigma}^{-1}(X)]^V=\bigcup_{P\in\P_V(X)}^V([\gamma_{r,m,\sigma}^{-1}(P)]^V\cap^V(\class_{V,P}(X)+V^\perp))$ from lemmas \ref{lem:simVgamma} and \ref{lem:simVgammadi} and corollary \ref{cor:gammacone}.
  \qed
\end{proof}

\begin{corollary}\label{cor:Mdetectable}
  Let $X$ be a Presburger-definable set and $V$ be an affine component of $\vecsaff(X)$. Any set $P\in\P_V(X)$ is detectable in $X$.
\end{corollary}
\begin{proof}
  Let us consider a pair $(\sigma_1,\sigma_2)$ of words in $\Sigma_{r,m}^*$ such that $\gamma_{r,m,\sigma_1}^{-1}(X)=\gamma_{r,m,\sigma_2}^{-1}(X)$. From lemma \ref{lem:Xsigma} we deduce that $\bigcup_{P\in\P_V(X)}^V([\gamma_{r,m,\sigma_1}^{-1}(P)]^V\cap^V(\class_{V,P}(X)+V^\perp))=\bigcup_{P\in\P_V(X)}^V([\gamma_{r,m,\sigma_2}^{-1}(P)]^V\cap^V(\class_{V,P}(X)+V^\perp))$. By intersecting the previous equality by $\class_{V,P}(X)+V^\perp$, we get $[\gamma_{r,m,\sigma_1}^{-1}(P)]^V\cap^V(\class_{V,P}(X)+V^\perp)=[\gamma_{r,m,\sigma_2}^{-1}(P)]^V\cap^V(\class_{V,P}(X)+V^\perp)$. From lemma \ref{lem:dimdim} we deduce that $\gamma_{r,m,\sigma_1}^{-1}(P)=\gamma_{r,m,\sigma_2}^{-1}(P)$.
  \qed
\end{proof}

Even if the following two corollaries are not used in this section, they become useful in the sequel.
\begin{corollary}\label{cor:wP}
  Let $X$ be a $(r,m,w)$-cyclic Presburger-definable set and let $V$ be an affine component of $\vecsaff(X)$. Any semi-$V$-pattern $P\in\P_V(X)$ is relatively prime with $r$ and included in the $V$-affine space $A=\xi_{r,m}(w)+V$.
\end{corollary}
\begin{proof}
  Since any $P\in\P_V(X)$ is $(r,m)$-detectable in $X$, we deduce that any $P\in\P_V(X)$ is $(r,m,w)$-cyclic. From lemma \ref{lem:cyclicpattern}, any $P\in\P_V(X)$ is relatively prime with $r$ and included in $A$.
  \qed
\end{proof}

\begin{corollary}\label{cor:detdet}
  The set $\Z^m\cap (\xi_{r,m}(w)+V)$ is $(r,m)$-detectable in $X$ for any Presburger-definable set $X\subseteq \Z^m$ and any affine component $V\in\comp(\saff(X))$.
\end{corollary}
\begin{proof}
  Let $A$ be the $V$-affine space $A=\xi_{r,m}(w)+V$. Let us consider $P\in\P_V(X)\moins\{\emptyset\}$. It is sufficient to prove that $\Z^m\cap A$ is $(r,m)$-detectable in $P$. Consider a pair $(\sigma_1,\sigma_2)$ of words in $\Sigma_{r,m}^*$ such that there exists $P'$ satisfying $\gamma_{r,m,\sigma_1}^{-1}(P)=P'=\gamma_{r,m,\sigma_2}^{-1}(P)$. Remark that if $P'=\emptyset$ then the dense pattern corollary \ref{cor:densepattern} shows that $\gamma_{r,m,\sigma_1}^{-1}(\Z^m\cap A)=\emptyset=\gamma_{r,m,\sigma_2}^{-1}(\Z^m\cap A)$. If $P'=\not\emptyset$, we deduce that $\saff(\gamma_{r,m,\sigma_i}^{-1}(P))=\Gamma_{r,m,\sigma_i}^{-1}(A)$. Therefore $\Gamma_{r,m,\sigma_1}^{-1}(A)=\Gamma_{r,m,\sigma_2}^{-1}(A)$. In particular, by intersecting the previous equality by $\Z^m$, we get $\gamma_{r,m,\sigma_1}^{-1}(\Z^m\cap A)=\gamma_{r,m,\sigma_2}^{-1}(\Z^m\cap A)$.
  \qed
\end{proof}


\begin{theorem}\label{thm:getdetectablepatterns}
  Let $\automaton$ be a FDVA, let $M$ be a $V$-vector lattice included in $\Z^m$, and let $B$ be a non empty finite subset of $\Z^m$. We can compute in polynomial time a partition $B_0$, $B_1$, ..., $B_n$ of $B$ such that a semi-$V$-pattern $P$ of the form $P=B'+M$ where $B'\subseteq B$ is represented by a FDVA of the form $\automaton^F$ if and only if there exists $J\subseteq\finiteset{1}{n}$ such that $B=\bigcup_{j\in J}B_j$.
\end{theorem}
\begin{proof}
  Let us denote by $\class$ the class of subsets of $X'\subseteq \Z^m$ that can be represented by the FDVA $\automaton^F$ where $F$ is any final function. Since $\class$ is stable by boolean operations in $\{\cup,\cap,\moins,\Delta\}$, we deduce that exists a unique partition $B_0$, $B_1$, ..., $B_n$ of a subset of $B$ satisfying the theorem. From proposition \ref{prop:critdetectable}, we deduce that there exists a finite set $U$ of pairs $(\sigma_1,\sigma_2)$ of words in $\Sigma_r^*$ computable in polynomial time such that $|\sigma_1|+m.\Z=|\sigma_2|+m.\Z$ for any $(\sigma_1,\sigma_2)\in U$, and such that a subset $X'\subseteq\Z^m$ is in in $\class$ if and only if $\gamma_{r,m,\sigma_1}^{-1}(X')=\gamma_{r,m,\sigma_2}^{-1}(X')$ for any $(\sigma_1,\sigma_2)\in U$. Let us consider the binary relation $\relation$ over $B$ defined by $b_1\relation b_2$ if and only if there exists $(\sigma_1,\sigma_2)\in U$ such that $\gamma_{r,m,\sigma_1}^{-1}(b_1+M)\cap \gamma_{r,m,\sigma_2}^{-1}(b_2+M)\not=\emptyset$. The symmetrical and transitive closure of $\relation$ denoted by $\relation'$ provides an equivalence relation of $B$. Let us consider the equivalence classes $B'_1$, ..., $B'_k$ of $\relation'$ such that the last classes $B'_{n'+1}$, ..., $B'_k$ are the equivalence classes such that $B'_i+M$ is not in $\class$. 

  Let us prove that $B_0=\bigcup_{i=n+1}^kB'_i$ and $B_1$, ..., $B_n$ are equal up to a permutation to $B'_1$, ..., $B'_{n'}$. Observe that $B_i+M$ is in $\class$ for any $i\geq 1$. Thus for any $(\sigma_1,\sigma_2)\in U$, we have $\gamma_{r,m,\sigma_1}^{-1}(B_i+M)=\gamma_{r,m,\sigma_2}^{-1}(B_i+M)$. In particular $b_1\relation b_2$ implies that there exists $i\geq 0$ such that $b_1,b_2\in B_i$. We have proved that for any equivalence class $B'$ of $\relation'$, there exists $i$ such that $B'\subseteq B_i$. Note that if $B'\subseteq B_0$ then $B'+M$ is not in $\class$ by definition of $B_0$. Next, assume that $B'\subseteq B_i$ with $i\geq 1$. Let us consider $(\sigma_1,\sigma_2)\in U$ and let $x\in \gamma_{r,m,\sigma_1}^{-1}(B'+M)$. There exists $b_1\in B'$ such that $\gamma_{r,m,\sigma_1}(x)\in b_1+M$. Since $B_i+M\in \class$, we get $\gamma_{r,m,\sigma_1}^{-1}(B_i+M)=\gamma_{r,m,\sigma_2}^{-1}(B_i+M)$. As $b_1\in B'\subseteq B_i$, we deduce that there exists $b_2\in B_i$ such that $\gamma_{r,m,\sigma_2}(x)\in b_2+M$. Thus $\gamma_{r,m,\sigma_1}^{-1}(b_1+M)\cap \gamma_{r,m,\sigma_2}^{-1}(b_2+M)\not=\emptyset$ and we have proved that $b_1\relation b_2$. Since $b_1\in B'$ we get $b_2\in B'$ and we have proved that $\gamma_{r,m,\\sigma_1}^{-1}(B'+M)\subseteq\gamma_{r,m,\sigma_2}^{-1}(B'+M)$. By symmetry, we get the equality $\gamma_{r,m,\sigma_1}^{-1}(B'+M)=\gamma_{r,m,\sigma_2}^{-1}(B'+M)$. We have proved that $B'+M\in\class$. Since $B'$ is non empty and included in $B_i$, we deduce that $B'=B_i$. We have proved that $B_0=\bigcup_{i=n+1}^kB'_i$ and $B_1$, ..., $B_n$ are equal up to a permutation to $B'_1$, ..., $B'_{n'}$.
  
  Therefore, it is sufficient to prove that we can decide in polynomial time if $\gamma_{r,m,\sigma_1}^{-1}(b_1+M)\cap \gamma_{r,m,\sigma_2}^{-1}(b_2+M)\not=\emptyset$ for any $b_1,b_2\in B$, and we can decide in polynomial time if $B'+M\in\class$ for any $B'\subseteq B$. Proposition \ref{prop:invsemipattern} prove that for any word $\sigma$ and for any finite subset $B'\subseteq\Z^m$, we can compute in polynomial time a finite subset $B_\sigma\subseteq\Z^m$ and a vector lattice $M_\sigma$ such that $|B_\sigma|\leq |B'|$ and $\gamma_{r,m,\sigma}^{-1}(B'+M)=B_\sigma+M_\sigma$. Therefore, it is sufficient to prove that given two vector lattices $M_1$ and $M_2$, two finite subsets $B_1$ and $B_2$ of $\Z^m$, and two vectors $b_1$ and $B_2$ in $\Z^m$, we can decide in polynomial time if $b_1+M_1\cap b_2+M_2\not=\emptyset$ and we can decide in polynomial time if $(B_1+M_1)=(B_2+M_2)$. From corollaries \ref{cor:intersectionaffinelattice} and \ref{cor:semiaffinelatticeequal}, we are done.
  \qed
\end{proof}


\section{Terminal components}
A \emph{terminal component}\index{Gen}{terminal component} $T$ of a FDVA $\automaton=(q_0,G,F_0)$ is a component of $G$ satisfying:
\begin{itemize}
\item $T$ is reachable (for $[G]$) from the initial state $q_0$, 
\item there exists a state $q\in T$ such that $[F_0](q)\not=\emptyset$, and 
\item any state $q'$ reachable (for $[G]$) from $T$ such that $[F_0](q')\not=\emptyset$ is in $T$.
\end{itemize}
 The set of terminal components of a FDVA $\automaton$ is denoted by $\T_\automaton$\index{Not}{$\T_\automaton$}. 

Observe that $V_G(T)$ is defined for any terminal component $T$ since the following proposition \ref{prop:terminalpropunt} show that such a $T$ is untransient.
\begin{proposition}\label{prop:terminalpropunt}
  A terminal component is untransient.
\end{proposition}
\begin{proof}
  Let $T$ be a terminal component of a FDVA $\automaton$. Consider a state $q\in T$ such that $[F_0](q)\not=\emptyset$, and let $s\in [F_0](q)$. Since $F_0$ is saturated for $G$ and $s\in [F_0](q)$ we deduce that $s\in [F_0](\delta(q,s^n))$ for any $n\in\Nat$. As $T$ is terminal, we have $\delta(q_0,s^n)\in T$. Moreover, as $Q$ is finite, there exits $n\in\Nat$ and $d\in\Nat\moins\{0\}$ such that $\delta(q_0,s^{n+d})=\delta(q_0,s^n)$. We have proved that there exists a loop of the state $q'=\delta(q,s^n)$. From $q'\in T$ we deduce that $T$ is untransient.
  \qed
\end{proof}
The \emph{terminal components} have a lot of applications in the sequel. In this section we show that $\saff(X_q)=a_G(q)+V_G(T)$ and we provide a geometrical characterization of the sets $X_q$.


\begin{lemma}[Destruction lemma]\label{lem:destruction}
  Let $\sigma\in\Sigma_{r,m}^+$ be a non-empty word and let $A$ be an affine space. There exists $k_0\in\Nat$ such that $\gamma_{r,m,\sigma^{k_0}}^{-1}(\Z^m\cap A)=\emptyset$ if and only if $\xi_{r,m}(\sigma)\not\in A$ or $\Z^m\cap A=\emptyset$.
\end{lemma}
\begin{proof}
  We can assume without loss of generality that $\Z^m\cap A\not=\emptyset$. In particular $\vec{A}$ is a vector space (because $A$ is non empty) and there exists a finite set $D\subseteq\Z^m\moins\{\unit_{0,m}\}$ such that $\vec{A}=\{x\in\Q^m;\;\bigwedge_{\alpha\in D}\scalar{\alpha}{x}=0\}$. 
  
  Assume first that $\xi_{r,m}(\sigma)\in A$. The set $\Z^m\cap A$ is equal to $\{x\in\Z^m;\;\bigwedge_{\alpha\in D}\scalar{\alpha}{x-\xi_{r,m}(\sigma)}=0\}$. Remark that $\gamma_{r,m,\sigma}^{-1}(\Z^m\cap A)=\{x\in\Z^m;\;\bigwedge_{\alpha\in D}\scalar{\alpha}{\gamma_{r,m,\sigma}(x)-\xi_{r,m}(\sigma)}=0\}=\Z^m\cap A$. In particular $\gamma_{r,m,\sigma^k}^{-1}(\Z^m\cap A)\not=\emptyset$ for any $k\in\Nat$.
  
  Next, assume that $\gamma_{r,m,\sigma^k}^{-1}(\Z^m\cap A)\not=\emptyset$ for any $k\in\Nat$. As $\Z^m\cap A\not=\emptyset$, there exists $a\in A$. For any $k\in\Nat$, we have:
  \begin{align*}
    &\gamma_{r,m,\sigma^{k}}^{-1}(\Z^m\cap A)\\
    &=\left\{x\in\Z^m;\;\bigwedge_{\alpha\in D}\scalar{\alpha}{\gamma_{r,m,\sigma^{k}}(x)-a}=0\right\}\\
    &=\left\{x\in\Z^m;\;\bigwedge_{\alpha\in D}\scalar{\alpha}{r^{k.|\sigma|_m}.(x-\xi_{r,m}(\sigma))+\xi_{r,m}(\sigma)-a}=0\right\}\\
    &=\left\{x\in\Z^m;\;\bigwedge_{\alpha\in D}\scalar{\alpha}{(r^{|\sigma|_m}-1).x+\gamma_{r,m,\sigma}(\unit_{0,m})}=(r^{|\sigma|_m}-1).\frac{\scalar{\alpha}{a-\xi_{r,m}(\sigma)}}{r^{k.|\sigma|_m}}\right\}
  \end{align*}
  Let us consider $k\in\Nat$ enough larger such that $|(r^{|\sigma|_m}-1).\frac{\scalar{\alpha}{a-\xi_{r,m}(\sigma)}}{r^{k.|\sigma|_m}}|<1$ for any $\alpha\in D$. As $\gamma_{r,m,\sigma^{k}}^{-1}(\Z^m\cap A)\not=\emptyset$, there exists $x$ in this set. From $\scalar{\alpha}{(r^{|\sigma|_m}-1).x+\gamma_{r,m,\sigma}(\unit_{0,m})}\in\Z$, we deduce that $(r^{|\sigma|_m}-1).\frac{\scalar{\alpha}{a-\xi_{r,m}(\sigma)}}{r^{k.|\sigma|_m}}$ is in the set $\{c\in\Z;\;|c|<1\}=\{0\}$. Therefore $\scalar{\alpha}{a-\xi_{r,m}(\sigma)}=0$ for any $\alpha\in D$. That means $\xi_{r,m}(\sigma)\in A$.
  \qed
\end{proof}

\begin{proposition}\label{prop:affeye}
  Let $\automaton=(q_0,G,F_0)$ by a FDVA that represents a set $X$, let $Y$ be an $s$-eye of a FDVG $G$ and let $T$ be a terminal component that contains $\ker_s(Y)$. We have $\saff(X_q^{F_{s,Y}}(G))=a_G(q)+V_G(T)$ for any principal state $q\in T$.
\end{proposition}
\begin{proof}
  Let us denote by $Z_q$ the set $Z_q=\{\xi_{r,m}(w);\;q\xrightarrow{w}q\}$. Recall that $\saff(Z_q)=a_G(q)+V_G(T)$.
  
  Let us first prove that $\saff(Z_q)\subseteq\saff(X_q^{F_{s,Y}})$. Consider a vector $x\in Z_q$. There exists a loop $q\xrightarrow{w}q$ with $w\in\Sigma_{r,m}^+$ such that $x=\xi_{r,m}(w)$. Let $q'\in\ker_s(Y)$. As $q$ and $q'$ are in the same component, there exists a path $q\xrightarrow{\sigma}q'$ with $\sigma\in\Sigma_{r,m}$. Remark that $\rho_{r,m}(w^k.\sigma,s)\in X_q^{F_{s,Y}}$ for any $k\in \Nat$. By developing $\rho_{r,m}(w^k.\sigma,s)$, we get $\rho_{r,m}(w^k.\sigma,s)=r^{k.|w|_m}.(\rho_{r,m}(\sigma,s)-x)+x$. From covering lemma \ref{lem:cov}, we get $\Q.(\rho_{r,m}(\sigma,s)-x)+x\subseteq\saff(X_q^{F_{s,Y}})$. In particular $x\in \saff(X_q^{F_{s,Y}})$ and we get $Z_q\subseteq \saff(X_q^{F_{s,Y}})$. By minimality of the semi-affine full, we deduce the inclusion $\saff(Z_q)\subseteq\saff(X_q^{F_{s,Y}})$.
  
  For the converse inclusion, let us consider a vector $x\in X_q^{F_{s,Y}}$. There exists a $(r,m)$-decomposition $(\sigma,s)$ of $x$ such that $\delta(q,\sigma)\in Y$. By replacing $\sigma$ by a word in $\sigma.s^*$, we can assume that $q'=\delta(q,\sigma)$ is in $\ker_s(Y)$. In particular, there exists $n_1\in\Nat\moins\{0\}$ such that $\delta(q',s^{n_1})=q'$. Proposition \ref{prop:untransient} shows that $\Gamma_{r,m,w}(\xi_{r,m}(s^{n_1}))\in \saff(\xi_{r,m,q})$. Remark that $\xi_{r,m}(s^{n_1})=\frac{s}{1-r}$, and we deduce that $x=\rho_{r,m}(w,s)\in \saff(Z_q)$. We have proved the inclusion $X_q^{F_{s,Y}}\subseteq\saff(Z_q)$. By minimality of the semi-affine hull, we deduce the other inclusion $\saff(X_q^{F_{s,Y}})\subseteq\saff(Z_q)$.
  \qed
\end{proof}

The following proposition shows that for any state $q$ in a terminal component of a FDVA that represents a set $X$, the semi-affine space $\saff(X_q)$ can be easily computed thanks to $a_G(q)$ and $V_G(T)$. 
\begin{proposition}\label{prop:terminalprop}
  Let $X$ be a set represented by a FDVA $\automaton$ and let $T$ be a terminal component. We have $\saff(X_q)=a_G(q)+V_G(T)$ for any state $q\in T$.
\end{proposition}
\begin{proof}
  Let us consider the class $\class_T$ of couple $(s,Y)\in S_{r,m}\times \partie(Q)$ such that $Y$ is an $s$-eye satisfying $\ker_s(Y)\subseteq T$ and $F_{s,Y}\subseteq F_0$. As $T$ is terminal, this class is non-empty. Proposition \ref{prop:affeye} shows that $\saff(X_q^{F_{s,Y}})=a_G(q)+V_G(T)$ for any $(s,Y)\in\class_T$. Let $F=\bigcup_{(s,Y)\in\class_T}F_{s,Y}$. As $q\in T$ and $T$ is terminal, we deduce that $X_q=X_q^{F}=\bigcup_{(s,Y)\in\class_T}X_q^{F_{s,Y}}$. From covering lemma \ref{lem:cov}, we get $\saff(X_q)=a_G(q)+V_G(T)$.
  \qed
\end{proof}

Remark that by definition of $\bound{V}{X}$, there exists a unique semi-$V$-pattern $P\in\P_V(X)$ such that $[C_{V,\#}]^V\subseteq^V\class_{V,P}(X)$ for any sequence $\#\in\{<,>\}^{\bound{V}{X}}$ such that $[C_{V,\#}]_V\not=[\emptyset]_V$. Let $X$ be a Presburger-definable set, let $V$ be an affine component of $\vecsaff(X)$, and let $P\in\P_V(X)$ be a semi-$V$-pattern. We denote by $\S_{V,P}(X)$ the set of sequences $\#\in\{<,>\}^{\bound{V}{X}}$ such that $[C_{V,\#}]^V\subseteq^V\class_{V,P}(X)$.

The following theorem provides a geometrical form of the set $X_q$ when $q$ is a state in a terminal component of a FDVA that represents a Presburger-definable set $X$. 
\begin{theorem}\label{thm:ultime}
  Let $X$ be a Presburger-definable set represented by a FDVA $\automaton$ and let $V$ be an affine component of $\vecsaff(X)$. For any state $q$ in a terminal component $T$ such that $V_G(T)$ is equal to $V$, there exists a vector $a_q\in\Q^m$ such that we have:
  $$X_{q}=\bigcup_{P\in\P_V(X)}\bigcup_{\#\in\S_{V,P}(X)}(P_q\cap (a_q+C_{V,\#}+V^\perp))$$
  such that for any $j\in\finiteset{1}{m}$, we have $-1<a_q[j]\leq 0$ if $V\subseteq \unit_{j,m}^\perp$ and we have $-1<a_q[j]<0$ otherwise.
\end{theorem}
\begin{proof}
  Let us first prove that there exists a loop $q\xrightarrow{w_j}q$ such that $w_j\not\in(\Sigma_{r,m}\cap \unit_{j,m}^\perp)^*$ for any $j\in\finiteset{1}{m}$ satisfying $V\not\subseteq\unit_{j,m}^\perp$. As $\vecsaff(X_q)=V$, from proposition \ref{prop:patternempty} we deduce that there exists $P\in\P_V(X)$ such that $P_q\not=\emptyset$. Let us consider a vector $x\in P_q$. As $V\not\subseteq\unit_{j,m}^\perp$, there exists a vector $v\in V$ such that $v[j]\not=0$ and by replacing $v$ by a vector in $(\Z\moins\{0\}).v$, we have proved that there exists a vector $v\in\inv_V(P_q)$ such that $v[j]>0$. In particular $x+\Z.v\subseteq P_q$. As $v[j]>0$, there exists $k\in\Nat$ enough larger such that $(x+k.v)[j]>0$. Let us consider a $(r,m)$-decomposition $(\sigma,s)$ of $x+k.n.v$. Naturally, as $(x+k.n.v)[j]>0$, we have $\sigma\not\in(\Sigma_{r,m}\cap \unit_{j,m}^\perp)^*$. Moreover, as $\rho_{r,m}(\sigma,s)\in P_q$, we get $P_{q'}\not=\emptyset$ where $q'=\delta(q,\sigma)$. Proposition \ref{prop:patternempty} shows that $X_{q'}\not=\emptyset$. As $T$ is terminal, we have proved that $q'\in T$. Hence, there exists a path $q'\xrightarrow{\sigma'}q$. Remark that the loop $q\xrightarrow{w_j}q$ where $w_j=\sigma.\sigma'$ satisfies $w_j\not\in(\Sigma_{r,m}\cap \unit_{j,m}^\perp)^*$.
  
  Let us consider the sequence $(C_{V,P})_{P\in\P_V(X)}$ of $V$-polyhedrons defined by $C_{V,P}=\bigcup_{\#\in\S_{V,P}(X)}C_{V,\#}$. Remark that $\class_{V,P}(X)=[C_{V,P}]^V$ for any $P\in\P_V(X)$. Hence, the set $Z=X\Delta(\bigcup_{P\in\P_V(X)}\bigcup_{\#\in\S_{V,P}(X)}(P_q\cap (a_q+C_{V,\#}+V^\perp)))$ is such that $[Z]^V=[\emptyset]^V$. Let us consider a path $q_0\xrightarrow{\sigma}q$ with $q$ in a terminal component $T$ such that $V_G(T)=V$. Thanks to the first paragraph, we can assume without loss of generality that $\sigma\not\in(\Sigma_{r,m}\cap \unit_{j,m}^\perp)^*$ for any $j\in\finiteset{1}{m}$ satisfying $V\not\subseteq\unit_{j,m}^\perp$. As $\vecsaff(X_q)=V$, and $[\gamma_{r,m,\sigma}^{-1}(Z)]^V=\gamma_{r,m,\sigma}^{-1}([\emptyset]^V)=[\emptyset]^V$, we deduce that $X_q$ is not included in $\saff(\gamma_{r,m,\sigma}^{-1}(Z))$. Hence, there exists a $(r,m)$-decomposition $(w_1,s)\in\rho_{r,m}^{-1}(X_q)$ such that $\rho_{r,m}(w_1,s)\not\in\saff(\gamma_{r,m,\sigma}^{-1}(Z))$. Destruction lemma \ref{lem:destruction} shows that by replacing $w_1$ by a word in $w_1.s^*$, we can assume that $\gamma_{r,m,\sigma.w_1}^{-1}(Z)=\emptyset$. Let $q'=\delta(q,w_1)$. As $s\in F_0(q')$ and $T$ is terminal, we deduce that $q'\in T$. As $q$ and $q'$ are in the strongly connected component $T$, there exists a path $q'\xrightarrow{w_2}q$. Let $w=w_1.w_2$ and let $a_q=\Gamma_{r,m,\sigma.w}^{-1}(\unit_{0,m})$. As $\sigma\not\in(\Sigma_{r,m}\cap \unit_{j,m}^\perp)^*$ for any $j\in\finiteset{1}{m}$ satisfying $V\not\subseteq\unit_{j,m}^\perp$, we deduce that for any $j\in\finiteset{1}{m}$, we have $-1<a_q[j]\leq 0$ if $V\subseteq \unit_{j,m}^\perp$ and we have $-1<a_q[j]<0$ otherwise. Remark that for any $V$-hyperplane $H$ such that $\vec{H}=H$ and for any $\#\in\{<,\leq,=,\geq,>\}$, we have $\Gamma_{r,m,\sigma.w}^{-1}(H^\#+V^\perp)=a_q+H^\#+V^\perp$. As $\gamma_{r,m,\sigma.w}^{-1}(Z)=\emptyset$ then $X_q=\bigcup_{P\in\P_V(X)}\bigcup_{\#\in\S_{V,P}(X)}(P_{q}\cap (a_q+C_{V,\#}+V^\perp))$.
  \qed
\end{proof}

%% file: chapter.geometricalextraction.tex
\chapter{Extracting Geometrical Properties}

\section{Semi-affine hull direction of a Presburger-definable FDVA}
In this section we prove that the semi-affine hull direction $\vec\saff(X)$ of a Presburger-definable set $X$ represented by a FDVA is computable in polynomial time.

This computation cannot be extended to $\saff(X)$. In fact, as shown by the following lemma \ref{lem:exexp}, the size of $\saff(X)$ can be exponentially larger than the size of a FDVA representing $X$.
\begin{lemma}\label{lem:exexp}
  There exist $\alpha,\beta\in\Q_+\moins\{0\}$, a sequence $(\automaton_n)_{n\in\Nat}$ of FDVA that represents a sequence $(X_n)_{n\in\Nat}$ of Presburger-definable sets in basis $r$, such that $\lim_{n\rightarrow+\infty}\size(\automaton_n)=+\infty$ and $\size(\saff(X_n))\geq \alpha.2^{\beta.\size(\automaton_n)}$.
\end{lemma}
\begin{proof}
  Consider the finite set $X_n=\finiteset{0}{r^n-1}^m$. Remark that $X_n$ is Presburger-definable and the FDVA $\automaton_{r,1}(X_n)$ that represents $X_n$ has $n+2$ principal states. Moreover, as $\comp(\saff(X_n))=\{\{x\};\;x\in X_n\}$, we deduce that $\size(\saff(X_n))=r^n$.
  \qed
\end{proof}

\begin{remark}
  The semi-affine hull of a set $X$ represented by a FDVA ($X$ is not necessarily Presburger-definable) can be computed in exponential time thanks to the algorithm provided in \cite{L-THESE03}. This result is not used in this paper.
\end{remark}

Our computation of $\vecsaff(X)$ is based on the following lemma \ref{lem:undervecsaff} that shows that an under-approximation of $\vecsaff(X)$ can be easily computed from a FDVA that represents a set $X$. In this section, we prove that this under-approximation is exact if $X$ is Presburger-definable.
\begin{lemma}\label{lem:undervecsaff}
  Let $X$ be a set represented by a FDVA. We have $\bigcup_{T\in \T_\automaton}V_G(T)\subseteq \vecsaff(X)$.
\end{lemma}
\begin{proof}
  Let us consider a FDVA $\automaton$ that represents a set $X$. Let us consider a terminal component $T\in \T_\automaton$ and let us prove that $V_G(T)\subseteq\vecsaff(X)$. Let us consider $q\in T$. As $T$ is reachable (for $[G]$) from the initial state, there exists a path $q_0\xrightarrow{\sigma\in\Sigma_{r,m}^*}q$. We have $X_q=\gamma_{r,m,\sigma}^{-1}(X)\subseteq \Gamma_{r,m,\sigma}^{-1}(X)$. Covering lemma \ref{lem:cov} shows that $\vecsaff(X_q)\subseteq\vecsaff(X)$. Moreover, as $q\in T$, proposition \ref{prop:terminalprop} shows that $\vecsaff(X_q)=V_G(T)$. Therefore $V_G(T)\subseteq\vecsaff(X)$ and we have proved the inclusion $\bigcup_{T\in T_\automaton}V_G(T)\subseteq \vecsaff(X)$.
  \qed
\end{proof}

\begin{proposition}\label{prop:ultime}
  Let $X$ be a Presburger-definable set represented by a FDVA $\automaton$ and let $V$ be an affine component of $\vecsaff(X)$. For any principal state $q$ reachable for $[G]$, there exists $P\in\P_V(X)$ such that $P_q\not=\emptyset$ if and only if there exists a terminal component $T\in \T_\automaton$ reachable from $q$ for $[G]$ such that $V_G(T)=V$.
\end{proposition}
\begin{proof}
  Assume first that there exists a terminal component $T\in\T_\automaton$ reachable from $q$ for $[G]$ such that $V_G(T)=V$ and let us prove that there exists $P\in\P_V(X)$ such that $P_q\not=\emptyset$. There exists $q'\in T$ and a path $q\xrightarrow{\sigma\in\Sigma_{r,m}^*}q'$. From theorem \ref{thm:ultime}, since $X_{q'}\not=\emptyset$, we deduce that there exists $P\in\P_V(X)$ such that $P_{q'}\not=\emptyset$. As $P_{q'}=\gamma_{r,m,\sigma}^{-1}(P_{q})$ we get $P_q\not=\emptyset$ and we have proved that there exists $P\in\P_V(X)$ such that $P_q\not=\emptyset$. Let us prove the converse. Assume that there exists $P\in\P_V(X)$ such that $P_q\not=\emptyset$ and let us prove that there exists a terminal component $T\in \T_\automaton$ reachable from $q$ for $[G]$ such that $V_G(T)=V$. Since $q$ is reachable for $[G]$ from the initial state, there exists a path $q_0\xrightarrow{\sigma_0}q$. Let us consider a sequence $(C_{V,P})_{P\in\P_V(X)}$ of $V$-polyhedrons such that $C_{V,P}\in\class_{V,P}(X)$. Let us consider $Z=X\Delta \bigcup_{P\in\P_V(X)}(P\cap (C_{V,P}+V^\perp))$. We have $[Z]^V=[\emptyset]^V$. That means $V$ is not included in $\vecsaff(Z)$.  Let $Z'=\gamma_{r,m,\sigma_0}^{-1}(Z)$. From covering lemma \ref{lem:cov}, we deduce that $V$ is not included in $\vecsaff(Z')$. Observe that if there exists $P\in\P_V(X)$ such that $P_q\not=emptyset$ then from $Z'=X_q\Delta \bigcup_{P\in\P_V(X)}(P_q\cap (C_{V,P}+V^\perp))$, we deduce that $V$ is included in $\vecsaff(X_q)$. Thus, there exists a $(r,m)$-decomposition $(\sigma,s)$ such that $\rho_{r,m}(\sigma,s)\in X_q$ and $\rho_{r,m}(\sigma,s)\not\in\saff(Z')$. Destruction lemma \ref{lem:destruction} proves that by replacing $\sigma$ by a word in $\sigma.s^*$, we can assume that $\gamma_{r,m,\sigma}^{-1}(Z')=\emptyset$. Let $q'=\delta(q,\sigma)$ and remark that $X_{q'}=\bigcup_{P\in\P_V(X)}(P_{q'}\cap (\Gamma_{V,r,m,\sigma}^{-1}(C_{V,P})+V^\perp))$. As $\rho_{r,m}(\epsilon,s)\in X_{q}$ then $s\in F_0(q')$. So there exists a terminal component $T$ reachable (for $[G]$) from $q'$. Let $q''\in T$. There exists a path $q'\xrightarrow{w\in\Sigma_{r,m}^*}q''$ such that $q''\in T$. We have $X_{q''}=\bigcup_{P\in\P_V(X)}(P_{q''}\cap(\Gamma_{V,r,m,\sigma.w}^{-1}(C_{V,P})+V^\perp))$. As $X_{q''}\not=\emptyset$, there exists $P\in\P_V(X)$ such that $P_{q''}\not=\emptyset$. In particular $P_{q''}$ is a non-empty semi-$V$-pattern. As $C_{V,P}$ is non-$V$-degenerate and $[C_{V,P}]_V=[\Gamma_{V,r,m,\sigma.w}^{-1}(C_{V,P})]_V$, we deduce that $\Gamma_{V,r,m,\sigma.w}^{-1}(C_{V,P})$ is non-$V$-degenerate. Lemma \ref{lem:dimdim} proves that $V$ is included in $\vecsaff(P_{q''}\cap (\Gamma_{V,r,m,\sigma.w}^{-1}(C_{V,P})+V^\perp))$. Therefore $V\subseteq\vecsaff(X_{q''})$. Moreover, as $\vecsaff(P_{q''})\subseteq V$ for any $P\in\P_V(X)$, we deduce that $\vecsaff(X_{q''})=V$. As $q''\in T$, recall that $V_G(T)=\vecsaff(X_{q''})$. Therefore, we have proved that there exists a terminal component $T$ such that $V_G(T)=V$.
  \qed
\end{proof}

From the previous proposition \ref{prop:ultime}, we deduce that $\vecsaff(X)$ can be easily computed in polynomial time from the sequence of vector spaces associated to the terminal components.
\begin{proposition}\label{prop:vechull}
  For any Presburger-definable set $X$ represented by a FDVA $\automaton$, we have:
  $$\vecsaff(X)=\bigcup_{T\in \T_\automaton}V_G(T)$$
\end{proposition}
\begin{proof}
  Lemma \ref{lem:undervecsaff} shows that $\bigcup_{T\in \T_\automaton}V_G(T)\subseteq \vecsaff(X)$. Now, let us prove the converse inclusion. Let $V$ be an affine component of $\vecsaff(X)$. Proposition \ref{prop:ultime} shows that there exists a terminal component $T$ such that $V_G(T)=V$. Therefore $V\subseteq \bigcup_{T\in T_\automaton}V_G(T)$. We deduce the other inclusion $\vecsaff(X)\subseteq\bigcup_{T\in T_\automaton}V_G(T)$.
  \qed
\end{proof}

From theorem \ref{thm:vgtpoly} and the previous proposition \ref{prop:vechull}, we get one of the \emph{main powerful theorem of this paper}.
\begin{theorem}\label{thm:semiaffinehull}
  The semi-affine hull direction of a Presburger-definable set represented by a FDVA is computable in polynomial time.
\end{theorem}

\subsection{An example}
Let us consider the set $X=X_1\cup X_2$ where $X_1=\{x\in\Nat^2;\;x[1]=2.x[2]\}$ and $X_2=\{x\in\Nat^2;\; x[2]=2.x[1]\}$. Naturally, the semi-vector space $\vecsaff(X_1)$ is equal to the vector space $V_1=\{x\in\Q^2;\;x[1]=2.x[2]\}$ and symmetrically the semi-vector space $\vecsaff(X_2)$ is equal to the vector space $V_2=\{x\in\Q^2;\;x[2]=2.x[1]\}$. As $\vecsaff(X)$ has two affine components $V_1$ and $V_2$, from proposition \ref{prop:vechull}, we deduce that whatever the FDVA $\automaton$ that represents $X$ we consider, for any terminal terminal components $T$, we have $V_G(T)\subseteq V_1$ or $V_G(T)\subseteq V_2$ (remark that we have implicitly used the insecable lemma \ref{lem:insecable}). Moreover, we also deduce that there exists at least one terminal component $T_1$ such that $V_G(T_1)=V_1$ and at least one terminal component $T_2$ such that $V_G(T_2)=V_2$.

This property can be verified in practice. Figure \ref{fig:dvadisj} represents the minimal FDVA $\automaton_{2,2}(X_1\cup X_2)$ where $X_1'=\{x\in\Nat;\;x[1]=2.x[2]+1\}$ and $X_2'=\{x\in\Nat;\;x[2]=2.x[1]+1\}$. Remark that this FDVA has 2 terminal components $T_1$ and $T_2$ defined by $T_1=\{X_1,X_1'\}$ and $T_2=\{X_2,X_2'\}$. We have $V_G(T_1)=\vec\aff(X_1)=\vec\aff(X_{1}')=V_1$ and $V_G(T_2)=\vec\aff(X_2)=\vec\aff(X_{2}')=V_2$.

\begin{figure}[htbp]
  \begin{center}
    \begin{picture}(100,80)(-10,-70)
      \put(-10,-70){\framebox(100,80){}}
      \node(n0)(10,-20){$X_1'$}
      \node[Nadjust=w,iangle=-120,Nmarks=i](n1)(40,-10){$X_1\cup X_2$}
      \node[Nadjust=wh,Nframe=n](n1p)(27,3){$\scriptstyle (0,0)$}
      \drawedge[dash={1.0 1.0 1.0 1.0}{0.0}](n1,n1p){}
      \node(n2)(70,-20){$X_2'$}
      \node(n3)(70,-44){$X_2$}
      \node[Nadjust=wh,Nframe=n](n3p)(70,-64){$\scriptstyle (0,0)$}
      \drawedge[dash={1.0 1.0 1.0 1.0}{0.0}](n3,n3p){}
      \node(n4)(10,-44){$X_1$}
      \node[Nadjust=wh,Nframe=n](n4p)(10,-64){$\scriptstyle (0,0)$}
      \drawedge[dash={1.0 1.0 1.0 1.0}{0.0}](n4,n4p){}
      \node(np)(40,-44){$\emptyset$}    
      \drawloop[loopangle=225](n4){$\scriptstyle (0,0)$}
      \drawloop[loopangle=-45](n3){$\scriptstyle (0,0)$}
      \drawloop(n0){$\scriptstyle (1,1)$}
      \drawloop[loopangle=45](n1){$\scriptstyle (0,0)$}
      \drawloop(n2){$\scriptstyle (1,1)$}
      \drawedge(n1,n2){$\scriptstyle (1,0)$}
      \drawloop[loopangle=-90](np){$\scriptstyle \Sigma_{2,2}$}
      \drawedge[ELside=r](n1,n0){$\scriptstyle (0,1)$}
      \drawedge[ELside=r,curvedepth=-8](n3,n2){$\scriptstyle (1,0)$}
      \drawedge[curvedepth=-8](n2,n3){$\scriptstyle (0,1)$}
      \drawedge[ELside=r,curvedepth=8](n0,n4){$\scriptstyle (1,0)$}
      \drawedge[curvedepth=8](n4,n0){$\scriptstyle (0,1)$}
      \drawedge(n4,np){$\scriptstyle \{1\}\times\Sigma_2$}
      \drawedge[ELside=r](n3,np){$\scriptstyle \Sigma_2\times\{1\}$}
      \drawedge(n0,np){$\scriptstyle \{0\}\times\Sigma_2$}
      \drawedge[ELside=r](n2,np){$\scriptstyle \Sigma_2\times\{0\}$}
      \drawedge[ELpos=25](n1,np){$\scriptstyle (1,1)$}
    \end{picture}
    \caption{The FDVA $\automaton_{2,2}(X_1\cup X_2)$\label{fig:dvadisj}}
  \end{center}
\end{figure}

\section{Polynomial time invariant computation}
Let $X$ be a Presburger-definable set and $V$ be an affine component of $\vecsaff(X)$. The $V$-vector lattice $\inv_V(X)$\index{Not}{$\inv_V(X)$} of invariants of $X$ is defined by the following equality: 
$$\inv_V(X)=\bigcap_{P\in\P_V(X)}\inv_V(P)$$
In this section we prove that the $V$-vector lattice of invariants $\inv_V(X)$ is computable in polynomial time from a \emph{cyclic} FDVA $\automaton$ that represents $X$ in basis $r$. We also prove that $|\Z^m\cap V/\inv_V(X)|$ is bounded by the number of principal states of $\automaton$.

Recall that corollary \ref{cor:wP} proves that any $P\in\P_V(X)$ is relatively prime with $r$ and included in the $V$-affine space $\xi_{r,m}(s)+V$. This $V$-affine space will be useful in the sequel. Our algorithm is based on the following proposition \ref{prop:invpolycyclic} and the remaining of this section is devoted to prove that all structures needed for applying this proposition are small and they can be computed efficiently. 
\begin{lemma}\label{lem:pourZrms}
  Let $A$ be a $V$-affine space and $s\in S_{r,m}$ be a $(r,m)$-sign vector such that $[Z_{r,m,s}\cap A]^V\not=[\emptyset]^V$. There exists a vector $v\in V$ such that $v[i]<0$ if $s[i]=r-1$ and $v[i]>0$ if $s[i]=0$ for any $i\in \finiteset{1}{m}$ such that $\unit_{i,m}\not\in V^\perp$.
\end{lemma}
\begin{proof}
  Since $A$ is a $V$-affine space, there exists $a\in A$. We denote by $\#_0$ the binary relation $\geq$ and by $\#_{r-1}$ the binary relation $<$, and we denote by $I$ the set of $i\in\finiteset{1}{m}$ such that $\unit_{i,m}\not\in V^\perp$. Remark that $Z_{r,m,s}\cap A=\Z^m\cap A\cap (C+\Pi_V(a)+V^\perp)$ where $C$ is the $V$-polyhedron $C=\bigcap_{i=1}^m\{x\in V;\;x[i]+a[i]\#_{s[i]}0\}$. As $\{x\in V;\;x[i]+a[i]\#_{s[i]}0\}=\{x\in V;\;\scalar{\Pi_V(\unit_{i,m})}{x}+a[i]\#_{s[i]}0\}$, we deduce that $\{x\in V;\;x[i]+a[i]\#_{s[i]}0\}$ is either empty or equal to $V$ for any $i\in \finiteset{1}{m}\moins I$. Moreover, as $[Z_{r,m,s}\cap A]^V\not=[\emptyset]^V$, we get $\{x\in V;\;x[i]+a[i]\#_{s[i]}0\}=V$ for any $i\in \finiteset{1}{m}\moins I$. Hence $C=\bigcap_{i\in I}\{x\in V;\;\scalar{\Pi_V(\unit_{i,m})}{x}+a[i]\#_{s[i]}0\}$. As $[\Z^m\cap A\cap (C+\Pi_V(a)+V^\perp)]^v\not=[\emptyset]^V$, lemma \ref{lem:dimdim} shows that $[C]^V\not=[\emptyset]^V$. From lemma \ref{lem:vdegpolyconv} we deduce that there exists a vector $v\in V$ such that $v[i]>0$ if $s[i]=0$ and $v[i]<0$ if $s[i]=r-1$ for any $i\in I$.
  \qed
\end{proof}

\begin{proposition}\label{prop:invpolycyclic}
  Let $X$ be a $(r,m,w)$-cyclic Presburger-definable set and let $V$ be an affine component of $\vecsaff(X)$. Assume that we have:
  \begin{itemize}
  \item  A $(r,m)$-sign vector $s\in S_{r,m}$ such that $[Z_{r,m,s}\cap (\xi_{r,m}(s)+V)]^V\not=[\emptyset]^V$, 
  \item A couple $(q_0,G)$ such that $q_0$ is a principal state of a FDVG $G$ such that $\delta(q_0,\sigma_1)=\delta(q_0,\sigma_2)$ if and only if $(\gamma_{r,m,\sigma_1}^{-1}(P))_{P\in\P_V(X)}=(\gamma_{r,m,\sigma_2}^{-1}(P))_{P\in\P_V(X)}$ for any $\sigma_1,\sigma_2\in\Sigma_{r,m}^*$,
  \item The set $Q'$ of principal states reachable for $[G]$ from $q_0$ such that  $(\emptyset)_{P\in\P_V(X)}\not=(\gamma_{r,m,\sigma}^{-1}(P))_{P\in\P_V(X)}$ if and only if $q'\in Q'$ for any path $q_0\xrightarrow{\sigma}q'$ with $\sigma\in\Sigma_{r,m}^*$,
  \item An integer $n_0\in\Nat\moins\{0\}$ relatively prime with $r$ such that $|\Z^m\cap V/\inv_V(X)|$ divides $n_0$,
  \item An integer $n\in\Nat\moins\{0\}$ such that $r^n\in 1+n_0.\Z$.
  \end{itemize}
  We denote by $U$ the set of pairs $u=(k,Z)\in K\times \Z/mn.\Z$ such that there exists a pair of words $(\sigma_u,\sigma'_u)$ in $\Sigma_r^*$ satisfying $|\sigma_u.\sigma'_u|\in m.n.\Z$, $(k,Z)=(\delta(q_0,\sigma_u),|\sigma_u|+m.n.\Z)$ and there exists an $s$-eye $Y'$ such that $\delta(k,\sigma'_u)\in \ker_s(Y')\subseteq Q'$. Given a sequence $(\sigma_u,\sigma'_u)_{u\in U}$ satisfying the previous conditions and such that $\sigma_{(q_0,m.n.\Z)}=\epsilon$, the vector lattice of invariants $\inv_V(X)$ is equal to the vector lattice generated by $n_0.\Z^m\cap V$ and the vectors $\rho_{r,m}(\sigma_{u_1}.b.\sigma'_{u_2},s)-\rho_{r,m}(\sigma_{u_2}.\sigma'_{u_2},s)$ where $u_1=(k_1,Z_1)\in U$, $b\in\Sigma_r$ and $u_2=(k_2,Z_2)\in U$ are such that $(k_2,Z_2)=(\delta(k_1,b),Z_1+1)$.
\end{proposition}
\begin{proof}
  Let us denote by $A$ the $V$-affine space $A=\xi_{r,m}(w)+V$.

  Since $\delta(q_0,\sigma_1)=\delta(q_0,\sigma_2)$ if and only if $(\gamma_{r,m,\sigma_1}^{-1}(P))_{P\in\P_V(X)}=(\gamma_{r,m,\sigma_2}^{-1}(P))_{P\in\P_V(X)}$ for any $\sigma_1,\sigma_2\in\Sigma_{r,m}^*$, for any principal state $q$ reachable for $[G]$ from $q_0$, there exists a unique sequence denoted by $(P_q)_{P\in\P_V(X)}$ such that $P_q=\gamma_{r,m,\sigma}^{-1}(P)$ for any $P\in\P_V(X)$ and for any $\sigma\in\Sigma_{r,m}^*$ such that $q=\delta(q_0,\sigma)$.
  
  We first prove that $\rho_{r,m}(\sigma',s)\in A$ for any word $\sigma'\in\Sigma_{r,m}^*$ such that there exists an $s$-eye $Y'$ satisfying $\delta(q_0,\sigma')\in \ker_s(Y')\subseteq Q'$. As the principal state $q'=\delta(q_0,\sigma')$ is in $Q'$, there exists $P\in\P_V(X)$ such that $P_{q'}\not=\emptyset$. As there exists a path $q'\xrightarrow{s^+}q'$ since $q'\in\ker_s(Y')$, we get $\saff(P_{q'})=\xi_{r,m}(s)+V$ from lemma \ref{lem:cyclicpattern}. Remark that $P_{q'}=\gamma_{r,m,\sigma'}^{-1}(P)$. Thus, from covering lemma \ref{lem:cov}, we get $\xi_{r,m}(s)+V\subseteq \Gamma_{r,m,\sigma'}(\saff(P))$ in particular from $\saff(P)=A$ and $\rho_{r,m}(\sigma,s)=\Gamma_{r,m,\sigma}(\xi_{r,m}(s))$, we deduce that $\rho_{r,m}(\sigma',s)\in A$. 
  
  Next, let us show that for any pair of integers $z_1,z_2\in\Nat$ such that $z_1+m.n.\Z=z_2+m.n.\Z$ and for any $x\in \Z^m$, we have $x'=\gamma_{r,m,0}^{z_2}(x)-\gamma_{r,m,0}^{z_1}(x)\in n_0.\Z^m$. Naturally, by symmetry, we can assume that $z_1<z_2$ and by replacing $x$ by $\gamma_{r,m,0}^{z_1}(x)$ and $(z_1,z_2)$ by $(0,z_2-z_1)$ we can assume that $z_1=0$. In this case $z=\frac{z_2}{m.n}$ is in $\Nat$ and $x'=(r^{n.z}-1).x$. Since $r^n-1$ divides $r^{n.z}-1$ and $n_0$ divides $r^n-1$, we have prove that $x'\in n_0.\Z^m$.
  
  Let us denote by $M$ the vector lattice generated by $n_0.\Z^m\cap V$ and the vectors $\rho_{r,m}(\sigma_{u_1}.b.\sigma'_{u_2},s)-\rho_{r,m}(\sigma_{u_2}.\sigma'_{u_2},s)$ where $u_1=(k_1,Z_1)\in U$, $b\in\Sigma_r$ and $u_2=(k_2,Z_2)\in U$ are such that $(k_2,Z_2)=(\delta(k_1,b),Z_1+1)$. 

  We first prove the inclusion $M\subseteq\inv_V(X)$.

  Let us show that $\rho_{r,m}(\sigma_2,s)-\rho_{r,m}(\sigma_1,s)\in\inv_V(X)$ for any pair of words $(\sigma_1,\sigma_2)$ in $(\Sigma_{r,m}^n)^*$ such that there exists a principal state $q'$ satisfying $\delta(q_0,\sigma_1)=q'=\delta(q_0,\sigma_2)$ and there exists an $s$-eye $Y'$ satisfying $q'\in\ker_s(Y')\subseteq Q'$. The previous paragraphs shows that $\rho_{r,m}(\sigma_1,s)$ and $\rho_{r,m}(\sigma_2,s)$ are both in $A$. Thus, from lemma \ref{lem:invdensepattern} we get $\gamma_{r,m,\sigma_i}^{-1}(P)=\xi_{r,m}(s)+P-\rho_{r,m}(\sigma_i,s)$ for any $i\in\{1,2\}$ and for any $P\in\P_V(X)$. In particular $\rho_{r,m}(\sigma_2,s)-\rho_{r,m}(\sigma_1,s)\in\inv_V(X)$.

  We can now easily prove that $M\subseteq\inv_V(X)$ since $n_0.\Z^m\cap V\subseteq \inv_V(X)$ (recall that $|\Z^m\cap V/\inv_V(X)|$ divides $n_0$) and from the previous paragraph we deduce that $\rho_{r,m}(\sigma_{u_1}.b.\sigma'_{u_2},s)-\rho_{r,m}(\sigma_{u_2}.\sigma'_{u_2},s)\in\inv_V(X)$ for any $u_1=(k_1,Z_1)\in U$, $b\in\Sigma_r$ and $u_2=(k_2,Z_2)\in U$ such that $(k_2,Z_2)=(\delta(k_1,b),Z_1+1)$.
  
  Next, let us prove the converse inclusion $\inv_V(X)\subseteq M$.

  Let us show that $\rho_{r,m}(\sigma_2.\sigma',s)-\rho_{r,m}(\sigma_1.\sigma',s)\in \rho_{r,m}(\sigma_2.\sigma'',s)-\rho_{r,m}(\sigma_1.\sigma'',s)+M$ for any pair of words $(\sigma_1,\sigma_2)$ in $\Sigma_r^*$ such that there exists $u=(k,Z)\in U$ satisfying $(\delta(q_0,\sigma_1),|\sigma_1|+m.n.\Z)=u=(\delta(q_0,\sigma_2),|\sigma_2|+m.n.\Z)$ and for any pair of words $(\sigma',\sigma'')$ in $\Sigma_r^*$ satisfying $Z+|\sigma'|=m.n.\Z=Z+|\sigma''|+m.n.\Z$ and there exists two $s$-eyes $Y'$ and $Y''$ satisfying $\delta(k,\sigma')\in\ker_s(Y')\subseteq Q'$ and $\delta(k,\sigma'')\in\ker_s(Y'')\subseteq Q'$. Let $x'=(\rho_{r,m}(\sigma_2.\sigma',s)-\rho_{r,m}(\sigma_1.\sigma',s))-(\rho_{r,m}(\sigma_2.\sigma'',s)-\rho_{r,m}(\sigma_1.\sigma'',s))$. This vector is in $V$ since the vectors $\rho_{r,m}(\sigma_1.\sigma',s)$, $\rho_{r,m}(\sigma_2.\sigma',s)$, $\rho_{r,m}(\sigma_1.\sigma'',s)$, and $\rho_{r,m}(\sigma_2.\sigma'',s)$ are in the $V$-affine space $A$ from the previous paragraphs. Moreover, let us remark that $x'=\gamma_{r,m,0}^{z_2}(x)-\gamma_{r,m,0}^{z_1}(x)$ where $z_1=|\sigma_1|$, $z_2=|\sigma_2|$ and $x=\rho_{r,m}(\sigma',s)-\rho_{r,m}(\sigma'',s)$. Thus, from the previous paragraphs, we get $x\in n_0.\Z^m$ and we have proved that $x'\in n_0.\Z^m\cap V\subseteq M$.
  
  Let us show that $\rho_{r,m}(\sigma_2,s)-\rho_{r,m}(\sigma_1,s)\in M$ for any  pair of words $(\sigma_1,\sigma_2)$ in $(\Sigma_{r,m}^n)^*$ such that there a principal state $q'$ satisfying $\delta(q_0,\sigma_1)=q'=\delta(q_0,\sigma_2)$ and there exists an $s$-eyes $Y'$ satisfying $q'\in\ker_s(Y')\subseteq Q'$. Since $M$ is a vector lattice, it is sufficient to prove that $\rho_{r,m}(\sigma,s)-\rho_{r,m}(\sigma_u,s)\in M$ for any word $\sigma\in(\Sigma_{r,m}^n)^*$ such that $u=(\delta(q_0,\sigma),m.n.\Z)$ is in $U$. Let us consider a sequence $b_1$, ..., $b_i$ of $r$-digits $b_j\in\Sigma_r$ such that $\sigma=b_1\ldots b_i$. We denote by $u_i$ the couple $u_i=(\delta(q_0,b_1\ldots b_j),j+m.n.\Z)$. Since $u_i=u$ is in $U$, we deduce that $u_j\in U$ for any $k\in\finiteset{0}{i}$. By definition of $M$, we have $\rho_{r,m}(\sigma_{u_{j-1}}.b_{j}.\sigma'_{u_{j}},s)-\rho_{r,m}(\sigma_{u_{j}}.\sigma'_{u_{j+1}},s)\in M$ for any $j\in\finiteset{1}{i}$. From the previous paragraph, we get $\rho_{r,m}(\sigma_{u_{j-1}}.b_j\ldots b_i,s)-\rho_{r,m}(\sigma_{u_j}.b_{j+1}\ldots b_i,s)\in M$ for any $j\in\finiteset{1}{i}$. By summing all the vectors, we deduce that $\rho_{r,m}(\sigma_{u_0}.b_1\ldots b_i)-\rho_{r,m}(\sigma_{u_i})\in M$. Now, just remark that $\sigma_{u_0}=\epsilon$ and $u_i=u$.
  
  Let us consider $v\in\inv_V(X)$ and let us prove that $v\in M$. Lemma \ref{lem:pourZrms} shows that there exists a vector $v_0\in V$ such that $v_0[i]<0$ if $s[i]=r-1$ and $v_0[i]>0$ if $s[i]=0$ for any $i\in \finiteset{1}{m}$ such that $\unit_{i,m}\not\in V^\perp$. By replacing $v_0$ by a vector in $(\Nat\moins\{0\}).v_0$, we can assume that $v_0\in \inv_V(X)$, $v_0[i]+v[i]<0$ if $s[i]=r-1$ and $v_0[i]+v[i]>0$ if $s[i]=0$ for any $i\in \finiteset{1}{m}$ such that $\unit_{i,m}\not\in V^\perp$. Since $[Z_{r,m,s}\cap A]^V\not=[\emptyset]^V$, there exists a vector $a\in Z_{r,m,s}\cap A$. Let $a_1=a+v_0$ and let $a_2=a+v_0+v$. Remark that $a_1,a_2\in Z_{r,m,s}$ since for any $i\in\finiteset{1}{m}$, if $\unit_i\not\in V^\perp$ then $a_1[i]=a[i]+v_0[i]$, $a_2[i]=a[i]+v_0[i]+v[i]$ and if $\unit_i\in V^\perp$ then $a_1[i]=a[i]$, $a_2[i]=a[i]$. As $a_1,a_2\in Z_{r,m,s}$, there exist $\sigma_1,\sigma_2\in\Sigma_{r,m}^*$ such that $a_1=\rho_{r,m}(\sigma_1,s)$ and $a_2=\rho_{r,m}(\sigma_2,s)$. By replacing $\sigma_1$ by a word in $\sigma_1.s^*$ and $\sigma_2$ by a word in $\sigma_2.s^*$ we can also assume that $|\sigma_1|$ and $|\sigma_2|$ are in $m.n.\Z$. Let $P\in\P_V(X)$. Since $\rho_{r,m}(\sigma_i,s)\in A$ and $r^{|\sigma_i|_m}\in 1+|\Z^m\cap V/\inv_V(P)|.\Z$, lemma \ref{lem:invdensepattern} proves that $\gamma_{r,m,\sigma_i}^{-1}(P)=\xi_{r,m}(s)+P-\rho_{r,m}(\sigma_i,s)$ for any $i\in\{1,2\}$. As $\rho_{r,m}(\sigma_2,s)-\rho_{r,m}(\sigma_1,s)=v_0\in\inv_V(X)$, we deduce that $\gamma_{r,m,\sigma_1}^{-1}(P)=\gamma_{r,m,\sigma_2}^{-1}(P)$ for any $P\in\P_V(X)$. Therefore there exists a state $q'\in Q'$ such that $\delta(q_0,\sigma_1)=q'=\delta(q_0,\sigma_2)$. Let us consider the $s$-eye $Y'$ that contains $q'$. Since $\gamma_{r,m,\sigma_1.s^{n}}^{-1}(P)=\xi_{r,m}(s)+P-\rho_{r,m}(\sigma_i.s^n,s)$ from lemma \ref{lem:invdensepattern}, we deduce that $(\gamma_{r,m,\sigma_1.s^n}^{-1}(P))_{P\in\P_V(X)}=(\gamma_{r,m,\sigma_1}^{-1}(P))_{P\in\P_V(X)}$. We have proved that $\delta(q',s^n)=q'$ and in particular $q'\in\ker_s(Y')$. By considering $P\in\P_V(X)\moins\{\emptyset\}$ let us remark that $\gamma_{r,m,\sigma_1}^{-1}(P)=\xi_{r,m}(s)+P-\rho_{r,m}(\sigma_i,s)$ is not empty. That means $q'\in Q'$. Moreover, as for any $q''\in\ker_s(Y')$ there exists a path $q''\xrightarrow{s^*}q'$ and $P_{q'}\not=\emptyset$ we get $P_{q''}\not=\emptyset$. Thus $\ker_s(Y')\subseteq Q'$. From the previous paragraph, we get $\rho_{r,m}(\sigma_2,s)-\rho_{r,m}(\sigma_1,s)\in M$. Now, just remark that $\rho_{r,m}(\sigma_2,s)-\rho_{r,m}(\sigma_1,s)=v$ and we have proved that $v\in M$.
  \qed
\end{proof}

The following proposition \ref{prop:ZrmsV} provides a simple algorithm for computing in polynomial time a $(r,m)$-sign vector $s\in S_{r,m}$ such that $[Z_{r,m,s}\cap (\xi_{r,m}(w)+V)]^V\not=[\emptyset]^V$ from a FDVA that represents a $(r,m,w)$-cyclic Presburger definable set $X$ in basis $r$.
\begin{proposition}\label{prop:ZrmsV}
  Let $X\subseteq\Z^m$ be a $(r,m,w)$-cyclic Presburger-definable set represented by a FDVA $\automaton$ in basis $r$, and let $V$ be an affine component of $\vecsaff(X)$. We have $[Z_{r,m,s}\cap (\xi_{r,m}(w)+V)]^V\not=[\emptyset]^V$ for any $(r,m)$-sign vector $s\in S_{r,m}$ such that $s\in [F_0](q)$ where $q$ is a principal state in a terminal component $T$ such that $V_G(T)=V$.
\end{proposition}
\begin{proof}
  Let us consider a terminal component $T$ of $\automaton$, a principal state $q\in T$ and a $(r,m)$-sign vector $s\in [F_0](q)$. Let $Y$ be the $s$-eye that contains $q$. As $T$ is terminal we deduce that $\ker_s(Y)\subseteq T$. From proposition \ref{prop:affeye}, we deduce that $\saff(X_q^{F_{s,Y}})=a_q(G)+V_G(T)$. From $X_q^{F_{s,Y}}\subseteq Z_{r,m,s}\cap X_q$, we deduce that $V\subseteq \vecsaff(Z_{r,m,s}\cap X_q)$. As $q$ is reachable, there exists a path $q_0\xrightarrow{w}q$ and we get $X_q=\gamma_{r,m,w}^{-1}(X)$. As $\gamma_{r,m,w}^{-1}(Z_{r,m,s}\cap X)=Z_{r,m,s}\cap X_q$, we have proved that $V\subseteq\vecsaff(Z_{r,m,s}\cap X)$ thanks to the covering lemma \ref{lem:cov}. Let $A$ be an affine component of $\saff(Z_{r,m,s}\cap X)$ such that $V\subseteq \vec{A}$. From $V\subseteq \vec{A}\subseteq \vecsaff(X)$ and as $V$ is an affine component of $\vecsaff(X)$, we deduce that $V=\vec{A}$. Moreover, as $Z_{r,m,s}\cap X$ is $(r,m,w)$-cyclic we deduce that $\xi_{r,m}(w)\in A$. Hence $A=\xi_{r,m}(w)+V$. From the dense component lemma \ref{lem:dense}, we get $\saff(Z_{r,m,s}\cap X\cap A)=A$. In particular $A\subseteq \saff(Z_{r,m,s}\cap A)$ and we have proved that $[Z_{r,m,s}\cap (\xi_{r,m}(w)+V)]^V\not=[\emptyset]^V$.
  \qed
\end{proof}

A couple $(q_0,G)$ and a set $Q'$ satisfying proposition \ref{prop:invpolycyclic} is obtained by a quotient of a FDVA $\automaton$ that represents $X$ in basis $r$ by the equivalence relation $\sim^V$ defined over the principal states of $\automaton$ by $q_1\sim^V q_2$ if and only if $X_{q_1} \sim^V X_{q_2}]^V$. Remark that $\sim^V$ is a polynomial time equivalence relation since $q_1\sim^Vq_2$ if and only $V$ is not included in $\vecsaff(X_{q_1}\Delta X_{q_2})$, and this last condition can be decided in polynomial because a FDVA that represents the Presburger-definable set $X_{q_1}\Delta X_{q_2}$ is computable in quadratic time and the semi-affine hull direction of this set is computable in polynomial time thanks to theorem \ref{thm:semiaffinehull}. The following propositions \ref{prop:saffpattern} and \ref{prop:patternempty} provides immediately the following corollary \ref{cor:qGQp}.

\begin{proposition}\label{prop:saffpattern}
  Let $X$ be a Presburger-definable set and let $V$ be an affine component of $\vecsaff(X)$. Given a pair $(\sigma_1,\sigma_2)$ of words in $\Sigma_{r,m}^*$, we have the equality $(\gamma_{r,m,\sigma_1}^{-1}(P))_{P\in\P_V(X)}=(\gamma_{r,m,\sigma_2}^{-1}(P))_{P\in\P_V(X)}$ if and only if $\gamma_{r,m,\sigma_1}^{-1}(X)\sim^V \gamma_{r,m,\sigma_2}^{-1}(X)$.
\end{proposition}
\begin{proof}
  Consider a pair $(\sigma_1,\sigma_2)$ of words in $\Sigma_{r,m}^*$. From lemma \ref{lem:Xsigma} we deduce that $[\gamma_{r,m,\sigma_i}^{-1}(X)]^V=\bigcup_{P\in\P_V(X)}^V([\gamma_{r,m,\sigma_i}^{-1}(P)]^V\cap^V(\class_{V,P}(X)+V^\perp))$ for any $i\in\{1,2\}$. As $(\class_{V,P}(X))_{P\in\P_V(X)}$
  is a polyhedral $V$-partition, we get $[\gamma_{r,m,\sigma_1}^{-1}(X)\Delta\gamma_{r,m,\sigma_2}^{-1}(X)]^V=\bigcup_{P\in\P_V(X)}^V([\gamma_{r,m,\sigma_1}^{-1}(P)\Delta\gamma_{r,m,\sigma_2}^{-1}(P)]^V\cap^V(\class_{V,P}(X)+V^\perp))$. 
  Remark that if $(\gamma_{r,m,\sigma_1}^{-1}(P))_{P\in\P_V(X)}=(\gamma_{r,m,\sigma_2}^{-1}(P))_{P\in\P_V(X)}$ then we have $[\gamma_{r,m,\sigma_1}^{-1}(X)\Delta\gamma_{r,m,\sigma_2}^{-1}(P)]^V=[\emptyset]^V$ and conversely if $[\gamma_{r,m,\sigma_1}^{-1}(X)\Delta\gamma_{r,m,\sigma_2}^{-1}(X)]^V=[\emptyset]^V$,
  by intersecting the following equality by
  $\class_{V,P}(X)+V^\perp$, we get $[\emptyset]^V=[\gamma_{r,m,\sigma_1}^{-1}(P)\Delta \gamma_{r,m,\sigma_2}^{-1}(P)]^V\cap^V(\class_{V,P}(X)+V^\perp)$:
$$[\gamma_{r,m,\sigma_1}^{-1}(X)\Delta \gamma_{r,m,\sigma_2}^{-1}(X)]^V=\bigcup_{P\in\P_V(X)}^V([\gamma_{r,m,\sigma_1}^{-1}(P)\Delta\Gamma_{r,m,\sigma_2}^{-1}(P)]^V\cap^V(\class_{V,P}(X)+V^\perp))$$
From lemma \ref{lem:dimdim} we get $\gamma_{r,m,\sigma_1}^{-1}(P)\Delta\gamma_{r,m,\sigma_2}^{-1}(P)=\emptyset$.
    \qed
\end{proof}

\begin{proposition}\label{prop:patternempty}
  Let $X$ be a Presburger-definable set and $V$ be an affine component of $\vecsaff(X)$. Given a word $\sigma\in\Sigma_{r,m}^*$, we have $(\gamma_{r,m,\sigma}^{-1}(P))_{P\in\P_V(X)}=(\emptyset)_{P\in\P_V(X)}$ if and only if $\gamma_{r,m,\sigma}^{-1}(X)\sim^V\emptyset$.
\end{proposition}
\begin{proof} 
  From lemma \ref{lem:Xsigma} we deduce that $[\gamma_{r,m,\sigma}^{-1}(X)]^V=\bigcup_{P\in\P_V(X)}^V([\gamma_{r,m,\sigma}^{-1}(P)]^V\cap^V(\class_{V,P}(X)+V^\perp))$. Remark that if $(\gamma_{r,m,\sigma}^{-1}(P))_{P\in\P_V(X)}=(\emptyset)_{P\in\P_V(X)}$ then $[\gamma_{r,m,\sigma}^{-1}(X)]^V=[\emptyset]^V$ and conversely if $[\gamma_{r,m,\sigma}^{-1}(X)]^V=[\emptyset]^V$, by intersecting the equality $[\gamma_{r,m,\sigma}^{-1}(X)]^V=\bigcup_{P\in\P_V(X)}^V([\gamma_{r,m,\sigma}^{-1}(P)]^V\cap^V(\class_{V,P}(X)+V^\perp))$ by $\class_{V,P}(X)+V^\perp$, we get $[\emptyset]^V=[\gamma_{r,m,\sigma}^{-1}(P)]^V\cap^V(\class_{V,P}(X)+V^\perp)$. From lemma \ref{lem:dimdim} we get $\gamma_{r,m,\sigma}^{-1}(P)=\emptyset$.
  \qed
\end{proof}

\begin{corollary}\label{cor:qGQp}
  Let $X$ be a $(r,m,w)$-cyclic Presburger-definable set represented by a FDVA $\automaton$ in basis $r$, and let $V$ be an affine component of $\vecsaff(X)$. We can compute in polynomial time a couple $(q_0,G)$ such that $q_0$ is a principal state of a FDVG $G$ such that $\delta(q_0,\sigma_1)=\delta(q_0,\sigma_2)$ if and only if $(\gamma_{r,m,\sigma_1}^{-1}(P))_{P\in\P_V(X)}=(\gamma_{r,m,\sigma_2}^{-1}(P))_{P\in\P_V(X)}$ for any $\sigma_1,\sigma_2\in\Sigma_{r,m}^*$, and we can compute in polynomial time the set $Q'$ of principal states reachable for $[G]$ from $q_0$ such that  $(\gamma_{r,m,\sigma}^{-1}(P))_{P\in\P_V(X)}\not=(\emptyset)_{P\in\P_V(X)}$ if and only if $q'\in Q'$ for any path $q_0\xrightarrow{\sigma}q'$ with $\sigma\in\Sigma_{r,m}^*$.
\end{corollary}

Let us consider a $(r,m,w)$-cyclic Presburger definable set $X$ represented by a FDVA $\automaton$ in basis $r$. The following proposition \ref{prop:invover} provides an algorithm for computing in polynomial time an integer $n_1\in\finiteset{1}{|\automaton|}$ such that there exists $z_0\in\Nat\moins\{0\}$ satisfying $n_1=z_0.|\Z^m\cap V/\inv_V(X)|$. Naturally the integer $n_1$ is not necessary relatively prime with $r$. However, let us remark that $n_0=h_r^\infty(n_1)$ is also computable in polynomial time (by an Euclid's algorithm) and it is also in $\finiteset{1}{n_1}\subseteq \finiteset{1}{|\automaton|}$. Moreover, as $\inv_V(X)$ is relatively prime with $r$ (recall that $X$ is cyclic), we deduce that $|\Z^m\cap V/\inv_V(X)|$ divides $n_0$. That means we have provided a polynomial time algorithm for computing an integer $n_0\in\finiteset{1}{|\automaton|}$ that satisfies proposition \ref{prop:invpolycyclic}. Now let us remark that an integer $n\in\finiteset{1}{n_0}$ satisfying proposition \ref{prop:invpolycyclic} can be easily computed in polynomial time. In fact, since $n_0$ is relatively prime with $r$, there exists an integer $n\in\finiteset{1}{n_0}$ such that $r^n\in 1+n_0.\Z$. By enumerating the integers in $\finiteset{1}{n_0}$ we compute in polynomial an integer $n$ satisfying  proposition \ref{prop:invpolycyclic}.

\begin{proposition}\label{prop:invover}
  Let $X$ be a cyclic Presburger-definable set and let $V$ be an affine component of $\vecsaff(X)$. There exists an integer $z_0\in\Nat\moins\{0\}$ such that for any $(q_0,G)$ and $Q'$ satisfying the same conditions as the one provided in proposition \ref{prop:invpolycyclic}, and for any $(r,m)$-sign vector $s\in S_{r,m}$ satisfying $[Z_{r,m,s}\cap (\xi_{r,m}(w)+V)]^V\not=[\emptyset]^V$, we have the following equality:
  $$|\Z^m\cap V/\inv_V(X)|=\frac{1}{z_0}(\sum_{\begin{array}{c}\scriptstyle Y\text{ $s$-eye of }[G]\\[-0pt] \scriptstyle \ker_s(Y)\subseteq Q'\end{array}}|\ker_s(Y)|)$$
\end{proposition}
\begin{proof}
  Let us recall that $A$ is the $V$-affine space $A=\xi_{r,m}(w)+V$. As $\bigcup_{P\in\P_V(X)}P$ is a non empty set included in $\Z^m\cap A$, there exists a vector $a_0$ in $\Z^m\cap A$. As $r$ and $|\Z^m\cap V/\inv_V(X)|$ are relatively prime, there exists an integer $z_1\in\Nat\moins\{0\}$ such that $r^{z_1}\in 1+|\Z^m\cap V/\inv_V(X)|.\Nat$. As $P-a_0$ is a relatively prime semi-$V$-pattern included in $V$ and $\rho_{r,m}(\unit_{0,m}^{z_1},\unit_{0,m})=\unit_{0,m}\in V$, lemma \ref{lem:invdensepattern} proves that $\gamma_{r,m,\unit_{0,m}}^{-z_1}(P-a_0)=P-a_0$ for any $P\in\P_V(X)$. In particular, there exists a minimal integer $z_0$ in $\Nat\moins\{0\}$ such that there exists a vector $v_0\in\Z^m\cap V$ satisfying $\gamma_{r,m,\unit_{0,m}}^{-z_0}(P-a_0)=P-a_0+v_0$ for any $P\in\P_V(X)$. Let us denote by $I$ the set of indexes $i\in\finiteset{1}{m}$ such that $\unit_{i,m}\not\in V^\perp$. Let us consider $s\in S_{r,m}$ such that $[A\cap (\xi_{r,m}(w)+V)]^V\not=[\emptyset]^V$. Let $Q_s$ be the union of the $s$-kernel $\ker_s(Y)$ where $Y$ is an $s$-eye of $G$ such that $\ker_s(Y)\subseteq Q'$.

  We are going to prove that there exists a one-to-one function from $Q_s$ to $\finiteset{0}{z_0-1}\times B_0$ by remarking that for any $z,z'\in\finiteset{0}{z_0-1}$ and for any $v,v'\in B_0$ such that $(\xi_{r,m}(s)+\gamma_{r,m,\unit_{0,m}}^{-z}(P-a_0+v))_{P\in\P_V(X)}$ and $(\xi_{r,m}(s)+\gamma_{r,m,\unit_{0,m}}^{-z'}(P-a_0+v'))_{P\in\P_V(X)}$ are equal, we have $v=v'$ and $z=z'$. Thanks to this one-to-one function we will obtain $|Q_s|=z_0.|\Z^m\cap V/\inv_V(X)|$ and concluded the proof of the proposition.
  
  Let us prove that for any state $q\in Q_s$, there exists $z\in \finiteset{0}{z_0-1}$ and $v\in B_0$ such that $P_q=\xi_{r,m}(s)+\gamma_{r,m,\unit_{0,m}}^{-z}(P-a_0+v)$ for any $P\in\P_V(X)$. Let $Y$ be the $s$-eye such that $q\in\ker_s(Y)\subseteq Q'$. As $q$ is reachable, there exists a path of the form $q_0\xrightarrow{\sigma}q$. As $q\in\ker_s(Y)$, there exists $n\in\Nat\moins\{0\}$ such that $q\xrightarrow{s^{n}}q$. By replacing $n$ by an integer enough larger in $n.(\Nat\moins\{0\})$, we can assume that there exists $\alpha,\beta\in\Nat$ and $z\in\finiteset{0}{z_0-1}$ such that $n=\alpha+z+\beta.z_0$ and $|\sigma|_m+\alpha\in z_1.\Nat$. Let $q'=\delta(q,s^{\alpha})$. As $(\emptyset)_{P\in\P_V(X)}$ is not in $\ker_s(Y)$, we deduce that there exists $P\in\P_V(X)$ such that $P_{q'}\not=\emptyset$. Moreover, as $P_{q'}$ is $(r,m,s^{n})$-cyclic and non-empty, from destruction lemma \ref{lem:destruction}, we get $\xi_{r,m}(s)\in \saff(P_{q'})$. From $P_{q'}=\gamma_{r,m,\sigma.s^{\alpha}}^{-1}(P)$, covering lemma \ref{lem:cov} proves that $\saff(P_{q'})\subseteq \Gamma_{r,m,\sigma.s^{\alpha}}^{-1}(\saff(P))$ and $P\subseteq A$, we deduce that $\xi_{r,m}(s)\in \Gamma_{r,m\sigma.s^{\alpha}}^{-1}(A)$. Therefore $\rho_{r,m}(\sigma.s^\alpha,s)\in A$. Moreover as $|\sigma.s^\alpha|_m\in 1+z_1.\Nat$, we deduce from lemma \ref{lem:invdensepattern} that $P_{q'}=\xi_{r,m}(s)+P-\rho_{r,m}(\sigma,s)$. Let $v'=a_0-\rho_{r,m}(\sigma,s)$. As $a_0$ and $\rho_{r,m}(\sigma,s)$ are both in $A$, we deduce that $v'\in\Z^m\cap V$. Remark that $P_q=\gamma_{r,m,s}^{-(z+\beta.z_0)}(\xi_{r,m}(s)+P-a_0+v')$ and we have proved that $P_q=\xi_{r,m}(s)+\gamma_{r,m,\unit_{0,m}}^{-(z+\beta.z_0)}(P-a_0+v')$ for any $P\in\P_V(X)$. Let us consider an integer $u\in\Nat$ such that $u.r\in 1+|\Z^m\cap V/\inv_V(X)|.\Nat$. An immediate induction over $\beta\in\Nat$ provides $\gamma_{r,m,\unit_{0,m}}^{-\beta.z_0}(P-a_0+v')=P-a_0+v$ where $v$ is the vector in $B_0$ satisfying $v\in u^{\beta.z_0}.v'+(u^{(\beta-1).z_0}+\cdots u^{0.z_0}).v_0+\inv_V(X)$. Hence $P_q=\xi_{r,m}(s)+\gamma_{r,m,\unit_{0,m}}^{-z}(P-a+v)$ for any $P\in\P_V(X)$.
  
  Now, let us prove that for any $z\in\finiteset{0}{z_0-1}$ and any $v\in B_0$, there exists a state $q\in Q_s$ such that $P_q=\xi_{r,m}(s)+\gamma_{r,m,\unit_{0,m}}^{-z}(P-a_0+v)$ for any $P\in\P_V(X)$. From lemma \ref{lem:pourZrms} we deduce that there exists a vector $v_0\in V$ such that $v_0[i]<0$ if $s[i]=r-1$ and $v_0[i]>0$ if $s[i]=0$ for any $i\in I$. By replacing $v_0$ by a vector in $(\Nat\moins\{0\}).v_0$, we can assume that $v_0\in\inv_V(X)$ and $(a-v+v_0)[i]>0$ if $s[i]=0$ and $(a-v+v_0)[i]<0$ if $s[i]=r-1$ for any $i\in I$. As $[Z_{r,m,s}\cap A]^V\not=[\emptyset]^V$, there exists a vector $a$ in $Z_{r,m,s}\cap A$. Remark that for any $i\in\finiteset{1}{m}$, if $i\in I$ the sign of $(a_0-v+v_0)[i]$ is $s[i]$ and if $i\not\in I$, as $\unit_{i,m}\in V^\perp$, we have $(a_0-v+v_0)[i]=a_0[i]=a[i]$ and form $a\in Z_{r,m,s}$, we also deduce that the sign of $(a_0-v+v_0)[i]$ is $s[i]$. Hence $a-v+v_0\in Z_{r,m,s}$. That implies there exists a word $\sigma\in\Sigma_{r,m}^*$ such that $\rho_{r,m}(\sigma,s)=a-v+v_0$. By replacing $\sigma$ by a word in $\sigma.s^*$, we can assume that $|\sigma|_m\in z_1.\Nat$. From $\rho_{r,m}(\sigma,s)\in A$ and $|\sigma|_m\in z_1.\Nat$, lemma \ref{lem:invdensepattern} shows that $\gamma_{r,m,\sigma}^{-1}(P)=\xi_{r,m}(s)+P-\rho_{r,m}(\sigma,s)=\xi_{r,m}(s)+P-a_0+v+v_0$. From $P+v_0=P$, we deduce that $\gamma_{r,m,\sigma}^{-1}(P)=\xi_{r,m}(s)+P-a_0+v$. Hence $\gamma_{r,m,\sigma.s^z}^{-1}(P)=\xi_{r,m}(s)+\gamma_{r,m,\unit_{0,m}}^{-1}(P-a_0+v)$. Let $q=\delta(q_0,\sigma.s^{z})$ and let $Y$ be the $s$-eye that contains $q$. As $\gamma_{r,m,s^{z_1}}^{-1}(P_q)=P_q$ for any $P\in\P_V(X)$, we deduce that $q\in\ker_s(Y)$. Moreover, as there exists $P\in\P_V(X)\moins\{\emptyset\}$ we deduce that $P_q\not=\emptyset$. Remark that for any $q'\in \ker_s(Y)$ there exists a path $q'\xrightarrow{s^*}q$ and $P_q\not=\emptyset$, we deduce that $P_{q'}\not=\emptyset$. Hence $\ker_s(Y)\subseteq Q'$. Therefore $q\in Q_s$.
  \qed
\end{proof}

\begin{theorem}\label{thm:polyinvcyclicmain}
  Given a cyclic Presburger-definable set $X\subseteq\Z^m$ represented by a FDVA $\automaton$ in basis $r$, and given an affine component $V$ of $\vecsaff(X)$ and given a full rank set of indices $I$ of $V$, the $I$-representation of $\inv_V(X)$ is computable in polynomial time. Moreover $|\Z^m\cap V/inv_V(X)|$ is bounded by the number of principal states of $\automaton$.
\end{theorem}

\section{Boundary of a Presburger-definable FDVA}
Let $X$ be a Presburger-definable set and $V$ be an affine component of $\vecsaff(X)$. The \emph{$V$-boundary} $\bound{V}{X}$\index{Not}{$\bound{V}{X}$} of $X$ is defined by the following equality:
$$\bound{V}{X}=\bigcup_{P\in\P_V(X)}\bound{V}{\class_{V,P}(X)}$$
In this section, we prove that $\bound{V}{X}\moins(\bigcup_{j=1}^m\{V\cap\unit_{j,m}^\perp\})$ is computable in polynomial time from a FDVA that represents $X$.

\begin{figure}[htbp]
  \setlength{\unitlength}{8pt}
  \pssetlength{\psunit}{8pt}
  \pssetlength{\psxunit}{8pt}
  \pssetlength{\psyunit}{8pt}
  \begin{center}
    \begin{picture}(20,20)(-2,-2)%
      \put(-2,-2){\framebox(20,20){}}
      \multido{\iii=0+1}{17}{\psline[linecolor=lightgray,linewidth=1pt](0,\iii)(16,\iii)}\multido{\iii=0+1}{17}{\psline[linecolor=lightgray,linewidth=1pt](\iii,0)(\iii,16)}\psline[linecolor=gray,arrowscale=1,linewidth=1pt]{->}(0,0)(0,17)\psline[linecolor=gray,arrowscale=1,linewidth=1pt]{->}(0,0)(17,0)
      \multido{\iii=0+1}{17}{\psdot*(0,\iii)}
      \multido{\iii=0+1}{17}{\psdot*(2,\iii)}
      \multido{\iii=0+1}{17}{\psdot*(4,\iii)}
      \multido{\iii=0+1}{17}{\psdot*(6,\iii)}
      \multido{\iii=0+1}{17}{\psdot*(8,\iii)}
      \multido{\iii=0+1}{17}{\psdot*(10,\iii)}
      \multido{\iii=0+1}{17}{\psdot*(12,\iii)}
      \multido{\iii=0+1}{17}{\psdot*(14,\iii)}
      \multido{\iii=0+1}{17}{\psdot*(16,\iii)}
    \end{picture}
    \begin{picture}(20,20)(-2,-2)%
      \put(-2,-2){\framebox(20,20){}}
      \multido{\iii=0+1}{17}{\psline[linecolor=lightgray,linewidth=1pt](0,\iii)(16,\iii)}\multido{\iii=0+1}{17}{\psline[linecolor=lightgray,linewidth=1pt](\iii,0)(\iii,16)}\psline[linecolor=gray,arrowscale=1,linewidth=1pt]{->}(0,0)(0,17)\psline[linecolor=gray,arrowscale=1,linewidth=1pt]{->}(0,0)(17,0)
      \multido{\iii=0+1}{17}{\psdot*(\iii,0)}
      \multido{\iii=0+1}{17}{\psdot*(\iii,2)}
      \multido{\iii=0+1}{17}{\psdot*(\iii,4)}
      \multido{\iii=0+1}{17}{\psdot*(\iii,6)}
      \multido{\iii=0+1}{17}{\psdot*(\iii,8)}
      \multido{\iii=0+1}{17}{\psdot*(\iii,10)}
      \multido{\iii=0+1}{17}{\psdot*(\iii,12)}
      \multido{\iii=0+1}{17}{\psdot*(\iii,14)}
      \multido{\iii=0+1}{17}{\psdot*(\iii,16)}
    \end{picture}
  \end{center}
  \vspace{0.1cm}
  \begin{center}
    \begin{picture}(20,20)(-2,-2)%
      \put(-2,-2){\framebox(20,20){}}
      \multido{\iii=0+1}{17}{\psline[linecolor=lightgray,linewidth=1pt](0,\iii)(16,\iii)}\multido{\iii=0+1}{17}{\psline[linecolor=lightgray,linewidth=1pt](\iii,0)(\iii,16)}
      \begin{psclip}{\pspolygon*[linecolor=lightgray,fillcolor=lightgray](0,0)(16,8)(16,16)(8,16)}%
	\multido{\iii=0+1}{17}{\psline[linecolor=gray,linewidth=1pt](0,\iii)(16,\iii)}\multido{\iii=0+1}{17}{\psline[linecolor=gray,linewidth=1pt](\iii,0)(\iii,16)}
      \end{psclip}%
      \psline[linecolor=gray,arrowscale=1,linewidth=1pt]{->}(0,0)(0,17)\psline[linecolor=gray,arrowscale=1,linewidth=1pt]{->}(0,0)(17,0)
      \psline(8,16)(0,0)(16,8)
      \psline[linewidth=1pt,linestyle=dashed,dash=3pt 2pt](0,0)(16,16)
      \put(12,9.3){$C_H$}
      \put(10,12){$H$}
    \end{picture}
    \begin{picture}(20,20)(-2,-2)%
      \put(-2,-2){\framebox(20,20){}}
      \multido{\iii=0+1}{17}{\psline[linecolor=lightgray,linewidth=1pt](0,\iii)(16,\iii)}\multido{\iii=0+1}{17}{\psline[linecolor=lightgray,linewidth=1pt](\iii,0)(\iii,16)}\psline[linecolor=gray,arrowscale=1,linewidth=1pt]{->}(0,0)(0,17)\psline[linecolor=gray,arrowscale=1,linewidth=1pt]{->}(0,0)(17,0)
      \psdot*(2,3)
      \psdot*(4,5)\psdot*(4,6)\psdot*(4,7)
      \psdot*(6,7)\psdot*(6,8)\psdot*(6,9)\psdot*(6,10)\psdot*(6,11)
      \psdot*(8,9)\psdot*(8,10)\psdot*(8,11)\psdot*(8,12)\psdot*(8,13)\psdot*(8,14)\psdot*(8,15)
      \psdot*(10,11)\psdot*(10,12)\psdot*(10,13)\psdot*(10,14)\psdot*(10,15)\psdot*(10,16)
      \psdot*(12,13)\psdot*(12,14)\psdot*(12,15)\psdot*(12,16)
      \psdot*(14,15)\psdot*(14,16)
      \psdot*(3,2)
      \psdot*(5,4)\psdot*(6,4)\psdot*(7,4)
      \psdot*(7,6)\psdot*(8,6)\psdot*(9,6)\psdot*(10,6)\psdot*(11,6 )
      \psdot*(9,8)\psdot*(10,8)\psdot*(11,8)\psdot*(12,8)\psdot*(13,8)\psdot*(14,8)\psdot*(15,8)
      \psdot*(11,10)\psdot*(12,10)\psdot*(13,10)\psdot*(14,10)\psdot*(15,10)\psdot*(16,10)
      \psdot*(13,12)\psdot*(14,12)\psdot*(15,12)\psdot*(16,12)
      \psdot*(15,14)\psdot*(16,14)
    \end{picture} 
    \caption{On top left a semi-$\Q^2$-pattern $P^<$, on top right a semi-$\Q^2$-pattern $P^>$, on bottom left an open convex $\Q^2$-polyhedron $C_H$ and a $\Q^2$-hyperplane $H$, on bottom right the set $(P^<\cap C_H\cap H^<)\cup (P^>\cap C_H\cap H^>)$.\label{fig:PP}}
  \end{center}
\end{figure}
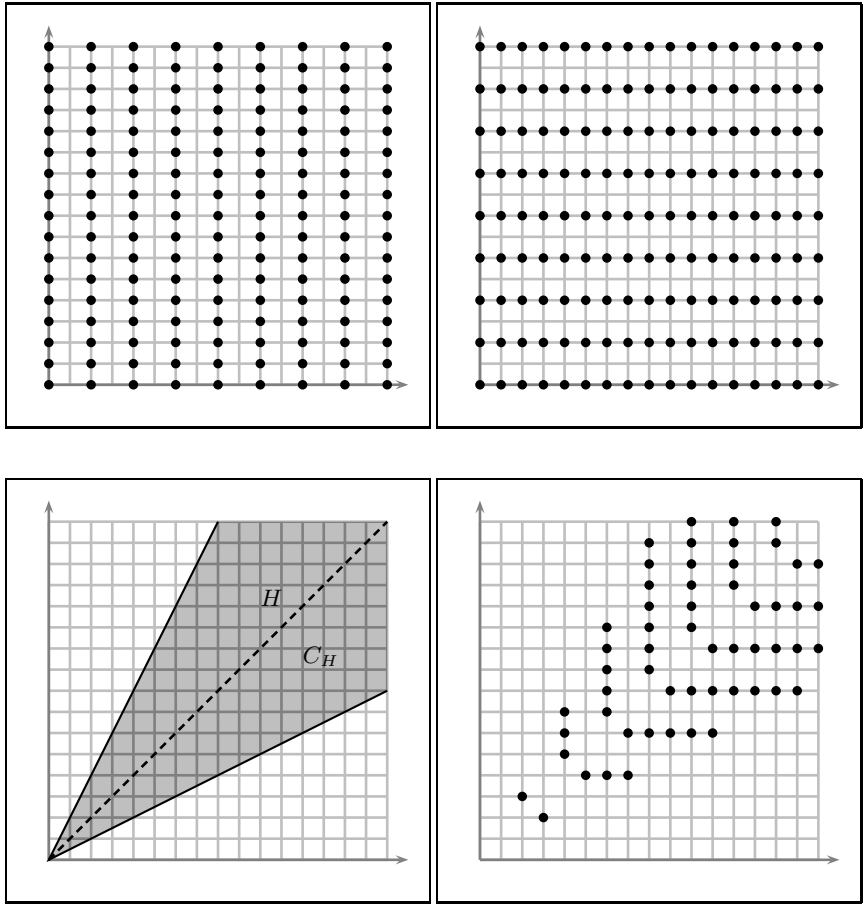

The set $\bound{V}{X}$ plays an important role as proved by the following proposition \ref{prop:patternH} (see also figure \ref{fig:PP}).
\begin{proposition}\label{prop:patternH}
  Let $X$ be a Presburger-definable set and let $V$ be an affine component of $\vecsaff(X)$. For any $H\in\bound{V}{X}$, there exist two different semi-$V$-patterns $P^<\not=P^>$ in $\P_V(X)$, an open convex $V$-polyhedron $C_H$ satisfying $[C_H\cap H^<]^V\not=[\emptyset]^V$, $[C_H\cap H^>]^V\not=[\emptyset]^V$ and such that:
  $$[X\cap (C_H+V^\perp)]^V=[P^<\cap ((C_H\cap H^<)+V^\perp)]^V\cup^V[P^>\cap ((C_H\cap H^>)+V^\perp)]^V$$
  Moreover, if $X$ is $(r,m,w)$-cyclic then one of these two sets is $(r,m)$-detectable in $X$:
  $$(P^<\cap (\xi_{r,m}(w)+H^<+V^\perp))\cup (P^>\cap (\xi_{r,m}(w)+H^\geq+V^\perp))$$
  $$(P^<\cap (\xi_{r,m}(w)+H^\leq+V^\perp))\cup (P^>\cap (\xi_{r,m}(w)+H^>+V^\perp))$$
\end{proposition}
\begin{proof}
  Let $H\in\bound{V}{X}$ and let us prove that there exist two different semi-$V$-patterns $P^<\not=P^>$ in $\P_V(X)$, an open convex $V$-polyhedron $C_H$ satisfying $[C_H\cap H^<]^V\not=[\emptyset]^V$, $[C_H\cap H^>]^V\not=[\emptyset]^V$ and such that $[X\cap (C_H+V^\perp)]^V=[P^<\cap ((C_H\cap H^<)+V^\perp)]^V\cup^V[P^>\cap ((C_H\cap H^>)+V^\perp)]^V$. From decomposition theorem \ref{thm:decomposition}, we have $[X]^V=\bigcup_{P\in\P_V(X)}([P]^V\cap(\class_{V,P}(X)+V^\perp))$. Let $H\in\bound{V}{X}$ and let $\Hrond'=\bound{V}{X}\moins\{H\}$. By definition of $\bound{V}{X}$, there exists $P_0\in\P_V(X)$ such that $H\in\bound{V}{\class_{V,P_0}(X)}$. Hence, there exist an open convex $V$-polyhedron $C$ and $\#_0\in\{<,>\}$ such that $[C\cap H^<]^V\not=[\emptyset]^V$, $[C\cap H^>]^V\not=[\emptyset]^V$ and $\class_{V,P_0}(X)\cap^V[C]^V =[C\cap H^{\#_0}]^V$. From lemma \ref{lem:alldir}, we deduce that there exists $\#'\in\{<,>\}^{\Hrond'}$ such that $[C\cap C_{V,\#'}\cap H^<]^V\not=[\emptyset]^V$, $[C\cap C_{V,\#'}\cap H^>]^V\not=[\emptyset]^V$. Let us denote by $C_H$ the open convex $V$-polyhedron $C_H=C\cap C_{V,\#'}$. Since $\Hrond'\cup\{H\}=\bound{V}{X}$ we deduce that $\class_{V,P}(X)\cap [C_{V,\#'}\cap H^\#]^V$ is either equal to $[\emptyset]^V$ or equal to $[C_{V,\#'}\cap H^\#]^V$ for any $P\in\P_V(X)$ and for any $\#\in\{<,>\}$. By definition of the sequence $(\class_{V,P}(X))_{P\in \P_V(X)}$ (a kind of partition of $[V]^V$) and since $[C_H\cap H^\#]^V\not=[\emptyset]^V$, there exists a unique $P^\#\in\P_V(X)$ such that $\class_{V,P^\#}(X) \cap^V [C_H\cap H^\#]^V=[C_H\cap H^\#]^V$. Since $\class_{V,P_0}(X)\cap^V [C_H]^V=[C_H\cap H^{\#_0}]^V$ we deduce that $\class_{V,P_0}(X)\cap^V [C_H\cap H^{\#_0}]^V=[C_H\cap H^{\#_0}]^V$ and $\class_{V,P_0}(X)\cap^V [C_H\cap H^{\#_1}]^V=[\emptyset]^V$ where $\#_1\in\{<,>\}\moins\{\#_0\}$. Hence $P^{\#_0}=P_0$ and $P^{\#_1}\not=P_0$. That means $P^<\not=P^>$ and we have proved that $[X\cap (C_H+V^\perp)]^V=[(P^<\cap ((C_H\cap H^<)+V^\perp))\cup(P^>\cap ((C_H\cap H^>)+V^\perp))]^V$ with $P^<\not=P^>$ in $\P_V(X)$.
  
  Now, assume that $X$ is $(r,m,w)$-cyclic, let $A$ be the $V$-affine space $A=\xi_{r,m}(w)+V$. Let $H\in\bound{V}{X}$, let $P^<$ and $P^>$ be two different semi-$V$-patterns in $\P_V(X)$, let $C_H$ be an open convex $V$-polyhedron such that $[C_H\cap H^<]^V\not=[\emptyset]^V$, $[C_H\cap H^>]^V\not=[\emptyset]^V$ and such that $[X\cap (C_H+V^\perp)]^V=[P^<\cap ((C_H\cap H^<)+V^\perp)]^V\cup^V[P^>\cap ((C_H\cap H^>)+V^\perp)]^V$, and let us prove that one of these two sets is $(r,m)$-detectable in $X$:
  $$(P^<\cap (\xi_{r,m}(w)+H^<+V^\perp))\cup (P^>\cap (\xi_{r,m}(w)+H^\geq+V^\perp))$$
  $$(P^<\cap (\xi_{r,m}(w)+H^\leq+V^\perp))\cup (P^>\cap (\xi_{r,m}(w)+H^>+V^\perp))$$
  Let $X'=X\cap A$. Corollary \ref{cor:detdet} shows that $\Z^m\cap A$ is $(r,m)$-detectable in $X$. 
  By replacing $C_H$ by $\vec{C_H}$, we can assume that $C_H=\vec{C_H}$. Let $X'=X\cap A$. Corollary \ref{cor:detdet} shows that $\Z^m\cap A$ is $(r,m)$-detectable in $X$ and in particular $X'$ is $(r,m)$-detectable in $X$. Since $X$ is $(r,m,w)$-cyclic and $P$ is $(r,m)$-detectable in $X$ from corollary \ref{cor:Mdetectable}, we deduce that any $P\in\P_V(X)$ is $(r,m,w)$-cyclic. From lemma \ref{lem:cyclicpattern}, we deduce that any $P\in\P_V(X)$ is relatively prime with $r$ and included in $A$. 
  
  Let us prove that by modifying $C_H$, we can assume that $X'\moins (\xi_{r,m}(w)+H)\cap (\xi_{r,m}(w)+C_H+V^\perp)=(P^<\cap (\xi_{r,m}(w)+C_H\cap H^<+V^\perp))\cup(P^>\cap (\xi_{r,m}(w)+C_H\cap H^>+V^\perp))$. Let $Z= (X'\moins(\xi_{r,m}(w)+V)\cap (\xi_{r,m}(w)+C_H+V^\perp))\Delta((P^<\cap (\xi_{r,m}(w)+C_H\cap H^<+V^\perp))\cup(P^>\cap (\xi_{r,m}(w)+C_H\cap H^>+V^\perp)))$. From $[X\cap (C_H+V^\perp)]^V=[(P^<\cap ((C_H\cap H^<)+V^\perp))\cup(P^>\cap ((C_H\cap H^>)+V^\perp))]^V$, we deduce that $[Z]^V=[\emptyset]^V$. Since $X'$, $P^<$ and $P^>$ are included in $A$, we deduce that $\saff(Z)\subseteq A$. In particular $\vecsaff(Z)\subseteq V$. Since $[Z]^V=[\emptyset]^V$, we deduce that $V$ is not included in $\vecsaff(Z)$. Assume by contradiction that $H\subseteq\vecsaff(Z)$. There exists an affine component $W$ of $\vecsaff(X)$ such that $H\subseteq W$. Since $H$ is a $V$-hyperplane, either $W=H$ or $W=V$. The last case is not possible since $V$ is not included in $\vecsaff(Z)$. Hence $W=H$ is an affine component of $\vecsaff(Z)$. Since $\saff(Z)=\xi_{r,m}(w)+\vecsaff(Z)$ we deduce that $\xi_{r,m}(w)+H$ is an affine component of $\saff(Z)$. From the dense component lemma \ref{lem:dense}, we deduce that $\saff(Z\cap (\xi_{r,m}(w)+H))=\xi_{r,m}(w)+H$. As $Z\cap (\xi_{r,m}(w)+H)=\emptyset$, we deduce a contradiction. Hence, there exists a finite set $\Hrond_0$ of $V$-hyperplane such that $\vec{\Hrond_0}=\Hrond_0$, $H\not\in \Hrond_0$ and such that $\vecsaff(Z)\subseteq \bigcup_{H_0\in\Hrond_0}H_0$. Thanks to lemma \ref{lem:alldir}, we deduce that there exists $\#\in\{<,>\}^{\Hrond_0}$ such that $[C_H\cap C_{V,\#}\cap H^>]\not=[\emptyset]^V$ and $[C_H\cap C_{V,\#}\cap H^<]\not=[\emptyset]^V$. Hence, by replacing $C_H$ by $C_H\cap C_{V,\#}$, since $Z\cap C_{V,\#}=\emptyset$, we can assume without loss of generality that $Z=\emptyset$. Thus $X'\moins(\xi_{r,m}(w)+V)\cap (\xi_{r,m}(w)+C_H+V^\perp)= (P^<\cap (\xi_{r,m}(w)+C_H\cap H^<+V^\perp))\cup(P^>\cap (\xi_{r,m}(w)+C_H\cap H^>+V^\perp))$.

  Assume first that $\Z^m\cap(\xi_{r,m}(w)+V)$ is $(r,m)$-detectable in $X$ and let us show that $X''= (P^<\cap (\xi_{r,m}(w)+H^<+V^\perp))\cup(P^>\cap (\xi_{r,m}(w)+H^>+V^\perp))$ is $(r,m)$-detectable in $X$. Let us consider a pair $(\sigma_1,\sigma_2)$ of words in $\Sigma_{r,m}^*$ such that $\gamma_{r,m,\sigma_1}^{-1}(X)=\gamma_{r,m,\sigma_2}^{-1}(X)$. Let us consider $x\in\gamma_{r,m,\sigma_1}^{-1}(X'')$. Then $\gamma_{r,m,\sigma_1}(x)\in (P^<\cap (\xi_{r,m}(w)+H^<+V^\perp))\cup(P^>\cap (\xi_{r,m}(w)+H^>+V^\perp))$. By definition of $v$, there exists an integer $k\in \Nat$ enough larger such that $\gamma_{r,m,\sigma_1}(x+k.v)$ is in $X''\cap (\xi_{r,m}(w)+C_H+V^\perp)$ and such that $\gamma_{r,m,\sigma_2}(x+k.v)\in \xi_{r,m}(w)+C_H+V^\perp$. Since $X''\cap (\xi_{r,m}(w)+C_H+V^\perp)=X'\moins (\xi_{r,m}(w)+H)\cap  (\xi_{r,m}(w)+C_H+V^\perp)$, we deduce that $\gamma_{r,m,\sigma_1}(x+k.v)\in X'\moins (\xi_{r,m}(w)+H)$. Since $X'$ and $\Z^m\cap(\xi_{r,m}(w)+H)$ are both $(r,m)$-detectable in $X$, we deduce that $\gamma_{r,m,\sigma_2}(x+k.v)\in X'\moins (\xi_{r,m}(w)+H)$. Moreover, as $\gamma_{r,m,\sigma_2}(x+k.v)(x)\in (\xi_{r,m}(w)+C_H+V^\perp)$, we have proved that $\gamma_{r,m,\sigma_2}(x+k.v)\in  X'\moins (\xi_{r,m}(w)+H)\cap (\xi_{r,m}(w)+C_H+V^\perp)$. Since this last set is equal to $X''\cap  (\xi_{r,m}(w)+C_H+V^\perp)$ we get $\gamma_{r,m,\sigma_2}(x+k.v)\in X''$. By definition of $v$, we get $\gamma_{r,m,\sigma_2}(x)\in X''$. Therefore $X''$ is $(r,m)$-detectable in $X$.
  
  We deduce that if $\Z^m\cap(\xi_{r,m}(w)+H)$ is $(r,m)$-detectable in $X$, since $P^<\cap (\xi_{r,m}(w)+H)$ and $P^>\cap (\xi_{r,m}(w)+H)$ are both $(r,m)$-detectable in $X$ as the intersection of $(r,m)$-detectable sets, the following two sets are $(r,m)$-detectable in $X$:
  $$(P^<\cap (\xi_{r,m}(w)+H^<+V^\perp))\cup (P^>\cap (\xi_{r,m}(w)+H^\geq+V^\perp))$$
  $$(P^<\cap (\xi_{r,m}(w)+H^\leq+V^\perp))\cup (P^>\cap (\xi_{r,m}(w)+H^>+V^\perp))$$
  
  Now, assume that $\Z^m\cap (\xi_{r,m}(w)+H)$ is not $(r,m)$-detectable in $X$
  
  Let us first show that there exists a pair $(\sigma_1,\sigma_2)$ of words such that $\gamma_{r,m,\sigma_1}^{-1}(X)=\gamma_{r,m,\sigma_2}^{-1}(X)$, $\Gamma_{r,m,\sigma_1}^{-1}(\xi_{r,m}(w)+V)$ and $\Gamma_{r,m,\sigma_2}^{-1}(\xi_{r,m}(w)+V)$ are equal, $\Z^m\cap \Gamma_{r,m,\sigma_1}^{-1}(\xi_{r,m}(w)+H)$ is not empty, and such that $\Gamma_{r,m,\sigma_1}^{-1}(\xi_{r,m}(w)+H+V^\perp)$ and $\Gamma_{r,m,\sigma_2}^{-1}(\xi_{r,m}(w)+H+V^\perp)$ have an empty intersection. Since $\Z^m\cap (\xi_{r,m}(w)+H)$ is not $(r,m)$-detectable in $X$, there exists a pair $(\sigma_1,\sigma_2)$ of words in $\Sigma_{r,m}^*$ such that $\gamma_{r,m,\sigma_1}^{-1}(X)=\gamma_{r,m,\sigma_2}^{-1}(X)$ and such that $\gamma_{r,m,\sigma_1}^{-1}(\Z^m\cap (\xi_{r,m}(w)+H))$ and $\gamma_{r,m,\sigma_2}^{-1}(\Z^m\cap (\xi_{r,m}(w)+H))$ are disjoint. Remark that $\gamma_{r,m,\sigma_i}^{-1}(\Z^m\cap (\xi_{r,m}(w)+H))=\Z^m\cap \Gamma_{r,m,\sigma_i}^{-1}(\xi_{r,m}(w)+H)$ for any $i\in\{1,2\}$. By replacing $(\sigma_1,\sigma_2)$ by $(\sigma_2,\sigma_1)$, we can assume that $\Z^m\cap \Gamma_{r,m,\sigma_1}^{-1}(\xi_{r,m}(w)+H)$ is not empty. Since $\Z^m\cap (\xi_{r,m}(w)+V)$ is $(r,m)$-detectable in $X$, we deduce that $\Z^m\cap \Gamma_{r,m,\sigma_1}^{-1}(\xi_{r,m}(w)+V)$ and $\Z^m\cap \Gamma_{r,m,\sigma_2}^{-1}(\xi_{r,m}(w)+V)$ are equal. Moreover, as $\Z^m\cap \Gamma_{r,m,\sigma_1}^{-1}(\xi_{r,m}(w)+H)$ is non empty and $H\subseteq V$, we deduce that the sets   $\Z^m\cap \Gamma_{r,m,\sigma_1}^{-1}(\xi_{r,m}(w)+V)$ and $\Z^m\cap \Gamma_{r,m,\sigma_2}^{-1}(\xi_{r,m}(w)+V)$ are non empty. Taking the semi-affine hull of these sets, we get $\Gamma_{r,m,\sigma_1}^{-1}(\xi_{r,m}(w)+V)=\Gamma_{r,m,\sigma_2}^{-1}(\xi_{r,m}(w)+V)$. Assume by contradiction that $\Gamma_{r,m,\sigma_1}^{-1}(\xi_{r,m}(w)+H+V^\perp)$ and $\Gamma_{r,m,\sigma_2}^{-1}(\xi_{r,m}(w)+H+V^\perp)$ have a non empty intersection and let $x$ be a vector in this intersection. From $x\in \Gamma_{r,m,\sigma_1}^{-1}(\xi_{r,m}(w)+H+V^\perp)$ we deduce that there exists $v_\perp\in V^\perp$ such that $x-v_\perp\in\Gamma_{r,m,\sigma_1}^{-1}(\xi_{r,m}(w)+H)\subseteq \Gamma_{r,m,\sigma_1}^{-1}(\xi_{r,m}(w)+V)$. From $\Gamma_{r,m,\sigma_1}^{-1}(\xi_{r,m}(w)+V)=\Gamma_{r,m,\sigma_2}^{-1}(\xi_{r,m}(w)+V)$, we deduce that $x-v\in \Gamma_{r,m,\sigma_2}^{-1}(\xi_{r,m}(w)+V)$. Moreover, since $x\in \Gamma_{r,m,\sigma_2}^{-1}(\xi_{r,m}(w)+H+V^\perp)$, we get $x-v\in \Gamma_{r,m,\sigma_2}^{-1}(\xi_{r,m}(w)+H)$. Therefore $\Gamma_{r,m,\sigma_1}^{-1}(\xi_{r,m}(w)+H)$ and $\Gamma_{r,m,\sigma_2}^{-1}(\xi_{r,m}(w)+H)$ are equal and we get a contradiction.
  
  Since $\Gamma_{r,m,\sigma_1}^{-1}(\xi_{r,m}(w)+H+V^\perp)$ and $\Gamma_{r,m,\sigma_2}^{-1}(\xi_{r,m}(w)+H+V^\perp)$ have an empty intersection, there exists $\#\in\{<,>\}$ such that $\Gamma_{r,m,\sigma_1}^{-1}(\xi_{r,m}(w)+H+V^\perp)\subseteq\Gamma_{r,m,\sigma_1}^{-1}(\xi_{r,m}(w)+H^\#+V^\perp)$. 
  
  Let us consider the $(r,m,w)$-cyclic Presburger definable set $X'_H=X'\cap (\xi_{r,m}(w)+H)$ and the semi-$H$-pattern $P^\#_H=P^\#\cap (\xi_{r,m}(w)+H)$, and let us prove the following equality:
  $$\gamma_{r,m,\sigma_1}^{-1}(X'_H\cap (\xi_{r,m}(w)+C_H\cap H))=\gamma_{r,m,\sigma_1}^{-1}(P_H^\#\cap (\xi_{r,m}(w)+C_H\cap H))$$
  Remark that $\gamma_{r,m,\sigma_2}^{-1}(X'\moins(\xi_{r,m}(w)+H))\cap \Gamma_{r,m,\sigma_1}^{-1}(\xi_{r,m}(w)+C_H\cap H+V^\perp)$ is equal to $\gamma_{r,m,\sigma_2}^{-1}(X')\moins \Gamma_{r,m,\sigma_2}^{-1}(\xi_{r,m}(w)+H+V^\perp)\cap \Gamma_{r,m,\sigma_1}^{-1}(\xi_{r,m}(w)+ C_H\cap H+V^\perp)$. Since $\Gamma_{r,m,\sigma_1}^{-1}(\xi_{r,m}(w)+ H+V^\perp)$ and $\Gamma_{r,m,\sigma_2}^{-1}(\xi_{r,m}(w)+H+V^\perp)$ have an empty intersection, and $\gamma_{r,m,\sigma_1}^{-1}(X')=\gamma_{r,m,\sigma_2}^{-1}(X')$, we deduce that $\gamma_{r,m,\sigma_2}^{-1}(X'\moins(\xi_{r,m}(w)+C_H\cap H+V^\perp))\cap \Gamma_{r,m,\sigma_1}^{-1}(\xi_{r,m}(w)+ C_H\cap H+V^\perp)$ is equal to $\gamma_{r,m,\sigma_1}^{-1}(X'_H\cap (\xi_{r,m}(w)+C_H\cap H+V^\perp))$. On the other hand, since $X'\moins(\xi_{r,m}(w)+H+V^\perp)\cap (\xi_{r,m}(w)+V_H+V^\perp)=\bigcup_{\#'\in\{<,>\}}(P^{\#'}\cap (\xi_{r,m}(w)+C_H\cap H^{\#'}+V^\perp)))$, we get $\gamma_{r,m,\sigma_2}^{-1}(X'\moins(\xi_{r,m}(w)+H+V^\perp)\cap (\xi_{r,m}(w)+V_H+V^\perp))\cap \Gamma_{r,m,\sigma_1}^{-1}(\xi_{r,m}(w)+H+V^\perp)=\bigcup_{\#'\in\{<,>\}}(\gamma_{r,m,\sigma_2}^{-1}(P^{\#'})\cap \Gamma_{r,m,\sigma_2}^{-1}(\xi_{r,m}(w)+C_H\cap H^{\#'}+V^\perp)\cap \Gamma_{r,m,\sigma_1}^{-1}(\xi_{r,m}(w)+ H+V^\perp))$. Remark that $\Gamma_{r,m,\sigma_2}^{-1}(\xi_{r,m}(w)+H^{\#'}+V^\perp)$ and  $\Gamma_{r,m,\sigma_1}^{-1}(\xi_{r,m}(w)+ H+V^\perp)$ have an empty intersection if $\#'$ is not equal to $\#$ and $\Gamma_{r,m,\sigma_2}^{-1}(\xi_{r,m}(w)+C_H\cap H^{\#}+V^\perp)\cap \Gamma_{r,m,\sigma_1}^{-1}(\xi_{r,m}(w)+ H+V^\perp)=\Gamma_{r,m,\sigma_2}^{-1}(\xi_{r,m}(w)+C_H+V^\perp)\cap \Gamma_{r,m,\sigma_1}^{-1}(\xi_{r,m}(w)+H+V^\perp)$. As $\Gamma_{r,m,\sigma_1}^{-1}(\xi_{r,m}(w)+V)=\Gamma_{r,m,\sigma_2}^{-1}(\xi_{r,m}(w)+V)$ and $\vec{C_H}=C_H$, we deduce that $\Gamma_{r,m,\sigma_2}^{-1}(\xi_{r,m}(w)+C_H+V^\perp)=\Gamma_{r,m,\sigma_1}^{-1}(\xi_{r,m}(w)+C_H+V^\perp)$. Moreover, since $\gamma_{r,m,\sigma_2}^{-1}(P^\#)=\gamma_{r,m,\sigma_1}^{-1}(P^\#)$, we have proved that $\gamma_{r,m,\sigma_2}^{-1}(X'\moins(\xi_{r,m}(w)+H+V^\perp))\cap \Gamma_{r,m,\sigma_1}^{-1}(\xi_{r,m}(w)+ C_H\cap H+V^\perp)=\gamma_{r,m,\sigma_1}^{-1}(P^\#\cap (\xi_{r,m}(w)+\C_H\cap H+V^\perp))$. Combining the two equalities proved in this paragraph, we are done.

  Let us prove that $C_H\cap H$ is a non $H$-degenerate $H$-polyhedron. The proof is obtained thanks to lemma \ref{lem:vdegpolyconv}. Since $[C_H\cap H^\#]^V\not=[\emptyset]^V$, there exists a vector $v_\#\in H^\#\cap C_H$. Now just remark that there exists $k_<$, $k_>$ in $\Q_+\moins\{0\}$ such that $v=k_<.v_<+k_>.v_>$ is in $H$. In particular $v\in H\cap C_H$. Thus $H\cap C_H$ is non-$H$-degenerate.
  
  Next, let us prove that $[X'_H\cap (\xi_{r,m}(w)+C_H\cap H)]^H=[P^\#_H\cap (\xi_{r,m}(w)+C_H\cap H)]^H$. Since $\gamma_{r,m,\sigma_1}^{-1}(\Z^m\cap (\xi_{r,m}(w)+H))$ is non empty, there exists a $(r,m)$-decomposition $(w,s)$ such that $\rho_{r,m}(w,s)$ is in this set. By replacing $w$ by a word in $w.s^*$, since $\inv_V(P^\#)$ is relatively prime with $r$, we can assume that $r^{|\sigma_1.w|_m}\in 1+|\Z^m\cap V/\inv_V(P^\#)|.\Z$. From lemma \ref{lem:invdensepattern} we get $\gamma_{r,m,\sigma_1.w}^{-1}(P^\#)=\xi_{r,m}(w)+P^\#-\rho_{r,m}(\sigma_1.w,s)$. In particular, if $[X'_H]^H=[\emptyset]^H$, then $[\gamma_{r,m,\sigma_1.w}^{-1}(X'_H)]^H=[\emptyset]^H$ and from the equality $\gamma_{r,m,\sigma_1}^{-1}(X'_H\cap (\xi_{r,m}(w)+C_H\cap H))=\gamma_{r,m,\sigma_1}^{-1}(P_H^\#\cap (\xi_{r,m}(w)+C_H\cap H))$ 
  we deduce that $[\gamma_{r,m,\sigma_1.w}^{-1}(P_H^\#\cap (\xi_{r,m}(w)+C_H\cap H))]^H=[\emptyset]^H$. Since $C_H\cap H$ is non $H$-degenerate, we get $[P^\#_H]^H=[\emptyset]^H$ 
  and we have proved that $[X'_H\cap (\xi_{r,m}(w)+C_H\cap H)]^H=[P^\#_H\cap (\xi_{r,m}(w)+C_H\cap H)]^H$. So, we can assume that $[X'_H]^H\not=[\emptyset]^H$. 
  In this case $H$ is an affine component of the $(r,m,w)$-cyclic Presburger definable set $X'_H$. In particular $\inv_H(X'_H)$ is relatively prime with $r$ and by replacing $w$ by a word in $w.s^*$ we can assume that $r^{|\sigma_1.w|_m}\in 1+|\Z^m\cap H/\inv_H(X)|.\Z$. Since $\rho_{r,m}(\sigma.w,s)\in \Z^m\cap (\xi_{r,m}(w)+H)$, 
  from lemma \ref{lem:invdensepattern} we deduce that $\gamma_{r,m,\sigma_1.w}^{-1}(P)=\xi_{r,m}(s)+P-\rho_{r,m}(\sigma_1.w,s)$. From $[X'_H]^H=\bigcup^H_{P\in\P_H(X'_H)}([P]^H\cap^H\class_{H,P}(X'_H)+H^\perp)$, we deduce that $[\gamma_{r,m,\sigma_1.w}^{-1}(X'_H)]^H=\bigcup^H_{P\in\P_H(X'_H)}([\gamma_{r,m,\sigma_1.w.s}^{-1}(P)]^H\cap^H\class_{H,P}(X'_H)+H^\perp)$. From the equality $[X'_H\cap (\xi_{r,m}(w)+C_H\cap H)]^H=[P^\#_H\cap (\xi_{r,m}(w)+C_H\cap H)]^H$, 
  decomposition theorem \ref{thm:decomposition} shows that there exists $P\in \P_H(X'_H)$ such that $[C_H\cap H]^H\subseteq^H\class_{H,P}(X'_H)$ and such that $\gamma_{r,m,\sigma_1.w}^{-1}(P)=\gamma_{r,m,\sigma_1.w}^{-1}(P^\#_H)$. Since $\gamma_{r,m,\sigma_1.w}^{-1}(P)=\xi_{r,m}(s)+P-\rho_{r,m}(\sigma_1.w,s)$ and $\gamma_{r,m,\sigma_1.w}^{-1}(P^\#_H)=\xi_{r,m}(s)+P^\#_H-\rho_{r,m}(\sigma_1.w,s)$ we get $P=P^\#_H$. Thus $[X'_H\cap (\xi_{r,m}(w)+C_H\cap H)]^H=[P^\#_H\cap (\xi_{r,m}(w)+C_H\cap H)]^H$ and we are done.
  
  Let us consider the set $E=(P^<\cap (\xi_{r,m}(w)+H^<+V^\perp))\cup (P^\#\cap (\xi_{r,m}(w)+H+V^\perp))\cup(P^>\cap (\xi_{r,m}(w)+H^\geq+V^\perp))$ and remark that these set is equal to one of the following two sets and it is such that $[Z]^H=[\emptyset]^H$ where $Z=(X'\Delta E)\cap (\xi_{r,m}(w)+C_H+V^\perp)$.
  $$(P^<\cap (\xi_{r,m}(w)+H^<+V^\perp))\cup (P^>\cap (\xi_{r,m}(w)+H^\geq+V^\perp))$$
  $$(P^<\cap (\xi_{r,m}(w)+H^\leq+V^\perp))\cup (P^>\cap (\xi_{r,m}(w)+H^>+V^\perp))$$
  Let us prove that $E$ is $(r,m)$-detectable in $X$. Consider a pair $(w_1,w_2)$ of words in $\Sigma_{r,m}^+$ such that $\gamma_{r,m,w_1}^{-1}(X)=\gamma_{r,m,w_2}^{-1}(X)$. Since $X'$ is $(r,m)$-detectable in $X$, we deduce that t$\gamma_{r,m,w_1}^{-1}(X')=\gamma_{r,m,w_2}^{-1}(X')$. From $Z=(X'\Delta E)\cap (\xi_{r,m}(w)+C_H+V^\perp)$, we deduce that $Z'=(\gamma_{r,m,w_1}^{-1}(E)\Delta \gamma_{r,m,w_2}^{-1}(E))\cap (C'+V^\perp)$ where $C'$ is the open convex $V$-polyhedron such that $C'+V^\perp=\Gamma_{r,m,w_1}^{-1}(\xi_{r,m}(w)+C_H+V^\perp)\cap \Gamma_{r,m,w_2}^{-1}(\xi_{r,m}(w)+C_H+V^\perp)$ and $Z'=(\gamma_{r,m,w_1}^{-1}(E)\Delta \gamma_{r,m,w_2}^{-1}(E))\cap (C'+V^\perp)$. Since $[Z]^H=[\emptyset]^H$, from covering lemma \ref{lem:cov}, we get $[Z']^H=[\emptyset]^H$. Moreover, as $[C']^V=[C_H]^V$, we deduce that $C'$ is non $V$-degenerate and such that $[C'\cap H^<]^V$ and $[C'\cap H^>]^V$ are both non equal to $[\emptyset]^V$. Let us remark that $\gamma_{r,m,w_1}^{-1}(E)\Delta\gamma_{r,m,w_2}^{-1}(E)$ is a semi-$H$-pattern and $C'\cap H$ is a non-$H$-degenerate $H$-polyhedron from lemma \ref{lem:Pleqgeq}. Since $[Z']^H=[\emptyset]^H$, we deduce from lemma \ref{lem:dimdim} that $\gamma_{r,m,w_1}^{-1}(E)=\gamma_{r,m,w}^{-1}(E)$. Thus $E$ is $(r,m)$-detectable. We are done. 
  \qed
\end{proof}

Recall that a semi-$V$-pattern $P$ detectable in a $(r,m,w)$-cyclic set $X$ are relatively prime with $r$. The following proposition will become useful in the last section in order to check that some sets that must be detectable in $X$ if $X$ is Presburger-definable are effectively detectable in $X$.
\begin{proposition}\label{prop:verifbound}
  Let $\automaton$ be a FDVA, let $P_1=B_1+M$, $P_2=B_2+M$ be two semi-$V$-patterns where $B_1$, $B_2$ are two finite subsets of $\Z^m$, and $M$ is a $V$-vector lattices included in $\Z^m$ relatively prime with $r$, let $H$ be a $V$-hyperplane, let $a_0\in \Q^m$ and let $(\#_1,\#_2)\in\{(<,\geq),(\leq,>)\}$. Assume that there exists a final function $F_i$ such that $P_i$ is represented by $\automaton^{F_i}$. We can decide in polynomial time if there exists a final function $F$ such that the following set is represented by $\automaton^F$:
  $$(P_1\cap (a_0+H^{\#_1}+V^\perp))\cup(P_2\cap (a_0+H^{\#_2}+V^\perp))$$
\end{proposition}
\begin{proof}
  From proposition \ref{prop:critdetectable} we deduce in polynomial time a set $U$ of pairs $(\sigma_a,\sigma_b)$ of words in $\Sigma_r^*$ such that $|\sigma_a|+m.\Z=|\sigma_b|+m.\Z$ for any $(\sigma_a,\sigma_b)\in U$ and such that a set $X'\subseteq \Z^m$ is represented by a FDVA of the form $\automaton^F$ if and only if $\gamma_{r,m,\sigma_a}^{-1}(X')=\gamma_{r,m,\sigma_b}^{-1}(X')$ for any $(\sigma_a,\sigma_b)\in U$. Let $X'$ be the set $X'=(P_1\cap (a_0+H^{\#_1}+V^\perp))\cup(P_2\cap (a_0+H^{\#_2}+V^\perp))$. Since $\gamma_{r,m,\sigma_a}^{-1}(P_i)=\gamma_{r,m,\sigma_b}^{-1}(P_i)$ for any $i\in\{1,2\}$, for proving the proposition, it is sufficient to show that given a pair $(\sigma_a,\sigma_b)$ of words such that $|\sigma_a|+m.\Z=|\sigma_b|+m.\Z$ and $\gamma_{r,m,\sigma_a}^{-1}(P_i)=\gamma_{r,m,\sigma_b}^{-1}(P_i)$ for any $i$, we can decide in polynomial time if $\gamma_{r,m,\sigma_a}^{-1}(X')=\gamma_{r,m,\sigma_b}^{-1}(X')$. In polynomial time, we can compute a vector $\alpha\in\Z^m\cap V$ such that $H^\#=\{x\in V;\;\scalar{\alpha}{x}\#0\}$ for any $\#\in\{<,\leq,=,\geq,>\}$. Let $z\in\finiteset{0}{m-1}$ such that $|\sigma_a|+m.\Z=z+m.\Z=|\sigma_b|+m.\Z$, let $\alpha_z$ be the vector in $\Z^m$ such that $\scalar{\alpha}{\gamma_{r,m,0}^z(x)}=\scalar{\alpha_z}{x}$ for any $x\in\Q^m$, and let $V_z$ be the vector space $V_z=\Gamma_{r,m,0}^{-z}(V)$. Proposition \ref{prop:invsemipattern} proves that we can compute in polynomial time two finite subsets $B_1'$ and $B_2'$ of $\Z^m$ such that $\gamma_{r,m,\sigma_a}^{-1}(P_i)=B_i'+\gamma_{r,m,0}^{-|\sigma|}(M)$. Since $M$ is relatively prime with $r$, we deduce that $\gamma_{r,m,0}^{-|\sigma|}(M)$ is equal to $M_z=\gamma_{r,m,0}^{-z}(M)$. Note that $\gamma_{r,m,\sigma_b}^{-1}(P_i)=B_i'+M_z$. Let $c_a=r^{\frac{z-|\sigma_a|}{m}}.\scalar{\alpha}{a_0}$ and $c_b=r^{\frac{z-|\sigma_b|}{m}}.\scalar{\alpha}{a_0}$. Observe that $x\in \Gamma_{r,m,\sigma_a}^{-1}(a_0+H^\#+V^\perp)$ if and only if $\Gamma_{r,m,\sigma_a}(x)\in a_0+H^\#+V$ if and only if $\scalar{\alpha}{\Gamma_{r,m,\sigma_a}(x)}\#\scalar{\alpha}{a_0}$ if and only if $\scalar{\alpha_z}{x}\#c_a$. We deduce the following equalities (the equality with $\sigma_b$ is obtained by symmetry):
  $$\begin{cases}
    \gamma_{r,m,\sigma_a}^{-1}(X')=\{x\in B_1'+M_z;\;\scalar{\alpha_z}{x}\#_1 c_a\}\cup \{x\in B_2'+M_z;\;\scalar{\alpha_z}{x}\#_2 c_a\}\\
    \gamma_{r,m,\sigma_b}^{-1}(X')=\{x\in B_1'+M_z;\;\scalar{\alpha_z}{x}\#_1 c_b\}\cup \{x\in B_2'+M_z;\;\scalar{\alpha_z}{x}\#_2 c_b\}
  \end{cases}$$
  If $c_a=c_b$ then $\gamma_{r,m,\sigma_a}^{-1}(X')=\gamma_{r,m,\sigma_2}^{-1}(X')$. Otherwise, by symmetry, we can assume that $c_a<c_b$. In this case, the set $\gamma_{r,m,\sigma_a}^{-1}(X')\Delta\gamma_{r,m,\sigma_b}^{-1}(X')$ is equal to the following set:
  $$\{x\in (B_1'+M_z)\Delta (B_2'+M_z);\; c_a (-\#_2) \scalar{\alpha_z}{x}\#_1 c_b\}$$
  Let us consider the set $B$ equal to the union of the set of vectors $b\in B_1$ such that there does not exist $b_2\in B_2$ such that $b-b_2\in M_z$ and the set of vectors $b\in B_2$ such that there does not exist $b_1\in B_1$ satisfying $b-b_1\in M_z$. Observe that $B$ is computable in polynomial time and $(B_1'+M_z)\Delta (B_2'+M_z)=B+M_z$. Thus we have reduced our problem to decide if there exists $b\in B$ such that the following set is non empty where $c_1'=c_a-\scalar{\alpha_z}{b}$, $c_2'=c_b-\scalar{\alpha_z}{b}$, and $(\#_1', \#_2')=(-\#_2,\#_1)$:
  $$\{x\in M_z;\; c_1' \#_1' \scalar{\alpha_z}{x}\#_2' c_2'\}$$
  From an Hermite representation of $M_z$, we deduce in linear time a $\Z$-basis $v_1$, ..., $v_d$ of $M_z$. Note that the set $\{\scalar{\alpha_z}{x};\;x\in M_z\}$ is equal to $\sum_{i=1}^d\Z.\scalar{\alpha_z}{v_i}$. Thus, considering the lattice generated by $\{\scalar{\alpha_z}{v_i};\;1\leq i\leq d\}$, we compute in polynomial time a rational number $\mu>0$ such that  $\{\scalar{\alpha_z}{x};\;x\in M_z\}$ is equal to $\Z.\mu$. We deduce that $\{x\in M_z;\; c_1' \#_1' \scalar{\alpha_z}{x}\#_2' c_2'\}$ is non empty if and only if there exists an integer $z\in\Z$ such that $\frac{c_1'}{\mu}\#_1'z\#_2'\frac{c_2'}{\mu}$. This property property is decidable in linear time. We are done.
  \qed
\end{proof}

\subsection{A polynomial time algorithm} 
As for any pair of serialized encoded FDVA $(\automaton_1,\automaton_2)$, we can compute in quadratic time a serialized encoded FDVA $\automaton$ that represents $X_1\Delta X_2$ where $X_i$ is the set represented by $\automaton_i$, the following proposition \ref{prop:boundpoly} shows that our computation problem can be effectively done in polynomial time thanks to the semi-affine hull direction computation.

\begin{proposition}\label{prop:boundpoly}
  Let $X$ be a Presburger-definable set represented by a FDVA $\automaton$ and let $V$ be an affine component of $\vecsaff(X)$. Consider $I_\automaton(V)$, the set of pairs of states $(q_1,q_2)\in T\times T$ where $T$ is a terminal component such that $V_G(T)=V$ and such that $q_1\sim^V q_2$\index{Not}{$\sim^V$}. We have the following equality:
  $$\bound{V}{X}\moins(\bigcup_{j=1}^m\{V\cap \unit_{j,m}^\perp\})=\comp(\bigcup_{(q_1,q_2)\in I_\automaton(V)}\vecsaff(X_{q_1}\Delta X_{q_2}))$$
\end{proposition}
\begin{proof}
  Let $J$ be the set of indices in $\finiteset{1}{m}$ such that $V\cap \unit_{j,m}^\perp$ is a $V$-hyperplane. As $\bound{V}{X}$ contains only $V$-hyperplanes, we deduce that $\bound{V}{X}\moins(\bigcup_{j=1}^m\{V\cap\unit_{j,m}^\perp\})$ and $\bound{V}{X}\moins(\bigcup_{j\in J}\{V\cap\unit_{j,m}^\perp\})$ are equal. We denote by $\Hrond_0$ this class. The semi-affine space $S=\bigcup_{H\in \Hrond_0}H$ satisfies $\comp(S)=\Hrond_0$. Consider the semi-affine space $S'=\bigcup_{(q_1,q_2)\in I_\automaton(V)}\vecsaff(X_{q_1}\Delta X_{q_2})$. We have to prove that $S=S'$.

  Let us first prove the inclusion $S'\subseteq S$. Let $(q_1,q_2)\in I_\automaton(V)$ and let $W=\vecsaff(X_{q_1}\Delta X_{q_2})$. Naturally, if $W=\emptyset$, we immediately have $W\subseteq S$. So we can assume that $W\not=\emptyset$. Let us consider an affine component $A_0$ of $W$.
  
  From theorem \ref{thm:ultime} there exists $a_1,a_2\in\Q^m$ satisfying the following equality (where $i\in\{1,2\}$) and such that $-1<a_i[j]<0$ for any $(i,j)\in\{1,2\}\times J$:
  $$X_{q_i}=\bigcup_{P\in\P_V(X)}\bigcup_{\#\in\S_{V,P}(X)}P_{q_i}\cap (a_i+C_{V,\#}+V^\perp)$$
  
  We denote by $v_i$ the vector $v_i=\Pi_V(a_i)$ for $i\in\{1,2\}$. Remark that $P_{q_1}=P_{q_2}$ for any $P\in\P_V$ since $q_1\sim^V q_2$. We denote by $P_{q_1,q_2}$ this semi-$V$-pattern.
  
  Let us prove that there exists $H\in\bound{V}{X}$, $\#\in\{<,>\}$ and a $V$-affine space $A$ such that $A_0\subseteq \vecsaff(\Z^m\cap A\cap (((v_1+H^\#)\Delta (v_2+H^\#))+V^\perp))$. The set $X_{q_1}\Delta X_{q_2}$ is included into the finite union of sets $P_{q_1,q_2}\cap (((v_1+C_{V,\#})\Delta (v_2+C_{V,\#}))+V^\perp)$ over $P\in\P_V(X)$ and $\#\in\S_{V,P}(X)$. As $C_{V,\#}=\bigcap_{H\in\bound{V}{X}}H^{\#_H}$, we deduce that $X_{q_1}\Delta X_{q_2}$ is included into the finite union of sets $P_{q_1,q_2}\cap (((v_1+H^\#)\Delta (v_2+H^\#))+V^\perp)$ over $P\in\P_V(X)$, $H\in\bound{V}{X}$ and $\#\in\{<,>\}$. From insecable lemma \ref{lem:insecable}, we deduce that there exists $P\in\P_V(X)$, $H\in\bound{V}{X}$ and $\#\in\{<,>\}$ such that $A_0\subseteq \vecsaff(P_{q_1,q_2}\cap (((v_1+H^\#)\Delta (v_2+H^\#))+V^\perp))$. As $P_{q_1,q_2}$ is a semi-$V$-pattern, it is included into a finite union of sets of the form $\Z^m\cap A$ where $A$ is a $V$-affine space. From insecable lemma \ref{lem:insecable} we deduce that there exists a $V$-affine space $A$ such that $A_0\subseteq \vecsaff(\Z^m\cap A\cap (((v_1+H^\#)\Delta (v_2+H^\#))+V^\perp))$.
  
  Let us show that $H\not\in\{V\cap \unit_{j,m}^\perp;\;j\in J\}$. As $A_0\not=\emptyset$, it is sufficient to show that otherwise, the set $\Z^m\cap A\cap (((v_1+H^\#)\Delta (v_2+H^\#))+V^\perp)$ is empty. Remark that this set is included in $(\Z^m\cap (a_1+H^\#+V^\perp))\Delta(\Z^m\cap (a_1+H^\#+V^\perp))$. If $H=V\cap \unit_{j,m}^\perp$ where $j\in J$, there exists $\epsilon\in\{-1,1\}$ such that $H^\#=\{x\in V;\;\epsilon.x[j]\#0\}$. Remark that $a_i+H^\#+V^\perp=\{x\in\Q^m;\;\epsilon.(x[j]-a_i[j])\#0\}$. As $a_1[j]$ and $a_2[j]$ are two rational numbers in $\{x\in\Q;\;-1<x<0\}$. We deduce that $\Z^m\cap (a_1+H^\#+V^\perp)$ and $\Z^m\cap (a_2+H^\#+V^\perp)$ are equal. Therefore $(\Z^m\cap (a_1+H^\#+V^\perp))\Delta(\Z^m\cap (a_1+H^\#+V^\perp))$ is empty. We have proved that $H\not\in\{V\cap \unit_{j,m}^\perp;\;j\in J\}$.
  
  Let us prove that $A_0\subseteq H$. Consider $\alpha\in \Z^m\cap V\moins\{\unit_{0,m}\}$ such that $H^\#=\{x\in V;\;\scalar{\alpha}{x}\#0\}$. Let $K=\{k\in\Z;\;k\leq \max\{|\scalar{\alpha}{v_1}|,|\scalar{\alpha}{v_2}|\}\}$ and remark that for any $x\in \Z^m\cap (((v_1+H^\#)\Delta (v_2+H^\#))+V^\perp)$, we have $\scalar{\alpha}{x}\in K$. Hence $\Z^m\cap A\cap  (((v_1+H^\#)\Delta (v_2+H^\#))+V^\perp)$ is included into $\bigcup_{k\in K}\{x\in A;\;\scalar{\alpha}{x}=k\}$. From insecable lemma \ref{lem:insecable}, we deduce that $A_0\subseteq H$.
  
  We have proved that $A_0\subseteq S$ for any affine component $A_0$ of $W$. Therefore $W\subseteq S$. We deduce that $S'\subseteq S$.
  
  Now, let us prove the converse inclusion $S\subseteq S'$. Consider a $V$-hyperplane $H_0\in\Hrond_0=\bound{V}{X}\moins (\bigcup_{j\in J}(V\cap \unit_{j,m}^\perp))$. Let $\H=\bound{V}{X}\moins\{H_0\}$. We denote by $\alpha_0\in V\moins\{\unit_{0,m}\}$ a vector such that $H_0^{\#_0}=\{x\in V;\;\scalar{\alpha_0}{x}\#_00$ for any $\#_0\in\{<,>\}$. Given $\#\in\{<,>\}^\H$ and $\#_0\in\{<,>\}$, we denote by $(\#,\#_0)$ the sequence in $\{<,>\}^{\bound{V}{X}}$ naturally defined. Remark that for any sequence $\#\in\{<,>\}^\H$ and for any $\#_0\in\{<,>\}$ such that $[C_{V,(\#,\#_0)}]_V\not=[\emptyset]_V$, there exists a unique $P_{\#,\#_0}\in\P_V(X)$ such that $(\#,\#_0)\in \S_{V,P_{\#,\#_0}}$. 
  
  Let us prove that there exists $\#\in\{<,>\}^\H$ such that $[C_{V,(\#,<)}]_V$ and $[C_{V,(\#,>)}]_V$ are both not equal to $[\emptyset]_V$, and such that $P_{\#,<}\not=P_{\#,>}$. By contradiction, if for any $\#\in\{<,>\}^\H$ such that $[C_{V,(\#,<)}]_V$ and $[C_{V,(\#,>)}]_V$ are both not equal to $[\emptyset]_V$, we have $P_{\#,<}=P_{\#,>}$, decomposition theorem \ref{thm:decomposition} shows that $\bound{V}{X}\subseteq \H$ which is impossible. Hence, there exists at least one sequence $\#\in\{<,>\}^\H$ such that  $[C_{V,\#}\cap H_0^{<}]_V$ and $[C_{V,\#}\cap H_0^{>}]_V$ are both not equal to $[\emptyset]_V$ and such that $P_{\#,<}\not=P_{\#,>}$.
  
  From the previous paragraph, we deduce that the semi-$V$-pattern $P_0=P_{\#,<}\Delta P_{\#,>}$ is not empty. Moreover, as $P_{\#,<}$ and $P_{\#,>}$ are both $(r,m)$-detectable in $X$, we deduce that $P_0$ is also $(r,m)$-detectable in $X$ and for any reachable state $q\in Q$, the set $(P_0)_q$ is well defined. 
  
  
  Let us prove that there exists a terminal component $T$ such that $V_G(T)=V$ and such that $(P_{\#,<})_q\not=(P_{\#,>})_q$ for any state $q\in T$. As $P_0$ is not empty, there exists a $(r,m)$-decomposition $(\sigma_0,s)\in\rho_{r,m}^{-1}(P_0)$. By replacing $\sigma_0$ by a word in $\sigma_0.s^*$, we can assume that there exists a loop labelled by a word in $s^+$ on the state $q_0'=\delta(q_0,\sigma_0)$. In particular $\inv_V((P_0)_{q_0'})$ is relatively prime with $r$ and $\xi_{r,m}(s)\in (P_0)_{q_0'}$. From destruction lemma \ref{lem:destruction}, we deduce that $P_{q_0'}\subseteq \xi_{r,m}(s)+\Z^m\cap V$ for any $P\in\P_V(X)$. As $(P_0)_{q_0'}$ is non empty, there exists $\#_0\in\{<,>\}$ such that $(P_{\#,\#_0})_{q_0'}\not=\emptyset$. From proposition \ref{prop:patternempty}, we deduce that $V$ is included in $\vecsaff(X_{q_0'})$. As $X_{q_0'}\subseteq\Gamma_{r,m,\sigma_0}^{-1}(X)$, covering lemma \ref{lem:cov} shows that $V$ is an affine component of $\vecsaff(X_{q_0'})$. Proposition \ref{prop:vechull} applied to $X_{q_0'}$ shows that there exists a terminal component $T$ reachable from $q_0'$ such that $V_G(T)=V$. Consider a state $q\in T$ and let us consider a path $q_0'\xrightarrow{\sigma_1} q$. From proposition \ref{prop:patternempty}, we deduce that there exists $P\in\P_V(X)$ such that $P_q\not=\emptyset$. Therefore $\gamma_{r,m,\sigma_1}^{-1}(P_{q_0'})\not\emptyset$. From $P_{q_0'}\subseteq \xi_{r,m}(s)+\Z^m\cap V$, we deduce that $\gamma_{r,m,\sigma_1}^{-1}(\xi_{r,m}(s)+\Z^m\cap V)$. From the dense pattern corollary \ref{cor:densepattern} we deduce that $\gamma_{r,m,\sigma_1}^{-1}((P_0)_{q_0'})\not=\emptyset$. That means $(P_{\#,<})_q\not=(P_{\#,>})_q$ for any $q\in T$.
  
  As there exists a loop on each state $q$ of $T$, we deduce that $P_q$ is relatively prime with $r$ for any $P\in\P_V(X)$ and for any $q\in T$. Hence, there exists an integer $n$ relatively prime with $r$ such that $\inv_V(P_q)\subseteq n.(\Z^m\cap V)$ for any $P\in\P_V(X)$ and for any $q\in T$.
  
  From an immediate induction and lemma \ref{lem:alldir}, we deduce that there exists a sharing of $J$ into $J=J_<\cup J_>$ such that $[C_{V,\#}\cap C\cap H_0^{\#_0}]_V\not=[\emptyset]_V$ for any $\#_0\in\{<,>\}$ where $C=\bigcap_{j\in J_<}\{x\in V;\;x[j]<0\}\bigcap_{j\in J_>}\{x\in V;\;x[j]>0\}$. In particular there exists a vector $v_{\#_0}\in C_{V,\#}\cap C\cap H_0^{\#_0}$ for each $\#_0\in\{<,>\}$. By replacing $v_{\#_0}$ by a vector in $(\Nat\moins\{0\}).v_{\#_0}$, we can also assume that $v_{\#_0}\in n.(\Z^m\cap V)$. 
  
  Let us show that there exists a $(r,m)$-sign vector $s\in S_{r,m}$ and a state $q\in T$ such that $\frac{s}{1-r}\in (P_0)_q$ and such that $s[j]=r-1$ for any $j\in J_<$ and such that $s[j]=0$ for any $j\in J_>$. Consider a state $q'\in T$. As $(P_0)_{q'}$ is not empty, there exists a vector $x$ is this set. As $v_{\#_0}\in\Z^m\cap V$ and $(P_0)_{q'}$ is a semi-$V$-pattern, we deduce that $x_k=x+k.n.v_{\#_0}$ is in $(P_0)_{q'}$ for any $k\in\Z$. As $v_{\#_0}[j]<0$ for any $j\in J_<$ and $v_{\#_0}[j]>0$ for any $j\in J$, we deduce that there exists $k\in\Nat$ enough larger such that $x_k[j]<0$ for any $j\in J_<$ and such that $x_k[j]>0$ for any $j\in J_>$. Let us consider a $(r,m)$-decomposition $(\sigma,s)$ of $x_k$ and remark that $s[j]=r-1$ for any $j\in J_<$ and $s[j]=0$ for any $j\in J_>$. Moreover, $\frac{s}{1-r}\in (P_0)_q$ where $q=\delta(q_0',\sigma)$. As $(P_0)_{q}\not=\emptyset$, proposition \ref{prop:patternempty} proves that $X_q\not=\emptyset$. As $T$ is a terminal component and $q$ is reachable from $T$, we deduce that $q\in T$.
  
  Consider a $(r,m)$-decomposition $(\sigma_{\#_0},s_{\#_0})$ of $\frac{s}{1-r}+v_{\#_0}$ for each $\#_0\in\{<,>\}$. By replacing $\sigma_{\#_0}$ by a word in $\sigma_{\#_0}.s_{\#_0}^*$, as $n$ is relatively prime with $r$, we can assume that $r^{|\sigma_{\#_0}|}\in 1+n.\Nat$ for any $\#_0\in\{<,>\}$. We denote by $w_{\#_0}$ the word $w_{\#_0}=\sigma_{\#_0}^n$.
 
  Let us show that $s_<=s=s_>$. For any $j\in\finiteset{1}{m}\moins J$, as $V\cap\unit_{j,m}^\perp$ is not a $V$-hyperplane, we deduce that $\unit_{j,m}\in V^\perp$. That means $v[j]=0$ for any $v\in V$. In particular $(\frac{s}{1-r}+v_{\#_0})[j]=\frac{s}{1-r}[j]$ and we deduce that $s_<[j]=s[j]=s_>[j]$. For any $j\in J_<$, as $s[j]=r-1$ and $v_{\#_0}[j]<0$, we get $s_{\#_0}[j]=r-1=s[j]$. Symmetrically, for any $j\in J_>$, we get $s_{\#_0}[j]=0=s[j]$. Therefore $s_<=s=s_>$.

  Let us prove that $\gamma_{r,m,w_{\#_0}}^{-1}(P_q)=P_q$ for any $P\in\P_V(X)$ and for any $\#_0\in\{<,>\}$. Let $P\in\P_V(X)$. Remark that $\gamma_{r,m,\sigma_{\#_0}}(x)\in x+\gamma_{r,m,\sigma_{\#_0}}(\unit_{0,m})+n.\Z^m$ for any $x\in\Z^m$. Hence $\gamma_{r,m,w_{\#_0}}(x)\in x+n.\Z^m$ for any $x\in\Z^m$. As $M_{q,P}$ is a $n$-mask, we deduce that $\gamma_{r,m,w_{\#_0}}^{-1}(M_{q,P})=M_{q,P}$. Moreover, from $\frac{s}{1-r}\in (P_0)_q$ we deduce that $\frac{s}{1-r}\in A_q$. Hence $A_q=\frac{s}{1-r}+V$. So $\Gamma_{r,m,\sigma_{\#_0}}^{-1}(A_q)=r^{-|\sigma_{\#_0}|}.(\frac{s}{1-r}-\gamma_{r,m,\sigma_{\#_0}}(\unit_{0,m}))+V$. Recall that $\rho_{r,m}(\sigma_{\#_0},s)=\frac{s}{1-r}+v_{\#_0}$ and remark that $\rho_{r,m}(\sigma_{\#_0},s)=\gamma_{r,m,\sigma_{\#_0}}(\unit_{0,m})+r^{|\sigma_{\#_0}|}.\frac{s}{1-r}$. We get $\Gamma_{r,m,\sigma_{\#_0}}^{-1}(A_q)=\frac{s}{1-r}-r^{-|\sigma_{\#_0}|}.v_{\#_0}+V=A_q$. An immediate induction show that $\Gamma_{r,m,w_{\#_0}}^{-1}(A_q)=A_q$. As $P_q=M_{q,P}\cap A_q$, we get $\gamma_{r,m,w_{\#_0}}^{-1}(P_q)=\gamma_{r,m,w_{\#_0}}^{-1}(M_{q,P})\cap \Gamma_{r,m,w_{\#_0}}^{-1}(A_q)=M_{q,P}\cap A_q=P_q$. We have proved that $\gamma_{r,m,w_{\#_0}}^{-1}(P_q)=P_q$ for any $P\in\P_V(X)$ and for any $\#_0\in\{<,>\}$.
  
  Let us prove that $\delta(q,w_{\#_0}^*)\subseteq T$ for any $\#_0\in\{<,>\}$. From the previous paragraph, we deduce that for any $k\in\Nat$, the set $\gamma_{r,m,w_{\#_0}}^{-k}((P_0)_q)=(P_0)_q$ is not empty. From proposition \ref{prop:patternempty}, we deduce that $\gamma_{r,m,w_{\#_0}}^{-k}( X_q)$ is also non empty. As $T$ is a terminal component, we deduce that $\delta(q,w_{\#_0}^k)\in T$.
  
  As $T$ is a finite set, there exists a state $q_{\#_0}\in T$ such that there exists a path $q\xrightarrow{w_{\#_0}^{r_{\#_0}}}q_{\#_0}$ and a loop $q_{\#_0}\xrightarrow{w_{\#_0}^{k_{\#_0}}}q_{\#_0}$ where $r_{\#_0}\in\Nat$ and $k_{\#_0}\in\Nat\moins\{0\}$.
  
  From theorem \ref{thm:ultime}, we deduce that there exists a vector $a\in\Q^m$ such that:
  $$X_{q}=\bigcup_{\begin{array}{@{}c@{}} P\in\P_V(X)\\[-0pt] \#'\in\S_{V,P}(X) \end{array}  }(P_q\cap (a+C_{V,\#'}+V^\perp))$$
  As $\gamma_{r,m,w_{\#_0}}^{-1}(P_q)=P_q$ for any $P\in\P_V(X)$ and for any $\#_0\in\{<,>\}$ we deduce the following equality for any $\#_0\in\{<,>\}$ and for any $k\in r_{\#_0}+\Nat.k_{\#_0}$:
  $$X_{q_{\#_0}}=\bigcup_{\begin{array}{@{}c@{}} P\in\P_V(X)\\[-5pt] \#'\in\S_{V,P}(X) \end{array}}(P_q\cap (\Gamma_{r,m,w_{\#_0}^k}^{-1}(a)+C_{V,\#'}+V^\perp))$$
  
  As $P_{q_<}=P_{q_>}$ for any $P\in\P_V(X)$, we deduce that $q_<\sim^V q_>$. Hence $(q_<,q_>)\in I_\automaton(V)$.
  
  Let us prove that $X_{q_{\#_0}}\cap (\frac{s}{1-r}+C_{V,\#}\cap C\cap H_0)=P_{q_{\#_0}}\cap (\frac{s}{1-r}+C_{V,\#}\cap C\cap H_0)$. Let us consider a vector $x\in (\frac{s}{1-r}+C_{V,\#}\cap C\cap H_0)$. By developing the expression $\Gamma_{r,m,w_{\#_0}^k}^{-1}(a)$, we deduce that $\lim_{k\rightarrow\infty}\Gamma_{r,m,w_{\#_0}^k}^{-1}(a)=\frac{s}{1-r}-\frac{v_{\#_0}}{r^{|\sigma_{\#_0}|}-1}$. As $v_{\#_0}\in C_{V,\#}\cap C\cap H_0^{\#_0}$ and $\frac{1}{r^{|\sigma_{\#_0}|}-1}\in \Q_+\moins\{0\}$, we deduce that there exists $k\in r_{\#_0}+\Nat.k_{\#_0}$ enough larger such that $x\in \Gamma_{r,m,w_{\#_0}^k}^{-1}(a)+C_{V,(\#,\#_0)}\cap C\cap H_0+V^\perp$. Therefore $X_q\cap \{x\}=P_{\#_0}\cap\{x\}$ and we have proved that $X_{q_{\#_0}}\cap (\frac{s}{1-r}+C_{V,\#}\cap C\cap H_0)=P_{\#_0}\cap (\frac{s}{1-r}+C_{V,\#}\cap C\cap H_0)$. 

  We deduce that $(P_0)_q\cap (\frac{s}{1-r}+H_0)\cap (\frac{s}{1-r}+C_{V,\#}\cap C\cap H_0)\subseteq X_{q<}\Delta X_{q_>}$. Since $[C_{V,\#}\cap C\cap H_0^<]_V$ and $[C_{V,\#}\cap C\cap H_0^>]_V$ are both not equal to $[\emptyset]_V$, lemma \ref{lem:Pleqgeq} shows that $C_{V,\#}\cap C\cap H_0$ is a non $H_0$-degenerate $H_0$-polyhedron. Moreover, since $(P_0)_q\cap (\frac{s}{1-r}+H_0)$ is a non-empty semi-$H_0$-pattern, from lemma \ref{lem:dimdim}, we deduce that $\vecsaff((P_0)_q\cap (\frac{s}{1-r}+H_0)\cap (\frac{s}{1-r}+C_{V,\#}\cap C\cap H_0))=H_0$.  Hence $H_0\subseteq \vecsaff(X_{q_<}\Delta X_{q_>})$. As $(q_{<},q_{>})\in I_\automaton(V)$, we also get $\vecsaff(X_{q_<}\Delta X_{q_>})\subseteq S'$. We deduce that $H_0\subseteq S'$. We have proved that $S\subseteq S'$.
  \qed
\end{proof}

From the previous proposition \ref{prop:boundpoly}, theorem \ref{thm:semiaffinehull} and theorem \ref{thm:vgtpoly}, we deduce the following main theorem of this paper.
\begin{theorem}\label{thm:boundary}
  Let $X$ be a Presburger-definable set represented by a serialized encoded FDVA, and let $V$ be an affine component of $\vecsaff(X)$. The boundary $\bound{V}{X}\moins(\bigcup_{j=1}^m\{V\cap\unit_{j,m}^\perp\})$ is computable in polynomial time. 
\end{theorem}

\subsection{An example}
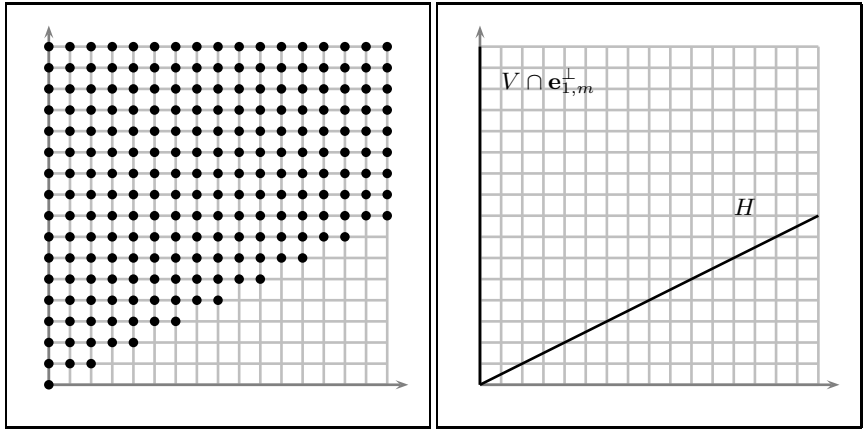
\begin{figure}[htbp]
  \setlength{\unitlength}{8pt}
  \pssetlength{\psunit}{8pt}
  \pssetlength{\psxunit}{8pt}
  \pssetlength{\psyunit}{8pt}
  \begin{center}
    \begin{picture}(20,20)(-2,-2)%
      \put(-2,-2){\framebox(20,20){}}
      \multido{\iii=0+1}{17}{\psline[linecolor=lightgray,linewidth=1pt](0,\iii)(16,\iii)}\multido{\iii=0+1}{17}{\psline[linecolor=lightgray,linewidth=1pt](\iii,0)(\iii,16)}\psline[linecolor=gray,arrowscale=1,linewidth=1pt]{->}(0,0)(0,17)\psline[linecolor=gray,arrowscale=1,linewidth=1pt]{->}(0,0)(17,0)
      \multido{\iii=0+1}{17}{\psdot*(0,\iii)}
      \multido{\iii=1+1}{16}{\psdot*(1,\iii)}
      \multido{\iii=1+1}{16}{\psdot*(2,\iii)}
      \multido{\iii=2+1}{15}{\psdot*(3,\iii)}
      \multido{\iii=2+1}{15}{\psdot*(4,\iii)}
      \multido{\iii=3+1}{14}{\psdot*(5,\iii)}
      \multido{\iii=3+1}{14}{\psdot*(6,\iii)}
      \multido{\iii=4+1}{13}{\psdot*(7,\iii)}
      \multido{\iii=4+1}{13}{\psdot*(8,\iii)}
      \multido{\iii=5+1}{12}{\psdot*(9,\iii)}
      \multido{\iii=5+1}{12}{\psdot*(10,\iii)}
      \multido{\iii=6+1}{11}{\psdot*(11,\iii)}
      \multido{\iii=6+1}{11}{\psdot*(12,\iii)}
      \multido{\iii=7+1}{10}{\psdot*(13,\iii)}
      \multido{\iii=7+1}{10}{\psdot*(14,\iii)}     
      \multido{\iii=8+1}{9}{\psdot*(15,\iii)}
      \multido{\iii=8+1}{9}{\psdot*(16,\iii)}
    \end{picture}
    \begin{picture}(20,20)(-2,-2)%
      \put(-2,-2){\framebox(20,20){}}
      \multido{\iii=0+1}{17}{\psline[linecolor=lightgray,linewidth=1pt](0,\iii)(16,\iii)}\multido{\iii=0+1}{17}{\psline[linecolor=lightgray,linewidth=1pt](\iii,0)(\iii,16)}\psline[linecolor=gray,arrowscale=1,linewidth=1pt]{->}(0,0)(0,17)\psline[linecolor=gray,arrowscale=1,linewidth=1pt]{->}(0,0)(17,0)
      \psline[linecolor=black,linewidth=1pt](0,0)(0,16)\put(1,14){$V\cap \unit_{1,m}^\perp$}
      \psline[linecolor=black,linewidth=1pt](0,0)(16,8)\put(12,8){$H$}
    \end{picture}
  \end{center}
  \caption{On the left the Presburger-definable set $X=\{x\in\Nat^2;\;(x[1]\leq 2.x[2])$. On the right $\bound{V}{X}$ where $V=\Q^2$ and $H=\{x\in V;\;x[1]=2.x[2]\}$.\label{fig:leq2}}
\end{figure}

Let us consider the set $X=\{x\in\Nat^2;\;x[1]\leq 2.x[2]\}$ given in figure \ref{fig:leq2}.

The minimal FDVA $\automaton_{2,2}(X)$ that represents $X$ is given in figure \ref{fig:DVAleq2}. We denote by $q_{-1}=\{x\in\Nat^2;\;x[1]\leq 2.x[2]-1\}$, $q_0=\{x\in\Nat^2;\;x[1]\leq 2.x[2]\}$ and $q_1=\{x\in\Nat^2;\;x[1]\leq 2.x[2]+1\}$ the states of this FDVA. 

\begin{figure}[htbp]
  \begin{center}
    \begin{picture}(131,60)(5,-65)
      \put(5,-65){\framebox(131,60){}}
      \node[NLangle=0.0,Nadjust=w](n0)(34.0,-44.0){$x\in\Nat^2;\;x[1]\leq 2.x[2]-1$}
      \node[NLangle=0.0,Nadjust=w,Nmarks=i](n1)(72.0,-24.0){$x\in\Nat^2;\;x[1]\leq 2.x[2]$}
      \node[NLangle=0.0,Nadjust=w](n2)(110.0,-44.0){$x\in\Nat^2;\;x[1]\leq 2.x[2]+1$}
      \drawedge[curvedepth=16.0](n1,n0){$\scriptstyle(1,0)$}
      \drawedge[curvedepth=16.0](n2,n1){$\scriptstyle(0,0),(1,0)$}
      \drawedge[curvedepth=16.0](n0,n1){$\scriptstyle(0,1),(1,1)$}
      \drawedge[curvedepth=16.0](n1,n2){$\scriptstyle(0,1)$}
      \drawloop[loopangle=-90.0](n0){ $\scriptstyle(0,0),(1,0)$}
      \drawloop[loopangle=-90.0](n2){$\scriptstyle(0,1),(1,1)$}
      \drawloop[loopangle=90.0](n1){$\scriptstyle(0,0),(1,1)$}
      \node[Nframe=n,Nadjust=wh](n5)(72,-60){$\scriptstyle \{(0,0)\}$}
      \drawedge[dash={1.0 1.0 1.0 1.0}{0.0}](n1,n5){}
      \node[Nframe=n,Nadjust=wh](n6)(110,-8){$\scriptstyle \{(0,0)\}$}
      \drawedge[dash={1.0 1.0 1.0 1.0}{0.0}](n2,n6){}
      \node[Nframe=n,Nadjust=wh](nA)(62,-30){$q_0$}
      \node[Nframe=n,Nadjust=wh](nB)(19,-50){$q_{-1}$}
      \node[Nframe=n,Nadjust=wh](nB)(125,-50){$q_{1}$}
    \end{picture}
  \end{center}
  \caption{The minimal FDVA $\automaton_{2,2}(\{x\in\Nat^2;\;x[1]\leq 2.x[2]\})$\label{fig:DVAleq2}}
\end{figure}

Remark that $T=\{q_{-1},q_0,q_1\}$ is the unique terminal component. Moreover, the algorithm that computes the vector space associated to an untransient component provides $V_G(T)=\Q^2$. Remark that from proposition \ref{prop:vechull}, we get $\vecsaff(X)=V_T(T)=\Q^2$. That means $V=\Q^2$ is the only affine component of $\vecsaff(X)$.

\begin{figure}[htbp]
  \begin{center}
    \begin{picture}(111,67)(15,-62)
      \put(15,-62){\framebox(111,67){}}
      \node[NLangle=0.0,Nadjust=w](n0)(34.0,-44.0){$X_{q_{-1}}\Delta X_{q_{-1}}$}
      \node[NLangle=0.0,Nadjust=w](n1)(72.0,-44.0){$X_{q_0}\Delta X_{q_0}$}
      \node[NLangle=0.0,Nadjust=w](n2)(110.0,-44.0){$X_{q_1}\Delta X_{q_1}$}
      \node[NLangle=0.0,Nadjust=w](n3)(51.0,-24.0){$X_{q_{-1}}\Delta X_{q_0}$}
      \node[NLangle=0.0,iangle=0.0,Nadjust=w](n4)(93.0,-24.0){$X_{q_0}\Delta X_{q_1}$}
      \node[NLangle=0.0,iangle=0.0,Nadjust=w](n5)(72.0,-4.0){$X_{q_{-1}}\Delta X_{q_1}$}
      \node[Nframe=n,Nadjust=wh](nsb)(102,-4.0){$\scriptstyle\{(0,0)\}$}
      \drawedge[dash={1.0 1.0 1.0 1.0}{0.0}](n5,nsb){}
      \drawedge[ELside=r](n5,n3){$\scriptstyle(0,0),(1,0)$}
      \drawedge(n5,n4){$\scriptstyle(0,1),(1,1)$}
      \drawedge[curvedepth=8.0](n3,n4){$\scriptstyle (0,1)$}
      \drawedge[ELside=r](n3,n0){$\scriptstyle(1,0)$}
      \drawedge(n3,n1){$\scriptstyle(1,1)$}
      \drawedge[ELside=r](n4,n1){$\scriptstyle(0,0)$}
      \drawedge(n4,n2){$\scriptstyle(0,1)$}
      \drawedge[curvedepth=8.0](n1,n0){$\scriptstyle(1,0)$}
      \drawedge[curvedepth=8.0](n2,n1){$\scriptstyle(0,0),(1,0)$}
      \drawloop[loopangle=120](n3){$\scriptstyle(0,0)$}
      \drawloop[loopangle=60](n4){$\scriptstyle(1,1)$}
      \drawedge[ELside=r](n4,n3){$\scriptstyle(1,0)$}
      \drawedge(n0,n1){$\scriptstyle(0,1),(1,1)$}
      \drawedge(n1,n2){$\scriptstyle(0,1)$}
      \drawloop[loopangle=-90.0](n0){ $\scriptstyle(0,0),(1,0)$}
      \drawloop[loopangle=-90.0](n2){$\scriptstyle(0,1),(1,1)$}
      \drawloop[loopangle=-90.0](n1){$\scriptstyle(0,0),(1,1)$}
      \node[Nframe=n,Nadjust=wh](nsa)(21.0,-24.0){$\scriptstyle\{(0,0)\}$}
      \drawedge[dash={1.0 1.0 1.0 1.0}{0.0}](n3,nsa){}
    \end{picture}
  \end{center}
  \caption{The Cartesian product $\automaton'$ of $\automaton_{q_{0}}$ and $\automaton_{q_{1}}$ that represents the symmetrical difference $X_{q_0}\Delta X_{q_1}$ where $X$ is represented by the FDVA $\automaton$ given in figure \ref{fig:DVAleq2}.\label{fig:DVAleqDelta}}
\end{figure}
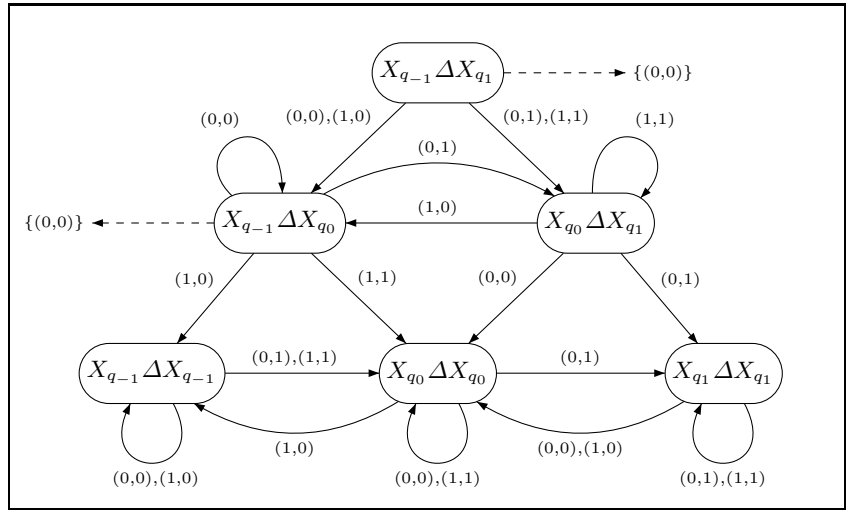

Let us prove that $\vecsaff(X_{q_i}\Delta X_{q_j})=H$ for any $i\not=j$. In figure \ref{fig:DVAleqDelta}, we have represented the FDVA Cartesian products of the FDVA $\automaton_{q_i}$ and the FDVA $\automaton_{q_j}$ that recognize the sets $X_{q_i}\Delta X_{q_j}$ where $i,j\in\{-1,0,1\}$. These FDVA (when $i\not=j$) have only one terminal component $T'=\{(X_{q_0}\Delta X_{q_1}), (X_{q_{-1}}\Delta X_{q_0})\}$ and we have $V_{G'}(T')=H$. Therefore $\vecsaff(X_{q_i}\Delta X_{q_j})=H$ for any $i\not=j$. 

Symmetrically, we get $\vecsaff(X_{q_i}\Delta X_{q_j})=H$ for any $i\not=j$. We deduce that $I_\automaton(V)=\{(q_i,q_j);\;i\not=j\}$ and $\bigcup_{(q_i,q_j)\in I_\automaton(V)}\vecsaff(X_{q_i}\Delta X_{q_j})=H$. From proposition \ref{prop:boundpoly}, we get $\bound{V}{X}\moins\{V\cap\unit_{1,m}^\perp,V\cap\unit_{2,m}^\perp\}=\{H\}$.

Now, just remark that the previous computation can be done in polynomial time from serialized encoded FDVA. Remark also that on this example $\bound{V}{X}=\{H,V\cap\unit_{1,m}^\perp\}$.

%% file: chapter.finalalgorithm.tex
\chapter{The polynomial time algorithm}

In this section we provide a polynomial time algorithm for deciding if the set represented by a FDVA is Presburger-definable and in this case we provide in polynomial time a Presburger formula that defines the same set. 

The algorithm is based on the fact that even if the set $X$ represented by a FDVA $\automaton$ is not Presburger-definable, the algorithms developed in the previous sections can be applied in order to extract from $\automaton$ sets of the form $P\cap H^\#$ where $P$ is a semi-$V$-pattern relatively prime with $r$ included in a $V$-affine space and $H$ is a $V$-hyperplane, and if $X$ is Presburger definable then these sets are $(r,m)$-detectable in $X$ and $X$ is equal to a boolean combination of these sets. 

In the remaining of this section we assume that $\automaton$ is a positive $(r,m,w)$-cyclic FDVA that represents a set $X_0\subseteq\Nat^m$ in basis $r$ and dimension $m$. Naturally these conditions are not restrictive thanks to the cyclic reduction provided by proposition \ref{prop:NDDtocyclic} and thanks to the positive reduction given by proposition \ref{prop:Sdetection}. 

Since a positive final function $F$ is such that $[F](q)\in\{\{\unit_{0,m}\},\emptyset\}$, without ambiguity such a function can be denoted as the set of principal states $q\in Q$ such that $[F](q)=\{\unit_{0,m}\}$. In the sequel, a positive final function $F$ is always denoted as a subset of $Q$.

The following proposition shows that given a set $X'\subseteq \Nat^m$ that can be represented by a FDVA of the form $\automaton^{F}$ where $F$ is an unknown final function, the computation of a positive final function $F'$ such that $X'$ is represented by $\automaton^{F'}$ can be reduced the membership problem for $X'$.
\begin{proposition}\label{prop:semsem}
  Let $\automaton$ be a FDVA. We denote by $\Y$ the set of $\unit_{0,m}$-eye $Y$ such that $Y$ is reachable for $[G]$ from the initial state. For any eye $Y\in\Y$, let us consider a word $\sigma_Y\in\Sigma_{r,m}^*$ such that $\delta(q_0,\sigma_Y)\in \ker_{\unit_{0,m}}(Y)$. Any set $X'\subseteq \Nat^m$ such that there exists a final function $F$ satisfying $X'$ is represented by $\automaton^F$ is represented by $\automaton^{F'}$ where $F'$ is the union of eyes $Y\in \Y$ such that $\rho_{r,m}(\sigma_Y,\unit_{0,m})\in X'$.
\end{proposition}
\begin{proof}
  Let $X$ be the set represented by $\automaton^{F'}$ and let us prove that $X=X'$. Consider $x\in X$. Let $(\sigma,\unit_{0,m})$ be a $(r,m)$-decomposition of $x$. There exists an eye $Y\in\Y$ such that $\delta(q_0,\sigma)\in Y$. Since $\delta(q_0,\sigma_Y)\in\ker_{\unit_{0,m}}(Y)$, by replacing $\sigma$ by a word in $\sigma.\unit_{0,m}^*$, we can assume without loss of generality that $\delta(q_0,\sigma)=\delta(q_0,\sigma_Y)$. Since there exists a final function $F$ such that $X'$ is represented by $\automaton^F$, we deduce that $\gamma_{r,m,\sigma}^{-1}(X')=\gamma_{r,m,\sigma_Y}^{-1}(X')$. From $\rho_{r,m}(\sigma_Y,\unit_{0,m})\in X'$ and the previous equality, we get $\rho_{r,m}(\sigma,\unit_{0,m})\in X'$. Therefore $x\in X'$ and we have proved the inclusion $X\subseteq X'$. For the converse inclusion, let $x\in X'$. Consider a $(r,m)$-decomposition $(\sigma,\unit_{0,m})$ of $x$ and let $Y\in \Y$ such that $\delta(q_0,\sigma)\in Y$. By replacing $\sigma$ by a word in $\sigma.\unit_{0,m}^*$ since $\delta(q_0,\sigma_Y)\in \ker_s(Y)$, we can assume that $\delta(q_0,\sigma)=\delta(q_0,\sigma_Y)$. As $\gamma_{r,m,\sigma}^{-1}(X')=\gamma_{r,m,\sigma_Y}^{-1}(X')$ and $\rho_{r,m}(\sigma,\unit_{0,m})\in X'$, we get $\rho_{r,m}(\sigma_Y,\unit_{0,m})\in X'$. We have proved that $\delta(q_0,\sigma_Y)\in F'$. Thus $\delta(q_0,\sigma)\in F'$ and we have proved that $x\in X$. We have proved the other inclusion $X'\subseteq X$.
  \qed
\end{proof}

Observe that we can decide in linear time if $X_0$ is empty. Thus, we can assume that $X_0$ is non-empty (otherwise we decide that $X_0$ is Presburger-definable and defined by the formula $\textrm{false}$). Theorem \ref{thm:semiaffinehull} proves that a non-empty semi-vector space $S$ such that $\saff(X_0)=\xi_{r,m}(w)+S$ if $X_0$ is Presburger-definable is computable in polynomial time.

Let us fix an affine component $V$ of $S$ and let $T_{V}$ be the finite union of terminal components $T\in\T_\automaton$ such that $V_G(T)=V$. By construction of the semi-affine space $S$, for any affine component $V$ of $S$, there exists at least one terminal component $T$ such that $V_G(T)=V$.

Observe that if $X_0$ is Presburger-definable then $\Z^m\cap (\xi_{r,m}(w)+V)$ is non empty from the dense component lemma \ref{lem:dense}. Since this property can be decided in polynomial time by proposition \ref{prop:heher}, we can assume that this set is non-empty (otherwise we decide that $X_0$ is not Presburger-definable) and from this same proposition we compute in polynomial time a vector $a_0\in \Z^m\cap (\xi_{r,m}(w)+V)$.

Theorem \ref{thm:polyinvcyclicmain} proves that we can compute in polynomial time a $V$-vector lattice $M$ included in $\Z^m$ such that if $X_0$ is Presburger-definable then $M=\inv_V(X_0)$ is relatively prime with $r$ and $|\Z^m\cap V/\inv_V(X_0)|$ is bounded by the number of principal states of $\automaton$. Theorem \ref{thm:inv1} proves that we can compute in polynomial time the characteristic sequence $n_1$, ..., $n_d$ of $M$ in $\Z^m\cap V$ and a $\Z$-basis $v_1$, .., $v_d$ of $\Z^m\cap V$ such that $n_1.v_1$, ..., $n_d.v_d$ is a $\Z$-basis of $M$. Observe that $|\Z^m\cap V/M|=n_1\ldots n_d$. We can assume that $n_1\ldots n_d$ is relatively prime with $r$ and it is bounded by the number of principal states of $\automaton$ (otherwise we decide that $X_0$ is not Presburger-definable). Let $B$ be the finite set $B=\{a_0+\sum_{i=1}^dk_i.v_i;\;0\leq k_1< n_1\wedge\ldots\wedge 0\leq k_d<n_d\}$. Observe that the cardinal of $B$ is equal to $n_1\ldots n_d$. Thus $B$ is computable in polynomial time. Moreover, by definition of $\inv_V(X_0)=M$, we deduce that if $X_0$ is Presburger-definable, for any semi-$V$-pattern $P\in\P_V(X_0)$, there exists a subset $B'\subseteq B$ such that $P=B'+M$.

Theorem \ref{thm:getdetectablepatterns} shows that we can compute in polynomial time a partition $B_0$, $B_1$, ..., $B_n$ of $B$ such that a semi-$V$-pattern $P$ of the form $P=B'+M$ where $B'\subseteq B$ is represented by a FDVA of the form $\automaton^F$ if and only if there exists $J\subseteq\finiteset{1}{n}$ such that $B'=\bigcup_{j\in J}B_j$. Let $i\geq 1$. Observe that there exists a final function $F$ such that $\Nat^m\cap (B_i+M)$ is represented by $\automaton^F$. Since we can decide in polynomial time if a vector $x$ is in $\Nat^m\cap(B_i+M)$, proposition \ref{prop:semsem} proves that we can compute in polynomial time a positive final function $Q_i$ such that $\Nat^m\cap (B_i+M)$ is represented by $\automaton^{Q_i}$. 

Note that $Z_i=X_0\cap (B_i+M)=X_0\cap (\Nat^m\cap (B_i+M))$ is represented by the FDVA $\automaton^{F_0\cap Q_i}$. Theorem \ref{thm:semiaffinehull} proves that a semi-vector space $S_i$ such that $\saff(Z_0)=\xi_{r,m}(w)+S_i$ if $X_0$ is Presburger-definable is computable in polynomial time. Let us consider the set $I$ of $i\in\finiteset{1}{n}$ such that $V\subseteq S_i$.

Let us show that if $X_0$ is Presburger-definable, then any state $q\in Q_i$ is co-reachable from $T_V$. Consider a state $q\in Q_i$, there exists a word $\sigma\in\Sigma_{r,m}^*$ such that $\delta(q_0,\sigma)=q$ and $\rho_{r,m}(\sigma,\unit_{0,m})\in B_i+M$. In particular $\rho_{r,m}(\sigma,\unit_{0,m})\in a_0+V$. Considering a semi-$V$-pattern $P\in\P_V(X)\moins\{\emptyset\}$ and recall that since $P$ is $(r,m)$-detectable in $X$ (from corollary \ref{cor:Mdetectable}), the semi-$V$-pattern $P$ is relatively prime with $r$ and included into the $V$-affine space $a_0+V$ (from lemma \ref{lem:cyclicpattern}). The dense pattern corollary \ref{cor:densepattern} proves that $\gamma_{r,m,\sigma}^{-1}(P)\not=\emptyset$. Proposition \ref{prop:ultime} proves that if $X_0$ is Presburger-definable, then $T_V$ is co-reachable from $q$. Therefore, we have proved that any state $q\in Q_i$ is co-reachable from $T_V$ if $X_0$ is Presburger-definable. Since this property is decidable in polynomial time, we can assume that it is verified (otherwise we decide that $X_0$ is not Presburger-definable).

Now, let us prove that if $X_0$ is Presburger-definable then $F_0\cap T_V\subseteq \bigcup_{i\in I}Q_i$. Consider $q\in F_0\cap T_V$. There exists a path $q_0\xrightarrow{\sigma}q$ with $\sigma\in\Sigma_{r,m}^*$. Since $q\in F_0$, we get $\rho_{r,m}(\sigma,\unit_{0,m})\in X_0$. Theorem \ref{thm:ultime} proves that there exists $P\in\P_V(X_0)\moins\{\emptyset\}$ such that $\rho_{r,m}(\sigma,\unit_{0,m})\in P$. Since there exists a $J\subseteq \finiteset{1}{n}$ such that $P=\bigcup_{j\in J}B_j+M$, we deduce that there exists $j\in\finiteset{1}{n}$ such that $\rho_{r,m}(\sigma,\unit_{0,m})\in B_j+M$. Theorem \ref{thm:ultime} proves that in this case $\vecsaff(Z_j)=V$. Thus $j\in J$ and $q\in \bigcup_{i\in I}Q_i$ and we have proved that $F_0\cap T_V\subseteq \bigcup_{i=1}^n Q_i$. Since this property is decidable in polynomial time, we can assume that it is true (otherwise we decide that $X_0$ is not Presburger-definable).

If $X_0$ is Presburger-definable then $Z_i$ is Presburger-definable and if $i\in I$ then $[Z_i]^V=V$ and in this case $\P_V(X_0)\moins\{\emptyset\}=\{B_i+M\}$ since for any semi-$V$-pattern $P\in\P_V(X_0)$, there exists $J\subseteq\finiteset{1}{n}$ such that $P=\bigcup_{j\in J}B_j+M$ (recall that corollary \ref{cor:Mdetectable} proves that any semi-$V$-pattern $P\in\P_V(X_0)$ is $(r,m)$-detectable in $X_0$). Theorem \ref{thm:boundary} provides a polynomial time algorithm for computing a finite set $\H_i$ of vector spaces such that if $X_0$ is Presburger-definable then $\bound{V}{Z_i}\moins\bigcup_{i=1}^m\{V\cap \unit_{i,m}^\perp\}=\H_i$. We can assume that $\H_i$ is a set of $V$-hyperplanes (otherwise we decide that $X_0$ is not Presburger-definable). Proposition \ref{prop:patternH} shows that if $X_0$ is Presburger-definable then for any $H\in \H_i$, there exists $\#_{i,H}\in\{\geq,>\}$ such that $(B_i+M)\cap (\xi_{r,m}(w)+H^{\#_{i,H}}+V^\perp)$ is represented by a FDVA of the form $\automaton^F$. Since we can decide this property in polynomial time thanks to proposition \ref{prop:verifbound}, we can assume that such a relation $\#_{i,H}$ exists. As we can decide in polynomial time if a vector $x$ is in $\Nat^m\cap (B_i+M)\cap (\xi_{r,m}(w)+H^{\#_{i,H}}+V^\perp)$, proposition \ref{prop:semsem} proves that we can compute in polynomial time a positive final function $Q_{i,H}$ such that $\Nat^m\cap (B_i+M)\cap (\xi_{r,m}(w)+H^{\#_{i,H}}+V^\perp)$ is represented by $\automaton^{Q_{i,H}}$.

Now observe that if $X_0$ is Presburger-definable, lemma \ref{lem:Xsigma} proves that there exists a boolean combination $Z_i'$ of the set $\Nat^m\cap (B_i+M)$ and the sets $\Nat^m\cap (B_i+M)\cap (\xi_{r,m}(w)+H^{\#_{i,H}}+V^\perp)$ such that $[X_0\Delta Z_i']^V=[\emptyset]^V$. Since any state in $Q_i$ is co-reachable from $T_V$, if such a boolean combination exists, there exists a boolean combination $Q_i'$ of the set $Q_i$ and the sets $Q_{i,H}$ where $H\in \H_i$ such that $Q_i'\cap T_V=F_0\cap T_V$. In particular $F_0\cap T_V$ is a boolean combination of the set $Q_i\cap T_V$ and the sets $Q_{i,H}\cap T_V$. Since this last property is decidable in polynomial time by the lemma \ref{lem:bool} we can assume that such a boolean combination exists (otherwise we decide that $X_0$ is not Presburger-definable). This same lemma \ref{lem:bool} also proves that we can compute in polynomial time a boolean formula $\psi_i$ such that $q\in F_0\cap T_V$ is defined by $\psi_i(q\in Q_i\cap T_V,(q\in Q_{i,H}\cap T_V)_{H\in\H_i})$. Observe that the set $Q_i'$ defined by $q\in Q_i'$ if $\psi_i(q\in Q_i,(q\in Q_{i,H})_{H\in\H_i})$ is computable in polynomial time. Moreover, the set $Z_i'$ represented by $\automaton^{Q_i'}$ is defined by the Presburger-formula $\phi_i$:
$$\phi_i(x):=(x\in \Nat^m\cap (B_i+M)) \wedge \psi_i(\textrm{true},(x\in a_0+H^{\#_{i,H}}+V^\perp)_{H\in\H_i})$$

Now, let us consider the Presburger formula $\phi':=\bigvee_{i\in I}\phi_i$ and the positive final function $Q'=\bigcup_{i\in I}Q_i'$. Remark that the set $Z'=\bigcup_{i\in I}Z'_i$ is represented by the FDVA $\automaton^{Q'}$ and it is defined by the Presburger formula $\phi'$. 

Note that $X_1=X\Delta Z'$ is the set represented by the FDVA $\automaton^{F_1}$ where $F_1=F_0\Delta F'$ and $X_0$ is Presburger-definable if and only if $X_1$ is Presburger-definable. Moreover, by construction of $F'$, any state $q\in F'$ is co-reachable from $T_V$ and $F'\cap T_V=F_0\cap T_V$. That means the set of strongly-connected components of $\automaton^{F_1}$ reachable from the initial state and co-reachable from a final state is strictly included in the strongly connected components of $\automaton^{F_0}$ satisfying this same property. 

Thus, by repeating the previous constructions we obtain a finite sequence $X_0$, $X_1$,..., $X_k$ where $k$ is bounded by the number of strongly connected components of $\automaton$, and a sequence $\phi_1$, ..., $\phi_k$ of Presburger-formulas $\phi_i$ defining $X_{i-1}\Delta X_i$ such that $X_k=\emptyset$. Note that $X_0$ is therefore Presburger-definable since we have the following equality:
$$X_0=(X_0\Delta X_1)\Delta\cdots\Delta(X_{n-1}\Delta X_k)$$
Moreover, from $\phi_1$, ..., $\phi_k$ we get a Presburger-formula $\phi$ that defines $X$.

We have proved the following theorem.
\begin{theorem}
  Let $X\subseteq \Z^m$ be the set represented by a FDVA $\automaton$ in basis $r$ and in dimension $m$. We can decide in polynomial time if $X$ is Presburger-definable. Moreover, in this case, we can compute in polynomial time a Presburger-formula $\phi$ that defines $X$. 
\end{theorem}